\documentclass[preprint,aps]{revtex4}
\usepackage{graphics}% Include figure files
\usepackage{dcolumn}% Align table columns on decimal point
\usepackage{bm}% bold math
\begin{document}
\preprint{Harvard/Frankfurt}

\title{The thermodynamics for a hadronic gas of fireballs 
with internal color structures and chiral fields}
\author{Ismail Zakout$^{123}$ and Carsten Greiner$^{1}$}
\affiliation{$^{1}$ Institut f\"ur Theoretische Physik,\\
$^2$ Frankfurt Institute for Advanced Studies, J. W. Goethe
Universit\"at, Max von Laue Stra$\beta$e 1,
D-60054 Frankfurt am Main, Germany}

\affiliation{
$^{3}$ Laboratory for Particle Physics and Cosmology,
Harvard University, Cambridge MA 02138, USA}

%%%%%%%%%%%%%%%%%%%%%%%%%%%%%%%%%
%%%%%%%%%%%%%%%%%%%%%%%%%%%%%%%%%

%%%%%%%%%%%%%%%%%%%%%%%%%%%%%%%%%
%%%%%%%%%%%%%%%%%%%%%%%%%%%%%%%%%

%%%%%%%%%%%%%%%%%%%%%%%%%%%%%%%%%
%%%%%%%%%%%%%%%%%%%%%%%%%%%%%%%%%
%%%%%%%%%%%%%%%%%%%%%%%%%%%%%%%%%
%%%%%%%%%%%%%%%%%%%%%%%%%%%%%%%%%
%%%%%%%%%%%%%%%%%%%%%%%%%%%%%%%%%%%%%%%%%%%%%%%%%%%%%%%%%%%%%%%%
%%%%%%%%%%%%%%%%%%%%%%%%%%%%%%%%%%%%%%%%%%%%%%%%%%%%%%%%%%%%%%%%
%%%%%%%%%%%%%%%%%%%%%%%%%%%%%%%%%%%%%%%%%%%%%%%%%%%%%%%%%%%%%%%%

%%%%%%%%%%%%%%%%%%%%%%%%%%%%%%%%%%%%%%%%%%%%%%%%%%%%%%%%%%%%%%%%
%%%%%%%%%%%%%%%%%%%%%%%%%%%%%%%%%%%%%%%%%%%%%%%%%%%%%%%%%%%%%%%%

\date{\today} % It is always \today, today,
\begin{abstract}
The thermodynamical partition function for a gas 
of color-singlet bags consisting of
fundamental and adjoint particles in both $U(N_c)$ and $SU(N_c)$ 
group representations is reviewed in detail. 
The constituent particle species 
are assumed to satisfy various thermodynamical statistics.
The gas of bags is probed to study the phase transition
for a nuclear matter in the extreme conditions.
These bags are interpreted as the Hagedorn states
and they are the highly excited hadronic states 
which are produced below the phase transition point 
to the quark-gluon plasma.
The hadronic density of states has
the Gross-Witten critical point and exhibits
a third order phase transition from a hadronic phase dominated 
by the discrete low-lying hadronic mass spectrum particles 
to another hadronic phase dominated 
by the continuous Hagedorn states.
The Hagedorn threshold production is found 
just above the highest known experimental  
discrete low-lying hadronic mass spectrum.   
The subsequent Hagedorn phase undergoes 
a first order deconfinement phase transition 
to an explosive quark-gluon plasma.
The role of the chiral phase transition in the phases 
of the discrete low-lying mass spectrum and
the continuous Hagedorn mass spectrum
is also considered.  
It is found crucial in the phase transition diagram.
Alternate scenarios are briefly discussed for the Hagedorn gas 
undergoes a higher order phase transition through multi-processes 
of internal color-flavor structure modification. 
%
%
%%%%%%%%%%%%%%%%%%%%%%%%%%%%%%%%%
\end{abstract}

%%%%%%%%%%%%%%%%%
\maketitle
%%%%%%%%%%%%%%%%%%%%%%%%%%%%%%%%%%%%%%%%%%%%%%%%%%%%%%%%%%%%%%%%%%%%%%
%%%%%%%%%%%%%%%%%%%%%%%%%%%%%%%%%%%%%%%%%%%%%%%%%%%%%%%%%%%%%%%%%%%%%%
%%%%%%%%%%%%%%%%%%%%%%%%%%%%%%%%%%%%%%%%%%%%%%%%%%%%%%%%%%%%%%%%%%%%%%
\section{Introduction}

%%%
%%%
%%%
%%%

%%%%%%%%%%%%%%%%%%%%%%%%%%%%%%%%%%%%

%%%
%%%
%%%
%%%

%%%%%%%%%%%%%%%%%%%%%%%%%%%%%%%%%%%%%%%%%%%%%%%%%%%%%%%%%%%%%%%%%%%%%%%%%
%%%%%%%%%%%%%%%%%%%%%%%%%%%%%%%%%%%%%%%%%%%%%%%%%%%%%%%%%%%%%%%%%%%%%%%%%

The thermodynamical description of the strongly interacting hadronic
matter can be approximated by a free gas with a modified level
density of hadronic states.
This new level density is given by the statistical bootstrap equation.
A solution to this equation exists only for some range
of parameters~\cite{Hagedorn:1965a,Frautschi:1971a}.
It has been argued that the statistical evaluation
of the mass spectrum in the bag made using
the micro-canonical ensemble
behaves the strong bootstrap model~\cite{Kapusta1981a,Kapusta:1982a}.
Naively, the thermodynamics implied by this hadronic mass-spectrum
has a limiting temperature $T_0$.
The bootstrap model with internal
symmetry~\cite{Redlich80a,Turko:1981a}
of the fireball provides subsidiary variables and allows
for new types of phase transitions.
The fireballs, described by
the bags of confined quark and gluon components, stand
for excited exotic hadrons. 
They are denoted as the hadronic bubbles 
or more definitely as the Hagedorn states. 
The Hagedorn states are defined by the fireballs 
which are consisting quarks and gluons
in the color-singlet states.
The confrontation with the experiment is made by projecting
the color-singlet states for the non-Abelian color gauge symmetry.
In the simple model, it is assumed that there
are no interactions among the color charges in the model
except for a color-singlet constraint from the confinement
assuming the results are stable to the
introduction of small interaction terms.
Gross and Witten~\cite{GrossWitten1980a}
have studied the action for fundamental
particles in the
large-$N_c$ limit of the two-dimensional $U(N_c)$
lattice gauge theory. Using the functional integral method
of Brezin, Itzykson, Parisi and Zuber~\cite{Brezin1978a}
in the large $N_c$-limit.
The action is explicitly evaluated
for all fixed weak-strong coupling 
$\lambda_{WS}=N_c g^2$ by steepest-descent methods.
In this limit, particular configuration
totally dominates the functional integral.
The Vandermonde determinant, which appears in the measure
of the lattice gauge theory also contributes a potential term.
Their solution suggests the existence of Gross-Witten point
for a third order phase transition from weak- to strong-coupling
$\lambda_{WS}$ interaction in the large-$N_c$ four dimensional
lattice gauge theory. They have also mentioned that
the occurrence of such a phase
transition would not mean that the large-$N_c$ theory does not confine
but more precisely that the occurrence of such a phase transition
are not described by the same analytic functions.
This means that the behavior of the mass spectral density
must be modified at the Gross-Witten point.

The Gross-Witten solution~\cite{GrossWitten1980a}
is also obtained for
quarks and gluons by solving the singular integral equation.
The third order phase transition point is determined
when the solution of
the singular integration is evaluated over
an open contour rather than a closed
one~\cite{Lang:1981a,Skagerstam:1984a,Azakov1987a}.
The method given by Azakov, Salomonson
and Skagerstam~\cite{Azakov1987a} leads to the singular
integration over either closed or open contour.
They have considered gluons and quarks
by solving the resultant singular
integral equation and they have obtained
a third order phase transition point.
They have derived the phase transition point 
considering two possibilities.
The first one is the integration over 
a closed contour while the other
possibility is the integration over an open contour.
The phase transition
takes place when the integration is performed
over an open contour rather
than the closed one.
These two possibilities complicate the solution
in the realistic case.
The order of the phase transition is assumed
to be the same even 
for arbitrary constituent mass~\cite{Hallin1998a}.
It is interested to note here that the authors
of Ref.~\cite{Azakov1987a} have shown the existence
of critical chemical potential
$\mu_c$ such as that for $T>0$ the physical properties for
the low-lying spectrum are unaffected by the chemical potential
$|\mu|<\mu_c$. This means that generalizing and extending the results
for zero chemical potential to the diluted nuclear matter
with a small chemical potential is satisfactory though
more thorough investigation is required in particular 
when the chemical potential exceeds the critical value.
The low temperature phase has free energy of $O(N^0_c)$,
interpreted as a gas of mesons and glueballs while
the high-temperature phase has free energy 
of order $O(N_c^2)$ which is interpreted
as exotic color-singlet bags of adjoint gluons
and of fundamental and anti-fundamental quarks.

The deconfining phase transition in $SU(N_c)$ gauge theories
have been studied at nonzero
temperature using a matrix model of Polyakov
loops~\cite{Dumitru2004a,Dumitru:2004b,Dumitru:2005b}.
This model has been also extended for nonzero density.
The effective action for loops starts with a potential term.
At large $N_c$, the action is dominated by the loop potential.
It has been demonstrated that the
Gross-Witten model represents an ultra-critical point
where the deconfining transition at $N_c=\infty$
is close but not at the Gross-Witten point.
Although masses vanish at the Gross-Witten point,
the transition is found
of first order and it has been suggested that at finite $N_c$,
the fluctuations can derive the theory much closer
to the Gross-Witten
point~\cite{Dumitru2004a,Dumitru:2004b,Dumitru:2005b}.
%%%%%%%%%%%%%%%%%%%%%%%%%%%%%%%%%%%%%%%%%%%%%%%%%%%%%
% here 
%%%%%%%%%%%%%%%%%%%%%%%%%%%%%%%%%%%%%%%%%%%%%%%%%%%%%

The thermodynamics of quark jets with an internal color structure 
has been considered in the context  of a one dimensional quark 
gas~\cite{Nambu:1982a,Bambah:1983a,GrossM:1983a}.
It is considered based on an exact Hamiltonian formalism 
where the quarks are treated as classical particles 
but their interactions through
the group theory $U(N_c)$ or $SU(N_c)$ gauge fields 
are treated exactly~\cite{Nambu:1982a,Bambah:1983a,
GrossM:1983a,Jaimungal:1997a,Gattringer:1996a}.
%%%%%%%%%%%%%%%%%%%%%%%%%%%%%%%%%%%

%%%
The class of theories on a compact spatial manifold
with fundamental flavors and adjoint gluons are also found
undergo a third order deconfinement phase
transition at a temperature $T_c$ proportional to the inverse
length scale of the compact manifold~\cite{Schnitzer2004a}.
The same thermodynamic behavior of the deconfinement transition
for large $N_c$ at zero and weakly coupling is found
for a wide class of toy models such as ${\cal N}=4$ super Yang-Mills theory.
It is argued that AdS black hole
thermodynamics is related to Hagedorn
thermodynamics~\cite{Sundborg:2000a}. Sundborg~\cite{Sundborg:2000a}
has calculated the partition sums
for ${\cal N}=4$ super Yang-Mills on $S^3$
and he has discussed the connections
with gravitational physics.
It have been speculated that on a dual string theory
interpretation that the deconfinement transitions
are always associated with black hole formation
and furthermore that the intermediate temperature phase associated
with second order deconfinement transitions would be dual
to a string theory in a background dominated by
a strange new type of stable 
black hole~\cite{Aharony2003a,Aharony2005a,Aharony:2005b}.
Indeed, the Gross-Witten point and the Hagedorn states
remain very exciting and rich physics.
For example, in the language of AdS/CFT duality,
the Horowitz-Polchinski point for 
a small black hole should correspond to a Gross-Witten transition.
This can be related to the phase transition for the small black hole to
a big black hole~\cite{Alvarez-Gaume2005}.
%%%
%%%
%%%
%
%
%

%%%%%%%%%%%%%%%%%%%%%%%%%%%%%%%%%%%%%%%%%%%%%%%%%%%%%%%%%%%%%%%%%%%%%%%
The internal color symmetry of the bound state
remains to be of a color-singlet state even for finite temperature
and chemical potential due to the color confinement.
The discrete low-lying hadronic mass spectrum is generated
by the broken chiral symmetry.
The internal color structure of the hadronic bag of quarks and gluons
for hadronic states above the known discrete low-lying hadronic mass
spectrum remains in a total color-singlet state.
In the realistic calculations, the internal color structure is imposed 
in the partition function
for the hadronic bubbles~\cite{Muller:1982b}.
When the temperature reaches the critical one,
color is expected to be liberated and chiral symmetry is restored.
QCD predicts a phase transition from the hadronic phase 
to a deconfined quark-gluon plasma.
%%%%%%%%%%%%%%%%%%%%%%%%%%%%%%%%%%%%%%%%%%%%%%%%%%%%%%%%%%%%%%%%%%%%%%%
The hadronic phase consists of the whole hadronic mass spectrum
including resonances of all known particles.
Hardcore repulsive forces can be represented by the excluded volume.
The effects from strong interactions are included by adding a free gas
of hadronic bubbles which are bags of constituent quarks
and gluons with a specific internal color-flavor structure.
The hadronic bubbles remain in overall color-singlet states, 
despite the complexity of the internal color-flavor structure.
%%% important restatement %%%%%%%%%%%%
The bubble's size grows up but retains 
its own internal color symmetry 
with increasing the baryonic density at low temperature.
The volume fluctuations are expected to be suppressed
whenever the bags start to overlap with each other
at large chemical potential.
When the temperature increases, the bag's surface
is deformed and smeared out until the bubbles dissociate eventually
at the critical temperature.
It is expected that the bubble's volume fluctuation increases
at high temperature in contrast to the cold dense hadronic 
matter where the bubbles tend to squeeze each other.
The strength of the volume fluctuations is presumed to modify
the details of the phase transition diagram~\cite{Zakout2006}.

%%%%%%%%%%%%%%%%%%%%%%%%%
%%%%%%%%%%%%%%%%%%%%%%%%%%%%%%%%%%%%%%%%%%%%%%%%%%%%%%%%%%%%%%%%%%%%%%%
The (grand-) canonical ensemble and its Laplace transform to
the micro-canonical ensemble for gluonic bags or glueballs was derived by
Kapusta~\cite{Kapusta1981a,Kapusta:1982a}
without imposing any color constraint.
The internal symmetry constraint was originally introduced for
the statistical bootstrap model~\cite{Redlich80a,Turko:1981a}.
The bootstrap density of states can be derived from
the MIT bag model.
In the hadronic phase, the highly excited fireballs derived
from the bootstrap equation are hadronic bubbles of
confined quarks and gluons in a net color-singlet
state~\cite{Elze:1983a,Elze84a,Elze985a,Elze86a,Elze86b,
Gorenstein:1983a}.
Gorenstein {\em et. al.}
~\cite{Gorenstein:1981a,Gorenstein:1982a,
Gorenstein:1984a,Gorenstein:1998a}
have studied the gas of bags using
the isobaric partition function.
Auberson {\em et. al.}
~\cite{Auberson:1986a}
have studied the phase transition to the deconfined quark-gluon plasma
by considering the asymptotic Gaussian volume fluctuations.
The order and the shape of the phase transition have been studied
in details~\cite{Zakout2006}.
It is shown that the color-singlet constraint imposed
on the bag states is not only the critical condition for the existence 
of the phase transition 
to an explosive quark-gluon plasma.
Therefore, the appearance of the deconfined phase transition
depends essentially on the volume fluctuations beside
the internal color structure constraints.
Recently, it has been suggested that the phenomenology
of hadronic bubble internal structure decides the
order of the phase transition for low chemical potential
and high temperature~\cite{Gorenstein:2005a,Zakout2006}.

In the present work the partition function is written
as a function of the thermal running parameter $\lambda$.
This thermal running parameter is related to the Gross-Witten
weak-strong coupling $\lambda_{WS}$ as follows
$\lambda/N^2_c=1/\lambda_{WS}$.
Therefore, the weakly interaction corresponds
the large thermal running $\lambda>\lambda_{\mbox{critical}}$
while the strong interaction is analogous to
the small thermal running $\lambda<\lambda_{\mbox{critical}}$.

%%%%%%%%%%%%%%%%%%%%%%%%%%%%%%%%%%%%%%%%%%%%%%%%%%%%%%%%%%%%%%%%%%%%%%%%%%%%%%%%
%%%%%%%
%%%%%%%
%%%%%%%
%%%%%%%
%%%%%%%%%%%%%%%%%%%%%%%%%%%%%%%%%%%%%%%%%%%%%%%%%%%%%%%%%%%%%%%%%%%%%%%%%%%%%%%%
%%%%%%%%%%%%%%%%%%%%%%%%%%%%%%%%%%%%%%%%%%%%%%%%%%%%%%%%%%%%%%%%%%%%%%%%%%%%%%%%

The outline of the present paper is as follows.
In section~\ref{sect_partition_ii},
we review the partition function
in the Hilbert space for various statistics 
for the gas of particles
with a specific internal color symmetry.
The partition function is derived for an ideal gas.
The sum of states for particles confined in the cavity 
is presented for the convenience in the realistic calculations.
The extension to a specific geometry and other degree of freedom
in the conformal fields such 
as the super-symmetry Yang-Mills theory is mentioned.
The invariance measure which is used to project
a specific internal color state is presented
in Section~\ref{sect_haar_measure}.
The invariance measure is given by the Vandermonde determinant
for the $U(N_c)$ and $SU(N_c)$ theoretic groups.
It is shown that the Vandermonde determinant have two different 
asymptotic approximations.
The first asymptotic approximation is appropriate when
the color charges are distributed uniformly over the entire circle.
While the second one is more appropriate when in the color charges 
become more dominated in a specific physical range.
In Section~\ref{sect_fundamental_spectral},
we review the spectral density 
of the color eigenvalues method. 
It has been studied by Gross and Witten 
to discover the critical Gross-Witten point.
This method is reviewed for the fundamental particles
and the phase transition point is given.
%%%%%%%%%
In Section~\ref{sect_fundamental_GSP},
we present an alternate method
to derive the critical point for the phase transition.
This method is based on the Gaussian-like saddle points method
and it is more appropriate in the theoretic nuclear 
physics and realistic calculations.
It has be considered by several
authors~\cite{Elze84a,Elze:1983a,
Gorenstein:1983a,Auberson:1986b,Zakout2006}.
%
%%%%%%%%%
In Section~\ref{sect_adj}, the gas of adjoint particles 
is reviewed in the context
of two different methods.
Since in the realistic situation we have a gas of 
fundamental as well as adjoint particles,
the physics of the phase transition for a gas
consisting of different particle species and
satisfying various statistics 
is studied in section~\ref{sect_fund_adj}.
At first, we demonstrate the calculation for
the Maxwell-Boltzmann's statistics for fundamental and adjoint particles.
The realistic confined quark and gluon bag is considered in detail.
The adjoint gluons which are obeying the Bose-Einstein statistics
and the fundamental quarks and antiquarks 
which are obeying the Fermi-Dirac statistics 
are considered in computing the partition function 
and finding the phase transition's critical point.
%%%%%%%%%
The micro-canonical ensemble as a density of states
is presented in section~\ref{sect_micro_canonical}.
The micro-canonical ensemble is derived from the inverse Laplace
transform of the grand-canonical one.
The critical mass for emerging the fireballs 
which are obeying the Maxwell-Boltzmann
statistics is demonstrated at first 
and then the realistic critical
mass for emerging the Hagedorn states is given.
%%%%%%%%%%
In section~\ref{sect_dens_statistics},
the density of states is derived
using the statistical evaluation
of the micro-canonical ensemble.
The derivation is given along the lines 
of Kapusta~\cite{Kapusta1981a} but the internal color symmetry
is considered explicitly in our calculations.
%%%%%%%%%%%
The thermodynamics for the gas of bags 
with the excluded volume repulsion
is presented in section~\ref{sect_VdW}.
We summarize the conditions for the phase transition
in the context of the isobaric partition function construction.
Furthermore, various scenarios for the phase transition are given.
%%%%%%%%%%%%
The role of the chiral symmetry restoration phase transition
in the hadronic matter and quark-gluon plasma phase diagram 
is given in section \ref{chiral-restoration}.
%%%%%%%%%%%%%
Finally, our conclusion is given in section~\ref{sect_conclusion}.
%%%%%%%%%%%%%%%%%%%%%%%%%%%%%%%%%%%%%%%%%%%%%%%%%%%%%%%%%%%%
%%%%%%%%%%%%%%%%%%%%%%%%%%%%%%%%%%%%%%%%%%%%%%%%%%%%%%%%%%%%

%%%%%%%%%%%%%%%%%%%%%%%%%%%%%%%%%%%%%%%%%%%%%%%%%%%%%%%%%%%%%%%
%%%%%%%%%%%%%%%%%%%%%%%%%%%%%%%%%%%%%%%%%%%%%%%%%%%%%%%%%%%%%%%
%%% section
\section{The canonical partition function\label{sect_partition_ii}}
The canonical ensemble for the gas of fundamental particles
in the Hilbert space is given by the tensor product of the Fock
spaces of particles and antiparticles
\begin{eqnarray}
Z_{q\overline{q}}(\beta)=\int_{SU(N_c),U(N_c)} d\mu(g)\left[
\mbox{Tr}_q \hat{U}_q(g)e^{ -\beta \hat{H}_q }
\right]
\left[
\mbox{Tr}_{\overline{q}}
\hat{U}_{\overline{q}}(g)
e^{ -\beta \hat{H}_{\overline{q}} }
\right].
\end{eqnarray}
On the other hand, the canonical ensemble for the gas of fundamental
and adjoint particles is given by the tensor product
of fundamental particles, antiparticles and adjoint particles' 
Fock spaces. This tensor product reads
\begin{eqnarray}
Z_{q\overline{q} g}(\beta)=\int_{SU(N_c),U(N_c)} d\mu(g)\left[
\mbox{Tr}_q \hat{U}_q(g)e^{ -\beta \hat{H}_q }
\right]
\left[
\mbox{Tr}_{\overline{q}}
\hat{U}_{\overline{q}}(g)
e^{ -\beta \hat{H}_{\overline{q}} }
\right]
\left[
\mbox{Tr}_{g}
\hat{U}_{g}(g)
e^{ -\beta \hat{H}_{g} }
\right],
\end{eqnarray}
where the notation $\mbox{Tr}_{q,\overline{q},g}$
is the trace in each Fock space. 
We adopt the notation $\mbox{Tr}$ in order to distinguish 
the single particle statistics from the trace $\mbox{tr}_c$
over the color degrees of freedom.
In each Fock space there exists basis that
diagonalizes both operators as long as
$\hat{H}_q$ and $\hat{U}_q(g)$ commute.
Let $|\alpha,j>$ be
the one-particle states of such a basis, where $\alpha$ labels
the eigenvalues of $\hat{H}_q$ and $j$ labels those
${\bf R}(g)$ of $\hat{U}_q(g)$, 
then the diagonalized eigenstates read
\begin{eqnarray}
<\alpha',j'|\hat{H}_q|\alpha,j>&=&
\delta_{\alpha,\alpha'}\delta_{jj'}
E_{\alpha},\nonumber\\
<\alpha',j'|\hat{U}_q(g)|\alpha,j>&=&
\delta_{\alpha,\alpha'}\delta_{jj'}
{\bf R}_{jj'}(g).
\end{eqnarray}
Any configuration of the system is defined by 
the set of occupation numbers $\{n_{\alpha,j}\}$.
The additivity of $\hat{H}_q$ simply means that
\begin{eqnarray}
<\{n_{\alpha,j}\}|\hat{H}_q|\{n_{\alpha,j}\}>=
\sum_{\alpha,j} n_{\alpha,j} E_{\alpha}.
\end{eqnarray}
Then, using the basis $|\{n_{\alpha,j}\}>$
to evaluate the trace, we readily obtain
\begin{eqnarray}
\mbox{Tr} \hat{U}_q(g)e^{-\beta \hat{H}_q}&=&
\sum_{\{n_{\alpha,j}\}}\prod_{\alpha,j} \left(
{\bf R}_{jj}(g)
e^{-\beta E_{\alpha}}
\right)^{n_{\alpha,j}}_{\mbox{Boson}}
\nonumber\\
&+&
\sum_{\{n_{\alpha,j}\}}\prod_{\alpha,j}
(-1)^{n_{\alpha,j}+1}
\left(
{\bf R}_{jj}(g)e^{-\beta E_{\alpha}}
\right)^{n_{\alpha,j}}_{\mbox{Fermion}},
\nonumber\\
&=&
\prod_{\alpha,j}
\sum_{\{n_{\alpha,j}\}}\left(
{\bf R}_{jj}(g)
e^{-\beta E_{\alpha}}
\right)^{n_{\alpha,j}}_{\mbox{Boson}}
\nonumber\\
&+&
\prod_{\alpha,j}
\sum_{\{n_{\alpha,j}\}}
(-1)^{n_{\alpha,j}+1}
\left(
{\bf R}_{jj}(g)e^{-\beta E_{\alpha}}
\right)^{n_{\alpha,j}}_{\mbox{Fermion}}.
\end{eqnarray}
Hence, the partition function in the each Fock space
decomposes to Bosonic and Fermionic single-particle functions.
It is reduced to
\begin{eqnarray}
\mbox{Tr}_{\mbox{states}} \hat{U}_q(g)e^{-\beta \hat{H}_q}&=&
\mbox{Tr}_{\mbox{FD}} \hat{U}_q(g)e^{-\beta \hat{H}_q}
+
\mbox{Tr}_{\mbox{BE}} \hat{U}_q(g)e^{-\beta \hat{H}_q},
\nonumber\\
&\rightarrow&
\mbox{Tr}_{\mbox{FD}} \hat{U}_q(g)e^{-\beta \hat{H}_q},
\nonumber\\
&\rightarrow&
\mbox{Tr}_{\mbox{BE}} \hat{U}_q(g)e^{-\beta \hat{H}_q},
\end{eqnarray}
where the subscripts $\mbox{FD}$ and $\mbox{BE}$ indicate
Fermi-Dirac and Bose-Einstein statistics, respectively.
The resultant ensemble decomposes either to
Fermionic or Bosonic Fock space or even the superposition of Fock spaces
with different statistics.
The Hilbert space of gas, which is consisting of particles satisfying
Fermi-Dirac and Bose-Einstein statistics, has the structure
of tensor product of Fock spaces obeying Fermi-Dirac statistics
and another one obeying Bose-Einstein statistics.
The trace over Fermi-Dirac statistics is given by
\begin{eqnarray}
\mbox{Tr}_{\mbox{FD}} \hat{U}_q(g)e^{-\beta \hat{H}_q}&=&
\prod_{\alpha}
\left[1+ {\bf R}(g)e^{-\beta E_{\alpha}}\right],
\nonumber\\
&=&\exp\left[+
\mbox{tr}_c
\sum_{\alpha}
\ln\left[1+ {\bf R}(g)e^{-\beta E_{\alpha}}
\right]\right],
\end{eqnarray}
while the trace over Bose-Einstein statistics 
is given by
\begin{eqnarray}
\mbox{Tr}_{\mbox{BE}} \hat{U}_q(g)e^{-\beta \hat{H}_q}
&=&
\prod_{\alpha}\frac{1}
{\left[1-{\bf R}(g)e^{-\beta E_{\alpha}}\right]},
\nonumber\\
&=&\exp\left[-\mbox{tr}_c
\sum_{\alpha}
\ln\left[1- {\bf R}(g)e^{-\beta E_{\alpha}}\right]\right],
\end{eqnarray}
where $\mbox{tr}_c$ is the trace over the color degree of freedom.
In the realistic physical situation, 
the single-particle partition
function satisfies either Fermi-Dirac or Bose-Einstein
statistics. 
The quarks and antiquarks satisfy Fermi-Dirac statistics
while gluons satisfy Bose-Einstein statistics.
The sum of states $\{jn\}$ for Fermi-Dirac statistics 
is calculated explicitly as follows
\begin{eqnarray}
\mbox{Tr}_{\mbox{FD}} \hat{U}_q(g)e^{-\beta \hat{H}_q}&=&
\exp\left[+
\mbox{tr}_c
\sum_{\alpha}
\ln\left[
1+ {\bf R}(g)e^{-\beta E_{\alpha}}
\right]\right],
\nonumber\\
&=&\exp\left[+
\mbox{tr}_c\sum_{\alpha}\sum_{m=1}
\frac{1}{m}(-1)^{m+1}
\left({\bf R}(g)e^{-\beta E_{\alpha}}\right)^m
\right],\nonumber\\
&=&\exp\left[+
\sum_{m=1}
\frac{1}{m}(-1)^{m+1}
\left(
\mbox{tr}_c{\bf R}(g^m)
\sum_{\alpha}e^{-m\beta E_{\alpha}}\right)
\right],
\nonumber\\
&=&\exp\left[+
\sum_{m=1}\frac{1}{m}(-1)^{m+1}
\mbox{tr}_c{\bf R}(g^m)
z_{\mbox{F}}(e^{-m\beta})
\right].
\end{eqnarray}
In Bose-Einstein statistics, the single particle
partition function becomes
\begin{eqnarray}
\mbox{Tr}_{\mbox{BE}} \hat{U}_q(g)e^{-\beta \hat{H}_q}&=&
\exp\left[-\mbox{tr}_c
\sum_{\alpha}
\ln\left[1- {\bf R}(g)e^{-\beta E_{\alpha}}\right]\right],
\nonumber\\
&=&\exp\left[+
\mbox{tr}_c\sum_{\alpha}\sum_{m=1}
\frac{1}{m}
\left({\bf R}(g)e^{-\beta E_{\alpha}}\right)^m
\right],\nonumber\\
&=&
\exp\left[+\sum_{m=1}\frac{1}{m}
\mbox{tr}_c\left({\bf R}(g^m)
\sum_{\alpha}e^{-m\beta E_{\alpha}}\right)
\right],
\nonumber\\
&=&\exp\left[+
\sum_{m=1}\frac{1}{m}
\mbox{tr}_c{\bf R}(g^m) z_{\mbox{B}}(e^{-m\beta})
\right],
\end{eqnarray}
where $z_{\mbox{F/B}}(e^{-\beta})=\sum_{\alpha}e^{-\beta E_{\alpha}}$
is the sum of energy states with a specific structure and/or geometry
either for Fermi- or Bose-particles.
Furthermore, in the high energy limit (i.e. temperature),
the Maxwell-Boltzmann statistics becomes 
an appropriate approximation under certain conditions.
The single particle partition function in
the Maxwell-Boltzmann statistics is given by
\begin{eqnarray}
\mbox{Tr}_{\mbox{MB}} \hat{U}_q(g)e^{-\beta \hat{H}_q}&=&
\sum_{\{n_{\alpha,j}\}}\prod_{\alpha,j}
\frac{1}{n_{\alpha,j}}
\left(
{\bf R}_{jj}(g)
e^{-\beta E_{\alpha}}
\right)^{n_{\alpha,j}},
\nonumber\\
&=&
\prod_{\alpha,j}
\exp\left(
{\bf R}_{jj}(g)
e^{-\beta E_{\alpha}}
\right),
\nonumber\\
&=&
\exp\sum_{\alpha}
\ln \det_c
\left[\exp\left(
{\bf R}(g)
e^{-\beta E_{\alpha}}
\right)
\right],
\nonumber \\
&=&
\exp
\sum_{\alpha}
\mbox{tr}_c
\ln\exp\left(
{\bf R}(g)
e^{-\beta E_{\alpha}}\right),
\nonumber\\
&=&
\exp\left[
\mbox{tr}_c {\bf R} (g)\sum_{\alpha}e^{-\beta E_{\alpha}}
\right].
\end{eqnarray}
The theoretical group such as $U(N_c)$ and $SU(N_c)$ has a special
importance in the realistic physical application
such as quark and gluon bubble.

The $U(N_c)$ or $SU(N_c)$ internal color group symmetry
is introduced in the canonical ensemble in order to project
the state with a specific internal color structure.
The color structure of the quark and anti-quark is introduced
by the fundamental representation.
The Fock space single-particle partition function with
fundamental representation in the Fermi-Dirac statistics
reads as follows
\begin{eqnarray}
\mbox{Tr}_{\mbox{FD}} \hat{U}_q(g)e^{-\beta \hat{H}_q}
&=&
\exp\left[+
\mbox{tr}_c
\sum_{\alpha}
\ln\left[1+ {\bf R}_{\mbox{fun}}(g)e^{-\beta E_{\alpha}}\right]\right],
\nonumber\\
&=&
\exp\left[+
\sum^{N_c}_{i=1}
\sum_{\alpha}
\ln\left[1+ e^{i\theta_i}e^{-\beta E_{\alpha}}\right]\right],
\end{eqnarray}
where the trace over group $U(N_c)$ (or $SU(N_c)$) 
fundamental representation is given by
\begin{eqnarray}
\mbox{tr}_c {\bf R}_{\mbox{fun}}(g^k)=
\sum^{N_c}_{i=1} e^{i k\theta_i}.
\end{eqnarray}
The partition function for the fundamental particle and antiparticle 
Hilbert space becomes
%% skip1 %%
\begin{eqnarray}
\left[\mbox{Tr}_q \hat{U}_q(g)
e^{-\beta \hat{H}_q}\right]
\left[
\mbox{Tr}_{\overline{q}} \hat{U}_{\overline{q}}(g)
e^{-\beta \hat{H}_{\overline{q}}}\right]
&=&
\exp\left[+
\sum^{N_c}_{i=1}
\sum_{\alpha}
\ln\left[1+ 2\cos(\theta_i)e^{-\beta E_{\alpha}}+
e^{-2\beta E_{\alpha}}
\right]\right],
\end{eqnarray}
%% skip2 %%
On the other hand, the gluons are assumed as gauge particles 
in the theory and they are introduced 
in the context of adjoint representation.
The Fock space single-particle partition function
for adjoint particles satisfying
Bose-Einstein statistics reads
\begin{eqnarray}
\mbox{Tr}_{\mbox{BE}} \hat{U}_g(g)e^{-\beta \hat{H}_g}
&=&
\exp\left[-\mbox{tr}_c
\sum_{\alpha}
\ln\left[1- {\bf R}_{\mbox{adj}}(g)
e^{-\beta E_{\alpha}}\right]\right],
\nonumber\\
&=&
\exp\left[- \Re\sum^{N_c}_{i\neq j}
\sum_{\alpha}
\ln\left[1-e^{i \left(\theta_i-\theta_j\right)}
e^{-\beta E_{\alpha}}\right]
-N_c\sum_{\alpha}
\ln\left[1-e^{-\beta E_{\alpha}}\right]
\right].
\end{eqnarray}
The trace over the adjoint representation becomes
\begin{eqnarray}
\mbox{tr}_c {\bf R}_{\mbox{adj}}(g^k)&=&
\sum_{i\neq j}\cos k
\left(\theta_i-\theta_j\right)+N_c, 
~~U(N_c),
\nonumber\\
&=&
\sum_{i\neq j}
\cos k
\left(\theta_i-\theta_j\right)+(N_c-1), 
~~SU(N_c),
\end{eqnarray}
for the $U(N_c)$ and $SU(N_c)$ groups, respectively.
%%%%%%%%%%%%%%%%%%%%%%%%%%%%%%%%%%%%%%%%%%%%%%%%%%%%
%
%
%
%%%%%%%%%%%%%%%%%%%%%%%%%%%%%%%%%%%%%%%%%%%%%%%%%%%%%
The density of states for single particle levels has
been studied extensively using the multiple reflection
expansion method~\cite{Balian:1970a,Balian70b,Balian71a}.
The sum runs over the number of states
which can fill the one particle states.
The sum over states for constituent particles
those are confined in a spherical cavity
is calculated as 
follows~\cite{Balian:1970a,Balian70b,Balian71a}
\begin{eqnarray}
E_{\alpha}&=&\sqrt{p^2+m}+V(r),
\nonumber\\
\sum_{\alpha} &=& {\cal D}_D \int_V d^3 r
\int d E \rho(E),
\nonumber\\
&=& {\cal D}_D
V\int d E \rho(E),
\nonumber\\
&=& {\cal D}_D \int\frac{Vd^3p}{(2\pi)^3}.
\end{eqnarray}
The volume $V$ is the spherical cavity volume where the constituent
particles are confined inside and it corresponds 
the total occupational space.
The one particle degeneracy is given
by ${\cal D}_D=N_c {\cal D}_d$
where $N_c$ is the number of colors and
${\cal D}_d$ is the particle species degeneracy.
Usually we set ${\cal D}_d=(2j+1)$ 
as the spin multiplicity 
where $j$ is the spin quantum number.

%%%%%%%%%%%%%%%%%%%%%%%%%%%%%%%%%%%%%%%
%%%%%%%%%%%%%%%%%%%%%%%%%%%%%%%%%%%%%%%
Nonetheless, it is interesting to note here that
in the case we have a thermodynamical system of particles
confined in a compactified space with specific geometry
and other degrees of freedom, the sum of states
becomes nontrivial.
It has been noted that in the conformal field theories,
the partition function for scalars and chiral Fermions on
$S^3\times R$ are given
by~\cite{Sundborg:2000a,Aharony2003a}
\begin{eqnarray}
z_{\mbox{B}}\left(e^{-\beta}\right)=
\frac{ e^{-\beta}+e^{-2\beta} }
{\left(1-e^{-\beta}\right)^3},
\end{eqnarray}
and
\begin{eqnarray}
z_{\mbox{F}}\left(e^{-\beta}\right)=
\frac{ e^{-\frac{4}{3}\beta} }
{\left(1-e^{-\beta}\right)^3}.
\end{eqnarray}
Hence, the model analysis for the partition function that is given
in the nuclear physics for the search for a quark-gluon plasma 
can be generalized to study the conformal
fields such as pure Yang-Mills theory and ${\cal N}=4$
Super-symmetry Yang-Mills theory. 
The understanding of the partition
function with various internal structures can shed the light
to a new phase of matter beyond the quark-gluon plasma.
This might be also useful to construct new models 
to explore new physics such as the searching 
for a dark matter.
%%%%%%%%%%%%%%%%%%%%%%%%%%%%%%%%%%%%%%%%%%%%%%%%%%%%%%%%%%%%%%%%%%
%%%%%%%%%%%%%%%%%%%%%%%%%%%%%%%%%%%%%%%%%%%%%%%%%%%%%%%%%%%%%%%%%%
\section{Invariance measure\label{sect_haar_measure}}

The invariance measure is essential in order to project
the state with a specific internal structure.
The invariance Haar measure reads
\begin{eqnarray}
\int d\mu(g)&=&
\frac{1}{N_c!}\left(\frac{1}{2\pi}\right)^{N_c}
\int^{\pi}_{-\pi}\left[
\prod_{m>n} 2\sin^2\left(\frac{\theta_n-\theta_m}{2}\right)
\right]
\prod^{N_c}_{k=1}d\theta_k,
~~U(N_c),
\nonumber\\
&=&
\frac{1}{N_c!}\left(\frac{1}{2\pi}\right)^{N_c}
\int^{\pi}_{-\pi}\left[
\prod_{m>n} 2\sin^2\left(\frac{\theta_n-\theta_m}{2}\right)
\right]
\prod^{N_c}_{k=1}
\left(2\pi\delta(\sum^{N_c}_{i=1}\theta_i)\right) 
d\theta_k,
~~SU(N_c),
\end{eqnarray}
for $U(N_c)$ and $SU(N_c)$
group representations, respectively.
Haar invariance measure for the system with 
homogeneous and uniform distribution
over the Fourier color variables,
$|\theta_i|\le\pi$, is written for $SU(N_c)$
as follows
\begin{eqnarray}
\int d\mu(g)=
\frac{1}{N_c!}\left(\frac{1}{2\pi}\right)^{N_c-1}
\left(2^{(N^2_c-N_c)/2}\right)
\prod^{N_c-1}_{k=1} \int^{\pi}_{-\pi} d\theta_k
\left. e^{\left[
\frac{1}{2}\sum^{N_c}_{n=1}\sum^{N_c}_{m=1}
\ln \sin^2\left(
\frac{\theta_n-\theta_m}{2}
\right)
\right]}\right|_{n\neq m}.
\end{eqnarray}
The Vandermonde determinant, which appears in the measure 
contributes to the action as an additional potential term.
The Vandermonde effective potential term is soft 
when the color eigenvalues
in the stationary condition distribute uniformly
over the entire circle $|\theta_i|\le \pi$.
However, this will not be the case 
for the extreme condition in particular 
when the color eigenvalues are distributed
in a narrow domain $|\theta_i|<<\pi$. 
Under this condition
the Vandermonde effective potential 
becomes virtually singular and the action must be
regulated in a proper way in order to remove
the Vandermonde determinant divergence.
The regulated Haar measure in the extreme conditions becomes
\begin{eqnarray}
\int d\mu(g)=
\frac{1}{N_c!}\left(\frac{1}{2\pi}\right)^{N_c}
\int \prod^{N_c}_{k=1}\left[
\prod_{m>n}\left(\theta_n-\theta_m\right)^2
\right]2\pi\delta\left(\sum^{N_c}_{i=1}\theta_i\right)d\theta_k.
\label{invariance_measure_saddle}
\end{eqnarray}
This version for the invariance measure accommodates
the canonical ensemble in the hot medium.
The similar results can be verified for 
the theoretic group $U(N_c)$ representation.
%%%%%%%%%%%%%%%%%%%%%%%%%%%%%%%%%%%%%%%%%%%%%%%%%%%%%%%%%%%
%%%%%%%%%%%%%%%%%%%%%%%%%%%%%%%%%%%%%%%%%%%%%%%%%%%%%%%%%%%
%%%%%%%%%%%%%%%%%%%%%%%%%%%%%%%%%%%%%%%%%%%%%%%%%%%%%%%%%%%
%
%  NEW SECTION
%
%%%%%%%%%%%%%%%%

\section{Fundamental particles:
Spectral density method\label{sect_fundamental_spectral}}
%%%%%%%%%%%%%%%%%%%%%%%%%%%%%%%%%%%%%%%%%%%%%%%%%%%%%%%%%%%%%%%%%%%%%

The canonical ensemble for fundamental particles 
in the $U(N_c)$ or $SU(N_c)$
representation, confined in a spherical
cavity and obeying Maxwell-Boltzmann statistics, reads
\begin{eqnarray}
Z(\beta)&=&
\int d\mu(g)
\exp\left[
{\cal D}_d\int_V d^3 r\int\frac{d^3p}{(2\pi)^3}
{\cal G}_{\mbox{fun}}(\theta) e^{-\beta E(p,r)}
\right],
\end{eqnarray}
%%%%%%%%%%%%%%%%%%%%%
%%%%%%%%%%%%%%%%%%%%%
where ${\cal D}_d$ is the single particle degeneracy due 
to the degree of freedom 
which is not specified explicitly in the calculation 
such as the spin multiplicity.
The trace over the color degree of freedom 
for particles and their antiparticle partners
in the fundamental representation is given by
\begin{eqnarray}
{\cal G}_{\mbox{fun}}(\theta)&=&
\frac{1}{N_c}
\mbox{tr}\left(
{\bf R}_{\mbox{fun}}(g)+{\bf R}_{\mbox{fun}}^*(g)
\right),
\nonumber\\
&=& \frac{2}{N_c}\sum^{N_c}_{i=1}\cos\theta_i.
\end{eqnarray}
The sum of states for a gas of particles confined 
in a spherical cavity is calculated as follows
\begin{eqnarray}
\lambda&=& {\cal D}_d
\int_V d^3 r 
\int\frac{d^3p}{(2\pi)^3}
e^{-\beta
E(p,r)},
\nonumber\\
&=&
{\cal D}_d
\int\frac{V d^3p}{(2\pi)^3} e^{-\beta E(p,r)}.
\end{eqnarray}
The color-singlet state for the canonical partition
function for a gas of fundamental particles
and antiparticles is projected as follows,
\begin{eqnarray}
Z(\lambda)&=&
\frac{1}{N_c!} 2^{(N^2_c-N_c)/2}
\prod^{N_c}_{k=1} \int \frac{d\theta_k}{2\pi}
\nonumber\\
&\times&
\exp\left[
\left.\frac{1}{2}\sum^{N_c}_{n=1}\sum^{N_c}_{m=1}
\ln \sin^2\left(
\frac{\theta_n-\theta_m}{2}
\right)\right|_{n\neq m}
+
\lambda
\frac{1}{N_c}
\mbox{tr}\left({\bf R}_{\mbox{fun}}(g)
+{\bf R}_{\mbox{fun}}^*(g)\right)
\right].
\end{eqnarray}
The multiple integrations in the partition function 
are evaluated using the steepest descent method.
The partition function is reduced to
\begin{eqnarray}
Z(\lambda)=
\frac{1}{N_c!} 2^{(N^2_c-N_c)/2}
\exp\left[
\frac{1}{2}\sum^{N_c}_{n=1}\sum^{N_c}_{m=1}\left.\ln \sin^2\left(
\frac{\overline{\theta}_n-\overline{\theta}_m}{2}
\right)\right|_{n\neq m}
+ 2\lambda \frac{1}{N_c}\sum^{N_c}_{n=1} \cos\overline{\theta}_n
\right].
\end{eqnarray}
The saddle points $\overline{\theta}_n$
are determined by the stationary condition
\begin{eqnarray}
2\lambda\frac{1}{N_c}\sin\overline{\theta}_i=
\sum^{N_c}_{n\neq i}
\cot\left| 
\frac{\overline{\theta}_i-\overline{\theta}_n}{2}
\right|.
\label{Eqstationary}
\end{eqnarray}
The above equation is turned out 
to be an eigenvalue problem for the large $N_c$ limit
and the saddle points are determined by the eigenvalues.
Hereinafter, we introduce $\lambda=N^2_c\widetilde{\lambda}$.
The stationary condition becomes
\begin{eqnarray}
2\widetilde{\lambda}\sin\overline{\theta}_i=
\frac{1}{N_c}\sum^{N_c}_{n\neq i}
\cot\left| 
\frac{\overline{\theta}_i-\overline{\theta}_n}{2}
\right|.
\label{Eqstationarynorm}
\end{eqnarray}
The solution for the stationary condition 
that is given by 
Eq.(\ref{Eqstationarynorm}) depends
basically on the value of the thermal 
running parameter $\widetilde{\lambda}$.
It can be shown numerically
that the solution of $\overline{\theta}_i$
for $\widetilde{\lambda}\le \frac{1}{2}$ 
is distributed uniformly over 
the entire circle range $|\overline\theta_i|\le \pi$ 
and
$\left|\overline{\theta}_i-\overline{\theta}_j\right|\le \frac{2\pi}{N_c}$.
However, this solution ceases to exist for the full range
$|\theta_i|\le \pi$ 
when the thermal running parameter 
$\widetilde{\lambda}$ becomes relatively large and 
runs over the range $\widetilde{\lambda}>\frac{1}{2}$.
The eigenvalues $\theta_i$ turn to be distributed 
uniformly only over the incomplete range $|\theta_i|\le \theta_c <\pi$. 
This means that the solution
for $\widetilde{\lambda}>\frac{1}{2}$ becomes problematic
and the action must be regulated thoroughly when $\theta_c$ becomes small.  
When that action is regulated, 
another analytical solution may emerge 
and the change in the analytical solution characteristic   
becomes the responsible for the existence of the phase transition.

%
%
%
%%%%%%%%%%%%%%%%%%%%%%%%%%%%%%%%%%%
%%
%% Figure : stationary solution
%%          and the ultimate limit
%% Fig.: westen_station_a.eps fig_w_station_a
%%
%%%%%%%%%%%%%%%%%%%%%%%%%%%%%%%%%%%
%%%%%%%%%%%%%%%%%%%%%%%%%%%%%%%%%%%

\begin{figure}
\includegraphics{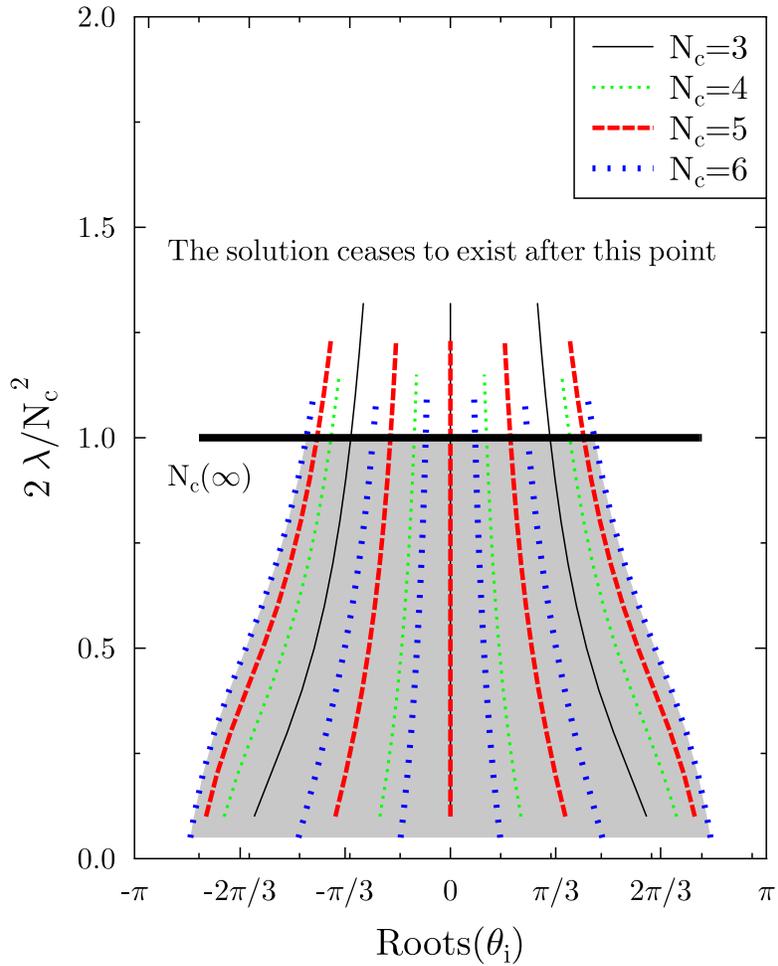}%
\caption{\label{fig_w_station_a}
(Color online)
The saddle points solution for particles
in the fundamental representation.
The saddle points are distributed uniformly over the range
$|\theta_i|\le \pi$ for a parameter runs over the range
$\lambda\le \lambda_0$. 
The numerical solution shows that saddle points
cease to exist when the thermal running parameter 
$\lambda$ exceeds the value
$\frac{\lambda_0}{N^2_c}=\frac{1}{2}$ in the large $N_c$ limit.}
\end{figure}

The numerical solution for the set of stationary conditions 
which are given by either Eq.(\ref{Eqstationary}) 
or Eq.(\ref{Eqstationarynorm})
versus the thermal running parameter $\widetilde{\lambda}$
for various color numbers $N_c$ with group symmetry $U(N_c)$
is displayed in Fig.\ref{fig_w_station_a}.
The resemblance between 't Hooft coupling
$\lambda_{W-S}=g^2_{YM} N_c$ and 
the thermal running parameter
$\widetilde{\lambda}=\lambda/N^2_c$ is given
by the relation $g^2_{YM} N_c=1/\widetilde{\lambda}$.
The weak 't Hooft coupling corresponds
the high-lying thermal running parameter 
$\widetilde{\lambda}\ge\widetilde{\lambda}_{\mbox{critical}}$
while the strong 't Hooft coupling 
is given by the low-lying $\widetilde{\lambda}
\le\widetilde{\lambda}_{\mbox{critical}}$.
The Vandermonde determinant contributes to the
effective action. 
Consequently, the saddle point locations
are computed using the stationary conditions 
where the Vandermonde determinant is included explicitly 
in the action as an additional potential term.
It is shown numerically that the saddle points are distributed
uniformly over the interval $-\pi\le\theta_i\le\pi$
for low-lying thermal running parameter $\widetilde{\lambda}$
(i.e. strong 't Hooft coupling).
The characteristic distribution of the saddle points
is found uniform for the small $\widetilde{\lambda}$
even for a small number of colors $N_c=2$.
This distribution becomes more dominant over a narrower range
$|\theta_i|\le \theta_0<\pi$
as $\widetilde{\lambda}$ increases and approaches the critical value.
However, the numerical calculation shows that the saddle points
cease to exist when the thermal running parameter
$\widetilde{\lambda}$ exceeds a critical value just
above $\widetilde{\lambda}\ge \frac{1}{2}$ (i.e. weak 't Hooft coupling).
For example, in the case of $N_c=3$,
the saddle points cease to exist when the thermal running
parameter exceeds the value $\widetilde{\lambda}>\frac{1.32}{2}$.
Furthermore, it is found that when the number of colors $N_c$
increases the saddle points cease to exist at smaller value of
$\widetilde{\lambda}$.
For example, in the case $N_c=5$, the solution for these points 
does not exist
beyond the critical point $\widetilde{\lambda}>\frac{1.23}{2}$.
In the limit of large $N_c$, the solution
for saddle points breaks down just above
the point $\widetilde{\lambda}=\frac{1}{2}$.
%
%%%%%%%%%%%%%%%%%%%%%%%%%%%%%%%%%%%%%%%
The Vandermonde term diverges when the color saddle points
become more dominant in a tiny domain around the origin. 
This divergence breaks the solution badly 
when these saddle points congregate and converge to the origin. 
It is found that the saddle points cease to exist
when $\widetilde{\lambda}>\widetilde{\lambda}_{\mbox{critical}}$.
This implies the stationary conditions must break down
and the action must be regulated
in a proper way and a new set of conditions must be re-formed.
This proves the existence of the phase transition in the system.
%%%%%%%%%%%%%%%%%%%%%%%%%%%%%%%%%%%%%%%%%%%
%
%
%%%%%%%%%%%%%%%%%%%%%%%%%%%%%%%%%%%%%%%%%%%%%%%%%%%%%%%%%%%%%%%
%%%%%%%%%%%%%%%%%%%%%%%%%%%%%%%%%%%%%%%%%%%%%%%%%%%%%%%%%%%%%%%

In the large-$N_c$ limit,
Gross and Witten have solved these equations
in the case of two-dimensional $U(N_c)$ lattice gauge theory
following Brezin {\em et. el.}~\cite{Brezin1978a}
by introducing the spectral density of eigenvalues.
It is shown by Brezin, Itzykson
and Parisi
and Zuber\cite{Brezin1978a} that the functional
integrals in the large-$N_c$ limit
can be calculated by steepest-descent methods.
In this limit, a particular configuration
totally dominates the functional integral.
The stationary equations in this limit
can be replaced
by their continuum
version by introducing a no decreasing function
\begin{eqnarray}
\overline{\theta}_n&=&\theta(x), \nonumber\\
x&=&\frac{n}{N_c}, \nonumber\\
n&=& 1,\cdots, N_c.
\end{eqnarray}
Hence, the stationary condition becomes
\begin{eqnarray}
2\widetilde{\lambda} \sin\theta(x_0)=\mbox{P}\int^1_0 dy
\cot\left(\frac{\theta(x_0)-\theta(y)}{2}\right),
\end{eqnarray}
where $\mbox{P}$ refers 
to the principle part of the integral.
By introducing the density of states
\begin{eqnarray}
\rho(\theta)=\frac{d x}{d \theta}\ge 0,
\end{eqnarray}
and the constraint
\begin{eqnarray}
\int^1_0 dx=
\int^{\theta_c}_{-\theta_c}d\theta \rho(\theta)=1,
\end{eqnarray}
where $|\theta_c|\le \pi$.
Hence the stationary condition becomes
\begin{eqnarray}
2\widetilde{\lambda}\sin\theta&=&
\mbox{P} \int^{\theta_c}_{-\theta_c}
d\theta'\rho(\theta')
\cot\left(\frac{\theta-\theta'}{2}\right).
\end{eqnarray}

%%%%%%%%%%%%%%%%%%%%%%%%%%%%%%%%%%%%%%%%%%%%%%%%%
%% subsection
%%%%%%%%%%%%%%%%%%%%%%%%%%%%%%%%%%%%%%%%%%%%%%%%%
\subsection{The highly thermal excited matter:
$\frac{\lambda}{N^2_c}\ge \frac{1}{2}$}

The solution of density of eigenvalues
in the large $N_c$ limit in $U(N_c)$ representation
has attracted much attention, recently.
Although the solution for the spectral density
of eigenvalues is difficult in the realistic situation,
the approximate solution for this density
has been derived for some simple specific
cases for fundamental 
and adjoint representations in the limit of large $N_c$.
In the following, we shall review
the partition function using the given solution 
of the spectral density of color eigenvalues.
Then in section~\ref{sect_fundamental_GSP}
we introduce another approximation
to evaluate the same partition function.
The other method is to approximate
the resultant integral to the Gaussian-like integral
and this approximation sound more efficient
when more complicate situations are involved.
The two methods have been shown to reproduce
the same results
for the large thermal running parameter
$\widetilde{\lambda}\ge\widetilde{\lambda}_0$
and $\widetilde{\lambda}=\lambda/N^2_c$.
%%%

The first method to evaluate the canonical
ensemble is based essentially
on the stationary equation solution
for the density of color eigenvalues
and it is written for
$\widetilde{\lambda}\ge \frac{1}{2}$
as follows
\begin{eqnarray}
\rho(\theta)=
\frac{2}{\pi}\widetilde{\lambda}\cos\frac{\theta}{2}
\left[
\sin^2\left(\frac{\theta_c}{2}\right)
-\sin^2\left(\frac{\theta}{2}\right)
\right]^{1/2},
\end{eqnarray}
where
\begin{eqnarray}
\sin^2\left(\frac{\theta_c}{2}\right)&=&
\frac{1}{2\widetilde{\lambda}}
\le 1 ~~~~~~~
\left(\lambda/N^2_c=\widetilde{\lambda}\ge\frac{1}{2}\right),
\end{eqnarray}
The color eigenvalues are not distributed uniformly
over the entire color circle range $-\pi$ to $\pi$ but instead,
in this solution they are distributed only over a narrow interval
$|\theta_i|\le\theta_c$ where $|\theta_c|<\pi$.
The canonical partition function after weighting the color density
of eigenvalues becomes
\begin{eqnarray}
Z(\lambda)&=& C e^{\left[ \frac{1}{2}N^2_c
\int^{\theta_c}_{-\theta_c} d\theta \rho(\theta)
\int^{\theta_c}_{-\theta_c} d\theta'\rho(\theta')
\ln\sin^2\left(\frac{\theta-\theta'}{2}\right)
+
2N^2_c\widetilde{\lambda} \int^{\theta_c}_{-\theta_c}
d\theta\rho(\theta)\cos\theta
\right]},
\end{eqnarray}
where the pre-exponential constant is given by $C=\frac{2^{(N^2_c-N_c)/2}}{N_c!}$.
Under this stationary approximation,
the canonical ensemble function is calculated as follows
\begin{eqnarray}
Z(\lambda)&\cong& \exp\left[
2N^2_c\widetilde{\lambda}-\frac{N^2_c}{2}\ln\left(2\widetilde{\lambda}
N^2_c\right)
+\frac{N^2_c}{2}\ln N^2_c
-\frac{3}{4}N^2_c
\right],\nonumber\\
&\cong&
\left(\frac{1}{2\widetilde{\lambda}}\right)^{\frac{N^2_c}{2}}
e^{\left(2N^2_c\widetilde{\lambda}-\frac{3}{4}N^2_c\right)},
~~~\mbox{where}~\widetilde{\lambda}\ge \frac{1}{2}.
\end{eqnarray}
Hence the asymptotic large running thermal coupling solution reads,
\begin{eqnarray}
Z(\lambda)&\equiv&
\left(\frac{N^2_c}{2\lambda}\right)^{\frac{N^2_c}{2}}
e^{\left(2\lambda-\frac{3}{4}N^2_c\right)}.
\label{DEV1}
\end{eqnarray}

%%%%%%%%%%%%%%%%%%%%%%%%%%%%%%%%%%%
%
% Small coupling constant
%
%%%%%%%%%%%%%%%%%%%%%%%%%%%%%%%%%%%
%%%%%%%%%%%%%%%%%%%%%%%%%%%%%%%%%%%%%%%%%%%%%%%%%
%% subsection
%%
\subsection{The diluted and relatively cold matter:
$\frac{\lambda}{N^2_c}\le\frac{1}{2}$}

At the low temperature, the stationary condition
for $\widetilde{\lambda}\le \frac{1}{2}$
produces the following density of color eigenvalues
\begin{eqnarray}
\rho(\theta)=
\frac{1}{2\pi}\left(1+2\widetilde{\lambda}\cos\theta\right),
~~ (-\pi\le\theta\le \pi).
\label{spectra_small_sat}
\end{eqnarray}
The chemical potential for the total particle number due to
the rotated $U(1)_B$ symmetry is not considered here and
it is left for the forthcoming work.
In the large $N_c$ limit,
the system is assumed highly compressed due
to the large number of colors
but this is not the case for a finite number of colors such as QCD.
In this case, the system is identified as a compressed one
when the chemical potential for particles numbers becomes relatively
large.
The density of eigenvalues given by Eq.(\ref{spectra_small_sat})
for $\frac{\lambda}{N^2_c}\le\frac{1}{2}$
produces the following partition function
\begin{eqnarray}
Z(\lambda)&=&e^{N^2_c\widetilde{\lambda}^2},\nonumber\\
&=&e^{\lambda^2/N^2_c}.
\end{eqnarray}
%
%%%%%%%%%%%%%%%%%%%%%%%%%%%%%%%%%%%%%%%%%%%%%%%%%
%
The occurrence of such a phase
transition from the diluted and relatively cold matter (i.e. low-lying phase) 
to the highly thermal excited matter (i.e. high-lying phase)
would not mean that the large-$N_c$ theory
does not confine. 
The highly excited thermal matter can be interpreted 
as an exotic hadronic phase dominated by the Hagedorn states.
In the lattice theory, this would imply
that the weak- and strong-$\lambda_{W-S}$ coupling
is not described by the same
analytic functions\cite{GrossWitten1980a}.
The weak- and strong-$\lambda_{W-S}$ are analogous to
large and small thermal running parameter $\lambda/N^2_c$,
respectively.
For the finite $N_c$, one could continue from
the strong to the weak coupling.
However, one would expect to see a sign of
phase transition for large enough $N_c$ (i.e. in the limit $N_c=\infty$)
whose manifestation would be a sharp transition
at $\lambda_{W-S}\approx \lambda_{W-S(\mbox{critical})}$
from the weak-coupling to the strong-coupling behavior.
As in the lattice theory, in the thermal and dense QCD, 
the strong- and weak-coupling transition is analogous to the phase transition
from the discrete low-lying mass spectrum particles to
the highly excited and massive Hagedorn states 
(i.e. continuous high-lying mass spectrum).
The Hagedorn states are the color-singlet (confined)
hadronic states. This phase should not be interpreted
as an immediate deconfined phase. 
The critical Gross-Witten point is the threshold point 
for the Hagedorn states to emerge in the system.
The deconfined phase transition can either take place
immediately when unstable Hagedorn states are produced
in the system or as a subsequent process when
the metastable Hagedorn phase undergoes 
a true deconfined phase transition.

%%%%%%%%%%%%%%%%%%%%%%%%%%%%%%%%
%
%  saddle point and Gaussian integration
%   our method
%
%%%%%%%%%%%%%%%%%%%%%%%%%%%%%%%%%
\section{Fundamental particles: Gaussian-like saddle points
method \label{sect_fundamental_GSP}}
%%%%%%%%%%%%%%%%%%%%%%%%%%%%%%%%%%%%%%%%%%%%%%%%%%%%%%%%%%%%%%%

We introduce an alternate novel method
to calculate the partition function
and to find the critical point for the phase transition.
This method accommodates the sophisticated physical problems
such as the internal color structure for a gas 
consisting particles with various statistics species 
in a specific space boundary
and with more complicate color-flavor correlations 
and chiral symmetries.
It may be useful to find the equation of state
for physics beyond QCD.
\subsection{The highly thermal excited matter:
$\frac{\lambda}{N^2_c}\ge \frac{1}{2}$}
In the realistic physical situation, 
more calculations are involved.
It becomes more difficult to calculate the density of
color eigenvalues under certain conditions 
in particular when the color eigenvalues populate
a tiny interval $|\theta_c|\ll 1$ under extreme conditions. 
The canonical ensemble has a special importance for 
the thermal running parameter 
$\widetilde{\lambda}$ over the range
$\widetilde{\lambda}>\frac{1}{2}$ 
and it corresponds the hot hadronic matter.
It is relevant to the relativistic heavy ion collisions
in RHIC and LHC, and the cosmology 
such as the physics of the early universe.

In following, we shall demonstrate a method
how to calculate the partition function
for the coupling parameter
$\frac{\lambda}{N^2_c}\ge \widetilde{\lambda}_0$
(in our case 
$\widetilde{\lambda}_0=
\widetilde{\lambda}_{\mbox{critical}}=\frac{1}{2}$).
The integral over the color invariance measure 
is approximated to the Gaussian-like
multi-integrals around the saddle points.
The color saddle points are dominant
in the narrow interval $|\theta|\le\theta_c$ 
around the origin where $|\theta_c|\ll 1$.
As far the saddle points are populated in a small domain
we can safely approximate all the saddle points
to be located near the origin.
The Vandermonde potential does not appear in the action 
but instead the invariance measure appears 
as a pre-factor power function of the exponential 
in the Gaussian integral.
The integration of the resultant multi-Gaussian integrals is straightforward.
We shall call this method as the Gaussian-like saddle points method.
In this limit the invariance Haar measure is approximated to,
\begin{eqnarray}
\int_{U(N_c)} d\mu(g)_{\mbox{saddle}}\approx
\prod^{N_c}_{k=1} \int d\theta_k
\frac{1}{N!}\left(\frac{1}{2\pi}\right)^{N_c}
\left[\prod^{N_c}_{n>m}(\theta_n-\theta_m)^2\right],
\end{eqnarray}
and
\begin{eqnarray}
\int_{SU(N_c)} d\mu(g)_{\mbox{saddle}}
&\approx&
\frac{1}{N!}\left(\frac{1}{2\pi}\right)^{N_c}
\prod^{N_c}_{k=1} \int d\theta_k
\left[\prod^{N_c}_{n>m}(\theta_n-\theta_m)^2\right]
2\pi\delta\left(\sum^{N_c}_i\theta_i\right),
\end{eqnarray}
for $U(N_c)$ and $SU(N_c)$ symmetries, respectively.
The partition ensemble is given by
\begin{eqnarray}
Z(\lambda)&=&\int d\mu(g) \exp\left[\frac{\lambda}{N_c}
\mbox{tr}_c \left(
{\bf R}_{\mbox{fun}}(g)+{\bf R}_{\mbox{fun}}^{*}(g)
\right)\right],
\nonumber\\
&=&\int d\mu(g) \exp\left(
2\frac{\lambda}{N_c}
\sum^{N_c}_{n=1}\cos\theta_n
\right).
\end{eqnarray}
When the thermal running parameter becomes a relatively large
and runs over the range 
$\widetilde{\lambda}\ge
\widetilde{\lambda}_{(II)\mbox{minimum}}$,
the partition ensemble is approximated  
around the saddle points which become dominant 
in a small range as follows 
\begin{eqnarray}
Z(\lambda)&\approx& Z_{(II)}(\lambda)
~~~
\left(@~~
\widetilde{\lambda}~ \mbox{runs over the range}~ 
\widetilde{\lambda}\ge
\widetilde{\lambda}_{(II)\mbox{minimum}}
\right), 
\nonumber\\
Z_{(II)}(\lambda)&=&\int d\mu(g)_{\mbox{saddle}} 
\exp\left(
2N_c\widetilde{\lambda}\sum^{N_c}_{n=1}\cos\theta_n
\right),\nonumber\\
&\approx&
e^{2N^2_c \widetilde{\lambda}}
\int d\mu(g)_{\mbox{saddle}}
\exp\left[-\frac{1}{2}\left(2N_c\widetilde{\lambda}\right)
\sum^{N_c}_{n=1}\theta^2_n\right],
\end{eqnarray}
where $\lambda=N^2_c\widetilde{\lambda}$.
In the case of $U(N_c)$ symmetry, the partition function 
becomes
\begin{eqnarray}
Z_{(II)}(\lambda)&=&
\frac{1}{N_c!}\left(\frac{1}{2\pi}\right)^{N_c}
\frac{e^{2N^2_c\widetilde{\lambda}}}
{\left(2N_c\widetilde{\lambda}\right)^{\frac{N^2_c}{2}}}
\left(\prod^{N_c}_{k=1} \int^{x_0}_{-x_0}
d x_k\right)
\left(\prod^{N_c}_{n>m}\left(x_n-x_m\right)^2\right)
e^{
-\frac{1}{2}
\sum^{N_c}_{n=1} x^2_n
}.
\end{eqnarray}
The Gaussian-like multi-integrals are evaluated
using the following standard formula
\begin{eqnarray}
\left(\prod^{N_c}_{k=1} \int^{\infty}_{-\infty}
d x_k\right)
\left(\prod^{N_c}_{n>m}\left(x_n-x_m\right)^2\right)
\exp\left[
-\frac{1}{2}
\sum^{N_c}_{n=1} x^2_n
\right]
&\equiv&
\left(2\pi\right)^{N_c/2}\prod^{N_c}_{n=1} n!.
\end{eqnarray}
Hence, the partition function is evaluated as follows
\begin{eqnarray}
Z_{(II)}(\lambda)&\cong&
\left(
\frac{1}{2N_c\widetilde{\lambda}}
\right)^{\frac{N^2_c-1}{2}}
e^{2N^2_c \widetilde{\lambda}}
\left[\frac{1}{(2\pi)^{\frac{N_c}{2}-1}}
\frac{1}{N_c!}
\frac{1}{\sqrt{2\pi N_c}} \prod^{N_c}_{n=1}n!
\right]
~~~@~~
\widetilde{\lambda}\ge
\widetilde{\lambda}_{(II)\mbox{minimum}},
\end{eqnarray}
and
\begin{eqnarray}
Z_{(II)}(\lambda)&\cong&
\left(\frac{1}{2N_c\widetilde{\lambda}}\right)^{\frac{N^2_c}{2}}
e^{2N^2_c \widetilde{\lambda}}
\left[\frac{1}{(2\pi)^{\frac{N_c}{2}}}
\frac{1}{N_c!}
\prod^{N_c}_{n=1}n!
\right]
~~~@~~
\widetilde{\lambda}\ge
\widetilde{\lambda}_{(II)\mbox{minimum}},
\end{eqnarray}
for $SU(N_c)$ and $U(N_c)$ symmetries, respectively.
In order to compare the results of the present method 
with those derived using the density of eigenvalues method 
(i.e. the spectral density method)
given by Eq.(\ref{DEV1}),
the partition function for $U(N_c)$ is written as follows
\begin{eqnarray}
Z_{(II)}(\lambda)=C_{N_c}
\left(\frac{1}{2\widetilde{\lambda}}\right)^{\frac{N^2_c}{2}}
e^{2N^2_c\widetilde{\lambda}}
~~~@~~
\widetilde{\lambda}\ge
\widetilde{\lambda}_{(II)\mbox{minimum}},
\end{eqnarray}
where the pre-factor constant is given by
\begin{eqnarray}
C_{N_c}=N^{-\frac{N^2_c}{2}}_c
\left[\frac{1}{(2\pi)^{\frac{N_c}{2}}}
\frac{1}{N_c!}
\prod^{N_c}_{n=1}n!
\right].
\label{Normc}
\end{eqnarray}
Using the Stirling's approximation
$\ln n!\approx n\ln n -n
+\frac{1}{2}\ln n +\frac{1}{2}\ln 2\pi$,
the pre-factor given by Eq.(\ref{Normc}) is simplified to
\begin{eqnarray}
\ln C_{N_c}&=& -\frac{N^2_c}{2}\ln N_c -\frac{N_c(N_c+1)}{2}
-\frac{1}{2}\sum^{N_c}_{n=1}\ln n
+\sum^{N_c}_{n=1}n \ln n,
\nonumber\\
&\approx&
-\frac{N^2_c}{2}\ln N_c -\frac{N_c(N_c+1)}{2}
-\frac{1}{2}\int^{N_c}_{1}dn \ln n
+\int^{N_c}_{1}dn n\ln n,
\nonumber\\
&=& -\frac{3}{4}N^2_c-\frac{1}{2}N_c\ln N_c.
\label{stirlingfund}
\end{eqnarray}
The above expression simplifies 
the partition function to
\begin{eqnarray}
\lim_{N_c\rightarrow \mbox{large}}
Z_{(II)}(\lambda)
\approx \left(\frac{1}{2\widetilde{\lambda}}\right)^{N^2_c/2}
\exp\left[2 N^2_c\widetilde{\lambda}
-\frac{3}{4}N^2_c\right]
~~~@~~
\widetilde{\lambda}\ge
\widetilde{\lambda}_{(II)\mbox{minimum}}.
\end{eqnarray}
The above solution is the approximate solution 
over the range 
\begin{eqnarray}
\widetilde{\lambda}\ge\widetilde{\lambda}_{\mbox{critical}}
\ge\widetilde{\lambda}_{(II)\mbox{minimum}}, 
~~~{\mbox{where}}~~
\widetilde{\lambda}_{\mbox{critical}}=\frac{1}{2}.
\end{eqnarray}
It is in agreement with Eq.(\ref{DEV1}).
This means that the Gaussian-like saddle points method
is consistent with the density of eigenvalues method
developed by Brezin {\em et. el.}~\cite{Brezin1978a}
for the energy range $\frac{\lambda}{N^2_c}\ge\frac{1}{2}$.

%%%%%%%%%%%%%%%%%%%%%%%%%%%%%%%%%%%%%%%%%%%%%%%%%
%% subsection
%%
%%%%%%%%%%%%%%%%%%%%%%%%%%%%%%%%%%%%%%%%%%%%%%%%%
%%%%%%%%%%%%%%%%%%%%%%%%%%%%%%%%%%%%%%%%%%%%%%%%%
%
%
\subsection{The diluted and relatively cold matter:
$\frac{\lambda}{N^2_c}\le \frac{1}{2}$}

The Fourier color angles are distributed uniformly
over the entire color circle range $-\pi\le \theta_i\le \pi$
in the low-lying energy domain
$\frac{\lambda}{N^2_c}\le\frac{1}{2}$ as follows
\begin{eqnarray}
Z(\lambda)&\approx&Z_{(I)}(\lambda) 
~
\left(@~~ \widetilde{\lambda}~
\mbox{runs over the range}~
\widetilde{\lambda}
\le\widetilde{\lambda}_{(I)\mbox{maximum}}
\right),
\nonumber\\
Z_{(I)}(\lambda)&=&\int^{\pi}_{-\pi} d\mu(g)
\exp\left[\lambda
\frac{1}{N_c}\mbox{tr}\left(
{\bf R}(g)+{\bf R}^{*}(g)
\right)\right].
\end{eqnarray}
The partition function is expanded 
with respect to orthogonal bases over 
$\frac{\lambda}{N^2_c}\le
\widetilde{\lambda}_{(I)\mbox{maximum}}$ 
and it reads
\begin{eqnarray}
Z_{(I)}(\lambda)&=&
\int^{\pi}_{-\pi} d\mu(g)
\exp\left[N_c\widetilde{\lambda}
\mbox{tr}\left(
{\bf R}(g)+{\bf R}^{*}(g)
\right)\right],\nonumber\\
&=&\int^{\pi}_{-\pi} d\mu(g)
\sum^{\infty}_{i=1}\sum^{\infty}_{j=1}
\frac{1}{i!} \left(N_c\widetilde{\lambda}\mbox{tr}{\bf R}(g)
\right)^i
\frac{1}{j!}
\left(N_c\widetilde{\lambda}\mbox{tr}{\bf R}^{\dagger}(g)
\right)^j.
 \label{ExpandSmall1}
\end{eqnarray}
Using the orthogonal relations over the $U(N_c)$ bases,
the partition function is reduced to\cite{GrossWitten1980a}
\begin{eqnarray}
Z_{(I)}(\lambda)&=&
\int^{\pi}_{-\pi} d\mu(g)
\sum^{\infty}_{i=1}
\left(\frac{1}{i!}\right)^2 
\left(N_c\widetilde{\lambda}\right)^{2i}
\left(\mbox{tr}{\bf R}(g)
\right)^i
\left(\mbox{tr}{\bf R}^{\dagger}(g)
\right)^i,
\nonumber\\
&=&\sum^{\infty}_{i=1} \frac{1}{i!}
\left(N^2_c\widetilde{\lambda}^2\right)^i,
\nonumber\\
&=&e^{\left(N^2_c\widetilde{\lambda}^2\right)},
\nonumber\\
&=&e^{\left(\frac{\lambda^2}{N^2_c}\right)}
~~~\left(
\mbox{over the range}~~
\widetilde{\lambda}\le
\widetilde{\lambda}_{\mbox{critical}}
\le\widetilde{\lambda}_{(I)\mbox{maximum}}
\right),
\label{ExpandSmall2}
\end{eqnarray}
where 
$\widetilde{\lambda}_{\mbox{critical}}=\frac{1}{2}$
and
$\widetilde{\lambda}_{(I)\mbox{maximum}}<1$.
%%%%%
%%%%%
%%%%%  NEW_Add
%%%%%
%%%%%

%%%%%%%%%%%%%%%%%%%%%%%%%%%%%%%%%%%%
The canonical ensemble with Maxwell-Boltzmann 
statistics single-particle partition function reads
\begin{eqnarray}
Z_{(I)}(\beta)&=&
\exp\left[
\frac{1}{N^2_c} (2j+1)^2
\left(\int\frac{V_{S^3}d^3 p}{(2\pi)^3}e^{-\beta E}
\right)^2
\right],
\nonumber\\
&=&
\exp\left[
\frac{ (2j+1)^2 V^2_{S^3} }{\pi^4 N^2_c \beta^6}
\right],
\nonumber\\
&\rightarrow&
\exp\left[
\frac{ (2j+1)^2 V^2_{S^3} T^6 }{\pi^4 N^2_c}
\right] ?
\end{eqnarray}
for an ideal gas of fundamental particles
confined in a spherical cavity in the three dimensional space 
where $V_{S^3}$ the cavity volume in the three dimensional space.
The free energy approaches the same value
as in the free theory for low temperature, 
whereas we have a mesonic-like and baryonic-like matter 
in the dilute hadronic phase.
The canonical ensemble 
in this range does not produce an explosive
phase transition.
The physics for the two phases will be discussed in rather details
when we present the micro-canonical ensemble 
as a density of states in section~\ref{sect_micro_canonical}.

%%%%%%%%%%%%%%%%%%%%%%%%%%%%%%%%%%%%%%%%%%
%%%%%%%%%%%%%%%%%%%%%%%%%%%%%%%%%%%%%%%%%%
%% Figures                             %%%
%% Figure: westen_fun.eps              %%%
%%  fig_w_fun                          %%%
%% fig_w_fun_smallb                    %%%
%% fig_w_fun_match                     %%%
%% Figure: westen_fun_smallb.eps       %%%
%% Figure: westen_fun_smallc.eps       %%%
%%%%%%%%%%%%%%%%%%%%%%%%%%%%%%%%%%%%%%%%%%
%%%%%%%%%%%%%%%%%%%%%%%%%%%%%%%%%%%%%%%%%%

\begin{figure}
\includegraphics{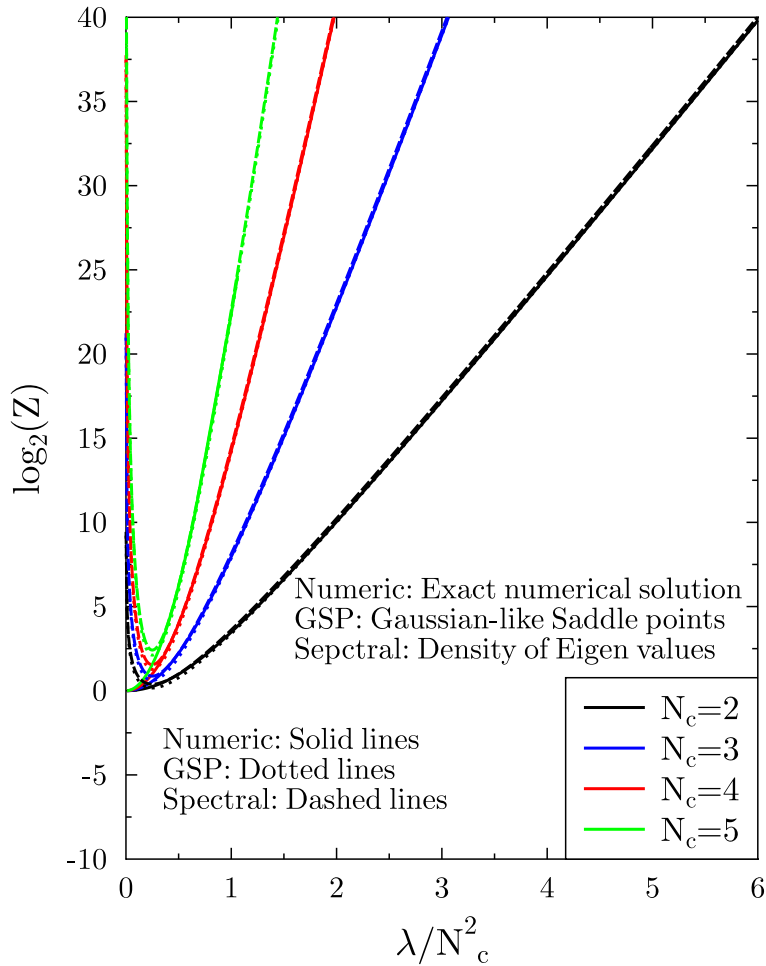}%
\caption{\label{fig_w_fun}
(Color online)
The density of states for the color-singlet fireballs
(i.e. Hagedorn states)
versus the thermal running parameter 
$\frac{\lambda}{N_c^2}$
with various color numbers $N_c$.
The high-lying solution is calculated 
for a bag consisting of fundamental
particles with a $\frac{\lambda}{N_c^2}$ 
runs over the range 
$\frac{\lambda}{N^2_c}\ge \frac{1}{2}$.
The results for the high-lying solution 
which is calculated
using the method of 
the spectral density of color eigenvalues
introduced by Brezin
{\em et. el.}~\cite{Brezin1978a}
in the limit of large color number $N_c$ 
and the results for 
the high-lying solution calculated using the Gaussian-like 
saddle points method are displayed together 
with the exact numerical solution. 
The high-lying solutions for the both methods 
and their extrapolations to $\frac{\lambda}{N_c^2}\le
\left(\frac{\lambda}{N_c^2}\right)_{\mbox{critical}}$
are compared with the exact numerical solution.}
\end{figure}

\begin{figure}
\includegraphics{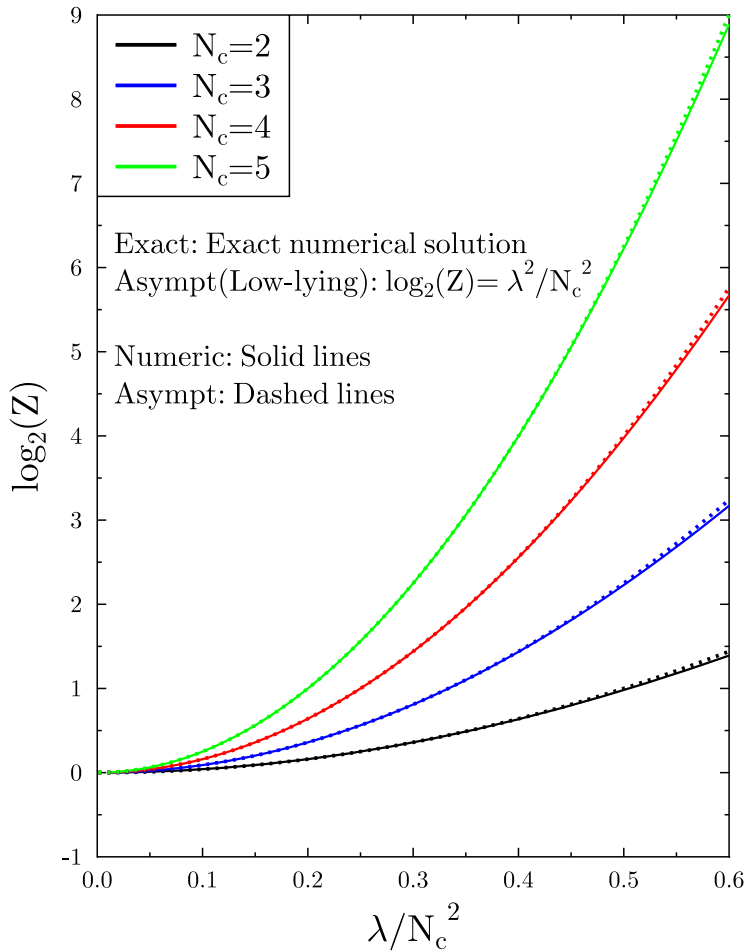}%
\caption{\label{fig_w_fun_smallb}
(Color online)
The density of states for the color-singlet fireballs
(i.e. Hagedorn states)
versus the thermal running parameter 
$\frac{\lambda}{N^2_c}$
for various color numbers $N_c$.
The low-lying density of states
is calculated for bags consisting of fundamental particles 
and with a thermal running parameter 
$\frac{\lambda}{N^2_c}$ runs over the range 
$\frac{\lambda}{N^2_c}\le \frac{1}{2}$.
The results of the density of states derived 
by the orthogonal expansion
given by Eqs.(\ref{ExpandSmall1}) and (\ref{ExpandSmall2})
are displayed together 
with the exact density of states evaluated
numerically.}
\end{figure}

\begin{figure}
\includegraphics{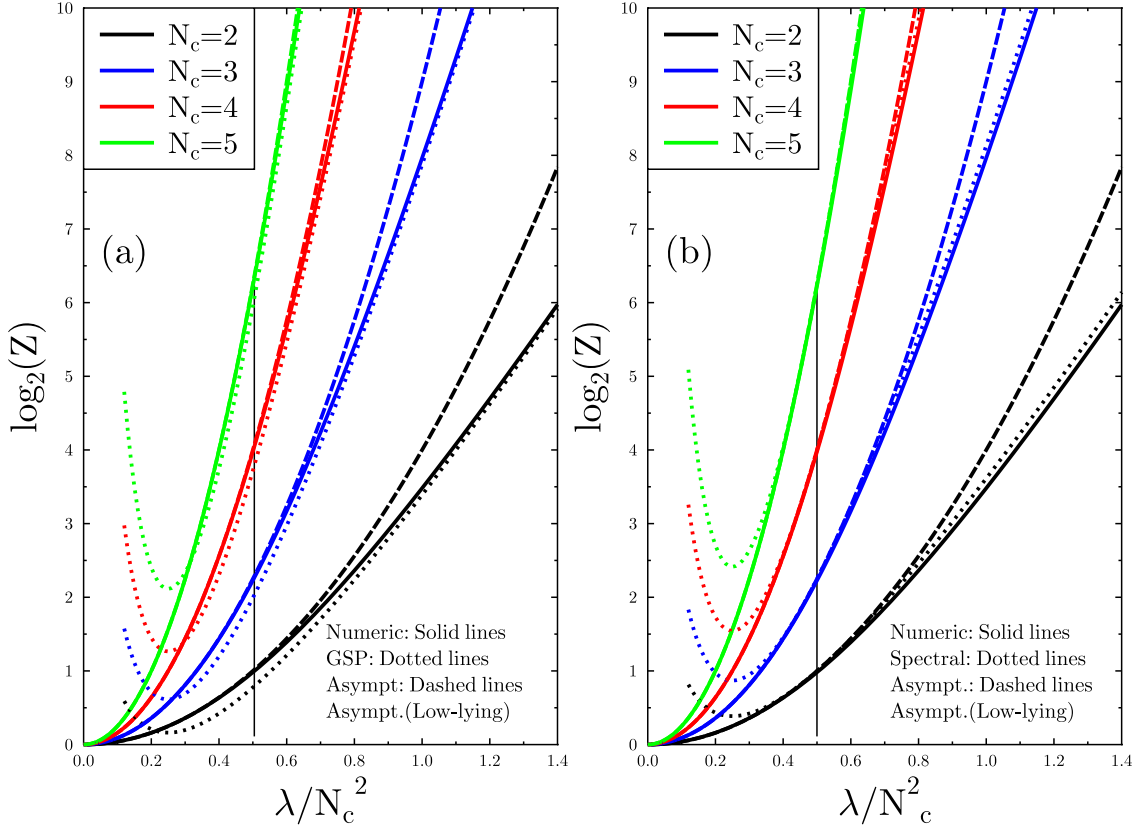}%
\caption{\label{fig_w_fun_match}
(Color online)
The match between the asymptotic
low-lying energy 
for $\frac{\lambda}{N^2_c}$ runs
over the range
$\frac{\lambda}{N^2_c}\le
\left(\frac{\lambda}{N^2_c}\right)_{\mbox{critical}}$
and the asymptotic high-lying energy solution
for $\frac{\lambda}{N^2_c}$ runs over the range 
$\frac{\lambda}{N^2_c}\ge
\left(\frac{\lambda}{N^2_c}\right)_{\mbox{critical}}$
for a color-singlet bag consisting of 
only fundamental particles.
The asymptotic low-lying solution is calculated
using the $\lambda$-power expansion
and the group orthogonality over
the entire color range $|\theta|\le\pi$.
a) The asymptotic high-lying energy solution 
is computed using the Gaussian-like
saddle points method. 
b) The asymptotic high-lying energy solution 
is computed using the spectral density 
of color eigenvalues method~\cite{Brezin1978a}.}
\end{figure}

We display the function 
$\ln Z(\lambda)$ versus $\frac{\lambda}{N^2_c}$ 
for the gas of fundamental particles 
in the color-singlet state
with various numbers of colors $N_c$
in Figs.(\ref{fig_w_fun}) and (\ref{fig_w_fun_smallb}).
The logarithm of the canonical ensemble
in the domain $\frac{\lambda}{N^2_c}\ge \frac{1}{2}$
is displayed in Fig.(\ref{fig_w_fun}).
The highly thermal excited partition parameter
$\frac{\lambda}{N^2_c}\ge\frac{\lambda_0}{N^2_c}$ 
corresponds the weak 't Hooft coupling 
and it describes a hot gas.
The partition function 
in the large $\widetilde{\lambda}$ regime
is evaluated exactly by
the exact numerical integration over the Fourier 
colors over the entire circle range $(-\pi\le\theta_i\le \pi)$.
The results for the numerical integration
are compared with the solution 
for the spectral density of eigenvalues 
method which is introduced 
by Brezin {\em et. el.}~\cite{Brezin1978a}
as well as the solution of the Gaussian-like saddle points method.
The two methods are found producing the same results
and they fit the exact numerical calculations when
the thermal partition parameter exceeds
the critical value $\frac{\lambda}{N^2_c}\ge \frac{1}{2}$.
In the regime below the critical point
${\widetilde{\lambda}}_0=\frac{\lambda_0}{N^2_c}$,
the two methods deviate from the exact numerical one
and this deviation increases as $\frac{\lambda}{N^2_c}$
decreases down below the critical thermal running parameter
$\frac{\lambda_0}{N^2_c}$.
The largest discrepancy is found
for the small number of colors $N_c=2$.
Despite the spectral density 
of color eigenvalues method
is originally derived
for the large $N_c\rightarrow\infty$ limit,
the solution is found satisfactory
for a finite number of colors
even for a relatively small one $N_c=2$.
Furthermore, the Gaussian-like saddle points method
is found to produce precisely the asymptotic solution
for the large $N_c\rightarrow\infty$ limit
although it is originally derived
for a finite number of colors.
However, the extrapolations of both methods 
to $\frac{\lambda}{N^2_c}<\frac{\lambda_0}{N^2_c}$ 
evidently fail to reproduce 
the exact numerical results 
in the low-lying energy regime 
in particular when $\frac{\lambda}{N^2_c}$
becomes sufficiently small
$\frac{\lambda}{N^2_c}\ll \frac{1}{2}$.
The failure of the large energy extrapolation 
to reproduce
the correct asymptotic low-lying
energy solution indicates the existence of
a possible phase transition near the point
$\frac{\lambda_0}{N^2_c}=\frac{1}{2}$.
This deviation from the exact numerical solution 
increases significantly
as $\frac{\lambda}{N^2_c}$ decreases
below the critical one
$\frac{\lambda_0}{N^2_c}$.
The low-lying energy solution for 
the small thermal running parameter domain 
$\frac{\lambda}{N^2_c}\le \frac{1}{2}$ 
with various color numbers $N_c$=2, 3, 4 and 5
is displayed in Fig.(\ref{fig_w_fun_smallb}).
The exact numerical solution is compared
with the asymptotic analytical solution 
$\ln Z=\lambda^2/N^2_c$ 
for the low-lying energy solution
which is derived for the small thermal 
running parameter $\frac{\lambda}{N^2_c}$.
The asymptotic analytical low-lying energy 
solution matches the exact numerical one 
in the range
$\frac{\lambda}{N^2_c}\le \frac{1}{2}$.
This agreement between the low-lying energy solution
and the exact numerical one is found satisfactory
even for a small color number $N_c=2$.
The critical point
$\widetilde{\lambda}_{\mbox{critical}}=\lambda_{\mbox{critical}}/N^2_c$
for the phase transition from the asymptotic
low-lying energy solution for the range
$\frac{\lambda}{N^2_c}\le\frac{\lambda_0}{N^2_c}$
to the asymptotic high-lying energy solution for
the range $\frac{\lambda}{N^2_c}\ge\frac{\lambda_0}{N^2_c}$
is determined when the both asymptotic  solutions 
match each other 
from left and right sides, respectively.  
Roughly speaking, 
they split only with a small redundant constant
due to several approximations considered in the derivation. 
This additive constant vanishes smoothly 
as the rough approximations are eliminated 
and the solution approaches the exact one. 
The solutions for the asymptotic small- and 
large-$\frac{\lambda}{N^2_c}$ energy domains are displayed 
in Fig.(\ref{fig_w_fun_match}). The both solutions 
are displayed with various color numbers $N_c$ 
and they are compared with the exact numerical one.
The phase transition is found a third order one.
This means that the asymptotic low-lying and large-lying energy 
solutions and their first and second derivatives are equal.
This implies that both of the solutions smoothly tangent
to each other and lie above each other. 
They look like that they do not really tend to be intersected 
but instead they look as they are approaching each other.
The critical point is the mid-point in the interpolation
between the low-lying and high-lying $\frac{\lambda}{N^2_c}$ 
energy solutions.
The matching of the asymptotic large $\frac{\lambda}{N^2_c}$
solution that is calculated using the Gaussian-like saddle points
method with the asymptotic small $\widetilde{\lambda}$ solution
is displayed in Fig.(\ref{fig_w_fun_match}.a).
It is found that at the critical point, 
the low-lying and high-lying energy solutions
are split to the minimum with a small redundant additive constant.
This redundant constant emerges 
due to consequent approximations adopted
in the analytical solutions with a finite $N_c$.
Although the two solutions deviate from each other 
by a small redundant constant, evidently, 
it is always possible to find a smooth interpolation 
between the two solutions. 
Since the order of phase transition is a third order 
then the tangent slope and its derivative are interpolated smoothly.
The exact numerical solution is shown to match the
low-lying energy solution and then
when $\frac{\lambda}{N^2_c}$ exceeds the critical point,
the exact numerical solution deviates 
the asymptotic low-lying energy solution. 
The exact numerical solution continues to match 
the asymptotic high-lying energy solution instead 
of the low-lying energy solution.
This tiny redundant constant should not worry us 
at all and the location of the critical point is realized 
in the midway of the interpolation between 
the two asymptotic solutions
and this constant should be vanished 
in order to preserve the continuation.
In this case, the interpolation between 
the two solutions is smooth and has a soft 
deflection characteristic at the critical point. 
On the other hand, the high-lying energy solution 
that is obtained by 
the spectral density of eigenvalues solution 
for the range 
$\frac{\lambda}{N^2_c}\ge\frac{\lambda_0}{N^2_c}$
is displayed in Fig.(\ref{fig_w_fun_match}.b).
It is shown that the low- and high-lying energy 
solutions smoothly tangent
to each other and located adjacent to each other
for a finite range where the Gross-Witten point
$\frac{\lambda_0}{N^2_c}=\frac{1}{2}$
is located in the midway.
The tangent slope and its derivative for the both 
left- and right- solutions are equal. 
Hence, the both low-lying and high-lying energy solutions
are scaled to be equal at the Gross-Witten point.

%
%
%
%%%%%%%%%%%%%%%%%%%%%%%%%%%%%%%%%%%%%%%%%%%%%%%%%%%%%%%%%%%%%%%%%%%%
%%see notation:
%%%%%%%%%%%%%%%%%%%%%%
%Exact: (exact) exact numerical solution
%GISp: (Gaussian) Gaussian-like Saddle points
%DoEV: (Spectral) Density of Eigen value
%%%%%%%%%%%%%%%%%%%%%
%%%%%%%%%%%%%%%%%%%%%%%%%%%%%%%%%%%%%%%%%%%%%%%%%%%%%%%%%%%%%%%%%%%
%%%%%%%%%%%%%%%%%%%%%%%%%%%%%%%

%
%
\subsection{The phase transition critical point}

The critical point for the phase transition is the solution
of the spectral density of eigenvalues method 
and it comes naturally
when the integration over the Fourier color variable
does not complete the entire circle range
but rather covers an incomplete circle range 
$|\theta_i|\le\theta_c$ where $|\theta_c|<\pi$. 
The spectral density method produces a solution
for the small thermal running parameter $\frac{\lambda}{N^2_c}\le\frac{1}{2}$
and another solution for the large parameter $\frac{\lambda}{N^2_c}\ge\frac{1}{2}$.
The critical point is determined by matching the two solutions.
It is interesting to point out that
the small- and large-$\widetilde{\lambda}$ parameters correspond
the strong- and weak-couplings in Gross-Witten scenario
\cite{GrossWitten1980a}, respectively.

On the other hand, the Gaussian-like saddle points method produces
the high-lying energy solution only for the thermal running parameter 
that runs over the interval $\frac{\lambda}{N^2_c}\ge\frac{1}{2}$.
Hereinafter, this is labeled as the solution (II).
In solution (II), 
the Fourier color variables $\theta_i$ are assumed
to be dominated in a narrow domain.
However, the low-lying energy solution for 
$\frac{\lambda}{N^2_c}\le\frac{1}{2}$
is known to be distributed uniformly over 
the entire circle with respect to the color angle.
Under this condition, the partition function
is evaluated trivially by expanding the integral 
with respect to the group basis powers and using the orthogonality
and then finally re-summing the resultant terms.
Hereinafter, we label the low-lying energy solution 
as the solution (I).
The extrapolation of both solutions fails to reproduce
the asymptotic results far away outside their domains.
The point of the phase transition 
is determined by examining the continuity 
when both solutions match each other
along the axis of the thermal running parameter
$\widetilde{\lambda}\equiv\frac{\lambda}{N^2_c}$.
The thorough study of the solution (II)
sheds more information about
the location of the critical point for the phase transition.
The solution (II) is concave up for the
small parameter $\frac{\lambda}{N^2_c}$.
The extremum left-hand point 
for the solution (II), namely, 
\begin{eqnarray}
\left(\lim_{N_c\rightarrow \infty}
\frac{\ln Z_{(II)}(\widetilde{\lambda})}{N^2_c}\right)
&=&
-\frac{1}{2}\ln 2\widetilde{\lambda}
+2\widetilde{\lambda}-\frac{3}{4},
\widetilde{\lambda}\ge\widetilde{\lambda}_{(II)\mbox{minimum}}
\end{eqnarray}
is determined by
\begin{eqnarray}
\left.\frac{\partial}{\partial \widetilde{\lambda}}
\left(\lim_{N_c\rightarrow \infty}
\frac{\ln Z_{(II)}(\widetilde{\lambda})}{N^2_c}
\right)
\right|_{\widetilde{\lambda}_{(II)\mbox{extremum}}}
&=&0
\rightarrow \widetilde{\lambda}_{(II)\mbox{extremum}}
=\frac{1}{4}.
\end{eqnarray}
This point is the
bottom of the concave up solution (II):
\begin{eqnarray}
\left.\frac{\partial^2}{\partial \widetilde{\lambda}^2}
\left(\lim_{N_c\rightarrow \infty}
\frac{\ln Z_{(II)}(\widetilde{\lambda})}{N^2_c}
\right)
\right|_{\widetilde{\lambda}_{(II)\mbox{extremum}}}
&\rightarrow& +
~~\mbox{at}~\widetilde{\lambda}_{(II)\mbox{minimum}}
=\frac{1}{4}.
\end{eqnarray}
The left-hand extremum point is found the minimum 
of the solution and the physical solution 
$Z_{(II)}(\widetilde{\lambda})$ is found for
$\widetilde{\lambda}$ runs over the range 
$\infty\ge\widetilde{\lambda}\ge
\widetilde{\lambda}_{(II)\mbox{minimum}}$.
This means that the solution (II) is 
the satisfactory solution over the range 
$\widetilde{\lambda}>\widetilde{\lambda}_{(II)\mbox{minimum}}$.
The critical point 
$\widetilde{\lambda}_{\mbox{critical}}$
is located somewhere above
$\widetilde{\lambda}_{(II)\mbox{minimum}}$,
\begin{eqnarray}
\widetilde{\lambda}\ge
\widetilde{\lambda}_{\mbox{critical}}\ge
\widetilde{\lambda}_{(II)\mbox{minimum}},
\mbox{where}~~ 
\widetilde{\lambda}_{(II)\mbox{minimum}}=\frac{1}{4}.
\end{eqnarray}
Therefore, when the low-lying energy solution 
$Z(\widetilde{\lambda})\approx 
Z_{(I)}(\widetilde{\lambda})$ over the range
$\widetilde{\lambda}<\widetilde{\lambda}_{\mbox{critical}}$
is not known, the Gaussian-like saddle points method still 
has the capability to shed much information 
about the critical point location and 
the continuous high-lying energy production threshold.

In order to determine the location of the critical 
point precisely,
we need the solution (I) for the energy domain 
$\widetilde{\lambda}<
\widetilde{\lambda}_{(I)_{\mbox{maximum}}}=1$.
The critical point is determined whenever
the solution (II) from above and the solution (I) 
from below match each other somewhere in 
the interval that is bounded by the interval
\begin{eqnarray}
\widetilde{\lambda}_{(II)\mbox{maximum}}\ge
\widetilde{\lambda}_{\mbox{critical}}\ge
\widetilde{\lambda}_{(I)\mbox{minimum}}.
\end{eqnarray}
In our case, the solution (I) reads
\begin{eqnarray}
\frac{\ln Z_{(I)}(\widetilde{\lambda})}{N^2_c}
=\widetilde{\lambda}^2 ~~\mbox{for}~~
\widetilde{\lambda}<\widetilde{\lambda}_{(I)\mbox{maximum}}.
\end{eqnarray}
Hence, the critical point for the phase transition
between the two solutions is computed by matching 
the extrapolation of the left- and right-hand solutions 
as follows
\begin{eqnarray}
\frac{\ln Z_{(I)}(\widetilde{\lambda})}{N^2_c}=
\frac{\ln Z_{(II)}(\widetilde{\lambda})}{N^2_c}
~~~@~~
\widetilde{\lambda}=
\widetilde{\lambda}_{\mbox{critical}}.
\end{eqnarray}
It is found that 
$\widetilde{\lambda}_{\mbox{critical}}=\frac{1}{2}$.

%%%%%%%%%%%%%%%%%%%%%%%%%%%%%%%%%%%%%%%%%%%%%%%%%%%%%%%%%%%%
%%%%%%%%%%%%%%%%%%%%%%%%%%%%%%%%%%%%%%%%%%%%%%%%%%%%%%%%%%%%
\subsection{Saddle points approximation
with the Vandermonde potential}
%%%%%%%%%%%%%%%%%%%%%%%%%%%%%%%%%%%%%%%%%%%%%%%%%%%%%%%%%%%%%%%%
%%%%%%%%%%%%%%%%%%%%%%%%%%%%%%%%%%%%%%%%%%%%%%%%%%%%%%%%%%%%%%%%
In order to understand the validity of Gaussian-like
approximation, it is essential to show that the color saddle
points are distributed in a narrow domain ($|\theta_c|\ll \pi$)
as the parameter $\widetilde{\lambda}$ runs over the range
$\widetilde{\lambda}\ge \widetilde{\lambda}_{\mbox{critical}}$.
The canonical partition function for the fundamental particles
obeying $U(N_c)$ can be written as follows,
\begin{eqnarray}
Z\left(\widetilde{\lambda}\right)&=&
\frac{1}{N_c!} \prod^{N_c}_k \int \frac{d\theta_k}{2\pi}
\exp\left[
\left.\frac{1}{2}\sum^{N_c}_{n=1}\sum^{N_c}_{m=1}
\ln \sin^2\left(
\frac{\theta_n-\theta_m}{2}
\right)\right|_{n\neq m}
+2 N_c \widetilde{\lambda} \sum^{N_c}_{n=1} \cos\theta_n
\right].
\end{eqnarray}
The above partition function is extremely difficult
to be evaluated exactly.
Fortunately, it can be integrated over the color
variables using the saddle points approximation.
The exponent under the integral can be written as follows,
\begin{eqnarray}
\ln Z\left(\widetilde{\lambda},\{\theta\}\right)
&=&\left[
\frac{1}{2}\sum^{N_c}_{n=1}\sum^{N_c}_{m=1}
\left.\ln \sin^2\left(
\frac{\theta_n-\theta_m}{2}
\right)\right|_{n\neq m}
+ 2 N_c \widetilde{\lambda} \sum^{N_c}_{n=1} 
\cos\theta_n
\right]+\mbox{constant}.
\end{eqnarray}
In order to avoid the singularity in the exponent,
the color saddle points are distributed uniformly
over the complete circle and do not approach each other
$(\overline{\theta}_i\neq\overline{\theta}_m)$.
However, when these points 
approach each other in the limit
$\overline{\theta}_i
\rightarrow \overline{\theta}_m$
the exponent blows up and diverges
and subsequently the saddle points integral approximation
is badly broken down and another analytic solution 
must emerge beyond this critical point.
%%%%%%%%%%%%%%%%%%%%%%%%%%%%%%%%%%%%%%%%%%%%%%%%%%%%%%%%%%%
The color saddle points are evaluated 
using the stationary conditions
\begin{eqnarray}
\left.\frac{\partial}{\partial \theta_i}
\ln Z\left(\widetilde{\lambda},\{\theta\}
\right)\right|_{\theta_i=\overline{\theta}_i}
&=&0, ~~~ i=1,2,\cdots N_c ~(\mbox{or} N_c-1),
\end{eqnarray}
for $U(N_c)$ or $SU(N_c)$.
Hence the saddle points are determined 
by the set of equations
\begin{eqnarray}
2N_c\widetilde{\lambda}\sin\overline{\theta}_i
&=&
\sum^{N_c}_{n\neq i}
\cot\left(\frac{\overline{\theta}_i
-\overline{\theta}_n}{2}\right),
~~\overline{\theta}_i\neq\overline{\theta}_m.
\end{eqnarray}
These points are satisfying 
the set of non-linear equations as follows
\begin{eqnarray}
2\widetilde{\lambda}\left(\frac{1}{N_c}\sum^{N_c}_{i=1}
\sin \overline{\theta_i}\right)
&=&
\frac{1}{N^2_c}\sum^{N_c}_{i}\sum^{N_c}_{n}
\cot\left(\frac{\overline{\theta}_i
-\overline{\theta}_n}{2}\right)_{i\neq n}.
\end{eqnarray}
This non-linear set is the root of
the spectral density 
that is derived using the spectral 
density of eigenvalues method 
in the large $N_c$ limit.
%
%%%
%%%%
%%%% important to emphasize this paragraph..........
%%%% ...............................................
%%%% ...............................................
The Gaussian-like integral around the saddle points
is approximated as follows
\begin{eqnarray}
Z\left(\widetilde{\lambda}\right)
&\sim&\prod_i\int^{\pi}_{-\pi}
\frac{d\theta_i}{2\pi}
\exp\ln Z\left(\widetilde{\lambda},\{\theta\}\right),
\nonumber\\
&\approx&
Z\left(\widetilde{\lambda},\{\overline{\theta}\}\right)
\prod_i\int^{\infty}_{-\infty}
\frac{d\theta_i}{2\pi} e^{-\sum_{ij}\frac{1}{2}
\Delta\left(\overline{\theta}_{i},\overline{\theta}_{j}\right)
\theta_i\theta_j},
\label{Gaussap1}
\end{eqnarray}
where the exponential elements are determined at the saddle
points $(\theta_i, i=1,2,\cdots,N_c)$ as follows
\begin{eqnarray}
\Delta\left(\overline{\theta}_{i},\overline{\theta}_{j}\right)
&=&-\left.\frac{\partial^2}{\partial \theta_i\partial\theta_j}
\ln Z\left(\widetilde{\lambda},\{\theta\}
\right)\right|_{\theta_i=\overline{\theta}_i}.
\end{eqnarray}
Eq.(\ref{Gaussap1}) is evaluated around the saddle points
using the Gaussian integral as follows
\begin{eqnarray}
Z(\widetilde{\lambda})
&\doteq& {\cal N}
\exp\left[
\ln Z(\widetilde{\lambda},\{\overline{\theta}\})
\right],
\end{eqnarray}
where the pre-exponential normalization satisfies 
the integral $\int\frac{d\theta_i}{2\pi}=1$ 
and it is determined by
\begin{eqnarray}
{\cal N}&=&
(2\pi)^{-N_c/2}
\left[\det
\Delta\left(
\overline{\theta}_{i},\overline{\theta}_{j}
\right)
\right]^{-1/2}\le 1.
\end{eqnarray}
However, it is not always possible 
to find real values 
for the saddle points distributed uniformly 
over the entire circle range
for the parameter $\widetilde{\lambda}$ 
that runs along the real axis.
These real saddle points cease 
to exist when the thermal running parameter 
$\widetilde{\lambda}$ reaches the critical value.
The Vandermonde potential characteristics 
is modified at the critical point. 
This point is the threshold 
for another solution with a different behavior.

%%%
%%%
%%%
%%%%%%%%%%%%%%%%%%%%%%%%%%%%%%%%%%%
%
% Small coupling constant
%
%%%%%%%%%%%%%%%%%%%%%%%%%%%%%%%%%%%

%%%%%%%%%%%%%%%%%%%%%%%%%%%%%%%%%%
%%%%%%%%%%%%%%%%%%%%%%%%%%%%%%%%%%
%%%%%%%%%%%%%%%%%%%%%%%%%%%%%%%%%%
%%%%%%%%%%%%%%%%%%%%%%%%%%%%%%%%%%
%%%%%%%%%%%%%%%%%%%%%%%%%%%%%%%%%%
%%%%%%%%%%%%%%%%%%%%%%%%%%%%%%%%%%

\section{Adjoint particles\label{sect_adj}}
The color-singlet bags consisting of a gas of fundamental
and anti-fundamental particles represent the mesonic
and baryonic states with no gluonic component.
On the other hand, the color-singlet bags consisting only of
the adjoint particles represent glueballs
states with no quark component.
The low-lying hadronic states are likely
mesons, baryons and glueballs.
The Hagedorn states with fundamental and adjoint 
constituent particles are essential to understand
the highly thermal excited hadronic states
near and just below 
the deconfinement phase transition diagram.

The canonical ensemble for the bags consisting 
only of the adjoint particles projected in the 
color-singlet state reads
\begin{eqnarray}
Z(\beta)&=&
\int d\mu(g)\exp
\left[
{\cal D}_{g}\int_V d^3 r\int\frac{d^3p}{(2\pi)^3}
{\cal G}_{\mbox{adj}}(\theta)
e^{-\beta E(p,r)}\right],
\end{eqnarray}
where ${\cal D}_g$ is the degeneracy 
such as the spin multiplicity 
${\cal D}_g=(2j+1)$.
The internal color structure 
for the adjoint constituents in the Fock space 
is given by
\begin{eqnarray}
{\cal G}_{\mbox{adj}}(\theta)&=&
\frac{1}{\mbox{dim}_{g}}
\mbox{tr}\left[{\bf R}_{\mbox{adj}}(g)\right],
\end{eqnarray}
where $\mbox{dim}_{g}=N^2_c$ and $(N^2_c-1)$
for $U(N_c)$ and $SU(N_c)$, respectively.
The partition function for the gas of free particles
occupying the volume $V$ is approximated to
\begin{eqnarray}
\lambda_g=
{\cal D}_{g}\int_V d^3 r\int\frac{d^3p}{(2\pi)^3}
e^{-\beta E(p,r)},
\end{eqnarray}
where $E(p,r)$ is the energy for each constituent particle species.
In order to simplify our notation, 
we define the following thermal running parameter
\begin{eqnarray}
\tilde{\lambda}_g=\frac{\lambda_g}{\mbox{dim}_{g}}.
\end{eqnarray}
Hence, the canonical ensemble is simplified as follows
\begin{eqnarray}
Z(\lambda_g)&=&\int d\mu(g)
\exp\left[ \tilde{\lambda}_g
\mbox{tr}_c\left[
{\bf R}_{\mbox{adj}}(g)
\right]
\right].
\end{eqnarray}
The adjoint group representation reads
\begin{eqnarray}
\mbox{tr}\left[{\bf R}_{\mbox{adj}}(g^k)\right]&=&
\mbox{tr}\left[{\bf R}_{\mbox{fund}}(g^k)\right]
\mbox{tr}\left[{\bf R}^*_{\mbox{fund}}(g^k)\right]-1,
\nonumber\\
&=&(N_c-1)+
\sum^{N_c}_{n=1}\sum^{N_c}_{m=1}
\left.\cos k\left(\theta_n-\theta_m\right)\right|_{n\neq m},
~~\left(\theta_{N_c}=\sum^{N_c-1}_{i=1} \theta_i\right),
\end{eqnarray}
and
\begin{eqnarray}
\mbox{tr}\left[{\bf R}_{\mbox{adj}}(g^k)\right]&=&
\mbox{tr}\left[{\bf R}_{\mbox{fund}}(g^k)\right]
\mbox{tr}\left[{\bf R}^*_{\mbox{fund}}(g^k)\right],
\end{eqnarray}
for the groups $SU(N_c)$ and $U(N_c)$, respectively.

We use the Gaussian-like saddle points method
developed in the previous section
in order to calculate 
the asymptotic high-lying energy solution
for the canonical ensemble
with large thermal running parameter 
$\tilde{\lambda}_g\ge
\left(\tilde{\lambda}_{g}\right)_{\mbox{critical}}$.
In this range, the Fourier color variables are assumed
to be dominant only in a narrow range.
The adjoint action near the saddle points
is approximated to
\begin{eqnarray}
\tilde{\lambda}_g
\mbox{tr}_c
\left[{\bf R}_{\mbox{adj}}(g)\right]&=&
\tilde{\lambda}_g\left\{ (N_c-1)
+\sum^{N_c}_{n=1}\sum^{N_c}_{m=1}
\left.\cos\left(\theta_n-\theta_m\right)\right|_{n\neq m}
\right\},\nonumber\\
&\approx&
\tilde{\lambda}_g
\left[
(N^2_c-1)
-\frac{1}{2}\sum^{N_c}_{n=1}\sum^{N_c}_{m=1}
\left(\theta_n-\theta_m\right)^2
\right].
\end{eqnarray}
The above approximation reduces
the canonical ensemble to
\begin{eqnarray}
\left.Z(\lambda_g)\right|_{SU(N_c)}
&\approx&
\left.Z_{(II)}(\lambda_g)\right|_{SU(N_c)}
~\left(@~~ 
\tilde\lambda_g ~\mbox{runs over the range}~
\tilde\lambda_g\ge
\tilde{\lambda}_{g(II)\mbox{minimum}}\right),
\nonumber\\
\left.Z_{(II)}(\lambda_g)\right|_{SU(N_c)}
&=&
e^{(N^2_c-1)\tilde{\lambda}_g}
\frac{1}{N!}
\int^{\pi}_{-\pi}d\theta_{N_c}
\delta\left(\sum^{N_c}_{i=1}\theta_i\right)
\left(\prod^{N_c-1}_k \int^{\infty}_{-\infty}
\frac{d\theta_k}{2\pi}\right)
\prod^{N_c}_{n>m}\left(\theta_n-\theta_m\right)^2
\nonumber\\
&\times&
\exp\left[
-\frac{\tilde{\lambda}_g}{2}
\sum^{N_c}_{n=1}\sum^{N_c}_{m=1}\left(\theta_n-\theta_m\right)^2
\right],
\nonumber\\
&\approx&
\left(\frac{1}{2N_c\tilde{\lambda}_g}\right)^{\frac{N^2_c-1}{2}}
e^{(N^2_c-1)\tilde{\lambda}_g}
\left[
\frac{1}{N_c!(2\pi)^{\frac{N_c}{2}-1}}\prod^{N_c}_{n=1}n!
\frac{1}{\sqrt{2\pi N_c}}
\right],
\end{eqnarray}
for $SU(N_c)$ group representation 
where the following relation has been adopted
\begin{eqnarray}
I_i&=&\left(\prod^{N_c}_k \int^{\infty}_{-\infty}
\frac{d\theta_k}{2\pi}\right)
2\pi\delta\left(\sum^{N_c}_{i=1}\theta_i\right)
\left[
\prod^{N_c}_{n\neq m}\left(\theta_n-\theta_m\right)^2
\right]^{\frac{1}{2}}
e^{\left[
-\frac{\tilde{\lambda}_g}{2}
\sum^{N_c}_{n=1}\sum^{N_c}_{m=1}\left(\theta_n-\theta_m\right)^2
\right]},\nonumber\\
&\equiv&\left(\prod^{N_c}_k \int^{\infty}_{-\infty}
\frac{d\theta_k}{2\pi}\right)
2\pi\delta\left(\sum^{N_c}_{i=1}\theta_i\right)
\left[
\prod^{N_c}_{n>m}\left(\theta_n-\theta_m\right)^2
\right]
e^{\left[
-\frac{2N_c\tilde{\lambda}_g}{2}
\sum^{N_c}_{n=1} \theta_n^2
\right]}.
\end{eqnarray}
The same procedure can be followed for $U(N_c)$ symmetry.
The canonical ensemble is reduced to
\begin{eqnarray}
\left.Z_{(II)}(\lambda_g)\right|_{U(N_c)}
&\approx&
\left(\frac{1}{2N_c\tilde{\lambda}_g}\right)^{\frac{N^2_c-1}{2}}
e^{N^2_c\tilde{\lambda}_g}
\left[
\frac{\sqrt{N_c}}{N_c!(2\pi)^{\frac{N_c-1}{2}}}\prod^{N_c}_{n=1}n!
\right]\int^{\infty}_{-\infty}\frac{d\theta_{N_c}}{2\pi},
\nonumber\\
&\approx&
e^{N^2_c\tilde{\lambda}_g}
\left(\frac{1}{2N_c\tilde{\lambda}_g}\right)^{\frac{N^2_c-1}{2}}
\left[
\frac{\prod^{N_c}_{n=1}n!}{N_c!(2\pi)^{\frac{N_c}{2}}}
\right]\sqrt{2\pi N_c}\int^{\pi}_{-\pi}
\frac{d{\theta}_{N_c}}{2\pi},
\nonumber\\
&~& \left(
@~~ 
\tilde\lambda_g 
~~\mbox{runs over the range}~~
\tilde\lambda_g\ge
\tilde{\lambda}_{g(II)\mbox{minimum}}\right).
\end{eqnarray}
Using the Stirling's approximation 
as done in Eq.(\ref{stirlingfund}) for the large $N_c$ limit,
the partition function becomes,
\begin{eqnarray}
\left.Z_{(II)}(\lambda_g)\right|_{U(N_c)}
&\approx&
\left(\frac{1}{2\tilde{\lambda}_g}\right)^{\frac{N^2_c-1}{2}}
e^{\left(N^2_c\tilde{\lambda}_g-\frac{3}{4}N^2_c
\right)},
\nonumber\\
&\rightarrow&
\left(\frac{1}{2\tilde{\lambda}_g}\right)^{\frac{N^2_c}{2}}
e^{\left(N^2_c\tilde{\lambda}_g-\frac{3}{4}N^2_c
+N^2_c C_{\mbox{adj}}
\right)},
\nonumber\\
&~& \left(
@~~ \tilde\lambda_g 
~~\mbox{runs over the range}~~
\tilde\lambda_g\ge
\tilde{\lambda}_{g(II)\mbox{minimum}}\right),
\end{eqnarray}
for the case of $U(N_c)$ representation.
The additional constant $C_{\mbox{adj}}$ insures
that the term 
$\ln Z_{(II)}(\lambda_g)/N^2_c$
is non-negative quantity and it should satisfy
the boundary near the critical point 
and furthermore it throws away any redundant constant.
This additional constant stems from
the normalization transformation
in the Gaussian-like integration
$\int^{\infty}_{-\infty}\frac{d\theta}{2\pi} e^{0}
\rightarrow
\int^{\pi}_{-\pi}\frac{d\theta}{2\pi}=1$.
Hence, the solution for the large parameter
$\tilde{\lambda}_g>\tilde{\lambda}_{g(II)\mbox{minimum}}$ 
reads
\begin{eqnarray}
\left.\left(
\lim_{N^2_c\rightarrow \mbox{large}}
\frac{\ln Z_{(II)}(\lambda_g)}{N^2_c}
\right)\right|_{U(N_c)}
&=&
-\frac{1}{2}\ln2\tilde{\lambda}_g
+\tilde{\lambda}_g-\frac{3}{4}+C_{\mbox{adj}},
\nonumber\\
&\ge&0,
\end{eqnarray}
in the large $N_c$-limit.
This function is concave up.
The value $\tilde{\lambda}_{g(II)\mbox{minimum}}$
is calculated as follows
\begin{eqnarray}
\left(\lim_{N^2_c\rightarrow \mbox{large}}
\frac{\partial}{\partial \lambda_g}
\frac{\ln Z_{(II)}(\lambda_g)}{N^2_c}
\right)_{\tilde{\lambda}_{g(II)\mbox{minimum}}}
&=&0
~~\rightarrow  
~~\tilde{\lambda}_{g(II)\mbox{minimum}}=\frac{1}{2},
\nonumber\\
\left(\lim_{N^2_c\rightarrow \mbox{large}}
\frac{\partial^2}{\partial \lambda_g^2}
\frac{\ln Z_{(II)}(\lambda_g)}{N^2_c}
\right)_{\tilde{\lambda}_{g(II)\mbox{minimum}}}
&\rightarrow& (+) ~~\rightarrow~~ \mbox{concave up}.
\end{eqnarray}
This means that our solution is valid only 
when $\tilde{\lambda}_{g}$ runs over the range
$\tilde{\lambda}_{g}\ge \tilde{\lambda}_{g(II)\mbox{minimum}}$.
The value $C_{\mbox{adj}}$ is determined as follows
\begin{eqnarray}
\left(\lim_{N^2_c\rightarrow \mbox{large}}
\frac{\ln Z_{(II)}(\lambda_g)}{N^2_c}\right)
&\ge&0,
\end{eqnarray}
for the point $\tilde{\lambda}_{g(II)\mbox{minimum}}$
in order to ensure that our solution is non negative in
the entire $\tilde{\lambda}$ range 
and also in order to preserve 
the analytic continuation of the solution. 
This leads to $C_{\mbox{adj}}=\frac{1}{4}$.
Hence, the canonical partition in the range
$\tilde{\lambda}_{g}\ge \tilde{\lambda}_{g(II)\mbox{minimum}}$
in the limit large $N_c$ limit reads
\begin{eqnarray}
Z_{(II)}(\lambda_g)
&\equiv&
\left(\frac{1}{2\tilde{\lambda}_g}\right)^{\frac{N^2_c}{2}}
e^{\left(N^2_c\tilde{\lambda}_g-\frac{N^2_c}{2}
\right)}.
\end{eqnarray}
%%%%%%%%%%%%%%
%

It will be shown below (e.g. in Eq.(\ref{smalladj1}))
that for 
$\tilde{\lambda}_g\le
\tilde{\lambda}_{g(I)\mbox{maximum}}$
where 
$\tilde{\lambda}_{g(I)\mbox{maximum}}<1$,
the canonical ensemble becomes
\begin{eqnarray}
\lim_{N^2_c\rightarrow\mbox{large}}
\ln Z(\lambda_g)&=&
\lim_{N^2_c\rightarrow\mbox{large}}
\ln Z_{(I)}(\lambda_g), 
\nonumber\\
&=&
-\frac{1}{N^2_c}\ln\left(1-\tilde{\lambda}_g\right),
\nonumber\\
&\approx&0,
\nonumber\\
&~&
\left(
@~~ \tilde{\lambda}_g~ \mbox{runs over the range}
\tilde{\lambda}_g\le 
\tilde{\lambda}_{g(I)\mbox{maximum}}
\right).
\end{eqnarray}
We are in the position to compute 
the Gross-Witten critical point
\begin{eqnarray}
\tilde{\lambda}_{g(I)\mbox{maximum}}\ge
\tilde{\lambda}_{g\mbox{critical}}
\ge
\tilde{\lambda}_{g(II)\mbox{minimum}}.
\end{eqnarray}
The critical point for the phase transition
is determined as follows
\begin{eqnarray}
\lim_{N^2_c\rightarrow \mbox{large}}
\frac{\ln Z_{(I)}(\lambda)}{N^2_c}
=
\lim_{N^2_c\rightarrow \mbox{large}}
\frac{\ln Z_{(II)}(\lambda)}{N^2_c}
~~~\left(
@~~
\mbox{the point}~
\tilde{\lambda}_g=
\tilde{\lambda}_{g\mbox{critical}}
\right).
\end{eqnarray}
It is found that the critical point is located at
$\tilde{\lambda}_{g\mbox{critical}}=\frac{1}{2}$.
This procedure is very useful to determine the point
for the phase transition for a finite number 
of colors $N_c$ and it will be more appropriate 
in the case of more complicated 
physical situations are involved.
Furthermore, it sounds that the Gaussian-like
saddle points method is a straightforward and easier 
than the spectral density method.

The solution of the canonical ensemble for the adjoint particles
in the range 
$\tilde{\lambda}_g\le\tilde{\lambda}_{g(I)\mbox{maximum}}$
is nontrivial because of the color structure for
$SU(N_c)$ and $U(N_c)$ group representations. 
Fortunately, the adjoint color structure
can be simplified to fundamental-like structure 
by introducing the Lagrange multiplier. 
The inclusion of the Lagrange multiplier trick 
reduces the partition function with $U(N_c)$ 
group structure to
\begin{eqnarray}
Z(\lambda_g)&=&\int d\mu(g) \exp\left(
\lambda_g \frac{1}{\mbox{dim}_g}\mbox{tr} {\bf R}_{\mbox{adj}}(g)
\right),\nonumber\\
&=&\int d\mu(g) \exp
\left(
\tilde{\lambda}_g\mbox{tr}{\bf R}_{\mbox{fun}}(g)
\mbox{tr}{\bf R}^*_{\mbox{fun}}(g)
\right),
\nonumber\\
&=&
\frac{N^2_c}{2\tilde{\lambda}_g}
\int^{\infty}_0 d\xi \xi e^{-\frac{N_c \xi^2}{4\tilde{\lambda}_g}}
\int d\mu(g) \exp\left[
\frac{N_c \xi}{2}
\left(
\mbox{tr} {\bf R}_{\mbox{fun}}(g)
+
\mbox{tr} {\bf R}^*_{\mbox{fun}}(g)
\right)\right].
\end{eqnarray}
The above equation is decomposed into two parts
\begin{eqnarray}
Z(\lambda_g)=Z_1(\lambda_g)+Z_2(\lambda_g),
\end{eqnarray}
where
\begin{eqnarray}
Z_1(\lambda_g)&=&
\frac{N^2_c}{2\tilde{\lambda}_g}
\int^{\xi_0}_0 d\xi \xi
e^{-\frac{N^2_c \xi^2}{4\tilde{\lambda}_g}}
e^{\frac{N^2_c\xi^2}{4}},
\label{eq1_lang}
\end{eqnarray}
and
\begin{eqnarray}
Z_2(\lambda_g)&=&
\frac{N^2_c}{2\tilde{\lambda}_g}
\int^{\infty}_{\xi_0} d\xi \xi e^{-\frac{N^2_c \xi^2}
{4\tilde{\lambda}_g}}
\left(\xi\right)^{-N^2_c/2}
e^{N^2_c\xi-3N^2_c/4}.
\label{eq2_lang}
\end{eqnarray}
The integration over the color-singlet state
is taken into account the splitting into small- 
and large-domains 
$0\le \xi\le \xi_0$ and $\xi_0\le\xi\le\infty$,
respectively, where $\xi_0=1$.
The integration over the range $(0\le\xi\le\xi_0)$,
reduces Eq.(\ref{eq1_lang}) to
\begin{eqnarray}
Z_1(\lambda_g)&=&
\frac{N^2_c}{4\tilde{\lambda}_g}
\int^{\xi_0}_0 d\xi^2
e^{
-\frac{N^2_c \xi^2}{4}\left(
\frac{1}{\tilde{\lambda}_g}-1\right)},
\end{eqnarray}
where $\tilde{\lambda}_g$ is assumed 
to run only over the range
$\tilde{\lambda}_g\le
\tilde{\lambda}_{g(I)\mbox{maximum}}<1$.
The extrapolation of the integration upper limit
$\xi_0\rightarrow\infty$ approximates 
the partition function to
\begin{eqnarray}
Z(\lambda_g)&=&Z_1(\lambda_g)+Z_2(\lambda_g),
\nonumber\\
&\approx&
\frac{N^2_c}{4\tilde{\lambda}_g}
\int^{\infty}_0 d\xi^2
e^{-\frac{ N^2_c\xi^2 (1-\tilde{\lambda}_g)}{4\tilde{\lambda}_g}},
\nonumber\\
&\approx&
\frac{N^2_c}{4\tilde{\lambda}_g}
\frac{4\tilde{\lambda}_g/N^2_c}
{\left(1-\tilde{\lambda}_g\right)}.
\label{smalladjz1}
\end{eqnarray}
This integration is trivial and is reduced to
\begin{eqnarray}
Z(\lambda_g)&\approx&Z_{1}(\lambda_g),\nonumber\\
Z_{1}(\lambda_g)&\approx&
\frac{1}{1-\tilde{\lambda}_g},
\nonumber\\
&=&\exp\left[-\ln(1-\tilde{\lambda}_g)
\right]
\sim \exp\left[\mbox{fun}(\tilde{\lambda}_g)\right].
\label{smalladj1}
\end{eqnarray}
%%%%%%%%%%5%%%%%%%%%%%%%%%%%%%%%%%%%%%%%%%%%%%%%%
The integration over $\xi_0\le\xi\le\infty$
where $\xi_0=1$
for the case 
$\tilde{\lambda}_g\le \frac{1}{2}
\le\tilde{\lambda}_{g(I)\mbox{maximum}}$
is approximated to
\begin{eqnarray}
Z_2(\lambda_g)&=&
\frac{N^2_c}{4\tilde{\lambda}_g}
\int^{\infty}_{\xi^2_0} d\xi^2
\left(\xi^2\right)^{-N^2_c/4}
e^{-\frac{N^2_c \xi^2}
{4\tilde{\lambda}_g}} e^{N^2_c\xi-\frac{3}{4}N^2_c},
\nonumber\\
&\le&
\frac{N^2_c}{4\tilde{\lambda}_g}
\int^{\infty}_{\xi^2_0} d\xi^2
e^{-\frac{N^2_c \xi^2}
{4\tilde{\lambda}_g}} e^{N^2_c\xi}
\le
\frac{N^2_c}{4\tilde{\lambda}_g}
\int^{\infty}_{x_0} dx
e^{-\frac{N^2(1-\lambda_g) x}{4\tilde{\lambda}_g}}.
\end{eqnarray}
This integral complements 
the integral given by Eq.(\ref{smalladjz1}) which runs
over the interval $0\le\xi\le\xi_0$.
In Eq.(\ref{smalladjz1}) when $\xi_0$
is extended and extrapolated to
$\infty$ in term $Z_1(\lambda_g)$, 
then the second term is suppressed.
%%%%%%%%%%%
%%%%%%%%%%%%%%%%%%%%%%%%%%%%%%%%%%%%%%%%%%%%%%%%%%%%%%%%%%%%%%%%%%%
%%%%%%%%%%%%%%%%%%%%%%%%%%%%%%%%%%%%%%%%%%%%%%%%%%%%%%%%%%%%%%%%%%%
%%%%%%%%%%%%%%%%%%%%%%%%%%%%%%%%%%%%%%%%%%%%%%%%%%%%%%%%%%%%%%%%%%%

On the other hand, the analytical solution behaves differently
in the energy domain 
$\tilde{\lambda}_g\ge 1
\ge\tilde{\lambda}_{g(II)\mbox{minimum}}$.
In this energy domain the integral over
$0\le\xi\le\xi_0$ is suppressed  and
$Z_1(\lambda_g)$ becomes negligible,
\begin{eqnarray}
Z_1(\lambda_g)\propto \frac{1}{\lambda_g}\sim 0.
\end{eqnarray}
The integration over the range
$\xi_0\le\xi\le\infty$ and $\tilde{\lambda}_g\ge 1$
approximates Eq.(\ref{eq2_lang}) to
\begin{eqnarray}
Z(\lambda_g)&\approx& Z_2(\lambda_g),
\nonumber\\
&=&\frac{N^2_c}{2\tilde{\lambda}_g}
\int^{\infty}_{\xi_0} d\xi \xi e^{-\frac{N^2_c \xi^2}
{4\tilde{\lambda}_g}}
\left(\xi\right)^{-N^2_c/2}
e^{N^2_c\xi-3N^2_c/4},
\nonumber\\
&=&\int^{\infty}_{\xi_0} d\xi e^{f(\xi)},
\end{eqnarray}
where
\begin{eqnarray}
f(\xi)\approx \left[-\frac{1}{4\tilde{\lambda}_g}\xi^2
-\frac{1}{2}\ln\xi+\xi-\frac{3}{4}\right].
\label{solextrm}
\end{eqnarray}
When the saddle point becomes $\xi_{\mbox{saddle}}\ge 1$,
the above equation is approximated to
\begin{eqnarray}
Z(\lambda_g)&\approx&
\int^{\infty}_{0} d\xi e^{f(\xi)}.
\end{eqnarray}
The saddle point can be found by the following solution 
\begin{eqnarray}
\left(\xi^2-2\tilde{\lambda}_g\xi+\tilde{\lambda}_g
\right)_{\xi=\xi_{\mbox{saddle}}}=0.
\label{constsol1}
\end{eqnarray}
The saddle point solution reads,
\begin{eqnarray}
\xi_{\mbox{saddle}}&=&
1+\sqrt{1-\frac{1}{\tilde{\lambda}_g}},
~~\left(
\mbox{with the constraint}~\tilde{\lambda}_g\ge 1
\right),
\nonumber\\
\xi_{\mbox{saddle}}&\approx& \tilde{\lambda}_g
\left[2-\frac{1}{2\tilde{\lambda}_g}\right],\nonumber\\
\xi_{\mbox{saddle}}&\approx& 2\tilde{\lambda}_g.
\end{eqnarray}
The saddle point can, alternatively, be derived from 
Eq.(\ref{constsol1}) as follows,
\begin{eqnarray}
&~&{\left(1-2\tilde{\lambda}_g\frac{1}{\xi}
+\tilde{\lambda}_g\frac{1}{\xi^2}
\right)}_{1/\xi=1/\xi_{\mbox{saddle}}}=0,
\nonumber\\
&~&~~\rightarrow~
1/\xi_{\mbox{saddle}}=
1-\sqrt{1-\frac{1}{\tilde{\lambda}_g}}
\approx\frac{1}{2\tilde{\lambda}_g}.
\end{eqnarray}
This result is analogous to Eq.(\ref{spectral-sin2}).
The above approximation softens
and extrapolates the constraint to
\begin{eqnarray}
\xi_{\mbox{saddle}}=2\tilde{\lambda}_g,
\left(\mbox{with constraint}~~
\tilde{\lambda}_g\ge \frac{1}{2}, ~~\mbox{and}~~
\xi_{\mbox{saddle}}\ge 1\right).
\label{constsol2}
\end{eqnarray}
Hence, by substituting the saddle point
into Eq.(\ref{solextrm}), we get
\begin{eqnarray}
f(\xi_{\mbox{saddle}})&=&
-\frac{1}{4\tilde{\lambda}_g}
\left[
\xi^2_{\mbox{saddle}}
-2\tilde{\lambda}_g\xi_{\mbox{saddle}}
+\tilde{\lambda}_g\right]
+\frac{\xi_{\mbox{saddle}}}{2}
-\frac{1}{2}\ln\xi_{\mbox{saddle}}-\frac{1}{2},
\nonumber\\
&=&
\frac{\xi_{\mbox{saddle}}}{2}
-\frac{1}{2}\ln\xi_{\mbox{saddle}}-\frac{1}{2},
\nonumber\\
&=&\frac{1}{2s^2_0}+\frac{1}{2}\ln s^2_0-\frac{1}{2}
=f(s^2_0),
\end{eqnarray}
where we have introduced $s^2_0=1/\xi_{\mbox{saddle}}$ 
for our convenience.
In terms of $s^2_0$, the partition function is reduced to
\begin{eqnarray}
Z_2\left(s^2_0\right)
&=&{\cal N}_{\xi}
\exp\left[N^2_c
\left(\frac{1}{2s^2_0}+\frac{1}{2}\ln s^2_0-\frac{1}{2}\right)
\right],
\nonumber\\
&\approx&
\exp\left[N^2_c
\left(\frac{1}{2s^2_0}+\frac{1}{2}\ln s^2_0-\frac{1}{2}
+{\cal O}\left(\frac{\ln N_c}{N^2_c}\right)
\right)
\right],
\end{eqnarray}
where
\begin{eqnarray}
{\cal N}_\xi&=&
\frac{N_c}{2\tilde{\lambda}_g}
\sqrt{\frac{2\pi}{-f''(\xi_0)}},
\nonumber\\
&=&\exp\left[N^2_c{\cal O}\left(
\frac{1}{N^2_c}\ln(N_c)
\right)\right].
\end{eqnarray}
With the approximation $s^2_0=\frac{1}{2\tilde{\lambda}_g}$ and
$\tilde{\lambda}_g\ge \frac{1}{2}$, 
the partition function reads
\begin{eqnarray}
Z(\tilde{\lambda}_g)=\exp\left[
N^2_c\left(\tilde{\lambda}_g-\frac{1}{2}\ln 2\tilde{\lambda}_g
-\frac{1}{2}
\right)\right].
\end{eqnarray}
This result is identical to the result 
Eq.(\ref{spectral-density-sol}) to be derived using
the spectral density method.
However, the saddle point derived from Eq.(\ref{constsol1})
can be approximated to
\begin{eqnarray} 
\xi_{\mbox{saddle}}=2\tilde{\lambda}_g-\frac{1}{2}.
\end{eqnarray}
This approximation in the large $N_c$ 
limit leads to
\begin{eqnarray}
Z\left(\tilde{\lambda}_{g}\right)&\approx&
\lim_{\tilde{\lambda}_g\rightarrow\mbox{Large}}
\exp\left[
N^2\left(\tilde{\lambda}_g
-\frac{1}{2}\ln\left( 2\tilde{\lambda}_g-\frac{1}{2}\right)
-\frac{3}{4}
\right)\right],\nonumber\\
&\approx&
\exp\left[
N^2\left(\tilde{\lambda}_g
-\frac{1}{2}\ln\left( 2\tilde{\lambda}_g\right)
-\mbox{constant}
\right)\right], ~\left(\mbox{constant}=\frac{3}{4}\right),
\end{eqnarray}
with an additive constant equivalent 
to the one in the Gaussian-like saddle points method. 
The discrepancy between the Gaussian-like saddle point method 
and the spectral density of the color eigenvalues method
comes from a redundant additive constant emerges due 
to the kind of approximation is considered. 
This redundant constant appears in the gas 
of only adjoint particles in the $U(N_c)$ representation.
The $SU(N_c)$ representation does not have 
this kind of problem.
%%%%%%%%%%%%%%%%%%%%%%%%%%%%%%%%%%%%%%%%%%%%%%%%%%%%%%%%%
%%%%%%%%%%%%%%%%%%%%%%%%%%%%%%%%%%%%%%%%%%%%%%%%%%%%%%%%%
%%%%%%%%%%%%%%%%%%%%%%%%%%%%%%%%%%%%%%%%%%%%%%%%%%%%%%%%%
%%%%%%%%%%%%%%%%%%%%%%%%%%%%%%%%%%%%%%%%%%%%%%%%%%%%%%%%%
In order to verify the results of the Gaussian-like
saddle points method, we compare them with the results
of the method of 
the spectral density of color eigenvalues\cite{Brezin1978a}.
The canonical ensemble in the adjoint representation
is computed in a similar way done 
in the fundamental representation.
Using a similar transformation,
the canonical ensemble in the term of 
the spectral density is reduced to
\begin{eqnarray}
Z(\lambda_g)&=&C \exp\left[
\frac{N^2_c}{2} \mbox{P}
\int^{\pi}_{-\pi} d\theta \rho(\theta)
\int^{\pi}_{-\pi} d\theta'\rho(\theta')
\ln \sin^2
\left(
\frac{\theta-\theta'}{2}\right)
\right.\nonumber\\
&+& \left.
\tilde{\lambda}_g
\left\{N_c+
N^2_c\mbox{P}
\int^{\pi}_{-\pi} d\theta \rho(\theta)
\int^{\pi}_{-\pi} d\theta' \rho(\theta')
\cos\left(\theta-\theta'\right)
\right\}
\right],\nonumber\\
&\approx&
C \exp\left[N^2_c\left(
\frac{1}{2}\mbox{P}
\int^{\pi}_{-\pi} d\theta \rho(\theta)
\int^{\pi}_{-\pi} d\theta'\rho(\theta')
\ln \sin^2
\left(
\frac{\theta-\theta'}{2}\right)
\right.\right.\nonumber\\
&+&
\left.\left.
\tilde{\lambda}_g
\int^{\pi}_{-\pi} d\theta \rho(\theta)
\int^{\pi}_{-\pi} d\theta' \rho(\theta')
\cos\left(\theta-\theta'\right)
\right)\right],
\end{eqnarray}
where the integral pre-factor $C$ is analogous 
to that one given in the fundamental representation.
Using the symmetry property of the spectral density
$\rho(\theta)$ in the range $-\pi\le\theta\le\pi$,
the stationary equation reads
\begin{eqnarray}
2\tilde{\lambda}_g\left[\int^{\pi}_{-\pi}
d\theta'\rho(\theta')\cos\theta'\right]\sin\theta
&=&\mbox{P}
\int^{\pi}_{-\pi}
d\theta'\rho(\theta')
\cot\left(\frac{\theta-\theta'}{2}\right).
\end{eqnarray}
The solution for the range $2\tilde{\lambda}_g\ge 1$
is already 
known~\cite{Brezin1978a,GrossWitten1980a,Sundborg:2000a,Aharony2003a}.
%%%%%%%%%%%%%
%
The density of eigenvalues reads
\begin{eqnarray}
\rho(\theta)&=&
\frac{1}{\pi}
\frac{\cos\frac{\theta}{2}}{\sin^2\frac{\theta_0}{2}}
\sqrt{
\sin^2\frac{\theta_0}{2}-\sin^2\frac{\theta}{2}
}, ~~~~ -\theta_0\le\theta\le\theta_0,\nonumber\\
&=& 0, ~~~~~~ \pi\ge|\theta|>\theta_0,
\label{spectral-sin2}
\end{eqnarray}
where
\begin{eqnarray}
\sin^2\left(\frac{\theta_0}{2}\right)&=&
1-\sqrt{1-\frac{1}{\tilde{\lambda}_g}}.
\end{eqnarray}
However, in the limit $\tilde{\lambda}_g\gg 1$, we get
$\sin^2\left(\theta_0/2\right)=
1/\left(2\tilde{\lambda}_g\right)$.
%%%%%%%%%%%%%%%%%%%%%
In the context of the spectral density method 
the canonical ensemble, in the large $N_c$ limit, becomes
\begin{eqnarray}
Z(\tilde{\lambda}_g)&\approx& \exp\left[
N_c^2 \tilde{\lambda}_g -
\frac{N^2_c}{2}\ln\left(2N_c\tilde{\lambda}_g\right)
+\frac{N^2_c}{2}\ln N_c
-\frac{N^2_c}{2}
\right],\nonumber\\
&\cong&
\left(\frac{1}{2\tilde{\lambda}_g}\right)^{\frac{N^2_c}{2}}
\exp\left[
N^2_c\tilde{\lambda}_g-\frac{N^2_c}{2}
\right],  ~~~~~
(\tilde{\lambda}_g\ge\frac{1}{2}).
\label{spectral-density-sol}
\end{eqnarray}
The solution for the energy domain
$\tilde{\lambda}_g\le\frac{1}{2}\ll 
\tilde{\lambda}_{g(I)\mbox{maximum}}<1$ 
reads
\begin{eqnarray}
\ln Z(\tilde{\lambda}_g)
&\approx&
\mbox{Constant}\times
\exp\left[\mbox{function}(\lambda_g)\right],
~~~~~\tilde{\lambda}_g\le \frac{1}{2},
\end{eqnarray}
for the entire color range $|\theta_i|\le\pi$, 
where the function of $\lambda_g$ 
is independent on $N_c$.
Hence, in the limit $N_c\rightarrow \infty$, 
we have
\begin{eqnarray}
\lim_{N_c\rightarrow\infty}
\frac{\ln Z(\tilde{\lambda}_g)}{N^2_c}&=&0.
\end{eqnarray}
%%%%%%%%%%
It means that in the limit $N_c\rightarrow \infty$,
the adjoint particle contribution vanishes 
and only the fundamental particles 
contribute in the low-lying energy solution 
for the energy range 
$\tilde{\lambda}_g\le \frac{1}{2}$.
This result is interpreted as follows:
the low-lying gluonic spectrum vanishes and 
the glueballs appear only as highly excited states 
in the high-lying mass spectrum.
Fortunately the existed hadronic mass spectrum agrees
this interpretation.

%%%%%%%%%%%%%%%%%%%%%%%%%%%%%%%%%%%%%%%%%%%%%%%%%%%%%%%%%%%%%%
%%%%%%%%%%%%%%%%%%%%%%%%%%%%%%%%%%%%%%%%%%%%%%%%%%%%%%%%%%%%%%
%%  Fig.
%%  Figure
%%  westen_adj_a.eps
%%  westen_adj_smallb.eps
%%  westen_adj_smallc.eps
%%  fig_w_adj_a
%%  fig_w_adj_smallb (0-0.5)
%%  fig_w_adj_smallc (0-1)
%%
%%%%%%%%%%%%%%%%%%%%%%%%%%%%%%%%%%%%%%%%%%%%%%%%%%%%%%%%%%%%%%
%%%%%%%%%%%%%%%%%%%%%%%%%%%%%%%%%%%%%%%%%%%%%%%%%%%%%%%%%%%%%%

\begin{figure}
\includegraphics{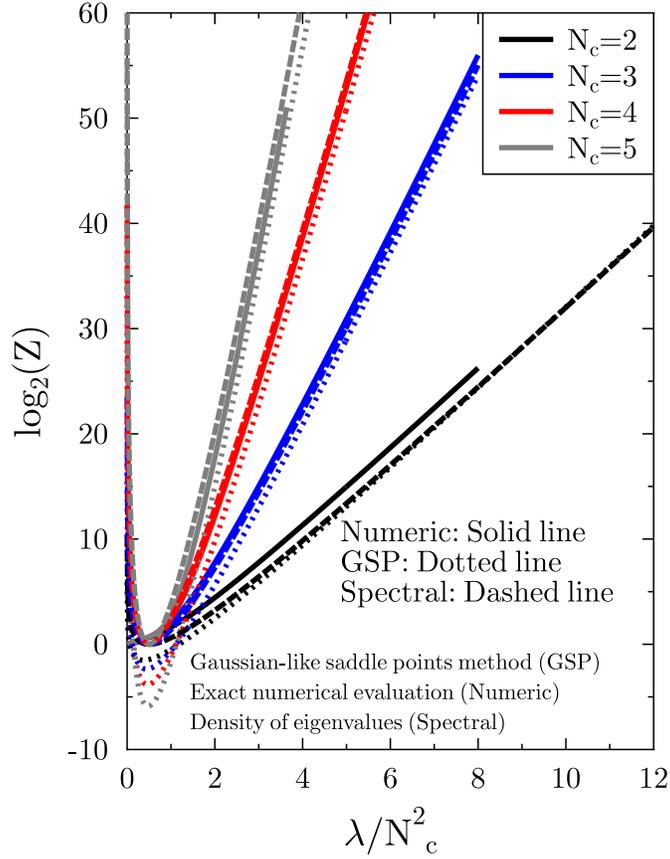}%
\caption{\label{fig_w_adj_a}
(Color online)
The high-lying density of states for 
the color-singlet bag consisting of only adjoint particles
versus the thermal running parameter 
$\frac{\lambda}{N^2_c}$ 
for various color numbers $N_c$.
The asymptotic high-lying energy solutions 
over the range
$\frac{\lambda}{N^2_c}\ge
\frac{\lambda_{(II)\mbox{min}}}{N^2_c}=\frac{1}{2}$
and their
extrapolation to the small 
$\frac{\lambda}{N^2_c}$ are displayed 
and compared with the exact numerical one.
The asymptotic high-lying energy solutions 
are computed using the Gaussian-like 
saddle points method and 
the spectral density of color eigenvalues.}
\end{figure}

The logarithm of partition function
$\ln Z(\lambda_{g})$
for the gas of color-singlet bags of adjoint particles
versus the thermal partition parameter
$\tilde{\lambda}_{g}$ 
with various color numbers $N_c$=2, 3, 4 and 5
is displayed in Fig.(\ref{fig_w_adj_a}).
The partition function for the high-lying energy
$\tilde{\lambda}_{g}\ge \tilde{\lambda}_0$
is solved using the spectral density 
of eigenvalues method
and the Gaussian-like saddle points method.
The both solutions are extrapolated to
the low-lying energy range
$\tilde{\lambda}_{g}<\tilde{\lambda}_0$.
It is shown that the solution
of the Gaussian-like saddle points method is adjacent
and almost parallel to the density of eigenvalues method
and they are splitting by a tiny additive constant
over the high-lying energy domain
$\tilde{\lambda}_{g}\ge \frac{1}{2}$.
Furthermore, the exact numerical solution is found
in the midway between the two solutions.
The tiny split between the exact numerical solution
and the asymptotic high-lying energy solution
for the two methods becomes less pronounced
as the number of colors $N_c$ increases.
The extrapolation of the asymptotic high-lying energy 
solution which is found by either methods 
deviates from the exact numerical solution over 
the thermal running parameter 
runs over the range $\tilde{\lambda}_g<\frac{1}{2}$.
This deviation increases significantly 
as $\tilde{\lambda}_g$
decreases far away from the critical point.
The extrapolation of the high-lying energy 
solution fails to fit the low-lying energy solution.
This characteristic behavior indicates 
a possible phase transition
to another analytic solution.
The high-lying energy solution is 
simply deflected at the Gross-Witten point.
%%%%%%%%%%%%%%%%%%%%%%%%%%%%%%%%%%%%%%%%%%%%%%%%%%%%%%%%%%%%%%%%%%%
%
%
%%%%%%%%%%%%%%%%%%%%%%%%%%%%%%%%%%%%%%%%%%%%%%%%%%%%%%%%%%%%%%%%%%

\begin{figure}
\includegraphics{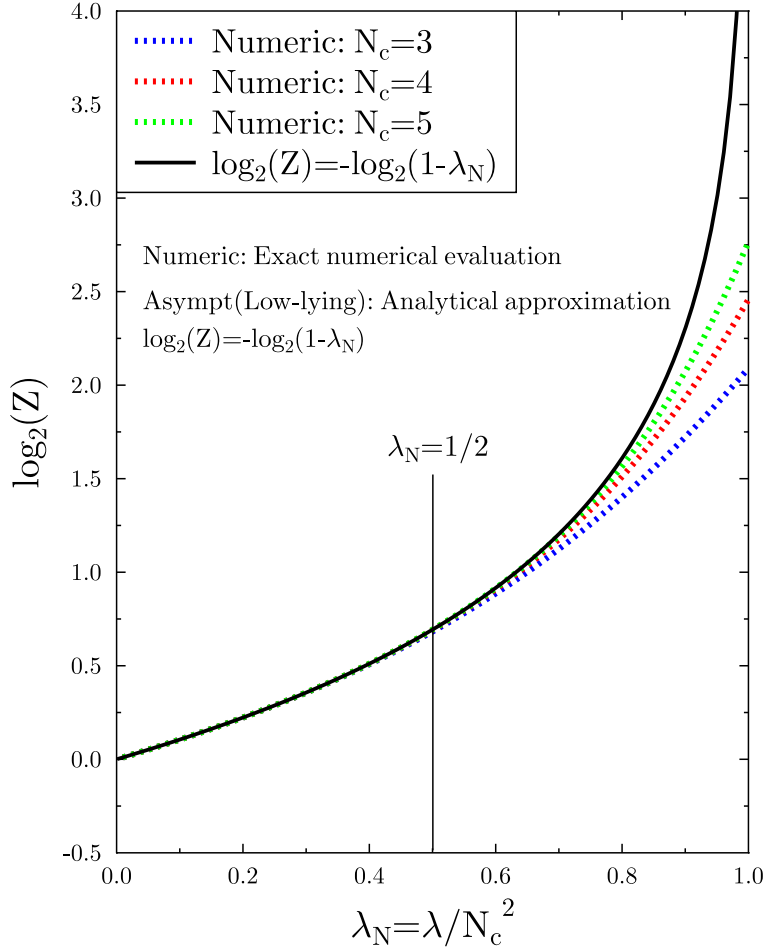}%
\caption{\label{fig_w_adj_smallc}
(Color online)
The asymptotic low-lying energy solution 
for a bag of adjoint particles in the color singlet state   
in the limit 
$N_c\rightarrow\infty$ 
versus the thermal running parameter $\frac{\lambda}{N^2_c}$.
The asymptotic low-lying energy solution 
versus $\frac{\lambda}{N^2_c}$
is compared with the exact numerical one 
for various color numbers $N_c$=3, 4 and 5.
The exact numerical solution is found to fit the
asymptotic low-lying energy solution over 
the range $\frac{\lambda}{N^2_c}\le 
\frac{\lambda_{(II)\mbox{min}}}{N^2_c}=\frac{1}{2}
\le\frac{\lambda_{\mbox{critical}}}{N^2_c}$.
The low-lying energy solution is found vanished
as $\lim_{N^2_c\rightarrow\infty}\frac{1}{N^2_c}
\ln\left(Z_{\mbox{Low-lying}}(\lambda/N_c^2)\right)
\rightarrow 0$. 
The asymptotic high-lying energy solution
$\lim_{N^2_c\rightarrow\infty}\frac{1}{N^2_c}
\ln\left(Z_{\mbox{High-lying}}(\lambda/N_c^2)\right)$
vanishes to match the low-lying solution
at the point of phase transition.}
\end{figure}

\begin{figure}
\includegraphics{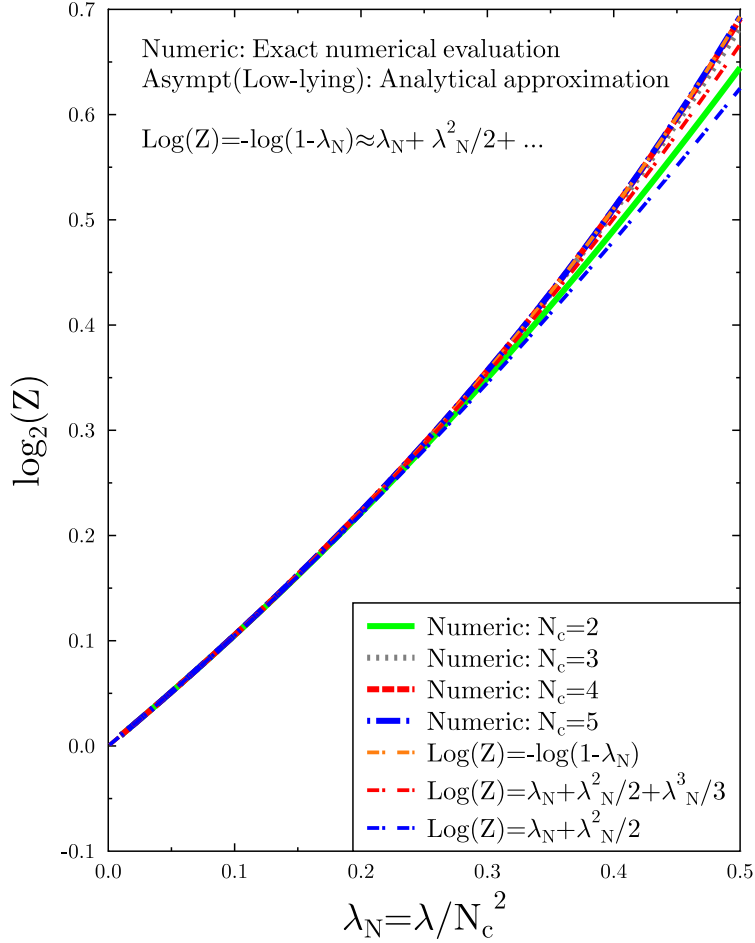}%
\caption{\label{fig_w_adj_smallb}
(Color online)
The density of states for the color-singlet fireballs 
(i.e. the Hagedorn states) versus the thermal running 
parameter $\frac{\lambda}{N_c^2}$ 
with various color numbers $N_c$. 
It is calculated for color-singlet bags 
consisting of only adjoint particles.
The exact numerical solution is compared 
with the asymptotic
low-lying energy solution 
with various analytic approximations
for the small thermal running parameter 
$\frac{\lambda}{N_c^2}\le 
\frac{\lambda_{(II)\mbox{min}}}{N^2_c}=\frac{1}{2}
\le\frac{\lambda_{\mbox{critical}}}{N^2_c}$. 
The exact numerical solution is found 
to fit the asymptotic analytic solution
$Z(\lambda)=1/\left(1-\frac{\lambda}{N^2_c}\right)$.}
\end{figure}

The asymptotic low-lying energy solution versus 
$\tilde{\lambda}_g$ in the energy domain
$\tilde{\lambda}_g\le \tilde{\lambda}_{g(\mbox{critical})}$
is displayed in Fig.(\ref{fig_w_adj_smallc}).
The asymptotic analytical solution for 
the low-lying energy states is basically obtained
by the power expansion over $\tilde{\lambda}_g$
and the orthogonal relations
in the fundamental group representation. 
The solution for adjoint representation 
is obtained by the convolution trick 
and introducing the Lagrange multiplier.
The exact numerical solution 
coincides with 
the asymptotic low-lying  energy solution 
in the energy domain
$\tilde{\lambda}_{g}\le
\tilde{\lambda}_{g(\mbox{critical})}$.
When the thermal running parameter reaches 
the Gross-Witten point 
$\tilde{\lambda}_{g(\mbox{critical})}
\rightarrow\tilde{\lambda}_{0}=\frac{1}{2}$, 
it deviates from the exact numerical one 
and this deviation grows up significantly
as $\tilde{\lambda}_{g}$ increases
and exceeds the Gross-Witten 
point 
$\tilde{\lambda}_0=\frac{1}{2}$.
The extrapolation of the low-lying energy 
solution fails to fit 
the high-lying energy solution 
and
subsequently this mechanism indicates that 
the low-lying energy solution is deflected 
and a phase transition 
to another analytical solution
takes place.
In order to demonstrate that the analytical 
low-lying energy solution is a correct one,
We expand it with respect 
to the thermal partition parameter
$\tilde{\lambda}_{g}$.
The results are displayed in Fig.(\ref{fig_w_adj_smallb}).
It is shown that the asymptotic analytic 
low-lying energy solution
coincides with the exact numerical one
and when the analytical solution 
is truncated to a lower order 
it deviates from the exact numerical 
one significantly.
%%%%%%%%%%%%%%%%%%%%%%%%%%%%%%%%%%%%%%%%%%%%%%%%%%%%%%%%%%%%%%
%%%%%%%%%%%%%%%%%%%%%%%%%%%%%%%%%%%%%%%%%%%%%%%%%%%%%%%%%%%%%%
%%%%%%%%%%%%%%%%%%%%%%%%%%%%%%%%%%%%%%%%%%%%%%%%%%%%%%%%%%%%%%

\section{Fundamental and Adjoint representation
for SU(3) and U(3)\label{sect_fund_adj}}

In the QCD the low-lying hadronic states
and the fireballs are treated as bags consisting of 
fundamental quark-antiquark particles 
and adjoint gluon particles as well.
Generally speaking, the adjoint particles are assumed 
to be the interaction among the fundamental particles, 
their anti-particles and the adjoint particles themselves.
It will be a reliable approximation to treat the hadronic states
as an ideal gas of fundamental and adjoint particles
in the color-singlet state.
It would be also a good approximation to ignore
the interaction among the constituent particles 
in the present model.

%%%%%%%%%%%%%%%%%%%%%%%%%%%%%%%%%%%%%%%%%%%%%%%%%%%%%%%%%%%%%
\subsection{Maxwell-Boltzmann statistics}

For the highly thermal excitations, the Maxwell-Boltzmann 
statistics becomes a reliable one.
Fortunately, the Maxwell-Boltzmann statistics is relatively 
a simple one.
In the context of this statistics,
the partition function for the fundamental 
and adjoint particles reads
\begin{eqnarray}
Z(\lambda_g,\lambda_{q\overline{q}})&=&\int d\mu(g)
\exp\left[\lambda_{g} \frac{1}{\mbox{dim}_g}\mbox{tr}_c
{\bf R}_{\mbox{adj}}(g)
+\lambda_{q\overline{q}} \frac{1}{N_c}
\mbox{tr}_c\left(
{\bf R}_{\mbox{fun}}(g)+{\bf R}^*_{\mbox{fun}}(g)
\right)
\right].
\end{eqnarray}
In order to simplify our notation
we rescale the thermal running parameters
$\tilde{\lambda}_g=\frac{\lambda_g}{\mbox{dim}_g}$, 
$\mbox{dim}_g=N_c^2-1$,
and
$\widetilde{\lambda}_{q\overline{q}}=
\frac{\lambda_{q\overline{q}}}{N^2_c}$.
With these redefinitions the canonical ensemble is simplified to
\begin{eqnarray}
\left.Z(\lambda_g,\lambda_{q\overline{q}})\right|_{SU(N_c)}
&=&\int d\mu(g)
e^{\left[\tilde{\lambda}_{g}
\left\{
(N^2_c-1)+\sum^{N_c}_{n=1}\sum^{N_c}_{m=1}
\cos\left(\theta_n-\theta_m\right)
\right\}
+
2N_c\widetilde{\lambda}_{q\overline{q}}
\sum^{N_c}_{n=1}\cos\theta_n\right]}.
\end{eqnarray}
For the large thermal running parameter limit
$\tilde{\lambda}_g\ge\tilde{\lambda}_{g(II)\mbox{min}}$
and
$\widetilde{\lambda}_{q\overline{q}}
\ge\widetilde{\lambda}_{q\overline{q}(II)\mbox{min}}$,
it is straightforward to evaluate the partition function 
using the Gaussian-like saddle points method.
In the asymptotic large $\tilde{\lambda}_g$ 
and
$\widetilde{\lambda}_{q\overline{q}}$,
the saddle points are dominated 
in a narrow interval around the origin.
The expansion around the saddle points 
$\theta_i\approx 0$ reduces
the partition function to
\begin{eqnarray}
\left.Z(\lambda_g,\lambda_{q\overline{q}})\right|_{SU(N_c)}
&=&
\left.Z_{(II)}(\lambda_g,\lambda_{q\overline{q}})
\right|_{SU(N_c)}, 
\nonumber\\
\left.
Z_{(II)}(\lambda_g,\lambda_{q\overline{q}})
\right|_{SU(N_c)}
&=&
Z^{(0)}_{SU(N_c)}
\int d\mu(g)
e^{\left[
-\frac{2N_c\tilde{\lambda}_g}{2}
\frac{1}{2N_c}
\sum^{N_c}_{n=1}\sum^{N_c}_{m=1}
\left(\theta_n-\theta_m\right)^2
-\frac{2N_c\widetilde{\lambda}_{q\overline{q}}}{2}
\sum^{N_c}_{n=1}\theta^2_n
\right]},
\nonumber\\
&\doteq&
Z^{(0)}_{SU(N_c)}
\int d\mu(g)
e^{
-\left(
2N_c\widetilde{\lambda}_{q\overline{q}}
+2N_c\tilde{\lambda}_g
\right)
\frac{1}{2}\sum^{N_c}_{n=1}\theta^2_n}
\nonumber\\
&~&
~\left(@~~ 
\left\{\begin{array}{c} \tilde{\lambda}_g 
\\ \widetilde{\lambda}_{q\overline{q}}
\end{array}\right\}
~\mbox{runs over the range}~ 
\left\{\begin{array}{c}
\tilde{\lambda}_g\ge
\tilde{\lambda}_{g(II)\mbox{min}}\\
\widetilde{\lambda}_{q\overline{q}}\ge
\widetilde{\lambda}_{q\overline{q}(II)\mbox{min}}
\end{array}\right\}
\right),
\nonumber\\
\label{MB-high-lying-approx}
\end{eqnarray}
where 
\begin{eqnarray}
Z^{(0)}_{SU(N_c)}
=\exp\left[
(N^2_c-1)\tilde{\lambda}_g
+2N^2_c\widetilde{\lambda}_{q\overline{q}}
\right].
\end{eqnarray}
The invariance measure is approximated to
\begin{eqnarray}
\frac{1}{2\pi}\int^{\pi}_{\pi} d\mu(g)
\rightarrow 
\frac{1}{2\pi}\int^{\infty}_{-\infty}d\mu_{\mbox{saddle}}.
\end{eqnarray}
The Gaussian-like integration over $SU(N_c)$ 
leads to
\begin{eqnarray}
\left.
Z_{(II)}(\lambda_g,\lambda_{q\overline{q}})
\right|_{SU(N_c)}
&=&\frac{1}{N_c!
(2\pi)^{N_c-1}}
\frac{(2\pi)^{N_c/2}\prod^{N_c}_{n=1}n!}{\sqrt{2\pi N_c}}
\left(
\frac{1}{2N_c(\tilde{\lambda}_g
+\widetilde{\lambda}_{q\overline{q}}) }
\right)^{\frac{N^2_c-1}{2}}
Z^{(0)}_{SU(N_c)}.
\end{eqnarray}
The same calculation can be carried 
in the same manner for $U(N_c)$ group.
In the asymptotic large thermal running parameters
$\widetilde{\lambda}_{q\overline{q}}\ge
\widetilde{\lambda}_{q\overline{q}(II)\mbox{min}}$
and 
$\tilde{\lambda}_g\ge\tilde{\lambda}_{g(II)\mbox{min}}$
the canonical ensemble for $U(N_c)$
becomes
\begin{eqnarray}
\left.
Z_{(II)}(\lambda_g,\lambda_{q\overline{q}})
\right|_{U(N_c)}
&=&\frac{(2\pi)^{N_c/2}\prod^{N_c}_{n=1}n!}{N_c!(2\pi)^{N_c-1}}
\frac{1}{\sqrt{2\pi}}
\left[
\int^{\pi}_{-\pi}\frac{d\theta}{2\pi}
e^{
-\frac{2N_c\widetilde{\lambda}_{q\overline{q}}}{2}
\theta^2}
\right]
\frac{Z^{(0)}_{U(N_c)}}
{\left(2N_c(\tilde{\lambda}_g
+\widetilde{\lambda}_{q\overline{q}})\right)^{\frac{N^2_c-1}{2}}
},
\nonumber\\
&=&\frac{
(2\pi)^{N_c/2}\prod^{N_c}_{n=1}n!}{N_c!(2\pi)^{N_c-1}}
\frac{1}{2\pi}
\frac{1}{\sqrt{2N_c\widetilde{\lambda}_{q\overline{q}}}}
\frac{Z^{(0)}_{U(N_c)}}
{
\left(2N_c(\tilde{\lambda}_g
+\widetilde{\lambda}_{q\overline{q}})
\right)^{\frac{N^2_c-1}{2}}},
\end{eqnarray}
where 
\begin{eqnarray}
Z^{(0)}_{U(N_c)}
=\exp\left[
N^2_c\tilde{\lambda}_g
+2N^2_c\widetilde{\lambda}_{q\overline{q}}
\right].
\end{eqnarray}
The canonical ensemble is simplified for a further analytical
investigation.
Using the Stirling's approximation
in the large $N_c\rightarrow\infty$ limit and approximating the
summation over $n$ to an integration over $n$,
the canonical ensemble can be approximated to
\begin{eqnarray}
\lim_{N_c\rightarrow\mbox{large}}
\left.
Z(\lambda_g,\lambda_{q\overline{q}})
\right|_{U(N_c)}
&=&
\lim_{N_c\rightarrow\mbox{large}}
\left.
Z_{(II)}(\lambda_g,\lambda_{q\overline{q}})
\right|_{U(N_c)}
\nonumber\\
&=&\frac{1}{\sqrt{2\widetilde{\lambda}_{q\overline{q}}}}
\frac{
\exp{\left[N^2_c\tilde{\lambda}_g
+2N^2_c\widetilde{\lambda}_{q\overline{q}}
-\frac{3N^2_c}{4}\right]}
}{
\left(2(\tilde{\lambda}_g
+\widetilde{\lambda}_{q\overline{q}})\right)^{\frac{N^2_c-1}{2}}
}
\nonumber\\
&~&
~\left(@~~ 
\left\{\begin{array}{c} \tilde{\lambda}_g 
\\ \widetilde{\lambda}_{q\overline{q}}
\end{array}\right\}
~\mbox{runs over the range}~ 
\left\{\begin{array}{c}
\tilde{\lambda}_g\ge
\tilde{\lambda}_{g(II)\mbox{min}}\\
\widetilde{\lambda}_{q\overline{q}}\ge
\widetilde{\lambda}_{q\overline{q}(II)\mbox{min}}
\end{array}\right\}
\right).
\nonumber\\
\label{fund_adj_mb1}
\end{eqnarray}
In contrary, it is hard to compute the spectral 
density in the large $\tilde{\lambda}_g$
and
$\widetilde{\lambda}_{q\overline{q}}$ limits
in the context of the spectral density of color eigenvalues method. 
Nonetheless, in that method, the canonical ensemble 
can be calculated only when the spectral density is known at first.

%%%%%%%%%%%%%%%%%%%%%%%%%%%%%%%%%%%%%%%%%%%%%%%%%%%%%%%%%%%%%%%%%%%%%%%%
%%%%%%%%%%%%%%%%%%%%%%%%%%%%%%%%%%%%%%%%%%%%%%%%%%%%%%%%%%%%%%%%%%%%%%%%
Under certain conditions when both
$\tilde{\lambda}_g$ and $\widetilde{\lambda}_{q\overline{q}}$
become small, the color eigenvalues are
distributed uniformly over the entire interval 
$-\pi\le\theta_i\le\pi$.
In this case, the partition function in the $U(N_c)$ 
representation is written as follows
\begin{eqnarray}
Z(\lambda_g,\lambda_{q\overline{q}})
&=&\int^{\pi}_{-\pi} d\mu(g)
e^{\lambda_g\frac{1}{\mbox{dim}_g}
\mbox{tr}{\bf R}(g)\mbox{tr}{\bf R}^*(g)}
e^{\lambda_{q\overline{q}}\frac{1}{N_c}
\mbox{tr}\left({\bf R}(g)+{\bf R}^*(g)\right)},
\nonumber\\
&=&\int^{\pi}_{-\pi} d\mu(g)
e^{\tilde{\lambda}_g
\mbox{tr}{\bf R}(g)\mbox{tr}{\bf R}^*(g)}
e^{N_c\widetilde{\lambda}_{q\overline{q}}
\mbox{tr}\left({\bf R}(g)+{\bf R}^*(g)\right)},
\nonumber\\
&=&\int^{\pi}_{-\pi} d\mu(g)
\left\{\frac{N^2_c}{2\tilde{\lambda}_g}
\int^{\infty}_0 d\xi \xi e^{-\frac{N^2_c\xi^2}
{4\tilde{\lambda}_g}}
e^{\frac{N_c\xi}{2}
\mbox{tr}\left({\bf R}(g)+{\bf R}^*(g)\right)}
\right\}
e^{N_c\widetilde{\lambda}_{q\overline{q}}
\mbox{tr}\left({\bf R}(g)+{\bf R}^*(g)\right)},
\nonumber\\
&=&
\frac{N^2_c}{2\tilde{\lambda}_g}
\int^{\infty}_0 d\xi \xi e^{-\frac{N^2_c\xi^2}
{4\tilde{\lambda}_g}}
\int^{\pi}_{-\pi} d\mu(g)
\left\{e^{N_c\left[\frac{\xi}{2}+
\widetilde{\lambda}_{q\overline{q}}\right]
\mbox{tr}\left({\bf R}(g)+{\bf R}^*(g)\right)}
\right\},
\label{fund_adj_small_exact}
\end{eqnarray}
where the Lagrange multiplier $\xi$ have been introduced 
in order to simplify the above equation.
By expanding the exponential 
and evaluating the series term by term
using the group bases orthogonality 
and then finally re-summing the resultant terms, 
the partition function is reduced to
\begin{eqnarray}
Z(\lambda_g,\lambda_{q\overline{q}})
&=&
\frac{N^2_c}{2\tilde{\lambda}_g}
\int^{\xi_0}_0 d\xi \xi e^{-\frac{N^2_c\xi^2}{4\tilde{\lambda}_g}}
e^{N^2_c\left(\frac{\xi}{2}
+\widetilde{\lambda}_{q\overline{q}}\right)^2}
\nonumber\\
&+&
\frac{N^2_c}{2\tilde{\lambda}_g}
\int^{\infty}_{\xi_0} d\xi \xi
e^{-\frac{N^2_c\xi^2}{4\tilde{\lambda}_g}}
\left(
\frac{
\exp\left[N^2_c
\left(\xi+2\widetilde{\lambda}_{q\overline{q}}\right)
-\frac{3N^2_c}{4}\right]}
{\left[
\xi+2\widetilde{\lambda}_{q\overline{q}}
\right]^{N^2_c/2}}\right),
\label{fun_adj_small_01}
\end{eqnarray}
where $\xi_0$ is determined by the constraint
\begin{eqnarray}
\left(\frac{\xi_0}{2}+\widetilde{\lambda}_{q\overline{q}}\right)
\ll 1 \longrightarrow \xi_0\ll
2\left(1-\widetilde{\lambda}_{q\overline{q}}\right),
~~~~\mbox{and}~~~~
\widetilde{\lambda}_{q\overline{q}}\le \frac{1}{2}.
\end{eqnarray}
This leads to an additional constraint $\xi_0\ll 1$.
It is possible to approximate 
Eq.(\ref{fun_adj_small_01})
for the energy domain
$\widetilde{\lambda}_{q\overline{q}}\le \frac{1}{2}$
in the following way
\begin{eqnarray}
Z(\lambda_g,\lambda_{q\overline{q}})
&\le&
\frac{N^2_c}{2\tilde{\lambda}_g}
\left\{
\int^{\xi_0}_0 d\xi 
\xi e^{-\frac{N^2_c\xi^2}{4\tilde{\lambda}_g}}
\exp\left[N^2_c\left(\frac{\xi}{2}
+\widetilde{\lambda}_{q\overline{q}}\right)^2\right]
\right.\nonumber\\
&+&
\left.
\int^{\infty}_{\xi_0} d\xi \xi
e^{-\frac{N^2_c\xi^2}{4\tilde{\lambda}_g}}
\exp\left[N^2_c
\left(\xi+2\widetilde{\lambda}_{q\overline{q}}\right)
-\frac{3N^2_c}{4}\right]\right\},
\nonumber\\
&\le&
\frac{N^2_c}{2\tilde{\lambda}_g}
\left\{
\int^{\xi_0}_0 d\xi 
\xi e^{-\frac{N^2_c\xi^2}{4\tilde{\lambda}_g}}
\exp\left[N^2_c\left(\frac{\xi}{2}
+\widetilde{\lambda}_{q\overline{q}}\right)^2\right]
\right.\nonumber\\
&+&
\left.
\int^{\infty}_{\xi_0} d\xi \xi
e^{-\frac{N^2_c\xi^2}{4\tilde{\lambda}_g}}
\exp\left[N^2_c
\left(\frac{\xi}{2}+\widetilde{\lambda}_{q\overline{q}}\right)^2
\right]
\right\}, \nonumber\\
&=&
\frac{N^2_c}{2\tilde{\lambda}_g}
\int^{\infty}_0 d\xi \xi 
\exp\left[-\frac{N^2_c\xi^2}{4\tilde{\lambda}_g}+
N^2_c\left(\frac{\xi}{2}
+\widetilde{\lambda}_{q\overline{q}}\right)^2\right].
\label{fun_adj_small1}
\end{eqnarray}
After evaluating the integral 
over the parameter $\xi$
in Eq.(\ref{fun_adj_small1}), 
the partition function is approximated to
\begin{eqnarray}
Z_{(I)}(\lambda_g,\lambda_{q\overline{q}})
=\frac{1}{1-\tilde{\lambda}_g}
\exp\left[
\frac{N^2_c\widetilde{\lambda}^2_{q\overline{q}}}
{1-\tilde{\lambda}_g}
\right]
\left(
2\int^{\infty}_0 dx x e^{-(x-b_0)^2}
\right),
\label{fun-adj-int}
\end{eqnarray}
where
\begin{eqnarray}
b_0&=&N_c \tilde{\lambda}_{q\overline{q}}
\sqrt{\frac{\tilde{\lambda}_g}{1-\tilde{\lambda}_g}}.
\end{eqnarray}
The integral on the right hand side 
of Eq.(\ref{fun-adj-int})
can be approximated to
\begin{eqnarray}
\int^{\infty}_0 dx x e^{-(x-b_0)^2}&=&
\frac{1}{2}\left[
e^{-b^2_0}+b_0\sqrt{\pi}\left(1+\mbox{erf}(b_0)\right)
\right],
\nonumber\\
&\approx&
\frac{1}{2} e^{-b^2_0} ~~~(\mbox{when}~ b_0\ll 1).
\label{fun-adj-int-approx}
\end{eqnarray}
Using the approximation given by 
Eq.(\ref{fun-adj-int-approx}),
the partition function is reduced to
\begin{eqnarray}
Z_{(I)}(\lambda_g,\lambda_{q\overline{q}})\approx
\frac{1}{1-\tilde{\lambda}_g}
\exp\left[
N^2_c\widetilde{\lambda}^2_{q\overline{q}}
\right].
\label{fun_adj_res1}
\end{eqnarray}
%%%%%%%%%%%%%%%%%%%%%%%%%%%%%%%%%%%%%%%%%%%%%%%
%
%%%%%%%%%%%%%%%%%%%%%%%%%%%%%%%%%%%%%%%%%%%%%%%
In order to verify the result, we assume that
$\tilde{\lambda}_g=0$ for a system
consisting only of fundamental particles. 
The partition function is reduced to
\begin{eqnarray}
\lim_{\tilde{\lambda}_g\rightarrow 0}
Z_{(I)}(\lambda_g,\lambda_{q\overline{q}})
&=&
\exp\left[
N^2_c\widetilde{\lambda}^2_{q\overline{q}}
\right].
\end{eqnarray}
On the other hand, in the limit 
$\widetilde{\lambda}_{q\overline{q}}=0$, 
Eq.(\ref{fun_adj_res1}) is converted to
\begin{eqnarray}
\lim_{\widetilde{\lambda}_{q\overline{q}}\rightarrow 0}
Z_{(I)}(\lambda_g,\lambda_{q\overline{q}})
&=&
\exp\left[-\ln(1-\tilde{\lambda}_{g})\right],
\end{eqnarray}
for a system consisting only of adjoint particles.
%%%%%%%%%%%%%%%%%%%%%%%%%%%%%%%%%%%%%%%%%%%%%%%%%%%%%%%%%%%%

%%%%%%%%%%%%%%%%%%%%%%%%%%%%%%%%%%%%%%%%%%%%%%%%%%%%%%%%%%%%%%%%%%%%%%%%
%%%%%%%%%%%%%%%%%%%%%%%%%%%%%%%%%%%%%%%%%%%%%%%%%%%%%%%%%%%%%%%%%%%%%%%%
We are in a position to determine 
the critical point for the phase transition.
This can be done by analyzing 
the canonical partition function.
The solution (II) is given for the thermal 
running parameters run over the ranges
$\left(\tilde{\lambda}_{g}
\ge
\tilde{\lambda}_{g\mbox{critical}}
\ge
\tilde{\lambda}_{g(II)\mbox{min}}\right)$
and
$\left(\widetilde{\lambda}_{q\overline{q}}
\ge
\widetilde{\lambda}_{q\overline{q}\mbox{critical}}
\ge
\widetilde{\lambda}_{q\overline{q}(II)\mbox{min}}\right)$.
Usually this solution corresponds the continuous 
high-lying hadronic mass spectrum.
The logarithm of solution (II) 
is given by Eq.(\ref{fund_adj_mb1}),
and it reads
\begin{eqnarray}
\lim_{N^2_c\rightarrow \mbox{Large}}
\frac{
\ln Z_{(II)}
(\tilde{\lambda}_g,\widetilde{\lambda}_{q\overline{q}})
}{N^2_c}
=\tilde{\lambda}_g
+2\widetilde{\lambda}_{q\overline{q}}-\frac{3}{4}
-\frac{1}{2}
\ln 2(\tilde{\lambda}_g+\widetilde{\lambda}_{q\overline{q}}).
\end{eqnarray}
This function is concave up where the minimum 
is located at the points 
$\tilde{\lambda}_{g(II)\mbox{min}}$
and
$\widetilde{\lambda}_{q\overline{q}(II)\mbox{\mbox{min}}}$.
However, the solution along the thermal running coupling axis 
in the range less than the minimum points
$\tilde{\lambda}_{g(II)\mbox{\mbox{min}}}$ and
$\widetilde{\lambda}_{q\overline{q}(II)\mbox{min}}$
is excluded since it will not be physical.
Usually, the running thermal parameters 
$\widetilde{\lambda}_{q\overline{q}}$ and
$\tilde{\lambda}_{g}$
are not strictly independent but rather they are  
functions of the variable $\zeta$.
They can be written as  
$\tilde{\lambda}_{g}(\zeta)$ 
and 
$\widetilde{\lambda}_{q\overline{q}}(\zeta)$. 
The variable $\zeta$ could be the hadronic mass
$m$ or $V/\beta^{3}$ etc.
The extreme left-side threshold of 
the solution's (II) range is determined 
by finding the solution's minimum location
as follows
\begin{eqnarray}
\lim_{N^2_c\rightarrow \mbox{Large}}
\left[
\left(
\frac{\partial}{\partial \tilde{\lambda}_g}
\frac{\ln Z_{(II)}
(\tilde{\lambda}_g,\widetilde{\lambda}_{q\overline{q}})}
{N^2_c}
\right)
\frac{\partial \tilde{\lambda}_g}
{\partial \zeta}
+
\left(
\frac{\partial}{\partial \tilde{\lambda}_{q\overline{q}}}
\frac{\ln Z_{(II)}
(\tilde{\lambda}_{g},\widetilde{\lambda}_{q\overline{q}})}
{N^2_c}
\right)
\frac{\partial\widetilde{\lambda}_{q\overline{q}}}
{\partial \zeta}
\right]_{\zeta=\zeta_{(II)\mbox{min}}}=0.
\nonumber\\
\end{eqnarray}
It leads to
\begin{eqnarray}
\tilde{\lambda}_{g}
+\widetilde{\lambda}_{q\overline{q}}
\ge
\frac{1}{2}
\left(
\frac{
\frac{
\partial \tilde{\lambda}_{g}
}{\partial \zeta}
+
\frac{
\partial\widetilde{\lambda}_{q\overline{q}}
}
{\partial \zeta} 
}
{
\frac{\partial \tilde{\lambda}_{g}}{\partial \zeta}
+
2\frac{\partial\widetilde{\lambda}_{q\overline{q}}}
{\partial \zeta} 
}\right)_{\zeta=\zeta_{(II)\mbox{min}}}.
\end{eqnarray}
The minimum thresholds are given by
\begin{eqnarray}
\widetilde{\lambda}_{q\overline{q}(II)\mbox{min}}
&=&\widetilde{\lambda}_{q\overline{q}}
\left(
\zeta_{(II)\mbox{min}}
\right),
\nonumber\\
\tilde{\lambda}_{g(II)\mbox{min}}
&=&\tilde{\lambda}_{g}
\left(
\zeta_{(II)\mbox{min}}
\right).
\end{eqnarray}
However, in the case that 
$\widetilde{\lambda}_{q\overline{q}}$
and $\tilde{\lambda}_g$  are independent parameters
then the minimum points are determined 
by the set of equations
\begin{eqnarray}
\lim_{N^2_c\rightarrow \mbox{Large}}
\frac{\partial}{\partial \tilde{\lambda}_{q\overline{q}}}
\left(
\frac{
\ln Z_{(II)}
(\tilde{\lambda}_{g},\widetilde{\lambda}_{q\overline{q}})
}{N^2_c}\right)_{\widetilde{\lambda}_{q\overline{q}}=
\widetilde{\lambda}_{q\overline{q}(II)\mbox{min}}}
&=&0,
\nonumber\\
\lim_{N^2_c\rightarrow \mbox{Large}}
\frac{\partial}{\partial \tilde{\lambda}_g}
\left(\frac{
\ln Z_{(II)}(\tilde{\lambda}_g,
\widetilde{\lambda}_{q\overline{q}})
}{N^2_c}
\right)_{\tilde{\lambda}_{g}=\tilde{\lambda}_{g(II)\mbox{min}}}
&=&0.
\end{eqnarray}
The set of constraints for the solution (II) is given by
\begin{eqnarray}
&~&\widetilde{\lambda}_{q\overline{q}}\ge
\widetilde{\lambda}_{q\overline{q}(II)\mbox{min}}=
\frac{1}{4},
\nonumber\\
&~&\mbox{or}
\nonumber\\
&~&
\tilde{\lambda}_{g}\ge
\tilde{\lambda}_{g(II)\mbox{min}}=
\frac{1}{2},
\nonumber\\
&~&
~\left(\mbox{The allowed range for solution (II)}
\right).
\end{eqnarray}
However, the number of the constraints 
must be kept minimum. 
In the case the thermal running parameters
$\widetilde{\lambda}_{q\overline{q}}$ and
$\tilde{\lambda}_{g}$
are strictly independent on each other
then the the thermal running
parameter $\widetilde{\lambda}_{q\overline{q}}$ 
for the fundamental particles 
is adopted as the master constraint for the minimum limit
while the constraint for $\tilde{\lambda}_{g}$ 
is restrained in order to get a feasible solution. 
Nevertheless, in the realistic QCD, the both parameters  
$\widetilde{\lambda}_{q\overline{q}}$ and
$\tilde{\lambda}_{g}$ depend on 
a characteristic variable such as the hadronic mass $m$ 
or even $V/\beta^3$ as mentioned above. 
%%%%%%%%%%%%%%%%%%%%%%%%%%%%%%%%%%%%%%%%%%%%%%%%%%%%%%%%%%%%%%%
%%%%%%%%%%%%%%%%%%%%%%%%%%%%%%%%%%%%%%%%%%%%%%%%%%%%%%%%%%%%%%%
%

On the other hand, the solution (I) that runs over the range
$\left(0\le\widetilde{\lambda}_{q\overline{q}}
\le
\widetilde{\lambda}_{q\overline{q}(I)\mbox{max}}\right)$ 
and
$\left(0\le\tilde{\lambda}_{g}
\le
\tilde{\lambda}_{g(I)\mbox{max}}\right)$,
is given by Eq.(\ref{fun_adj_res1})
\begin{eqnarray}
\lim_{N^2_c\rightarrow \mbox{Large}}
\frac{
\ln Z_{(I)}
(\tilde{\lambda}_g,\widetilde{\lambda}_{q\overline{q}})
}{N^2_c} &\approx& \widetilde{\lambda}^2_{q\overline{q}},
\label{approx-low-lying-log}
\end{eqnarray}
where $\widetilde{\lambda}_{q\overline{q}}$
and 
$\tilde{\lambda}_{g}$ are again assumed strictly 
independent on each other. 
The solution (I), usually, corresponds 
the discrete low-lying hadronic mass spectrum. 
Furthermore, the set of constraints which 
are associated with the solution (I) reads
\begin{eqnarray}
&~&\widetilde{\lambda}^2_{q\overline{q}}
\ll 1,
\nonumber\\
&\rightarrow&
\widetilde{\lambda}^2_{q\overline{q}}
\le
\widetilde{\lambda}^2_{q\overline{q}(I)\mbox{max}}
<1,
\nonumber\\
&~&
~\left(
\mbox{The allowed range for the solution (I)}
\right).
\end{eqnarray}
The critical points $\tilde{\lambda}_{q\overline{q}}$
and $\widetilde{\lambda}_g$
for the phase transition are determined
at the point where the two asymptotic solutions 
match each other
\begin{eqnarray}
\lim_{N^2_c\rightarrow \mbox{Large}}
\frac{
\ln Z_{(I)}
(\tilde{\lambda}_g,\widetilde{\lambda}_{q\overline{q}})
}{N^2_c}
&=&
\lim_{N^2_c\rightarrow \mbox{Large}}
\frac{
\ln Z_{(II)}
(\tilde{\lambda}_g,\widetilde{\lambda}_{q\overline{q}})
}{N^2_c} ~\rightarrow
\left(\tilde{\lambda}_g,
\widetilde{\lambda}_{q\overline{q}}\right)_{\mbox{critical}}.
\label{critical-point-match}
\end{eqnarray}
The both solutions are found match each other 
at the critical points
\begin{eqnarray}
\widetilde{\lambda}_{q\overline{q}\mbox{critical}}
&=&\frac{1}{2},
\nonumber\\
\tilde{\lambda}_{g\mbox{critical}}&=&0.
\end{eqnarray}
%%%%%%%%%%%%%%%%%%%%%%%%%%%%%%%%%%%%%%%%%%%%%%%%%%%%%%%%%%%%%%%
%%%%%%%%%%%%%%%%%%%%%%%%%%%%%%%%%%%%%%%%%%%%%%%%%%%%%%%%%%%%%%%
This solution is very interesting and is interpreted 
that the low-lying hadronic mass spectrum does not 
mix the fundamental particles
and the adjoint ones significantly. 
The hadronic states with the constituent
fundamental particles (e.g. quarks) 
has no significant adjoint components 
in the low energy domain.
This scenario explains why the low-lying hadronic mass spectrum
is likely consisting of mesons and baryons 
and is unlikely consisting of exotic hadronic states mixed 
with significant gluonic components
(e.g. hybrid: quarks, antiquarks and glueballs 
are not common in the low-lying mass spectrum).
The second example is the QCD with $N_c=3$ 
$n_{\mbox{flavor}}=2$. According to 
Eqs.(\ref{lambda-qq}) and (\ref{lambda-g}), 
we have 
$\tilde{\lambda}_g=\frac{16}{21} 
\widetilde{\lambda}_{q\overline{q}}$.
In this example, we get from the constraint
given by Eq.(\ref{critical-point-match}) 
the critical point 
$\widetilde{\lambda}_{q\overline{q}\mbox{critical}}\approx 
0.6\ge \widetilde{\lambda}_{q\overline{q}(II)\mbox{min}}$
where 
$\widetilde{\lambda}_{q\overline{q}(II)\mbox{min}}\approx 0.18$.
%%%%%%%%%%%%%%%%%%%%%%%%%%%%%%%%%%%%%%%%%%%%%%%%%%%%%%%%%%%%
%%%%%%%%%%%%%%%%%%%%%%%%%%%%%%%%%%%%%%%%%%%%%%%%%%%%%%%%%%%%
It is shown that the low-lying solution 
given by Eq.(\ref{approx-low-lying-log})
has not an adjoint component and subsequently 
it leads to an over estimation for 
$\widetilde{\lambda}_{q\overline{q}\mbox{critical}}$.
However for a rather large adjoint component 
the solution given by Eq.(\ref{approx-low-lying-log})
is not appropriate to determine precise the critical point.
In order to determine the threshold limit 
of the (pre-)critical point
we can return to Eq.(\ref{MB-high-lying-approx}).
It is noticed that Eq.(\ref{MB-high-lying-approx})
can be written as a semi-fundamental representation 
as follows 
\begin{eqnarray}
\left.
Z_{(II)}(\lambda_g,\lambda_{q\overline{q}})
\right|_{SU(N_c)}
&\doteq&
Z^{(0)}_{SU(N_c)}
\int d\mu(g)
e^{
-\left(
2N_c\widetilde{\lambda}_{q\overline{q}}
+2N_c\tilde{\lambda}_g
\right)
\frac{1}{2}\sum^{N_c}_{n=1}\theta^2_n},
\nonumber\\
&\rightarrow& 
Z^{(0)}_{SU(N_c)}
\int d\mu(g)
e^{-\frac{1}{2}\left(2N_c\tilde{\lambda}\right)
\sum^{N_c}_{n=1}\theta^2_n}.
\end{eqnarray}
The (pre-)critical point is determined by
\begin{eqnarray} 
\tilde{\lambda}\equiv\frac{1}{2N_c}
\left(
2N_c\widetilde{\lambda}_{q\overline{q}}
+2N_c\tilde{\lambda}_g\right)_{\mbox{critical}}&=&
\frac{1}{2}.
\end{eqnarray}
This result is analogous the critical point 
derived in section \ref{sect_fund_adj}.
The advantage of this method,
the explicit expression for the low-lying solution 
is not needed and it gives the threshold limit
of the critical point 
$\tilde{\lambda}_{(\mbox{pre}-)\mbox{critical}}\le
\tilde{\lambda}_{\mbox{critical}}$.
Hereinafter, we shall adopt the above procedure 
to estimate the location of the critical point
and set the correspondence
$\tilde{\lambda}_{\mbox{(pre-)critical}}
\rightarrow
\tilde{\lambda}_{\mbox{critical}}$.

%%%%%%%%%%%%%%%%%%%%%%%%%%%%%%%%%%%%%%%%%%%%
%%%%%%%%%%%%%%%%%%%%%%%%%%%%%%%%%%%%%%%%%%%%
%%%
%%%  Figure
%%%  Fig
%%%
%%%  westen_fun_adj.eps
%%%  westen_fun_adj_match.eps
%%%  westen_fun_adj_small.eps
%%%  fig_w_fun_adj
%%%  fig_w_fun_adj_small
%%%  fig_w_fun_match
%%%
%%%
%%%%%%%%%%%%%%%%%%%%%%%%%%%%%%%%%%%%%%%%%%%%%
%%%%%%%%%%%%%%%%%%%%%%%%%%%%%%%%%%%%%%%%%%%%%

\begin{figure}
\includegraphics{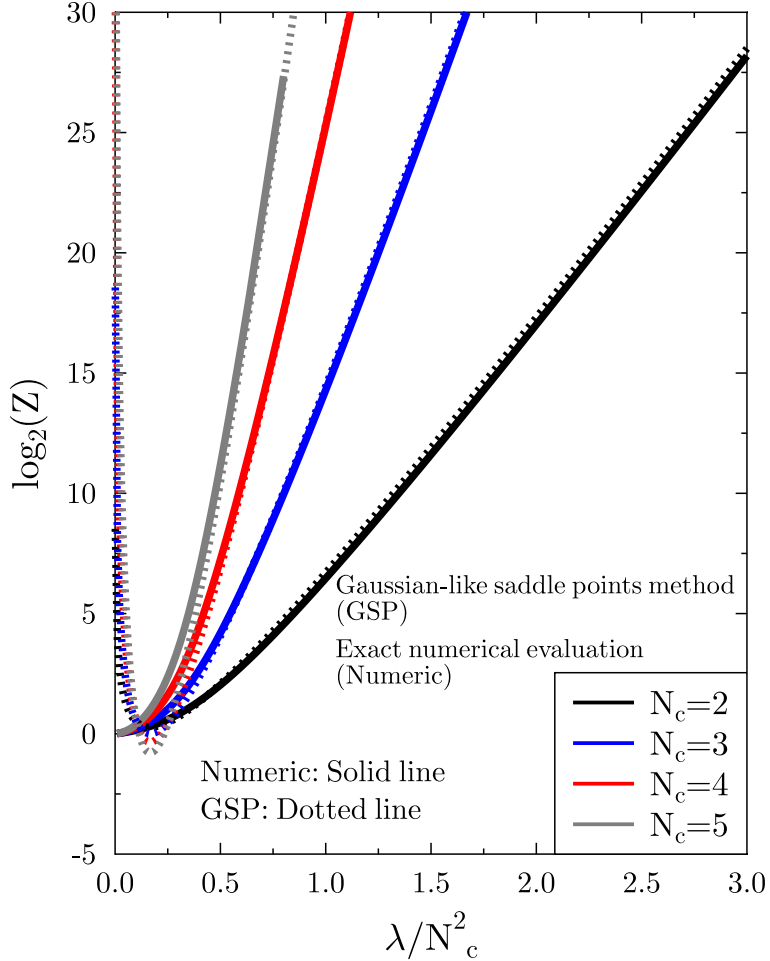}%
\caption{\label{fig_w_fun_adj}
(Color online)
The high-lying density of states and its extrapolation 
for the color-singlet bag consisting of fundamental 
and adjoint particles with 
the same thermal running parameter 
versus the thermal running parameter 
$\frac{\lambda}{N^2_c}$ 
with various color numbers $N_c$.
The high-lying energy solution
is the asymptotic solution for  
the thermal running parameter runs
over the range $\frac{\lambda}{N^2_c}
\ge 
\frac{\lambda_{\mbox{critical}}}{N^2_c}
\ge 
\frac{\lambda_{\mbox{(pre-)critical}}}{N^2_c}
\ge
\frac{\lambda_{(II)\mbox{min}}}{N^2_c}$.
The asymptotic high-lying energy solutions are computed 
using the Gaussian-like saddle points method 
and the spectral density of color eigen-values
and they are compared with the exact numerical one.}
\end{figure}

The grand canonical potential for the color-singlet bags
of fundamental and adjoint particles 
versus the thermal running parameter
$\tilde{\lambda}$  with various color numbers 
$N_c$= 2, 3, 4 and $5$ is displayed in Fig.(\ref{fig_w_fun_adj}).
In this toy calculations, we have set
$\tilde{\lambda}=\widetilde{\lambda}_{q\overline{q}}=\tilde{\lambda}_g$.
The asymptotic analytical high-lying energy solution is displayed for
the Gaussian-like saddle points method. 
The asymptotic high-lying energy solution is
compared with the exact numerical solution.
It is shown that the asymptotic analytical solution fits precisely the exact
numerical one for the energy domain 
$\tilde{\lambda}\ge\tilde{\lambda}_{\mbox{critical}}$
even with a small number of colors $N_c=2$.
The extrapolation of the high-lying energy solution
to the range 
$\tilde{\lambda}<\tilde{\lambda}_{\mbox{critical}}$
deviates from the exact numerical solution.
This deviation becomes significant as $\tilde{\lambda}$ decreases.
It is evident that the extrapolation of the high-lying energy solution 
to the low-lying solution is unphysical.
Moreover, it is evident that the grand potential is deflected 
at the Gross-Witten point.
Hence, the high-lying energy solution undergoes a phase transition
to another analytical function that describes the low-lying energy
solution more successfully.

\begin{figure}
\includegraphics{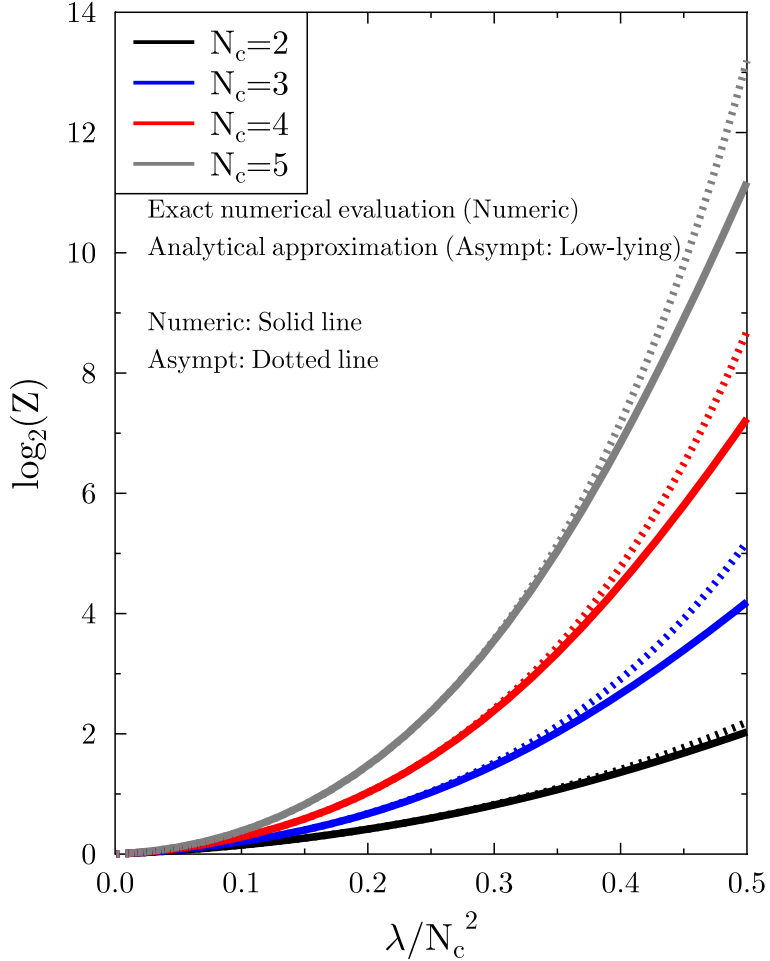}%
\caption{\label{fig_w_fun_adj_small}
(Color online)
The asymptotic low-lying energy solution
for color-singlet bags consisting 
of fundamental and adjoint particles
with the same thermal running parameter 
$\frac{\lambda}{N_c^2}$ 
for the both fundamental and adjoint particles
versus $\frac{\lambda}{N_c^2}$ 
with various color numbers $N_c$.
The exact numerical solution is compared 
with the analytic approximation of the low-lying energy 
solution for the thermal running parameter runs over
the range $\frac{\lambda}{N_c^2}\le\frac{1}{2}$.
It is found that the exact numerical solution fits 
the approximate analytical one for small values 
of $\frac{\lambda}{N_c^2}$. 
When the thermal running parameter exceeds
the (pre-)critical one 
$\frac{\lambda}{N^2_c}
\ge 
\left(\frac{\lambda}{N^2_c}
\right)_{\mbox{(pre-)critical}}$, 
the exact numerical solution starts 
to deviate slightly from 
the approximate analytical solution
and this deviation becomes 
noticeable beyond the point 
$\frac{\lambda_{\mbox{critical}}}{N^2_c}
=\left(
\frac{\lambda}{N^2_c}
\right)_{\mbox{match}}=0.34$.}
\end{figure}

The asymptotic analytical low-lying energy solution
is displayed in Fig.(\ref{fig_w_fun_adj_small}).
The low-lying energy solution is approximated by expanding the canonical
ensemble as a power expansion over $\tilde{\lambda}$
and then integrating it over the entire color circle and using
the Lagrange multiplier and the convolution 
for the adjoint particles.
The low-lying energy solution fits 
the exact numerical one over the range 
$\tilde{\lambda}<\tilde{\lambda}_{\mbox{critical}}$
even with a small number of colors $N_c=2$.
It deviates from the exact numerical solution 
when  $\tilde{\lambda}$ increases and exceeds
the critical point 
$\widetilde{\lambda}_{q\overline{q}}+\tilde{\lambda}_{g}=
2\tilde{\lambda}\ge 2\tilde{\lambda}_{\mbox{critical}}$.
The numerical calculations shows that the split between 
the exact numerical solution and 
the low-lying energy solution 
(i.e. solution (I)) becomes 
noticeable at the point 
$\tilde{\lambda}=\frac{\lambda}{N^2_c}\approx 0.34\ge 
\tilde{\lambda}_{\mbox{(pre-)critical}}=\frac{1}{4}$
in agreement with the result of 
Eq.(\ref{critical-point-match}).
Furthermore, the extrapolation of 
the low-lying energy solution
to a larger thermal running parameter
$\tilde{\lambda}>\tilde{\lambda}_{\mbox{critical}}$
deviates from the exact numerical solution significantly. 
This failure for the low energy extrapolation
indicates that the analytic solution is deflected 
in the middle between
the low-lying and high-lying energy solutions 
and proves the existence of the phase transition.
%%%%%%%%%%%%%%%%%%%%%%%%%%%%%%%%%%%%%%%%%%%%%%%%%%%%%%%%%%%%%

\begin{figure}
\includegraphics{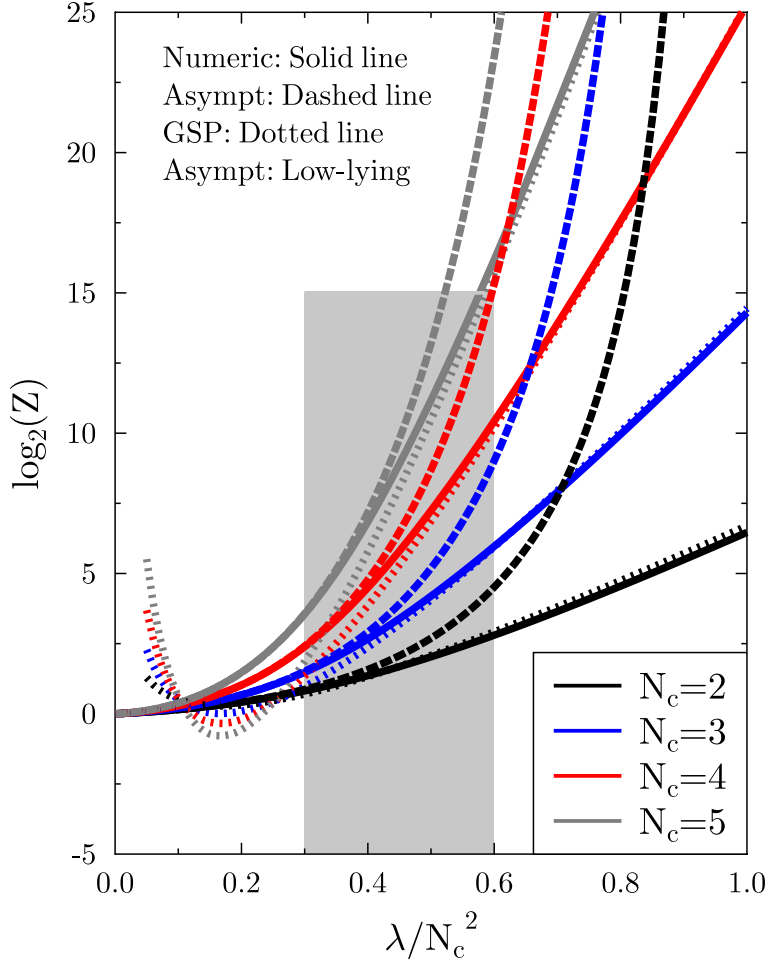}%
\caption{\label{fig_w_fun_adj_match}
(Color online)
The asymptotic low-lying and high-lying energy solutions
and their extrapolations for 
the color-singlet bag of fundamental and adjoint
particles with the same thermal running parameter 
versus $\frac{\lambda}{N_c^2}$ for various color numbers $N_c$. 
The exact numerical solution fits the asymptotic
low-lying energy solution over the range
$\frac{\lambda}{N_c^2}\le\frac{\lambda_{\mbox{critical}}}{N_c^2}$
and fits the high-lying energy solution over the range
$\frac{\lambda}{N_c^2}\ge\frac{\lambda_{\mbox{critical}}}{N_c^2}$.
The asymptotic low-lying and high-lying energy solutions split 
by a small redundant constant at the critical point 
of the phase transition. 
This redundant small constant emerges due to the kind  
of approximations are considered.}
\end{figure}

In order to demonstrate the phase transition and the deflection point
for the analytic solution, we display the low and high-lying energy
solutions and their extrapolations 
in Fig.(\ref{fig_w_fun_adj_match}).
The asymptotic analytical solutions and their extrapolations
for the entire energy domain are compared with the exact numerical one.
It is shown that the low-lying energy solution fits the numerical one
for the small thermal running parameter
$\tilde{\lambda}\le\tilde{\lambda}_{\mbox{critical}}$.
When the thermal running parameter $\tilde{\lambda}$
reaches the Gross-Witten critical point it starts to deviate
from the exact numerical one significantly as $\tilde{\lambda}$
increases and runs far away from that point.
On the other hand, the high-lying energy solution 
fits the exact numerical one for the large thermal running 
parameter over the range
$\tilde{\lambda}\ge\tilde{\lambda}_{\mbox{critical}}$
while its extrapolation to small values deviates
from the exact numerical one and 
this deviation increases significantly
as the thermal running parameter $\tilde{\lambda}$ decreases
down below the Gross-Witten point and approaches the origin.
Evidently, the extrapolations of both solutions
fail to fit the exact numerical one 
when they go far away beyond the critical point 
in the opposite directions.
This demonstrates how the analytical solution 
is smoothly deflected at the critical point.
This soft deflection causes the asymptotic 
analytic low and high-lying solutions to
lie just above each other and reduces 
the intersection possibility
to the minimum in any realistic rough approximations.
Therefore, Gross-Witten point is then determined 
in the midway between the smooth interpolation 
from the low to high-lying energy domain 
or from a small to the large thermal running parameter 
$\tilde{\lambda}$.
The exact numerical solution matches
the asymptotic low-lying energy solution from the left and
then it matches the asymptotic high-lying energy solution 
from the right beneath this small interval around 
the Gross-Witten point.
However, in the case that the low-lying energy solution
is not known,
the extreme left hand critical point is determined 
by the minimum point of the high-lying energy 
solution extrapolation.
This point is the extreme left hand threshold 
of the high-lying energy solution.
This threshold point can shed the light 
on the existence of the phase transition even
for a complicated realistic physical situation 
can not be solved using the spectral density method.
The characteristic solutions of the low-lying 
and high-lying energy domains
are found not restricted 
to large $N_c\rightarrow \infty$ limit 
and is satisfactory even for a small number 
of colors $N_c=2$.

%%%%%%%%%%%%%%%%%%%%%%%%%%%%%%%%%%%%%%%%%%%%%%%%%%%%%%%
%%%%%%%%%%%%%%%%%%%%%%%%%%%%%%%%%%%%%%%%%%%%%%%%%%%%%%%
\subsection{Quark and gluon fireball\label{sec_saddles}}
The hadronic states are bound states of quarks and gluons 
confined by a color-singlet state.
The excited hadronic states are assumed to be Hagedorn states.
The Hagedorn states are approximated as bags 
of ideal quark and gluon gases
and each bag is in the color-singlet state.
In the realistic QCD, 
the quarks are fundamental particles and are satisfying 
the Fermi-Dirac statistics
while the gluons are adjoint particles and are satisfying 
the Bose-Einstein statistics.
The canonical potential in the Fock space for quark 
and anti-quark with an internal color structure 
is formalized as follows
\begin{eqnarray}
\ln Z_{q\overline{q}}(\beta,V)&=&
\frac{1}{N_c}\mbox{tr}_c\sum_{\mbox{Flavors}}\left(
\sum_{\alpha}\ln
\left[1+{\bf R}(g)e^{-\beta E_{q\alpha}}\right]
+
\sum_{\alpha}\ln\left[1+{\bf R}^*(g)
e^{-\beta E_{q\alpha}}\right]\right),
\nonumber\\
&=&
\frac{1}{N_c}
\mbox{tr}_c
\sum_{\mbox{Flavors}}
\sum_{\alpha}\ln
\left[1+\left({\bf R}(g)+{\bf R}^*(g)\right)e^{-\beta E_{q\alpha}}
+e^{-2\beta E_{q\alpha}}
\right].
\end{eqnarray}
For the highly thermal excited hadronic states, 
the color saddle points become
dominant over a tiny interval around the origin.
This highly thermal excited states correspond 
the energy domain 
$\lambda\ge
\lambda_{\mbox{critical}}\ge
\lambda_{(II)\mbox{minimum}}$ and 
the analogue solution $Z_{(II)}(\beta,V)$.
In the Hagedorn Hilbert space, 
the quark-antiquark canonical 
potential is approximated to
\begin{eqnarray}
\ln Z_{q\overline{q}}(\beta,V)
&\approx&
\sum_{\mbox{Flavors}}\sum_{\alpha}\ln
\left[1+2 e^{-\beta E_{q\alpha}}+e^{-2\beta E_{q\alpha}}
\right]
\nonumber\\
&~&
-\frac{1}{2}\frac{1}{N_c} \sum^{N_c}_{i}\theta^2_i
\left[
\sum_{\mbox{Flavors}}
\sum_{\alpha}
\frac{2 e^{-\beta E_{q\alpha}}}
{\left[1+2 e^{-\beta E_{q\alpha}}+e^{-2\beta E_{q\alpha}}
\right]}\right],
\nonumber\\
&=&
\Phi^{(0)}_{q\overline{q}}-
\frac{1}{2}\Phi^{(2)}_{q\overline{q}} \sum^{N_c}_{i}\theta^2_i,
\end{eqnarray}
where the following parameters are introduced 
in order to simplify our notations
\begin{eqnarray}
\Phi^{(0)}_{q\overline{q}}&=&
\sum_{\mbox{Flavors}}
\sum_{\alpha}\ln
\left[1+2 e^{-\beta E_{q\alpha}}+e^{-2\beta E_{q\alpha}}
\right],\nonumber\\
\Phi^{(2)}_{q\overline{q}}&=&
\frac{1}{N_c}
\sum_{\mbox{Flavors}}
\sum_{\alpha}
\frac{2 e^{-\beta E_{q\alpha}}}
{\left[1+2 e^{-\beta E_{q\alpha}}+e^{-2\beta E_{q\alpha}}
\right]}.
\end{eqnarray}
The sum over states for an ideal gas
of massless quarks and antiquarks confined in a cavity
with a sharp boundary is approximated by the integration
over the density of states as follows
\begin{eqnarray}
\Phi^{(0)}_{q\overline{q}}&=&
(2j+1)\cdot N_c n_{\mbox{flavor}}
\int \frac{V d^{3} k}{(2\pi)^{3}}
2\ln
\left[1 + e^{-\beta p}\right],
\nonumber\\
&=&
\frac{7\pi^2}{360}(2j+1) N_c n_{\mbox{flavor}}
\frac{V}{\beta^3},
\label{lambda-qq}
\end{eqnarray}
and
\begin{eqnarray}
\Phi^{(2)}_{q\overline{q}}&=&
(2j+1)\cdot N_c n_{\mbox{flavor}}
\frac{1}{N_c}\int \frac{Vd^3 p}{(2\pi)^3}
\frac{2e^{-\beta p}}{\left(1+e^{-\beta p}\right)^2}
\nonumber\\
&=&\frac{1}{6}(2j+1) n_{\mbox{flavor}}\frac{V}{\beta^3},
\label{lambda-qq-2}
\end{eqnarray}
where $V$ is the bag's volume and $(2j+1)$ comes from 
the spin degeneracy.
%%%%%%%%%%%%%%%%%%%%%%%%%%%%%%%%%%%%%%%%%%%%%%%%%%%%%%%%%%%%%%%%%%%%%%%%%%
Since the constituent adjoint gluons 
are satisfying Bose-Einstein statistics,
the canonical potential for the gas of gluons reads
\begin{eqnarray}
\ln Z_{g}(\beta,V)&=&
-\frac{1}{\mbox{dim}_g}\mbox{tr}_c\sum_{\alpha}\ln
\left[1-{\bf R}_{\mbox{adj}}(g)e^{-\beta E_{g\alpha}}\right].
\end{eqnarray}
%%%%%%%%%%%%%%%%%%%%%%%%%%%%%%%%%%%%%
The internal color structure for $SU(N_c)$ (or $U(N_c)$) symmetry
group is separated and evaluated as follows
\begin{eqnarray}
\mbox{tr}_c \left\{ \ln[1-\lambda {\bf R}_{\mbox{adj}}(g)] \right\}
&=&
\sum^{\infty}_{k=1} \lambda^k \mbox{tr}_c {\bf R}_{\mbox{adj}}(g^k),
\nonumber\\
&=& \sum^{\infty}_{k=1} \lambda^k
\left\{\left[\sum^{N_c}_{i\neq j}\cos k\left(\theta_i-\theta_j\right)
\right]
+(N_c-1)\right\},\nonumber\\
&=&
\Re \sum^{\infty}_{k=1} \lambda^k
\left\{\left[
\sum^{N_c}_{i\neq j}
e^{i k\left(\theta_i-\theta_j\right)}\right]
+(N_c-1)\right\},
\nonumber\\
&=&
\Re \left(\sum^{N_c}_{i\neq j} \ln\left[1
-\lambda
e^{i \left(\theta_i-\theta_j\right)}\right]\right)
+(N_c-1)\ln[1-\lambda].
\end{eqnarray}
%%%%%%%%%%%%%%%%%%%%%%%%%%%%%%%%%%%%%%%%%
The approximation of the gluon canonical potential
around the color saddle points, which are dominated
near the origin, becomes
\begin{eqnarray}
\ln Z_{g}(\beta,V)
&\approx&
-\sum_{\alpha}\ln
\left[1-e^{-\beta E_{g\alpha}}\right]
-\frac{1}{2}\sum^{N_c}_{n=1}\sum^{N_c}_{m=1}
\left(\theta_n-\theta_m\right)^2
\left[\frac{1}{\mbox{dim}_g}
\sum_{\alpha}\frac{e^{-\beta E_{g\alpha}}}{
\left[1-e^{-\beta E_{g\alpha}}\right]^2}\right],
\nonumber\\
&=&
\Phi^{(0)}_{g}
-\frac{1}{2}
\Phi^{(2)}_{g}
\sum^{N_c}_{n=1}\sum^{N_c}_{m=1}
\left(\theta_n-\theta_m\right)^2.
\end{eqnarray}
The following terms
\begin{eqnarray}
\Phi^{(0)}_{g}&=&\sum_{\alpha}\ln
\left[1-e^{-\beta E_{g\alpha}}\right],
\end{eqnarray}
and
\begin{eqnarray}
\Phi^{(2)}_{g}&=&
\frac{1}{\mbox{dim}_g}\sum_{\alpha}\frac{e^{-\beta E_{g\alpha}}}{
\left[1-e^{-\beta E_{g\alpha}}\right]^2},
\end{eqnarray}
are introduced for our convenience in order to simplify 
our notations.
Furthermore, we have the dimension $\mbox{dim}_g=(N^2_c-1)$
and the color constraint 
$\theta_{N_c}=-\sum^{N_c-1}_{i=1}\theta_i$
for the $SU(N_c)$ group representation.
Since the quarks and gluons are assumed to be confined in
a cavity with sharp boundary then 
the summation over the energy states
is performed as follows 
\begin{eqnarray}
\Phi^{(0)}_{g}&=&(2j+1)\mbox{dim}_g
\int \frac{V d^{3} k}{(2\pi)^{3}}
\ln
\left[1 - e^{-\beta p}\right],
\nonumber\\
&=&
\frac{\pi^2}{90}
(2j+1)\mbox{dim}_g
\frac{V}{\beta^3},
\label{lambda-g}
\end{eqnarray}
and
\begin{eqnarray}
\Phi^{(2)}_{g}&=&
(2j+1)\int \frac{V d^{3} k}{(2\pi)^{3}}
\frac{e^{-\beta k}}{\left[1-e^{-\beta k}\right]^2},
\nonumber\\
&=&\frac{1}{6}(2j+1)\frac{V}{\beta^3}.
\label{lambda-g-2}
\end{eqnarray}
%%%%%%%%%%%%%%%%%%%%%%%%%%%%%%%%%%%%%%%%%%%%%%%%%%%%
%%%%%%%%%%%%%%%%%%%%%%%%%%%%%%%%%%%%%%%%%%%%%%%%%%%%
%
The partition function for an ideal quark and gluon gas, which is
in the color-singlet state in the Hilbert space 
of Hagedorn state, is the tensor
product of Fock spaces for the quark 
and gluon partition functions.
The resultant tensor product is then projected into 
a color-singlet state as follows
\begin{eqnarray}
Z_{q\overline{q} g}(\beta,V)&=&
\int d\mu(g) Z_g(\beta,V) Z_{q\overline{q}}(\beta,V).
\end{eqnarray}
The integration over the Fourier color variables 
is evaluated by the Gaussian quadrature integration 
around the  saddle points.
In the extreme condition in the limit $\beta\rightarrow 0$, 
the color saddle points become dominant 
in a tiny range around the origin.
Hence, the canonical ensemble for the highly 
thermal excited states is then approximated
by the quadratic expansion around the saddle points as follows
\begin{eqnarray}
Z_{q\overline{q} g}(\beta,V)&\approx&
Z_{q\overline{q} g(II)}(\beta,V),
\nonumber\\
Z_{q\overline{q} g(II)}(\beta,V)&=&
\exp\left[
\Phi^{(0)}_{q\overline{q}}+\Phi^{(0)}_{g}\right]
\nonumber\\
&~&
\times\int d\mu(g)_{\mbox{saddle}}
\exp\left[
-\frac{1}{2}\Phi^{(2)}_{q\overline{q}}
\sum^{N_c}_{i}\theta^2_i
-\frac{1}{2}
\frac{\left(2N_c\Phi^{(2)}_{g}\right)}{2N_c}
\sum^{N_c}_{n=1}\sum^{N_c}_{m=1}
\left(\theta_n-\theta_m\right)^2
\right],\nonumber\\
\label{qgp_canon}
\end{eqnarray}
where the approximate Haar measure near
the saddle points is embedded in the Gaussian integration.
%%%%%%%%%%%%%%%%%%%%%%%%%%%%%%%%%%%%%%%%%%%%%%%%%%%%%%
%
After evaluating the Gaussian-like integral 
over the Fourier color variables,
the canonical ensemble becomes
\begin{eqnarray}
Z_{q\overline{q} g(II)}(\beta,V)&=&
{\cal N}_{q\overline{q} g}
\left(
\frac{1}{ 2 N_c\Phi^{(2)}_g+\Phi^{(2)}_{q\overline{q}} }
\right)^{\frac{N^2_c-1}{2}}
\exp\left[
\Phi^{(0)}_{q\overline{q}}
+\Phi^{(0)}_{g}\right],
\nonumber\\
\end{eqnarray}
where the pre-factor normalization is given by
\begin{eqnarray}
{\cal N}_{q\overline{q} g}=
\frac{1}{N_c!
(2\pi)^{N_c-1}}
\frac{(2\pi)^{N_c/2}\prod^{N_c}_{k=1}k!}
{\sqrt{2\pi N_c}}.
\label{norm_qqg}
\end{eqnarray}
In order to be in a position to compute 
the micro-canonical ensemble, it is appropriate 
to extract the thermodynamical ensembles 
$(V,\beta^3)$ in the following way,
\begin{eqnarray}
Z_{q\overline{q} g(II)}(\beta,V)&=&
{\cal N}_{q\overline{q} g}
\left(
\frac{\beta^3/V}{ 2 N_c\tilde{\Phi}^{(2)}_g+
\tilde{\Phi}^{(2)}_{q\overline{q}} }
\right)^{\frac{N^2_c-1}{2}}
\exp\left[\frac{V}{\beta^3}\left(
\tilde{\Phi}^{(0)}_{q\overline{q}}
+\tilde{\Phi}^{(0)}_{g}\right)\right].
\nonumber\\
\label{partqqg}
\end{eqnarray}
The functions
$\tilde{\Phi}^{(0)}_{q\overline{q}}$,
$\tilde{\Phi}^{(0)}_{g}$,
$\tilde{\Phi}^{(2)}_{q\overline{q}}$,
and
$\tilde{\Phi}^{(2)}_g$ are independent on $V$ and $\beta$
and they are calculated by dividing the un-tilde terms
by $\frac{V}{\beta^3}$, 
\begin{eqnarray}
\tilde{\Phi}^{(0)}_{q\overline{q}}&=&
\Phi^{(0)}_{q\overline{q}}/\frac{V}{\beta^3},\nonumber\\
\tilde{\Phi}^{(2)}_{q\overline{q}}&=&
\Phi^{(2)}_{q\overline{q}}/\frac{V}{\beta^3},\nonumber\\
\tilde{\Phi}^{(0)}_{g}&=&
\Phi^{(0)}_{g}/\frac{V}{\beta^3},\nonumber\\
\tilde{\Phi}^{(2)}_{g}&=&
\Phi^{(2)}_{g}/\frac{V}{\beta^3}.
\end{eqnarray}
%%%%%%%%%%%%%%
%

The micro-canonical ensemble is found by calculating 
the inverse Laplace transformation as follows, 
\begin{eqnarray}
Z_{q\overline{q} g(II)}(W,V)
&=&
\frac{1}{2\pi i}
\int^{\beta_c+i \infty}_{\beta_c-i \infty} d\beta e^{\beta W}
Z_{q\overline{q} g(II)}(\beta,V),
\nonumber\\
&=&
\frac{{\cal N}_{q\overline{q} g}}{2\pi i}
\int^{\beta_c+i \infty}_{\beta_c-i \infty} d\beta e^{\beta W}
\beta^{\frac{3}{2}(N^2_c-1)}
\frac{\exp\left[\frac{V}{\beta^3}
\left(\tilde{\Phi}^{(0)}_{q\overline{q}}
+
\tilde{\Phi}^{(0)}_{g}\right)\right]}
{ V^{\frac{N^2_c-1}{2}}
\left(2 N_c\tilde{\Phi}^{(2)}_g
+\tilde{\Phi}^{(2)}_{q\overline{q}}
\right)^{\frac{N_c^2-1}{2}} }.
\end{eqnarray}
It is appropriate to define the following constants
explicitly which are independent on $\beta$ and $V$,
\begin{eqnarray}
\tilde{\Phi}^{(0)}_{q\overline{q} g}&=&
\left[\tilde{\Phi}^{(0)}_{q\overline{q}}
+\tilde{\Phi}^{(0)}_{g}\right],
\nonumber\\
&=&(2j+1)\left[
\frac{7\pi^2}{360} N_c n_{\mbox{flavor}}
+
\frac{\pi^2}{90} (N^2_c-1)\right],
\end{eqnarray}
and
\begin{eqnarray}
\tilde{\Phi}^{(2)}_{q\overline{q}}
+2N_c\tilde{\Phi}^{(2)}_g
&=&
\frac{1}{6}(2j+1)
\left[n_{\mbox{flavor}}+N_c\right].
\end{eqnarray}
The Laplace transform is evaluated 
using the steepest descent method.
The Laplace saddle point is determined 
at the following point
\begin{eqnarray}
\beta_0=\left[\frac{3V}{W}
\tilde{\Phi}^{(0)}_{q\overline{q}g}
\right]^{\frac{1}{4}}.
\label{saddle_micro_qqg}
\end{eqnarray}
The micro-canonical ensemble for the Hagedorn states
corresponds the high-lying energy solution as follows
\begin{eqnarray}
Z_{\mbox{Hagedorn}}(W,V)
&\leftrightarrow&
Z_{q\overline{q}g(II)}(W,V).
\end{eqnarray}
The micro-canonical ensemble for quark and gluon
bag in the color-singlet state reads
\begin{eqnarray}
Z_{q\overline{q}g(II)}(W,V)&\approx&
\frac{1}{2\sqrt{2\pi}}
{\cal N}_{q\overline{q} g}
N_{\widetilde{\Phi}}\frac{1}{W}
\left(\frac{1}{VW^3}\right)^{\frac{N_c^2-2}{8}}
\exp\left[
\frac{4}{3}
\left(
3\tilde{\Phi}^{(0)}_{q\overline{q} g}\right)^{\frac{1}{4}}
W^{3/4} V^{1/4}
\right],
\label{qgp_micro}
\end{eqnarray}
where the pre-factor constant $N_{\tilde{\Phi}}$ is given by
\begin{eqnarray}
N_{\widetilde{\Phi}}=
\frac{
\left(
3\tilde{\Phi}^{(0)}_{q\overline{q} g}
\right)^{\frac{3N^2_c-2}{8}}
}{
\left(
2N_c\tilde{\Phi}^{(2)}_g
+\tilde{\Phi}^{(2)}_{q\overline{q}}
\right)^{\frac{N^2_c-1}{2}}
}.
\end{eqnarray}
After scaling the bag energy with respect to its volume
$x=W/V$ which is denoted as the bag energy density,
the micro-canonical ensemble becomes
\begin{eqnarray}
Z_{q\overline{q}g (II)}(x,V)
&\equiv&
\frac{1}{2\sqrt{2\pi}}
{\cal N}_{q\overline{q} g}
N_{\widetilde{\Phi}}
V^{-\frac{N^2_c}{2}} x^{-\frac{3N^2_c+2}{8}}
\exp\left[
\frac{4}{3}
\left(
3\tilde{\Phi}^{(0)}_{q\overline{q} g}
\right)^{\frac{1}{4}}
x^{3/4} V
\right].
\end{eqnarray}
%%%%%%%%%%%%%%%%%%%%%%%%%%%%%%%%%%%%%%%%%%%%%%%%%%%%%%%%%%%%%%%
%%%%%%%%%%%%%%%%%%%%%%%%%%%%%%%%%%%%%%%%%%%%%%%%%%%%%%%%%%%%%%%
%%%%%%%%%%%%%%%%%%%%%%%%%%%%%%%%%%%%%%%%%%%%%%%%%%%%%%%%%%%%%%%

%%%%%%%%%%%%%%%%%%%%%%%%%%%%%%%%%%%%%%%%%%%%%%%%%%%%%%%%%%%%%
%%%%%%%%%%%%%%%%%%%%%%%%%%%%%%%%%%%%%%%%%%%%%%%%%%%%%%%%%%%%%
%%%%%%%%%%%%%%%%%%%%%%%%%%%%%%%%%%%%%%%%%%%%%%%%%%%%%%%%%%%%%
%
% Here
%
%
%
%%%%%%%%%%%%%%%%%%%%%%%%%%%%%%%%%%%%%%%%%%%%%%%%%%%%%%%%%%%%%

In the real world of the relativistic heavy ion collisions,
the story is more complicated.
The masses for the constituent quarks even for the light flavors
do not identically vanish to zero under the Gross-Witten
phase transition line to the Hagedorn phase 
due the chiral bound interaction.
Although the constituent zero mass approximation
for the light flavors is satisfactory 
in particular for the hadronic phase 
just below the deconfinement phase transition point,
the constituent strange flavor mass
still deviates significantly from zero.
Therefore, the inclusion of massive flavors such 
as strangeness degree of freedom will drastically 
modify the numerical results quantitatively.
The calculation of the micro-canonical ensemble
becomes more complicated for the massive flavors.
The major trouble comes from the analytical calculation
for the location of the Laplace saddle point 
in the steepest descent method.
In this case, after the integration by parts
the quark and antiquark 
partition function namely,
\begin{eqnarray}
\Phi^{(0)}_{q\overline{q}}&\equiv&
\Phi^{(0)}_{q\overline{q}}(\beta,V),
\nonumber\\
\Phi^{(0)}_{q\overline{q}}(\beta,V)&=&
(2j+1) N_c n_{\mbox{flavor}}
\int \frac{V d^{3} k}{(2\pi)^{3}}
2\ln
\left[1 + e^{-\beta \sqrt{k^2+m^2}}
\right],
\end{eqnarray}
is reduced to
\begin{eqnarray}
\Phi^{(0)}_{q\overline{q}}(\beta,V)
&=&
(2j+1)N_c n_{\mbox{flavor}}
\frac{V}{\beta^3}\frac{1}{\pi^2}
\int^{\infty}_0 dk
\frac{1}{3}
\frac{\beta^4 k^3}{\sqrt{\beta^2 k^2+\beta^2 m^2}}
\frac{1}{\left[e^{\sqrt{\beta^2 k^2+\beta^2 m^2}}+1\right]},
\nonumber\\
&=&
(2j+1) N_c n_{\mbox{flavor}}
\frac{V}{\beta^3}\frac{1}{\pi^2}
I_{\Phi^{(0)}_{q\overline{q}}}(m\beta),
\end{eqnarray}
where the integration term reads
\begin{eqnarray}
I_{\Phi^{(0)}_{q\overline{q}}}(m\beta)=
\int^{\infty}_0 dx
\frac{1}{3}
\frac{x^3}{\sqrt{x^2+\beta^2 m^2}}
\frac{1}
{\left[e^{\sqrt{x^2+\beta^2 m^2}}+1\right]}.
\end{eqnarray}
The same thing can also be done for the second order function
$\Phi^{(2)}_{q\overline{q}}$.
When the mixed-grand canonical ensemble is transformed
to the micro-canonical ensemble, the massive flavor
produces a complicated transcendental function for the stationary
Laplace saddle point solution.
Generally speaking,
it is possible to solve the transcendental equation by the iteration.
At first, the solution for the zero$^{\mbox{th}}$ iteration
is assumed to be $\beta^{(\underline{0})}_0 m=0$.
Then in the first iteration, we solve the stationary
saddle point $\beta_0=\beta^{(\underline{1})}_0$.
In the second iteration, the massive flavor is considered
explicitly and then we repeat the iteration
to the higher order until the convergence is achieved. 
This procedure is summarized as follows,
\begin{eqnarray}
\beta m&\rightarrow& 0,\nonumber\\
\beta m&\rightarrow& \beta^{(\underline{0})}_0 m,\nonumber\\
\beta m&\rightarrow& \beta^{(\underline{1})}_0 m,\nonumber\\
\vdots .
\end{eqnarray}
Nonetheless, it is expected that the convergence
is achieved very rapidly and subsequently 
the first order iteration
is an adequate and sufficient approximation for light flavors
including the strangeness.
Therefore, the first order iteration truncation
$\beta m\rightarrow \beta^{(\underline{1})}_0 m$
is an appropriate one 
for the relativistic heavy ion collisions
\begin{eqnarray}
\Phi^{(0)}_{q\overline{q}}(\beta,V)&=&
(2j+1) N_c n_{\mbox{flavor}}
\frac{V}{\beta^3}\frac{1}{\pi^2}
I_{\Phi^{(0)}_{q\overline{q}}}
\left(\beta^{(\underline{0})}_0 m\right).
\end{eqnarray}
%%%%%%%%%%%%%%%%%%%%%%%%%%%%%%%%%%%%%%%%%%%%%%%%%%%%
The value of the point $\beta^{(\underline{0})}_0$
can be simplified 
in the context of the standard MIT bag model. 
After considering
the bag's total energy $m=W+BV$ 
and the volume $V=m/4B$ constraints,
the stationary point becomes
\begin{eqnarray}
\beta^{(\underline{0})}_0&=&\left[
\left(\tilde{\Phi}^{(0)}_{q\overline{q}}
+
\tilde{\Phi}^{(0)}_{g}\right)/B\right]^{\frac{1}{4}},
\nonumber\\
&=&\frac{1}{T_{\beta}}.
\end{eqnarray}
The term $T_{\beta}$ can be interpreted 
as an effective temperature for the screening mass.
Alternatively, in order to overcome the possible complication 
in the micro-canonical ensemble, 
it is possible to assume the following ansatz
\begin{eqnarray}
\beta m\rightarrow \frac{m}{T},
~~\mbox{and}~~ T\beta\sim {\cal O}(1),
\end{eqnarray}
for the high temperature approximation.
This assumption simplifies the mixed-grand 
canonical ensemble significantly,
\begin{eqnarray}
\Phi^{(0)}_{q\overline{q}}(\beta,V)&=&
(2j+1)N_c n_{\mbox{flavor}}
\frac{V}{\beta^3}\frac{1}{\pi^2}
I_{\Phi^{(0)}_{q\overline{q}}}\left(\frac{m}{T}\right).
\end{eqnarray}
An alternate assumption regarding 
the highly compressed hadronic matter reads
\begin{eqnarray}
\beta m\propto \frac{1}{\mu} m.
\end{eqnarray}
Nonetheless, the density of states 
for the highly compressed matter 
at low temperature needs more consideration.
In this regime the physics is rich  due to
the configuration space, flavor and/or flavor-color correlations.
Furthermore, a new equation of state is expected to emerge 
due to the formation of the color superconductivity. 
%%%%%%%%%%%%%%%%%%%%%%%%%%%%%%%%%%%%%%%%%%%%%%%%%%%%
%%%%%%%%%%%%%%%%%%%%%%%%%%%%%%%%%%%%%%%%%%%%%%%%%%%%%%%%%%%%%%%%%%%%%%%%%%%%%
%%%%%%%%%%%%%%%%%%%%%%%%%%%%%%%%%%%%%%%%%%%%%%%%%%
%%%%%%%%%%%%%%%%%%%%%%%%%%%%%%%%%%%%%%%%%%%%%%%%%%
%%%%%%%%%%%%%%%%%%%%%%%%%%%%%%%%%%%%%%%%%%%%%%%%%%

%%%%%%%%%%%%%%%%%%%%%%%%%%%%%%%%%%%%%%%%%%%%%%%%%%%%%%%%%%%%%%%
On the other hand in the limit of low-lying energy 
$\beta\gg 1$
(i.e. the diluted and relatively cold matter 
which is analogous to $Z_{(I)}(\lambda)$ solution
where
$\lambda\le
\lambda_{\mbox{critical}}\le\lambda_{(I)\mbox{max}}$),
the partition function for an ideal 
gas of quarks and antiquarks can be approximated to
\begin{eqnarray}
\ln Z_{q\overline{q}}(\beta,V)&=&
\mbox{tr}_c
\sum_{\mbox{flavor}}
\sum_{\alpha}\ln
\left[1+{\bf R}(g)e^{-\beta E_{q\alpha}}\right]
+
\mbox{tr}_c
\sum_{\mbox{flavor}}
\sum_{\alpha}\ln\left[1+{\bf R}^*(g)
e^{-\beta E_{q\alpha}}\right],\nonumber\\
&\approx&
\sum_{\mbox{flavor}}
\left[\mbox{tr}_c{\bf R}(g)
\sum_{\alpha}
e^{-\beta E_{q\alpha}}
+
\mbox{tr}_c{\bf R}^*(g)
\sum_{\alpha}
e^{-\beta E_{q\alpha}}\right],\nonumber\\
&=&
\sum_{\mbox{flavor}}
\left(\sum_{\alpha}e^{-\beta E_{q\alpha}}\right)
\mbox{tr}_c\left[{\bf R}(g)+{\bf R}^*(g)\right].
\end{eqnarray}
The low-lying energy limit for the gluon gas reads
\begin{eqnarray}
\ln Z_{g}(\beta,V)&=&
-\mbox{tr}_c\sum_{\alpha}\ln
\left[1-{\bf R}_{\mbox{adj}}(g)e^{-\beta E_{g\alpha}}\right],
\nonumber\\
&\approx&
\mbox{tr}_c\sum_{\alpha}
{\bf R}_{\mbox{adj}}(g)e^{-\beta E_{g\alpha}},
\nonumber\\
&=&
\left(\sum_{\alpha}e^{-\beta E_{g\alpha}}\right)
\mbox{tr}_c{\bf R}_{\mbox{adj}}(g).
\end{eqnarray}
By substituting the following thermal running parameters
\begin{eqnarray}
\lambda_{q\overline{q}}&=&
\sum_{\mbox{flavor}}
\sum_{\alpha}
e^{-\beta E_{q\alpha}},\nonumber\\
\lambda_g&=&
\sum_{\alpha}e^{-\beta E_{g\alpha}},
\end{eqnarray} 
we get the partition function
\begin{eqnarray}
Z_{q\overline{q} g(I)}(\lambda_{q\overline{q}},\lambda_g)=
\int^{\pi}_{-\pi} d\mu(g)
e^{\lambda_g\frac{1}{\mbox{dim}_g}
\mbox{tr}_c{\bf R}(g)\mbox{tr}_c{\bf R}^*(g)}
e^{\lambda_{q\overline{q}}\frac{1}{N_c}
\mbox{tr}_c\left({\bf R}(g)+{\bf R}^*(g)\right)},
\end{eqnarray}
for the $SU(N_c)$ group representation.
This equation is similar to Eq.(\ref{fund_adj_small_exact}).
The result reads
\begin{eqnarray}
Z_{q\overline{q}g(I)}(\lambda_{q\overline{q}},\lambda_g)=
\frac{1}{1-\frac{\lambda_{g}}{(N^2_c-1)}}
\exp\left[\lambda^2_{q\overline{q}}/N^2_c\right],
\label{low-qqg-canonical}
\end{eqnarray}
where
$\lambda_{q\overline{q}}=
\frac{N_c}
{\pi^2}\frac{(2j+1)V}{\beta^3} n_{\mbox{flavor}}$
for massless flavors
and
$\lambda_{g}=\frac{(N^2_c-1)}{\pi^2}\frac{(2j+1)V}{\beta^3}$.
This approximation is an appropriate solution over
the energy domain 
$\lambda_{q\overline{q}}\le
\lambda_{q\overline{q}\mbox{critical}}\le
\lambda_{q\overline{q}(I)\mbox{max}}$
and
$\lambda_{g}\le
\lambda_{g\mbox{critical}}\le
\lambda_{g(I)\mbox{max}}$.
This result is analogous to the result 
given in Eq.(\ref{fun_adj_res1}).
Furthermore, we have for massive flavors 
the following thermal $q\overline{q}$-running 
parameter
\begin{eqnarray}
\lambda_{q\overline{q}}&=&
N_c n_{\mbox{flavor}} \frac{(2j+1)V}{\beta^3}
\frac{m_q^2\beta^2}{2\pi^2} K_2(m_q\beta),
\end{eqnarray}
where $K_2(x)$ is Bessel function of second kind.
In the case of the low-lying energy with massive flavors 
in the limit $\beta m_q\gg 1$, 
the $q\overline{q}$-running parameter becomes
\begin{eqnarray}
\lambda_{q\overline{q}}&\approx&
N_c n_{\mbox{flavor}}
\frac{(2j+1)V}{\beta^3}
\left(\frac{m_q\beta}{2\pi}\right)^{3/2} e^{-\beta m_q}.
\end{eqnarray}
Nonetheless, in the realistic case, 
the density of states derived 
from the micro-canonical transformation of the low-lying 
mixed-canonical ensemble 
given by Eq.(\ref{low-qqg-canonical})
will be found in Section \ref{sect_micro_canonical} 
to be replaced by the discrete low-lying mass spectrum.

%%%%%%%%%%%%%%%%%%%%%%%%%%%%%%%%%%%%%%%%%%%%%%%%%%%%%%%%%%%%%%%%
%%%%%%%%%%%%%%%%%%%%%%%%%%%%%%%%%%%%%%%%%%%%%%%%%%%%%%%%%%%%%%%%

%%%%%%%%%%%%%%%%%%%%%%%%%%%%%%%%%%%%%%%%%%%%%%%%%
%%%%%%%%%%%%%%%%%%%%%%%%%%%%%%%%%%%%%%%%%%%%%%%%%
%%%%%%%%%%%%%%%%%%%%%%%%%%%%%%%%%%%%%%%%%%%%%%%%%
%%%%%%%%%%%%%%%%%%%%%%%%%%%%%%%%%%%%%%%%%%%%%%%%%
\section{Micro-canonical Ensemble as a Density of
states\label{sect_micro_canonical}}
The micro-canonical ensemble becomes known after calculating
the inverse Laplace transform of the mixed-canonical ensemble,
\begin{eqnarray}
Z(W,V)&=&\frac{1}{2\pi i}
\int^{\beta_c+i \infty}_{\beta_c-i \infty}d\beta
e^{\beta W} Z(\beta,V),\nonumber\\
&=&\frac{1}{2\pi i}
\int^{\beta_c+i \infty}_{\beta_c-i \infty}d\beta
\exp\left[
\beta W+\ln Z(\beta,V)
\right].
\end{eqnarray}
The above integral can be computed
by the steepest descent method. In this method the integral
is approximated to the Gaussian-like integral around the saddle point.
The general result reads
\begin{eqnarray}
Z(W,V)&=&\frac{1}{2}\frac{1}{\sqrt{2\pi}}
\frac{
\exp\left[
\overline{\beta} W+\ln Z(\overline{\beta},V)
\right]
}
{\left[
\left(\frac{1}{Z(\beta,V)}
\frac{\partial Z(\beta,V) }{\partial \beta}
\right)^2-
\frac{1}{Z(\beta,V)}
\frac{\partial^2 Z(\beta,V) }{\partial \beta^2}
\right]^{1/2}_{\beta=\overline{\beta}}},
\end{eqnarray}
where the saddle point $\overline{\beta}$ of the stationary
condition is determined at the extremum point
\begin{eqnarray}
W+\frac{1}{Z(\overline{\beta},V)}
\left.\frac{\partial Z(\beta,V) }{\partial \beta}
\right|_{\beta=\overline{\beta}}
=0.
\end{eqnarray}
In the standard MIT bag model, the density of states
is determined for constituent particles confined 
in a cavity with a sharp boundary.
In a realistic model, it is possible to deform the cavity's
boundary. The $\delta$-function for the sharp boundary in the standard
MIT bag model can be smeared by the Gaussian smoothing function.
The extreme conditions in the relativistic heavy ion collisions
may smooth the sharp boundary for the quark and gluon bag.
Therefore, the micro-canonical ensemble for the gas of bags
with extended boundaries is approximated to
\begin{eqnarray}
Z(W,V)&=&\int d v
\delta\left(v-V\right) Z(W,v),
\nonumber\\
&=&\int d v
f_{\mbox{smooth}}\left(v-V\right) Z(W,v),
\end{eqnarray}
where the boundary surface is extended by
the Gaussian smoothing function
\begin{eqnarray}
f_{\mbox{smooth}}(v)
&=&\sqrt{\frac{\pi}{\Delta/W}}
e^{-\frac{\Delta}{W} v^2(1-\frac{v_0}{v})^2}.
\end{eqnarray}
The bag's boundary becomes more extended with respect 
to the energy
when
the bag's energy increases with respect to temperature,
while on the contrary, the surface turns to be sharper
for the low energy. 
The bag volume is proportional to its energy
$v_0\propto W$. In the low energy limit,
the bag's boundary is reduced to a $\delta$-function
\begin{eqnarray}
\lim_{W\rightarrow 0} f_{\mbox{smooth}}(v)=\delta(v-v_0).
\end{eqnarray}
Hence, the Hagedorn's density of states for the 
standard MIT bag with a sharp surface boundary reads
\begin{eqnarray}
\rho_{(II)}(m,v)&=&\delta\left(v-\frac{m}{4B}\right) Z(W,v),
\end{eqnarray}
where $W=m-Bv$. 
The micro-canonical ensemble $Z(W,v)$ is given by 
Eq.(\ref{qgp_micro}).
The density of states can be re-written as follows
\begin{eqnarray}
\rho_{(II)}(m,v)&=&\delta\left(m-4Bv\right) Z(m,v).
\end{eqnarray}
The generalization of the micro-canonical
ensemble to take into account the inclusion
of volume variation reads
\begin{eqnarray}
\rho_{(II)}(W,v)=f_{\mbox{smooth}}(v) Z(W,v).
\end{eqnarray}
The volume fluctuation effect is studied in details
in Ref.\cite{Zakout2006}
and the bag stability needs a further investigation.
%
%
%

%%%%%%%%%%%%%%%%%%%%%%%%%%%%%%%%%%%%%%%%%%%%%%%%%%%%%%%%%%%%%%%%%%%%%%%%%%%%%%
%%%%%%%%%%%%%%%%%%%%%%%%%%%%%%%%%%%%%%%%%%%%%%%%%%%%%%%%%%%%%%%%%%%%%%%%%%%%%%
\subsection{The critical mass for the bag consisting of
fundamental particles obeying Maxwell-Boltzmann statistics}

The canonical ensemble for a gas with the internal color 
symmetry $U(N_c)$ reads
\begin{eqnarray}
Z(\beta,V)&\approx&Z_{(I)}(\beta,V),
\nonumber\\
Z_{(I)}(\beta,V)&=&
e^{N^2_c \widetilde{\lambda}^2}, \widetilde{\lambda}\le  
\frac{1}{2}\le\lambda_{(I)\mbox{max}}\ll 1,
\end{eqnarray}
and
\begin{eqnarray}
\nonumber\\
Z(\beta,V)&\approx&Z_{(II)}(\beta,V),
\nonumber\\
Z_{(II)}(\beta,V)&=&
\left(\frac{1}{2\widetilde{\lambda}}\right)^{\frac{N^2_c}{2}}
e^{\left(2N^2_c \widetilde{\lambda}-\frac{3}{4}N^2_c\right)}
\left(@~~
\widetilde{\lambda}\ge 
\frac{1}{2}\ge\widetilde{\lambda}_{(II)\mbox{min}}\right),
\end{eqnarray}
for the low- and high-lying energy solutions, respectively.
The effective thermal running parameter
$\widetilde{\lambda}$ for an ideal gas
of fundamental particles embedded in the thermal bath and
satisfying Maxwell-Boltzmann statistics reads
\begin{eqnarray}
\widetilde{\lambda}&=&\lambda/N^2_c,
\nonumber\\
&=&\frac{1}{N_c}\left(2j+1\right)
\left.\int \frac{Vd^3p}{(2\pi)^3}e^{-\beta
\sqrt{p^2+m^2}}\right|_{m=0},\nonumber\\
       &=&\frac{1}{\pi^2}\frac{(2j+1)}{N_c} \frac{V}{\beta^3}.
\end{eqnarray}
The low temperature $1/\beta$ phase has a free energy
of order one $O(N^0_c)$.
The high temperature phase has 
a free energy of order $O(N_c^2)$.
It is  characterized by the Hagedorn growth
in its density of states.

The micro-canonical ensemble is computed by calculating
the inverse Laplace transform as follows
\begin{eqnarray}
Z_{(II)}(W,V)&=&\lim_{W\rightarrow \infty}\frac{1}{2\pi i}
\int^{\beta_c+i \infty}_{\beta_c-i \infty}d\beta
e^{\beta W} Z_{(II)}(\beta,V)
~~\left(@~~ 
\widetilde{\lambda}\ge\frac{1}{2}\ge
\widetilde{\lambda}_{(II)\mbox{min}}\right),
\nonumber\\
\nonumber\\
&=&
\frac{1}{2\sqrt{2\pi}}\frac{1}{\sqrt{12}N_c}
\left[\frac{\pi^2 N_c}{2(2j+1)}\right]^{\frac{N^2_c+1}{2}}
\beta_0^{(3N^2_c+5)/2} V^{-(N^2_c+1)/2}
\nonumber\\
&~&
~~~~~~~\times
\exp\left[
\frac{4}{3}\left(\frac{6(2j+1)N_c}{\pi^2}\right)^{1/4}
V^{1/4}W^{3/4}
-\frac{3}{4}N^2_c
\right].
\end{eqnarray}
The stationary point for the steepest descent 
method is found at
\begin{eqnarray}
\beta_0&=&\left(\frac{6(2j+1)N_c}{\pi^2}
\frac{V}{W}\right)^{\frac{1}{4}}.
\end{eqnarray}
In the micro-canonical representation,
the effective thermal running parameter 
and the energy constraint
for the highly excited states are reduced to the following
\begin{eqnarray}
\widetilde{\lambda}&=&
\frac{1}{6N^2_c}
\left(
\frac{6 (2j+1) N_c}{\pi^2}
\right)^{1/4}
W^{3/4}V^{1/4},\nonumber\\
&\ge&\frac{1}{2}\ge
\widetilde{\lambda}_{(II)\mbox{min}}.
\label{mass-critical-MB}
\end{eqnarray}
In the standard MIT bag model,
the effective bag's mass and volume are given by 
the relations
\begin{eqnarray}
m&=&W+BV,\\
m&=&4BV,\\
W&=&\frac{3}{4}m,\\
V&=&\frac{m}{4B}.
\end{eqnarray}
In the present model, the Hagedorn states 
appear in the hadronic mass spectrum
when the energy reaches the threshold 
$\widetilde{\lambda}\ge 
\widetilde{\lambda}_{\mbox{(pre-)critical}}
=\frac{1}{2}
\ge\widetilde{\lambda}_{(II)\mbox{min}}$
and subsequently the Hagedorn mass threshold exceeds the limit
\begin{eqnarray}
m&\ge&  2\sqrt{2\pi}N_c^{(1+3/4)}
\left[\frac{2 B}{(2j+1)}\right]^{1/4},
\nonumber\\
        &\ge& 6.2 \mbox{GeV}~~
\mbox{with}~~ B^{1/4}= 180 \mbox{MeV}.
\end{eqnarray}
This threshold mass is relatively large due to 
the classical Maxwell-Boltzmann statistics 
that is considered in the present scenario.

On the other hand, the situation is rather different for
the low-lying energy 
for the small thermal running parameter
$\widetilde{\lambda}\le\frac{1}{2}
\le\widetilde{\lambda}_{(I)\mbox{max}}$ 
(usually $\widetilde{\lambda}_{(I)\mbox{max}}<1$), 
where the micro-canonical ensemble is evaluated as follows
\begin{eqnarray}
Z_{(I)}(W,V)&=&\frac{1}{2\pi\imath}
\int^{\beta_c+i \infty}_{\beta_c-i \infty}d\beta
e^{\beta W} Z_{(I)}(\beta,V),\nonumber\\
&=&
\frac{1}{2\pi\imath}
\int^{\beta_c+i \infty}_{\beta_c-i \infty}d\beta
e^{\beta W}
e^{\left[d^2_{q}\frac{V^2}{\beta^6}\right]},
\end{eqnarray}
where $d_{q}=\frac{(2j+1)}{\pi^2}$.
In the standard MIT bag, we have the effective bag energy
$W\approx \frac{3}{4} m$.
For the energy excitation less than the threshold
$\widetilde{\lambda}\le\frac{1}{2}
\le\widetilde{\lambda}_{(I)\mbox{max}}$,
we have the low-lying hadronic
mass spectrum  and this low-lying spectrum ends 
when the energy scale reaches the critical point
\begin{eqnarray}
m&\le&  
m\left(
\widetilde{\lambda}_{\mbox{(pre-)critical}} 
\right): 6.2 \mbox{GeV},
~~~ \mbox{with}~~ 
B^{1/4}= 180 \mbox{MeV},
\nonumber\\
&\ll&m_{(I)\mbox{max}}: 10 \mbox{GeV},
\end{eqnarray}
where the upper limit $m_{(I)\mbox{max}}$
will be given 
by Eq.(\ref{m_low_lying_max_constraint}).
When the energy scale exceeds this point,
the micro-canonical ensemble
for the low-lying hadronic states becomes an inappropriate.
The extrapolation of the steepest descent approximation leads
to the density of states,
\begin{eqnarray}
Z_{(I)}(W,V)&=&
\frac{1}{2\sqrt{2\pi}}
\frac{{\left[6 d^2_q\right]}^{1/14}}
{\sqrt{7} V^{-1/7} W^{4/7}}
\exp\left[\frac{7}{6}[6d^2_{q}]^{1/7} V^{2/7}W^{6/7}
\right],
\nonumber\\
&\sim& (\cdots) V^{1/7} W^{-4/7} 
\exp\left[(\cdots) V^{2/7}W^{6/7}\right]
\left(
m\le m_{(I)\mbox{max}}
\right).
\label{low-lying-hagedorns}
\end{eqnarray}
This approximation is not valid
for small $W$ or $m\le m_{(I)\mbox{max}}$ 
if the low-lying spectrum is a discrete one as will be found
in the realistic physical situation
in section~\ref{mass_fire_critical}.
We shall assume that 
$m_{(I)\mbox{max}}\ge m_{\mbox{Critical}}
\sim m_{\mbox{physical}}$
where $m_{\mbox{physical}}\approx 2.0$ GeV 
is the maximum mass for the known physical 
particles found experimentally 
without strangeness\cite{databook2004}.
For the low-lying hadronic mass spectrum, 
we have considered the density of states 
for the known mass spectrum particles
\begin{eqnarray}
Z_{\mbox{Low-Lying}}(W,V)
&\leftrightarrow&Z_{(I)}(W,V),
\nonumber\\
\rho_{\mbox{Low-Lying}}(W,V)&\propto&
Z_{\mbox{Low-Lying}}(W,V),
\nonumber\\
&\approx&
\sum^{m_{\mbox{Critical}}}_{m_i,v_i}
\delta(m-m_i)\delta(v-v_i).
\end{eqnarray}

Nonetheless, the micro-canonical extrapolation 
given in Eq.(\ref{low-lying-hagedorns}) 
is useful to show the discrete hadronic
low-lying mass spectrum is essential.
If the standard MIT bag model
with a sharp surface boundary is applied 
to Eq.(\ref{low-lying-hagedorns}), no phase transition 
to an explosive quark-gluon plasma takes place.
To see this let us adopt the standard MIT bag approximation 
where $W\propto V$. 
This consideration leads to a continuous density of states 
$\rho\sim(\cdots)  v^{-3/7} e^{(\cdots) v^{8/7}}$. 
This density leads to a divergence 
in the isobaric partition function 
and this standard approximation 
of the MIT bag for the low-lying mass spectrum fails.
It is evident that the Hagedorn states do not appear 
in the low-lying mass spectrum limit.
Furthermore, if we argue that the continuous density 
of states grows exponentially as a linear exponential 
growth with respect to the hadron's excluded volume,
it is possible to extrapolate 
$V^{2/7} W^{6/7}\propto V$ or $W\propto V^{5/6}$.  
With this approximation, the density of states 
is reduced roughly speaking to 
$\rho\sim(\cdots) v^{-1/3} e^{(\cdots) v}$.
This density gives no phase transition 
to the quark-gluon plasma.
The both above approximations show that 
continuous low-lying mass spectrum 
can not generate explosive bags. 
%%%%%%%%%%%%%%%%%%%%%%%%%%%%%%%%%%%%%%%%%%%%%%%%%%%%%%%%%%%%%%%%
%%%%%%%%%%%%%%%%%%%%%%%%%%%%%%%%%%%%%%%%%%%%%%%%%%%%%%%%%%%%%%%%
%
%
%
%
%
%

\subsection{The critical mass for the quark and gluon fireball
(e.g. Hagedorn state threshold)
\label{mass_fire_critical}}

In QCD, Hagedorn states are assumed to be
a bag of weakly interacting quarks and gluons
confined in a color-singlet state. 
The canonical ensemble for a quark and gluon gas
in the color-singlet state is given by Eq.(\ref{partqqg}).
The logarithm of the canonical partition function is simplified
as follows
\begin{eqnarray}
\log\left( Z_{q\overline{q} g}\left(V/\beta^3\right)\right)
&\approx&
\log\left( Z_{q\overline{q} g(II)}\left(V/\beta^3\right) \right)
~~\left(\mbox{for the energy domain}~ 
\lambda_{q\overline{q}g}\ge
\lambda_{q\overline{q}g(II)\mbox{min}}\right),
\nonumber\\
\log\left( Z_{q\overline{q} g(II)}
\left(V/\beta^3\right) \right)
&=&
\frac{V}{\beta^3}
\left[\tilde{\Phi}^{(0)}_{q\overline{q}}
+\tilde{\Phi}^{(0)}_{g}\right]
-\frac{N^2_c-1}{2}\log\frac{V}{\beta^3}
+\mbox{constant},
\end{eqnarray}
for the energy domain 
$\lambda_{q\overline{q}g}\ge
\lambda_{q\overline{q}g(II)\mbox{min}}$ where 
$\lambda_{q\overline{q}g}\propto V/\beta^3$.
The critical point for the phase transition is determined
by finding the extremum left-hand point as follows
\begin{eqnarray}
\frac{\partial}{\partial \left(\frac{V}{\beta^3}\right)}
\log Z_{q\overline{q} g(II)}
\left(V/\beta^3\right)=0.
\end{eqnarray}
The parameter $\left(\frac{V}{\beta^3}\right)$ is related to
the running coupling parameter $\lambda$ and it determines 
the threshold point of
a possible phase transition to the Hagedorn phase.
The extreme left-hand side location for 
the prospective phase transition is located at
\begin{eqnarray}
\frac{1}{N^2_c-1}\left(\frac{V}{\beta^3}\right)
\left[\tilde{\Phi}^{(0)}_{q\overline{q}}
+\tilde{\Phi}^{(0)}_{g}\right]\ge \frac{1}{2}.
\end{eqnarray}
Below the minimum left-hand side point, the solution must
be already deflected to another analytical solution
and in this sense the critical point is bounded from below
by the extreme  left-hand side point.
The minimum limit for the threshold Hagedorn mass 
(i.e. $m_{(II)\mbox{min}}$ where 
$m_{\mbox{critical}}\ge m_{(II)\mbox{min}}$) 
for the phase transition
from the low-lying mass spectrum
to the highly excited spectrum of fireballs
or more precisely the continuous Hagedorn states
in the context of the micro-canonical ensemble
is given by the condition
\begin{eqnarray}
\frac{1}{3(N^2_c-1)}\left[3
\left(\tilde{\Phi}^{(0)}_{q\overline{q}}
+\tilde{\Phi}^{(0)}_{g}\right)
VW^3\right]^{\frac{1}{4}}\ge\frac{1}{2}.
\end{eqnarray}
The stationary point of the steepest descent method 
for the inverse Laplace transform
to the micro-canonical ensemble
is determined by Eq.(\ref{saddle_micro_qqg}).
In the context of standard MIT bag model with
a sharp boundary surface, the point of the critical point
is determined by the constraint
\begin{eqnarray}
\left[3VW^3\right]^{1/4}=\frac{3m}{4B^{1/4}}.
\end{eqnarray}
The minimum limit for the threshold Hagedorn 
mass production in
the context of the MIT bag with the sharp surface boundary
approximation is given by the following constraint
\begin{eqnarray}
&~&\frac{(2j+1)^{1/4}}{2(N^2_c-1)}\left[
\frac{7\pi^2}{360} N_c n_{\mbox{flavor}}
+\frac{\pi^2}{90}(N^2_c-1)\right]^{1/4}\frac{m}{B^{1/4}}
\ge 1,
\nonumber\\
&~&\rightarrow m\ge m_{(II)\mbox{min}}
~~(\mbox{the resultant constraint}).
\label{mass-minimum-constraint}
\end{eqnarray}
Hence, the threshold Hagedorn mass production
for $N_c=3$ and $B^{1/4}=180 \mbox{MeV}$
is given by
\begin{eqnarray}
m&\ge&m_{(II)\mbox{min}},\nonumber\\
m&\ge& 2206 \mbox{MeV}(\mbox{one massless flavor}),
\nonumber\\
&\ge& \underline{2029 \mbox{MeV}(\mbox{two massless flavors}}),
\nonumber\\
&\ge& 1906 \mbox{MeV}(\mbox{three massless flavors}).
\label{mass-minimum-limit}
\end{eqnarray}
The critical point for the phase transition 
from the discrete low-lying mass spectrum 
to the continuous Hagedorn density of states
is located just above the maximum mass 
of the known hadronic particles found 
in the Data book\cite{databook2004}.
The highest experimental non-strange hadron state
is roughly estimated to 
$m_{\mbox{max}}\sim 2.0-2.3$ GeV.
 The critical point exists above the minimum point solution
of the Hagedorn threshold 
$m_{\mbox{critical}}\ge m_{(II)\mbox{min}}$.
The estimation of the Hagedorn's critical mass threshold 
will be given in Eqs.(\ref{mass-critical-constraint})
and (\ref{mass-critical-limit}).

The low-lying mass spectrum is considered
a discrete density of states
that includes all the known mass spectrum particles
\begin{eqnarray}
Z_{\mbox{Low-lying}}(W,V)&=&
Z_{q\overline{q}g (I)}(W,V),
\nonumber\\
\rho_{\mbox{Low-lying}}(m,v)&\sim&\sum^{m_{\mbox{max}}}_{m_i,v_i} 
\delta(m-m_i)\delta(v-v_i).
\end{eqnarray}
Using the following connections between 
the canonical ensemble and the density of the states
\begin{eqnarray}
\rho(W,V)&\propto&
Z_{q\overline{q} g(I)}(W,V)+Z_{(q\overline{q} g(II)}(W,V),
\nonumber\\
&\propto&
Z_{\mbox{Low-Lying}}(W,V)+Z_{\mbox{Fireballs}}(W,V),
\end{eqnarray}
the density of states for 
the entire hadronic states is approximated to
\begin{eqnarray}
\rho(m,v)&\approx&
\sum^{m_{\mbox{max}}}_{m_i,v_i} \delta(m-m_i)\delta(v-v_i)
\nonumber\\
&~&
+Z_{\mbox{Fireballs}}(W,v)\times
\left(\mbox{volume-mass fluctuation}\right),
\end{eqnarray}
where the micro-canonical ensemble 
$Z_{q\overline{q}g(II)}(W,V)$ 
is the continuous mass spectrum 
for the Hagedorn states (i.e. $Z_{\mbox{Fireballs}}(W,V)$).
In the standard MIT bag model with the sharp surface boundary
the volume-mass fluctuation is determined by the relation
$\delta(m-4BV)$. 
For any fuzzy bag model, the volume-mass fluctuation 
becomes nontrivial and may change 
the order of phase transition.

%%%%%%%%%%%%%%%%%%%%%%%%%%%%%%%%%%%%%%%%%%%%%%%%%%%%%%%%%%%%%%%%
%%%%%%%%%%%%%%%%%%%%%%%%%%%%%%%%%%%%%%%%%%%%%%%%%%%%%%%%%%%%%%%%
%
%
%

\section{Density of states\label{sect_dens_statistics}}

The density of states can be evaluated by using 
the phase-space integral.
The calculation for an ideal gas of particles in finite size 
has been evaluated
and studied without the internal color 
structure\cite{Kapusta1981a}. 
We use the same procedure to re-derive the density of states 
reached in the previous sections
in particular section~\ref{sec_saddles}.
The density of states from the phase-space integral reads,
\begin{eqnarray}
\rho(W,V)=\sum^{\infty}_{n=2}
\frac{N^n_c(2j+1)^n}{n!}
\left[\prod^{n}_{i=1}\int \frac{Vd^3p_i}{(2\pi)^3}\right]
\delta\left(W-\sum^{n}_{i=1}\epsilon(\vec{p}_i)\right),
\end{eqnarray}
where the constituent particle's energy
$\epsilon(\vec{p}_i)=\sqrt{\vec{p}^2_i+m^2}$ 
and it is reduced to
$\epsilon(\vec{p}_i)\doteq |\vec{p}|$ 
for the massless one.
Here $j$ is the particle's quantum number and
$V$ and $W$ are the bag's volume and energy,
respectively.
The internal color symmetry is embedded in the phase-space
integral as follows,
\begin{eqnarray}
\rho(W,V)&=&\sum^{\infty}_{n=2}
\frac{(2j+1)^n}{n!}
\left[\prod^{n}_{i=1}\mbox{tr}
\left[{\bf R}(g)+{\bf R}^*(g)\right]
\int \frac{Vd^3p_i}{(2\pi)^3}\right]
\delta\left(W-\sum^{n}_{i=1}|\vec{p}_i|\right),
\nonumber\\
&=&
\sum^{\infty}_{n=2}
\frac{(2j+1)^n}{n!}
\left(\mbox{tr}\left[{\bf R}(g)+{\bf R}^*(g)\right]\right)^n
\left(\frac{V}{2\pi^2}\right)^n \frac{2^n W^{3n-1}}{(3n-1)!},
\nonumber\\
&=&\frac{1}{W}
\sum^{\infty}_{n=2}
\frac{1}{n!(3n-1)!} \left(a_{\{W^3V\}}({\bf R})\right)^n.
\label{phase-space-color}
\end{eqnarray}
In Eq.(\ref{phase-space-color}),
the following function has been introduced
\begin{eqnarray}
a_{\{W^3V\}}({\bf R})&=&
\tilde{a}_{\{W^3V\}}\mbox{tr}
\left[{\bf R}(g)+{\bf R}^*(g)\right],\nonumber\\
&=&
\pi^2
\tilde{a}_{\{W^3V\}}\mbox{tr}
\left({\bf R}(g)\int \frac{dx}{2\pi^2} x^2 e^{-x}
+{\bf R}^*(g)\int \frac{dx}{2\pi^2} x^2 e^{-x}\right),
\label{aw3v}
\end{eqnarray}
where
\begin{eqnarray}
\tilde{a}_{\{W^3V\}}=\frac{(2j+1)}{\pi^2}
\left[W^3 V\right].
\end{eqnarray}
The summation over the number $n$ is approximated
with the help of the Stirling's formula
to an integration over the variable $n$
\begin{eqnarray}
\rho(W,V)=\frac{1}{W}\int^{\infty}_0 dn
\exp\left[
n\ln\left(a_{\{W^3V\}}({\bf R})\right)-\ln n! -\ln(3n-1)!
\right].
\end{eqnarray}
The above integration is evaluated using
the saddle point approximation.
the saddle point is found as follows
\begin{eqnarray}
n\approx \left(
\frac{a_{\{W^3V\}}({\bf R})}{27}
\right)^{\frac{1}{4}}.
\end{eqnarray}
In this approximation, the density of states is reduced to
\begin{eqnarray}
\rho(W,V)=
\frac{\sqrt{3}}{2\pi}
\frac{1}{W}\exp\left[
\left(\frac{256}{27} a_{\{W^3V\}}({\bf R})\right)^{1/4}
\right] \times C_n(W,V),
\end{eqnarray}
where
\begin{eqnarray}
C_n(W,V)&=&
\int^{\infty}_0 dn \exp\left[
-\frac{1}{2}
\left(\frac{256\times 27}{a_{\{W^3V\}}({\bf R})}
\right)^{1/4} n^2
\right],\nonumber\\
&=&\frac{1}{2}\sqrt{2\pi}
\left(\frac{a_{\{W^3V\}}({\bf R})}
{256\times 27} \right)^{1/8}.
\end{eqnarray}
Hence the density of states reads
\begin{eqnarray}
\rho(W,V)=
\frac{1}{4\sqrt{2\pi}}
\left(3 a_{\{W^3V\}}({\bf R})\right)^{\frac{1}{8}}
\frac{1}{W}\exp\left[\frac{4}{3}
\left(3 a_{\{W^3V\}}({\bf R})\right)^{1/4}
\right].
\end{eqnarray}
%%%%%%%%%%%%%%%%%%%%%%%%%%%%%%%%%%%%%%%%%%%%%%%%%%%%%%
%%%%%%%%%%%%%%%%%%%%%%%%%%%%%%%%%%%%%%%%%%%%%%%%%%%%%%
%
After tracing over the color index for $U(N_c)$ (or $SU(N_c)$),
Eq.(\ref{aw3v}) becomes
\begin{eqnarray}
a_{\{W^3V\}}({\bf R})=
2\tilde{a}_{\{W^3V\}}
\sum^{N_c}_{i=1}\cos\theta_i.
\end{eqnarray}
The projection over the color-singlet state reduces
the density of states to
\begin{eqnarray}
\rho_{\mbox{Singlet}}=\int^{\pi}_{-\pi} d\mu(g) \rho(W).
\end{eqnarray}
The density of states under the integral 
is approximated to the Gaussian-like integral 
over the color variables 
since it dominates a tiny range around the origin
along the real axis of Fourier color variables
\begin{eqnarray}
\rho(W,V)&=&
\frac{1}{4\sqrt{2\pi}}
\left(
6\tilde{a}_{\{W^3V\}}\sum^{N_c}_i\cos\theta_i
\right)^{\frac{1}{8}}
\frac{1}{W}\exp\left[\frac{4}{3}
\left(
6 \tilde{a}_{\{W^3V\}}\sum^{N_c}_{i=1}\cos\theta_i
\right)^{1/4}
\right],\nonumber\\
&=&
{\cal C}_{\rho}
\times
\exp\left[-\frac{1}{2}
{\cal D}_{\rho}
\sum^{N_c}_{i=1}\theta^2_i
\right],
\label{rho-singlet1}
\end{eqnarray}
where the pre-exponential coefficient reads
\begin{eqnarray}
{\cal C}_{\rho}=
\frac{1}{ 4\sqrt{2\pi} }
\left(
6 N_c\tilde{a}_{\{W^3V\}}
\right)^{\frac{1}{8}}
\frac{1}{W}\exp\left[
\frac{4}{3}
\left(
6 N_c \tilde{a}_{\{W^3V\}}
\right)^{1/4}
\right].
\end{eqnarray}
The exponential term is given by
\begin{eqnarray}
{\cal D}_{\rho}&=&
\frac{4}{3}
\frac{1}{4 N^{3/4}_c}
\left(6 \tilde{a}_{\{W^3V\}}\right)^{1/4},
\nonumber\\
&=&
\frac{1}{3N_c}\left[
\frac{6 (2j+1)N_c}{\pi^2}W^3V
\right]^{1/4}.
\end{eqnarray}
The color-singlet density of states  
for the groups $U(N_c)$ and $SU(N_c)$
is, respectively, approximated to
\begin{eqnarray}
\rho_{\mbox{Singlet}}(W,V)&=&{\cal C}_{\rho}
\int^{\theta_c}_{-\theta_c} d\mu  \exp\left[-\frac{1}{2}
{\cal D}_{\rho}
\sum^{N_c}_{i=1}\theta^2_i
\right],
\nonumber\\
&\approx&
\frac{\prod^{N_c}_{n=1} n!}{N_c!(2\pi)^{N_c/2}}
\times\left(\frac{1}{{\cal D}_{\rho}}\right)^{N^2_c/2}
\times
{\cal C}_{\rho}, ~~~\mbox{for}~ U(N_c),
\nonumber\\
&\approx&
\frac{\prod^{N_c}_{n=1} n!}{N_c!(2\pi)^{N_c/2}}
\sqrt{\frac{2\pi}{N_c}}
\times
\left(\frac{1}{{\cal D}_{\rho}}\right)^{(N^2_c-1)/2}
\times{\cal C}_{\rho}, ~~~\mbox{for}~ SU(N_c),
\nonumber\\
&~&\left(@~~ {\cal D}_{\rho}/N_c\ge 1\right).
\label{rho-singlet-res1}
\end{eqnarray}
Furthermore, the approximation which is carried out 
in Eqs.(\ref{rho-singlet1}) 
and (\ref{rho-singlet-res1}) requires the constraint
${\cal D}_{\rho}/N_c\ge 1$. 
This is derived trivially using
the analogous behavior 
$\rho(\tilde{\lambda})\propto 
e^{2 N^2_c \tilde{\lambda}}
\int d\mu(g) \exp\left[-\frac{1}{2}
(2N_c\tilde{\lambda})\sum_i\theta_i^2\right]$ 
where $\tilde{\lambda}\ge 1/2$.
The constraint ${\cal D}_{\rho}/N_c\ge 1$ 
is exactly the same constraint given 
by Eq.(\ref{mass-critical-MB}).
The asymptotic color-singlet density of states 
for the $U(N_c)$ and $SU(N_c)$ group representations
behaves, respectively, 
as follows
\begin{eqnarray}
\rho_{\mbox{Singlet}}(W,V)&\propto&
\frac{1}{W}
\left[
\frac{1}{W^3V}\right]^{\frac{N^2_c-1}{8}}
\exp\left[
\mbox{constant} \left(W^3V\right)^{1/4}
\right], ~~~\mbox{for}~ U(N_c),
\nonumber\\
&\propto&
\frac{1}{W}\left[
\frac{1}{W^3V}\right]^{\frac{N^2_c-2}{8}}
\exp\left[
\mbox{constant} \left(W^3V\right)^{1/4}
\right], ~~~\mbox{for}~ SU(N_c),
\nonumber\\
&~&\left(@~~ {\cal D}_{\rho}/N_c\ge 1\right).
\end{eqnarray}
%%%%%%%%%%%%%%%%%%%%%%%%%%%%%%%%%%%%%%%%%%%%%%%%%%%%%%%%%%%%%%%%%%
%%%%%%%%%%%%%%%%%%%%%%%%%%%%%%%%%%%%%%%%%%%%%%%%%%%%%%%%%%%%%%%%%%
%%%%%%%%%%%%%%%%%%%%%%%%%%%%%%%%%%%%%%%%%%%%%%%%%%%%%%%%%%%%%%%%%%

In order to analyze the results for the low energy limit,
it would be worthy to do the analysis in the language 
of the canonical-like ensemble rather 
than the micro-canonical ensemble.
The density of states in term of the canonical-like 
ensemble can be defined as follows
\begin{eqnarray}
\rho_{\mbox{Singlet}}(W,V)&=&
\frac{1}{2}\frac{1}{W}\frac{1}{2\sqrt{2\pi}}
\left(3 a_{\{W^3V\}}({\bf R})\right)^{\frac{1}{8}}
\exp\left[\frac{4}{3}
\left(3 a_{\{W^3V\}}({\bf R})\right)^{1/4}\right],
\nonumber\\
&\sim&
\frac{1}{2}\frac{1}{W}
\frac{1}{2\pi i}
\int^{\zeta_0+i \infty}_{\zeta_0-i \infty}
d\zeta e^{\zeta y} \exp\left[
\frac{1}{\zeta^3}
\left(a_{\{W^3V\}}({\bf R})\right)
\right]_{y=1}.
\end{eqnarray}
The projection of the color-singlet state 
reduces the density of states to
\begin{eqnarray}
\rho_{\mbox{Singlet}}(W,V)&\cong&
\frac{1}{2}\frac{1}{W}
\int d\mu(g)
\frac{1}{2\pi i}
\int^{\zeta_0+i \infty}_{\zeta_0-i \infty}
d\zeta e^{\zeta y} \exp\left[
\frac{1}{\zeta^3}
\left(a_{\{W^3V\}}({\bf R})\right)\right]_{y=1},
\nonumber\\
&\cong&
\frac{1}{2}\frac{1}{W}
\frac{1}{2\pi i}
\int^{\zeta_0+i \infty}_{\zeta_0-i \infty}
d\zeta e^{\zeta y}
\int d\mu(g) \exp\left[
\frac{1}{\zeta^3}
\left(a_{\{W^3V\}}({\bf R})\right)\right]_{y=1}.
\end{eqnarray}
Hence, the analysis of the above equation becomes 
similar to the Gross-Witten critical point solution.
In the case 
$\frac{\tilde{a}_{\{W^3V\}}}{N_c\zeta^3}\ll 1$
the color saddle points will be distributed uniformly
over the invariance measure $\int^{\pi}_{-\pi} d\mu(g)$.
It is evaluated as follows
\begin{eqnarray}
\rho_{\mbox{Low-Lying}}(W,V)&\cong&
\frac{1}{2}\frac{1}{W}
\frac{1}{2\pi i}
\int^{\zeta_0+i  \infty}_{\zeta_0-i \infty}
d\zeta e^{\zeta y}
\int^{\pi}_{-\pi} d\mu(g) \exp\left[
\frac{\tilde{a}_{\{W^3V\}}}{\zeta^3}
\left(\mbox{tr}{\bf R}(g)
+\mbox{tr}{\bf R}^{*}(g)\right)\right]_{y=1},
\nonumber\\
&=&
\frac{1}{2}\frac{1}{W}
\frac{1}{2\pi i}
\int^{\zeta_0+i \infty}_{\zeta_0-i \infty}
d\zeta e^{\zeta y}\exp\left[
\left(
\frac{\tilde{a}_{\{W^3V\}}}{\zeta^3}
\right)^2\right]_{y=1},
\nonumber\\
&~&~~ 
(\mbox{i.e.: continuous low-lying spectrum}).
\end{eqnarray}
Under the constraint 
$\frac{\tilde{a}_{\{W^3V\}}}{N_c\zeta^3}\ll 1$,
the integration over the Laplace transform 
is evaluated using
the steepest descent method. 
It is reduced to
\begin{eqnarray}
\rho_{(I)}(W,V)&=&\rho_{\mbox{Low-Lying}}(W,V),
\nonumber\\
&=&\frac{1}{2}\frac{1}{W}
\frac{1}{\sqrt{2\pi}}\frac{1}{\sqrt{7}}
\left(6\tilde{a}^2_{\{W^3V\}}\right)^{1/14}
\exp\left[
\frac{7}{6}
\left(6\tilde{a}^2_{\{W^3V\}}\right)^{1/7}
\right], \nonumber\\
&\propto&
\frac{1}{2}\frac{1}{W}
\left(W^3V\right)^{1/7} \exp\left[\mbox{constant}
\left(W^3V\right)^{2/7}
\right],
\end{eqnarray}
where the saddle point of the stationary 
condition is found at
\begin{eqnarray}
\overline{\zeta}=
\left[6\tilde{a}^2_{\{W^3V\}}\right]^{1/7}.
\end{eqnarray}
The location of the saddle point 
must satisfy the energy constraint
\begin{eqnarray}
\frac{\tilde{a}_{\{W^3V\}}}{N_c\overline{\zeta}^3}
&=&\frac{1}{6^{4/7}N_c^{8/7}}
\left(6 N_c\tilde{a}_{\{W^3V\}}\right)^{1/7}\ll 1,
\nonumber\\
&\rightarrow&  \frac{1}{6 N^2_c} 
\left(6 N_c\tilde{a}_{\{W^3V\}}\right)^{1/4}\ll 1,
\nonumber\\
&\rightarrow&  \frac{1}{6 N^2_c} 
\left(\frac{6(2j+1)N_c}{\pi^2}\right)^{1/4}
\left[W^{3/4}V^{1/4}\right]
\ll 1,
\nonumber\\
&\rightarrow&  \left(\frac{1}{6 N^2_c} 
\left(
\frac{6(2j+1)N_c n_{\mbox{flavors}}}{\pi^2}
\right)^{1/4}
\left[W^{3/4}V^{1/4}\right]\right)
\ll 1,
\nonumber\\
&\rightarrow& m\ll 10 \mbox{GeV}~~~ 
(\mbox{for 2 flavors}).
\label{m_low_lying_max_constraint}
\end{eqnarray}
%%%%%%%%%%%%%%%%%%%%%%%%%%%%%%%%%%%%%%%%%%%%%%%%%%
%%%%%%%%%%%%%%%%%%%%%%%%%%%%%%%%%%%%%%%%%%%%%%%%%%
%%%%%%%%%%%%%%%%%%%%%%%%%%%%%%%%%%%%%%%%%%%%%%%%%%

In the realistic QCD, the quarks
obey Fermi-Dirac statistics while the gluons obey
Bose-Einstein statistics.
The partition-like function for a gas of massless 
quarks and anti-quarks confined in a finite 
size cavity with the internal color structure 
$SU(N_c)$ (or $U(N_c)$) reads
\begin{eqnarray}
a_{q\overline{q}}({\bf R})&=&
\frac{(2j+1)}{\pi^2} (W^3V) n_{\mbox{flavor}}\cdot 
\mbox{tr}\left(\int^{\infty}_0
dx x^2 \ln\left[1+{\bf R}_{\mbox{fun}}(g)e^{-x}\right]
\right.
\nonumber\\
&~&
\left.~~~~~~~~~
+\int^{\infty}_0
dx x^2 \ln\left[1+{\bf R}_{\mbox{fun}}^*(g)e^{-x}\right]
\right).
\end{eqnarray}
Similarly, the partition-like function 
for a gas of gluons is reduced to
\begin{eqnarray}
a_{g}({\bf R})=
-\frac{(2j+1)}{\pi^2} (W^3V)\cdot 
\mbox{tr}\left(\int^{\infty}_0
dx x^2\ln\left[1-{\bf R}_{\mbox{adj}}(g)e^{-x}\right]
\right).
\end{eqnarray}
It would be interested to note here that when
$\tilde{a}_{q\overline{q}}/\zeta^3\le O(1)_{\mbox{critical}}$
and
$\tilde{a}_{g}/\zeta^3\le O(1)_{\mbox{critical}}$,
the color saddle points will be distributed uniformly 
over the entire circle. 
In this case the Maxwell-Boltzmann statistics 
can be produced as follows
\begin{eqnarray}
&~&\ln\left[1+{\bf R}_{\mbox{fun}}^*(g)e^{-x}\right]
\approx {\bf R}_{\mbox{fun}}^*(g)e^{-x},\nonumber\\
&-&\ln\left[1-{\bf R}_{\mbox{adj}}(g)e^{-x}\right]
\approx {\bf R}_{\mbox{adj}}(g)e^{-x}.
\end{eqnarray}

On the other hand, in the case
$\tilde{a}_{q\overline{q}}/\zeta^3> O(1)_{\mbox{critical}}$
and
$\tilde{a}_{g}/\zeta^3> O(1)_{\mbox{critical}}$ 
the color saddle points become
more dominant in a narrow domain around the origin.
In this domain, the partition-like function is expanded 
around the color saddle points.
The coefficients of the expansion around the saddle points
for the quark and anti-quark partition-like function reads
\begin{eqnarray}
\left. a_{q\overline{q}}({\bf R})\right|_{\{\theta\}=0}
&=&
\frac{7\pi^2}{360}
N_c n_{\mbox{flavor}}
(2j+1)\cdot(W^3V),
\nonumber\\
\left.\frac{\partial^2 a_{q\overline{q}}({\bf R})}
{\partial \theta^2_i}
\right|_{\{\theta\}=0}&=&
-\frac{1}{6}
n_{\mbox{flavor}}
(2j+1)\cdot(W^3V).
\end{eqnarray}
Moreover, the same thing can be calculated for the gluons
\begin{eqnarray}
\left. a_{g}({\bf R})\right|_{\{\theta\}=0}
&=&
\frac{\pi^2}{90}(N^2_c-1) (2j+1) (W^3V),
\nonumber\\
\left.\frac{\partial^2 a_{g}({\bf R})}
{\partial (\theta_i-\theta_j)^2}
\right|_{\{\theta\}=0}
&=&-\frac{1}{6}(2j+1)(W^3V).
\end{eqnarray}
The coefficient of the quark, antiquark and gluon gas 
is reduced to
\begin{eqnarray}
a_{\{W^3V\}}({\bf R})&=&
a_{q\overline{q}}({\bf R})+a_{g}({\bf R}).
\end{eqnarray}
The coefficient of the zero$^{th}$ approximation reads
\begin{eqnarray}
\left.a_{\{W^3V\}}({\bf R})\right|_{\{\theta\}=0}
&=&d_{q\overline{q} g} W^3V,
\end{eqnarray}
where 
\begin{eqnarray}
d_{q\overline{q} g}&=&(2j+1)\left(
\frac{7\pi^2}{360}
N_c n_{\mbox{flavor}}
+
\frac{\pi^2}{90}(N^2_c-1)\right).
\end{eqnarray}
The density of states is approximated
to the Gaussian-like function around the saddle 
points as follows
\begin{eqnarray}
\rho_{(II)}(W,V)&=&\rho_{\mbox{High-Lying}}(W,V),
\nonumber\\
&=&\frac{1}{4\sqrt{2\pi}}\frac{1}{W}
\left(3 a_{\{W^3V\}}({\bf R})\right)^{1/8}
\exp\left[
\frac{4}{3}\left(
3 a_{\{W^3V\}}({\bf R})
\right)^{1/4}\right],\nonumber\\
&\approx&
{\cal C}_{q\overline{q} g}(W,V)
\exp\left[
-\frac{1}{2} \lambda^{(2)}_{q\overline{q}}
\sum^{N_c}_{i=1}\theta^2_i
-\frac{1}{2}
\frac{2N_c\lambda^{(2)}_{g}}{2N_c}
\sum^{N_c}_{i=1}\sum^{N_c}_{j=1}(\theta_i-\theta_j)^2
\right],
\end{eqnarray}
where the pre-exponential coefficient reads
\begin{eqnarray}
{\cal C}_{q\overline{q} g}(W,V)
&=&
\frac{1}{4\sqrt{2\pi}}\frac{1}{W}
\left(3
d_{q\overline{q} g} W^3V
\right)^{1/8}
\exp\left[
\frac{4}{3}
\left(3
d_{q\overline{q} g} W^3V
\right)^{1/4}\right].
\end{eqnarray}
The quadratic terms in the exponential read
\begin{eqnarray}
\lambda^{(2)}_{q\overline{q}}&=&
-\left(\frac{1}
{
3 d_{q\overline{q} g} W^3V
}\right)^{3/4}
\left.\frac{\partial^2 a_{q\overline{q}}({\bf R})}
{\partial\theta^2_i}\right|_0,\nonumber\\
&=&
\frac{1}
{\left(3 d_{q\overline{q} g}\right)^{3/4}
}
\cdot \frac{1}{6}(2j+1) 
n_{\mbox{flavor}}\left(W^3V\right)^{1/4},
\end{eqnarray}
and
\begin{eqnarray}
\lambda^{(2)}_{g}
&=&
\left(
\frac{1}
{
3 d_{q\overline{q} g} W^3V
}
\right)^{3/4}
\left.
\frac{\partial^2 a_{g}({\bf R})}
{\partial(\theta_i-\theta_j)^2}
\right|_0,
\nonumber\\
&=&
\frac{1}
{
\left(3 d_{q\overline{q} g}\right)^{3/4}
}
\cdot
\frac{1}{6}(2j+1)
\left(W^3V\right)^{1/4}.
\end{eqnarray}
The color-singlet state for the density of states
is projected as follows
\begin{eqnarray}
\rho_{(II)}(W,V)&=&
{\cal C}_{q\overline{q} g}(W,V)
\int^{\theta_c}_{-\theta_c} d\mu(g)
\exp\left[
-\frac{1}{2} \lambda^{(2)}_{q\overline{q}}
\sum^{N_c}_{i=1}\theta^2_i
-\frac{1}{2}
\frac{2N_c\lambda^{(2)}_{g}}{2N_c}
\sum^{N_c}_{i=1}\sum^{N_c}_{j=1}(\theta_i-\theta_j)^2
\right],
\nonumber\\
&\doteq&
{\cal C}_{q\overline{q} g}(W,V)
\int^{\theta_c}_{-\theta_c} d\mu(g)
\exp\left[
-\frac{1}{2} 
\left(\lambda^{(2)}_{q\overline{q}}
+ 2N_c\lambda^{(2)}_{g}
\right)
\sum^{N_c}_{i=1}\theta^2_i
\right].
\label{densityofstates-rho1}
\end{eqnarray}
The result of Eq.(\ref{densityofstates-rho1}) 
resembles Eq.(\ref{qgp_canon}) 
and its evaluation leads to
\begin{eqnarray}
\rho_{(II)}(W,V)
&=&
c_{q\overline{q}g}
\frac{1}{W}
\left(3
d_{q\overline{q} g} W^3V
\right)^{1/8}
\left(
\frac{1}{ \lambda^{(2)}_{q\overline{q}}
+2N_c\lambda^{(2)}_{g} }
\right)^{\frac{N^2_c-1}{2}}
\exp\left[
\frac{4}{3}
\left(3
d_{q\overline{q} g} W^3V
\right)^{1/4}\right],
\nonumber\\
\label{densityofstates-rho-4}
\end{eqnarray}
where the constant $c_{q\overline{q}g}$ reads 
\begin{eqnarray}
c_{q\overline{q}g}=
\frac{1}{4 N_c!(2\pi)^{N_c-1/2}}
\frac{(2\pi)^{N_c/2}
\prod^{N_c}_{k=1}k!}{\sqrt{2\pi N_c}}.
\end{eqnarray}
In the terms of $\tilde{\Phi}^{(2)}_{q\overline{q}}$ 
and $\tilde{\Phi}^{(2)}_{g}$, 
Eq.(\ref{densityofstates-rho-4}) becomes
\begin{eqnarray}
\rho_{(II)}(W,V)
&=&
c_{q\overline{q}g}
\left(
\frac{
3d_{q\overline{q} g}
}{ \tilde{\Phi}^{(2)}_{q\overline{q}}
+2N_c\tilde{\Phi}^{(2)}_{g} }
\right)^{\frac{N^2_c-1}{2}}
\frac{1}{W}
\left(\frac{1}{
3d_{q\overline{q} g} W^3V}
\right)^{\frac{N^2-2}{8}}
\exp\left[
\frac{4}{3}
\left(3
d_{q\overline{q} g} W^3V
\right)^{1/4}\right],
\nonumber\\
\label{densityofstates-rho2}
\end{eqnarray}
where
\begin{eqnarray}
\lambda^{(2)}_{q\overline{q}}&=&
{\left(W^{3} V\right)}^{1/4}
\tilde{\Phi}^{(2)}_{q\overline{q}}/
\left(3 d_{q\overline{q}g}\right)^{3/4},
\nonumber\\
\lambda^{(2)}_{g}&=&
{\left(W^{3} V\right)}^{1/4}
\tilde{\Phi}^{(2)}_{g}/
\left(3 d_{q\overline{q}g}\right)^{3/4}.
\end{eqnarray}
The Hagedorn density of states
is given by $\rho_{(II)}(W,V)=
\rho_{\mbox{High-Lying}}(W,V)\delta(m-4B V)$
for the bags with the sharp surface boundary.
It is found that Eq.(\ref{densityofstates-rho2}) 
is identical to the result given 
in Eq.(\ref{qgp_micro}). 
The constraint of the thermal running parameter for 
Eq.(\ref{densityofstates-rho1}) validity is given by 
\begin{eqnarray}
\tilde{\lambda}\equiv 
\frac{1}{2N_c}\left(
\lambda^{(2)}_{q\overline{q}}+2N_c\lambda^{(2)}_{g}
\right)\ge \frac{1}{2}  
~~~\left(\mbox{where}~
\tilde{\lambda}_{\mbox{(pre-)critical}}
=\frac{1}{2}\right). 
\end{eqnarray}
Using the standard MIT bag model we get
\begin{eqnarray}
\frac{1}{2N_c}\frac{1}{24 d_{q\overline{q}g}^{3/4} 
B^{1/4}}(2j+1)\left(n_{\mbox{flavor}}+2N_c\right) 
m \ge \frac{1}{2},
\label{mass-critical-constraint}
\end{eqnarray}
where $(W^3V)^{1/3}=3^{3/4} m/(4B^{1/4})$ 
and $B^{1/4}=180$ MeV.
Eq.(\ref{mass-minimum-constraint}) gives the minimum
mass limit $m_{(II)\mbox{min}}$
while the (pre-)critical mass threshold 
$m_{\mbox{critical}}$ is determined 
by Eq.(\ref{mass-critical-constraint})
with the condition 
$m_{\mbox{critical}}\ge m_{(II)\mbox{min}}$.
The Hagedorn mass constraint is bounded 
from below 
by the (pre-)critical mass $m\ge m_{\mbox{critical}}$.
Nonetheless, Eq.(\ref{mass-critical-constraint}) 
gives the following estimations
\begin{eqnarray}
m\ge m_{\mbox{critical}}&=&
2060 \mbox{MeV}~~ 
(\mbox{massless flavors}:~n_{\mbox{flavor}}=1),
\nonumber\\
m\ge m_{\mbox{critical}}&=&
2315 \mbox{MeV}~~
\underline{
(\mbox{massless flavors}:~n_{\mbox{flavor}}=2)},
\nonumber\\
m\ge m_{\mbox{critical}}&=&
2482 \mbox{MeV}~~
(\mbox{massless flavors}:~n_{\mbox{flavor}}=3).
\label{mass-critical-limit}
\end{eqnarray}
It is worth to note that 
Eq.(\ref{mass-minimum-constraint}) gives the constraint
for the minimum mass limit for the Hagedorn threshold production
where the (pre-)critical mass determined by 
Eq.(\ref{mass-critical-constraint}) 
must lie above this minimum mass limit
$m_{\mbox{critical}}\ge m_{(II)\mbox{min}}$.
According to Eqs.(\ref{mass-minimum-limit}) and 
(\ref{mass-critical-limit}),
the minimum mass threshold and the (pre-)critical mass
for the two-flavor hadronic mass spectrum are found  
$m_{\mbox{critical}}=2315 \mbox{MeV}\ge
m_{(II)\mbox{min}}=2029 \mbox{MeV}$.
The experimental data book indicates that
the maximum limit for the discrete mass spectrum 
is estimated to be around 2250-2300 MeV.
The critical mass for the production 
of fireballs has to be just above the 
the known discrete mass spectrum.
The present work predict that maximum limit for the
two flavor discrete hadronic mass spectrum 
is 2315 MeV and above this limit 
the continuous Hagedorn spectrum is produced.

%%%%%%%%%%%%%%%%%%%%%%%%%%%%%%%%%%%%%%%%%%%%%%%%%%%%%%%%%
%%%%%%%%%%%%%%%%%%%%%%%%%%%%%%%%%%%%%%%%%%%%%%%%%%%%%%%%%
%%%%%%%%%%%%%%%%%%%%%%%%%%%%%%%%%%%%%%%%%%%%%%%%%%%%%%%%%

%%%%%%%%%%%%%%%%%%%%%%%%%%%%%%%%%%%%%%%%%%%%%%%%%%%%%%%%%
%%%%%%%%%%%%%%%%%%%%%%%%%%%%%%%%%%%%%%%%%%%%%%%%%%%%%%%%%%%%
%%%%%%%%%%%%%%%%%%%%%%%%%%%%%%%%%%%%%%%%%%%%%%%%%%%%%%%%%%%%
%%%%%%%%%%%%%%%%%%%%%%%%%%%%%%%%%%%%%%%%%%%%%%%%%%%%%%%%%%%%
%%%%%%%%%%%%%%%%%%%%%%%%%%%%%%%%%%%%%%%%%%%%%%%%%%%%%%%%%%%%
%%%%%%%%%%%%%%%%%%%%%%%%%%%%%%%%%%%%%%%%%%
%%%%%%%%%%%%%%%%%%%%%%%%%%%%%%%%
%%%%%%%%%%%%%%%%%%%%%%%
%%%%%%%%%%%%%%%%%%%
%%%%%%%%%%%%%%%%%%%%%%%%%%%%%%%%%%%%%%%%%%%%%%%%%%%%%%%%%%%%
%%%%%%%%%%%%%%%%%%%%%%%%%%%%%%%%%%%%%%%%%%%%%%%%%%%%%%%%%%%%
%%%%%%%%%%%%%%%%%%%%%%%%%%%%%%%%%%%%%%%%%%%%%%%%%%%%%%%%%%%%
%%%%%%%%%%%%%%%%%%%%%%%%%%%%%%%%%%%%%%%%%%%%%%%%%%%%%%%%%%%%
%%%%%%%%%%%%%%%%%%%%%%%%%%%%%%%%%%%%%%%%%%
%%%%%%%%%%%%%%%%%%%%%%%%%%%%%%%%
%%%%%%%%%%%%%%%%%%%%%%%
%%%%%%%%%%%%%%%%%%%
%%%%%%%%%%%%%%%%%%%%%%%%%%%%%%%%%%%%%%%%%%%%%%%%%%%%%%%%%%%%%%%
%%%%%%%%%%%%%%%%%%%%%%%%%%%%%%%%%%%%%%%%%%%%%%%%%%%%%%%%%%%%%%%
%%%%%%%%%%%%%%%%%%%%%%%%%%%%%%%%%%%%%%%%%%%%%%%%%%%%%%%%%

\section{The gas of bags with van der Waals
repulsion\label{sect_VdW}}

\subsection{Isobaric partition function}

The partition function for a gas of bags
in the Maxwell-Boltzmann statistics reads
\begin{eqnarray}
Z(T,V)&=&\sum_{N\ge 0}
\frac{1}{N!} \prod^N_{i=1} Q_i(\Lambda),
\label{partition_bags}
\end{eqnarray}
where
\begin{eqnarray}
Q_i(\Lambda)&=&
\int d m_i d v_i
\int\left(V-\sum^N_{k=1} v_k\right)\frac{d^3 p_i}{(2\pi)^3}
\rho(\Lambda;m_i,v_i) e^{- E_i/T}
\Theta\left(V-\sum^N_{k=1} v_k\right),
\label{partition_bags2}
\end{eqnarray}
and $v_i, m_i$ and $E_i$ are the hadron's volume, mass 
and total energy, respectively.
The fugacity $\Lambda$ is the constraint
of all possible charge conservations such as
baryonic, strangeness, $\cdots$ etc
and is given by 
\begin{eqnarray}
\Lambda=\{e^{i\theta_B} \cdots\}
\equiv
\left\{ e^{\frac{\mu_B}{T}}\cdots\right\}.
\end{eqnarray}
The density of states given by 
Eq.(\ref{partition_bags}) consists all the discrete 
known mass spectrum particles and the continuous
Hagedorn density of states as follows
\begin{eqnarray}
\rho(\Lambda;m_i,v_i\cdots)&=&
\rho_{(I)}(m,v,T,\Lambda)+\rho_{(II)}(m,v,T,\Lambda),
\end{eqnarray}
where the first term $\rho_{(I)}(m,v,T,\Lambda)$ 
corresponds the discrete low-lying hadronic mass spectrum 
while the second term $\rho_{(II)}(m,v,T,\Lambda)$ 
corresponds the continuous Hagedorn mass spectrum.
The discrete low-lying density of states 
of the known mass spectrum particles reads
\begin{eqnarray}
\rho_{(I)}(m,v,T,\Lambda)&=&
\sum^{\mbox{Baryons}}_i D_{\mbox{FD}}
\left(m,v,T,\Lambda_i\right)
\delta(m-m_i)\delta(v-v_H)
\nonumber\\
&+&
\sum^{\mbox{Mesons}}_i D_{\mbox{BE}}
\left(m,v,T,\Lambda_i\right)
\delta(m-m_i)\delta(v-v_H).
\end{eqnarray}
The sum runs over the baryon and meson mass spectra
those are satisfying Fermi-Dirac and Bose-Einstein
statistics, respectively. 
The terms
$D_{\mbox{FD}}\left(m,v,T,\Lambda_i\right)$
and
$D_{\mbox{BE}}\left(m,v,T,\Lambda_i\right)$ 
are the degeneracies as well as the single-particle 
statistic ensemble functions 
for Fermi- and Bose- particles, respectively.
The continuous part $\rho_{(II)}(m,v,T,\Lambda)$
is the Hagedorn density of states 
and these states correspond the hadronic bubbles
with relatively large hadronic masses 
could emerge as fireballs 
just above the highest mass 
of the known hadronic particle 
without strangeness $m\ge 2.0 \mbox{GeV}$ 
represented by the discrete low-lying mass spectrum. 
In the simplest approximation there is 
no reason to prefer the Fermi-Dirac  
or Bose-Einstein statistics for the exotic 
hadronic states such as Hagedorn states.
The Hagedorn states are assumed to obey simply 
the classical Maxwell-Boltzmann statistics due 
to their relatively large masses. 
The grand canonical ensemble  for a gas of 
non-interacting multi-particle species obeys 
the tensor product of their 
Fock spaces~\cite{Gorenstein:1998a},
\begin{eqnarray}
Z(T,V;\Lambda)=
\prod_{\mbox{Baryons}}   Z(T,V)
\prod_{\mbox{Mesons}}    Z(T,V)
\prod_{\mbox{Hagedorns}} Z(T,V)
\cdot\theta\left(V-\sum_i v_i\right).
\end{eqnarray}
In order to overcome the volume step function 
problem in Eqs.(\ref{partition_bags}) 
and (\ref{partition_bags2}), 
the isobaric ensemble trick is introduced.
%%%%%%%%%%%%%%
The isobaric partition function is calculated
by taking the Laplace transformation of the grand partition
function~\cite{Gorenstein:1981a,Gorenstein:1982a,
Gorenstein:1984a,Gorenstein:1998a,Auberson:1986a}
\begin{eqnarray}
\hat{Z}(T,s;\Lambda)
&\equiv&\int_{0}^{\infty}dV
\exp\left(-sV\right)Z(T,V;\Lambda),
\nonumber\\
&=&
1/\left[s-
\int_i
\exp\left(-v_i s\right)
\varphi_{\cal S}(T;m_{i},\Lambda_{i})\right],
\nonumber\\
&=&
1/\left[s-f_{\mbox{Hadrons}}(T,s)\right],
\label{isobaric_z1}
\end{eqnarray}
where $v_i$  is the hadron's Van der Waals excluded volume. 
It is reduced to a continuous Van der Waals 
variable $v$ for the continuous mass spectrum. 
The isobaric function found in the denominator
in the right hand side of Eq.(\ref{isobaric_z1}) 
is decomposed as follows
\begin{eqnarray}
f_{\mbox{Hadrons}}(T,s)&=&
\int_i
\exp\left(-v_i s\right)\rho(m,v,T,\Lambda)
\varphi_{\cal S}(T;m_{i},\Lambda_{i}),
\nonumber\\
&=&
~\sum_i\int
\exp\left(-v_{i} s\right)\rho_{(I)}(m_{i},v_{i},T,\Lambda)
\varphi_{\cal S}(T;m_{i},\Lambda_{i}) 
~(\mbox{discrete Low-Lying})
\nonumber\\
&+&
\int dv d m
\exp\left(-v s\right)\rho_{(II)}(m,v,T,\Lambda)
\varphi_{\cal S}(T;m,\Lambda) 
~(\mbox{continuous Hagedorn})
\nonumber\\
&=& 
f_{\mbox{Low-Lying}}(T,s)+f_{\mbox{Fireballs}}(T,s).
\label{isobaric_den}
\end{eqnarray}
The first term $f_{\mbox{Low-Lying}}(T,s)$ 
denotes the isobaric function consisting
all the known hadronic mass spectrum particles 
and resonances as well as  their antiparticles
embedded in the hot and dense medium.
The masses of these particles are taken from
the particle data book~\cite{databook2004}.
The non-strange hadronic spectrum consists
76 mesons and 64 baryons. 
This isobaric function is given as follows
\begin{eqnarray}
f_{\mbox{Low-lying}}(T,s)
&=&
\sum_i^{\mbox{Mesons}}
\left[\varphi_{\mbox{BE}}(T;m_{i},\Lambda^{\star}_{i})
+
\varphi_{\mbox{BE}}(T;m_{i},{\overline{\Lambda}}^{\star}_{i})\right]
\nonumber\\
&+&
\sum_i^{\mbox{Baryons}}
\left[\varphi_{\mbox{FD}}(T;m_{i},\Lambda^{\star}_{i})
+
\varphi_{\mbox{FD}}
(T;m_{i},{\overline{\Lambda}}^{\star}_{i})\right].
\end{eqnarray}
%%%%%%%%%%%%%%%%%%%%%%%%%%%%%%%%%%%%%%%%%%%%%%%%%%%%%%%%%%
%
%%%%%%%%%

The ensemble functions for single-particle species obeying the
Maxwell-Boltzmann, Fermi-Dirac
and Bose-Einstein statistics, respectively, read
\begin{eqnarray}
\varphi_{\mbox{MB}}(T;m_i,\Lambda_i)=
(2 J_i+1)\Lambda_i
\int\frac{d^{3}k}{(2\pi)^{3}}
e^{-\frac{1}{T}\sqrt{k^{2}+m_i^{2}}},
\label{quant_stats_mb}
\end{eqnarray}
\begin{eqnarray}
\varphi_{\mbox{FD}}(T;m_i,\Lambda_i)=
(2 J_i+1)\int\frac{d^{3}k}{(2\pi)^{3}}
\ln\left[1+\Lambda_i
e^{-\frac{1}{T}\sqrt{k^{2}+m_i^{2}}}\right],
\label{quant_stats_fd}
\end{eqnarray}
and
\begin{eqnarray}
\varphi_{\mbox{BE}}(T;m_i,\Lambda_i)= -
(2 J_i+1)\int\frac{d^{3}k}{(2\pi)^{3}}
\ln\left[1-\Lambda_i
e^{-\frac{1}{T}\sqrt{k^{2}+m_i^{2}}}\right].
\label{quant_stats_be}
\end{eqnarray}
where $J_i$ is the quantum number stemming
from the internal degrees of freedom (e.g. spin multiplicity).
The effective chemical potentials read
\begin{eqnarray}
\Lambda^{\star}_{i}=e^{-v s}\Lambda_{i},~~
{\overline{\Lambda}}^{\star}_{i}=
e^{-v s}{\overline{\Lambda}}_{i},
\end{eqnarray}
where the isobaric pressure times 
Van der Waals volume $\left(s\cdot v\right)$ enters 
as an exponential pre-factor of the particle fugacity.
%%%%%%%%%%%%%%%%%%%%%%%%%%%%%%%%%%%%%%%%%%%%%%%%%%%%%%%%%%%%%%%%%%%%%%%%%%%%%
%%%%%%%%%%%%%%%%%%%%%%%%%%%%%%%%%%%%%%%%%%%%%%%%%%%%%%%%%%%%%%%%%%%%%%%%%%%%%
The second term in Eq.(\ref{isobaric_den}) corresponds 
the isobaric pressure of Hagedorn bubbles.
The isobaric function for a gas of Hagedorn states reads
\begin{eqnarray}
f_{\mbox{Fireballs}}(T,s)
&=&
\int^{\infty}_{v_0} dv
\int^{\infty}_{m_0} dm e^{-vs} \rho_{(II)}(m,v,T,\Lambda)
\varphi_{{\mbox{MB}}}(T;m,\Lambda),
\label{isobaric_f1}
\end{eqnarray}
where the first integration is over the hadron's mass 
while the second
one is over the hadron's excluded volume.
The asymptotic behavior of the function
$\varphi_{\mbox{MB}}(T;m,\Lambda)$ 
in the limit of the large bag's mass $m\gg T$ reads
\begin{eqnarray}
\varphi_{\mbox{MB}}(T;m,\Lambda)&=&
\int \frac{d^3 k}{(2\pi)^3}
\exp\left(-\sqrt{k^2+m^2}/T\right),
\nonumber\\
&=& \left[\frac{m^{2}T}{2\pi^{2}}\right]K_{2}(m/T),
\nonumber\\
&\approx&
\left(\frac{mT}{2\pi}\right)^{3/2}e^{-m/T}.
\label{kinetic_app}
\end{eqnarray}
By introducing Eq.(\ref{kinetic_app}) in Eq.(\ref{isobaric_f1}),
Eq.(\ref{isobaric_f1}) is reduced to
\begin{eqnarray}
f_{\mbox{Fireballs}}(T,s)&=&
\int^{\infty}_{v_0} dv
\int^{\infty}_{m_0} dm e^{-vs} \rho_{(II)}(m,v,T,\Lambda)
\left(\frac{mT}{2\pi}\right)^{3/2}e^{-m/T},
\end{eqnarray}
where $\rho_{(II)}(m,v,T,\Lambda)$
measures the mass spectral density {\em and}
the volume fluctuation {\em as well}
of the continuous Hagedorn states.

%%%%%%%%%%%%%%%%%%%%%%%%%%%%%%%%%%%%%%%%%%%%%%%%%%%%%%%%%%%%%%%%
%%%%%%%%%%%%%%%%%%%%%%%%%%%%%%%%%%%%%%%%%%%%%%%%%%%%%%%%%%%%%%%%
It is known from the property of the Laplace transformation that
the asymptotic behavior of $Z(T,V)$
as $V\rightarrow \infty$ is determined by the extreme right-hand
singularity of $\hat{Z}(T,s;\Lambda)$
with respect to the isobaric variable 
$s$ on the real axis.
We denote this singularity point by $s^*$.
The Laplace parameter $s^*$ plays the role of 
the isobaric pressure.
In the thermodynamic limit $V\rightarrow \infty$, 
the pressure reads
\begin{eqnarray}
p&=&T \lim_{V\rightarrow \infty}\frac{1}{V} \ln Z(T,V),
\nonumber\\
&=& T s^*.
\end{eqnarray}
The isobaric partition function with the isobaric 
ensembles $\left(T,s\right)$ is convenience 
for a system characterized by
the external pressure $p=Ts$ 
rather than the fixed volume $V$.
%%%%%%%%%%%
The isobaric partition function has another singularity 
beside the first singular point $s_0=s^*$. 
This singularity is determined
by the nonlinear equation 
$s^*=f_{\mbox{Hadrons}}(T,s^*)$ given by Eq.(\ref{isobaric_z1}).
The second pole $s_c$ is the singularity 
when the isobaric function 
$f_{\mbox{Fireballs}}(T,s_c)$ diverges 
because the fireball's internal pressure exceeds 
the external pressure of the gas of hadrons
\begin{eqnarray}
f_{\mbox{Fireballs}}(T,s_c)\rightarrow \infty.
\end{eqnarray}
%%%%%%%%%%%%%%%%%%%%%%%%%%%%%%%%%%%%%%%%%%%%%%%%%%%%%%%%
%%%%%%%%%%%%%%%%%%%%%%%%%%%%%%%%%%%%%%%%%%%%%%%%%%%%%%%%
%
%

The density of states for the Hagedorn states
described by the standard MIT bag model
with a sharp surface boundary reads
\begin{eqnarray}
\rho_{(II)}(m,v)=
\delta\left(m-4Bv\right) Z_{(II)q\overline{q}g}(W,v),
\label{mit_sharp1}
\end{eqnarray}
where $W=m-Bv$ and the micro-canonical ensemble
\begin{eqnarray}
Z_{(II)q\overline{q}g}(W,v)=
C v^{-N^2_c/2}x^{-\left(\frac{3N^2_c+2}{8}\right)}
\exp\left[\frac{4}{3} a x^{3/4} v\right] 
\nonumber\\
\left(
\mbox{with}~ 
\left(W,v\right)\rightarrow \left(x=W/v,v\right)
\right),
\end{eqnarray}
is the micro-canonical ensemble for a quark and gluon gas
projected on the color-singlet state
and confined in a spherical cavity 
with a specific volume $v$.
This micro-canonical ensemble has been derived 
in sections~\ref{sect_micro_canonical} 
and \ref{sect_dens_statistics}.
Hence the isobaric pressure 
for the continuous spectrum of the hadronic bubbles 
becomes
\begin{eqnarray}
f_{\mbox{Fireballs}}(T,s)
&=&
\int^{\infty}_{v_0} dv
\int^{\infty}_{m_0} dm e^{-vs} \delta(m-4Bv)
Z_{(II)q\overline{q}g}(W,v)
\left(\frac{mT}{2\pi}\right)^{3/2}e^{-m/T},
\nonumber\\
&=&
C'(T)\int^{\infty}_{v_0} dv
v^{-N^2_c/2+3/2}
\exp\left[-v\left(s-s_0\right)\right],
\end{eqnarray}
where the bag's isobaric internal pressure is given by
\begin{eqnarray}
s_0&=&\left[\frac{4}{3}
\left(\tilde{\Phi}^{(0)}_{q\overline{q} g}
\right)^{\frac{1}{4}}
(3B)^{3/4}-\frac{4B}{T}\right].
\end{eqnarray}
The integral pre-factor,
which is independent on the volume fluctuation, 
reads
\begin{eqnarray}
C'(T)&=&C \left[3B\right]^{-\left(\frac{3N^2_c+2}{8}\right)}
\left(\frac{4BT}{2\pi}\right)^{3/2}.
\end{eqnarray}
In the case of QCD, we have $N_c=3$, 
$C^{\star}(T)=\left.C'(T)\right|_{N_c=3}$
and the isobaric pressure 
for Hagedorn's gas reads
\begin{eqnarray}
f_{\mbox{Fireballs}}(T,s)&=&
C^{\star}(T)\int^{\infty}_{v_0} dv
v^{-3}
\exp\left[-v\left(s-s_0\right)\right].
\label{fireball-isobaric-fun}
\end{eqnarray}
%%%%%%%%%%%%%%%
The pressure  $p=T s^*$ is typically calculated 
from the isobaric function extreme right singularity.
The external isobaric pressure of Hadronic gas and 
the pressure of Hagedorn's gas, respectively, read
\begin{eqnarray}
s^*&=&f_{\mbox{Hadrons}}(T,s^*),
\nonumber\\
p_{\mbox{Fireballs}}&=& T\cdot f_{\mbox{Fireballs}}(T,s),
\end{eqnarray}
where
\begin{eqnarray}
f_{\mbox{Fireballs}}(T,s)
&=&
C^{\star}(T)\left(\frac{z_0}{v_0}\right)^{2}
\int^{\infty}_{z_0} dz
z^{-3} e^{-z},
\end{eqnarray}
and
\begin{eqnarray}
z_0=v_0\left(s-s_0\right).
\label{z0singular}
\end{eqnarray}
%%%%%%%%%%%%%%%%%%%%%%%%%%%%%%%%%%%%%%%%%%%%%%%%%%%%
%%%%%%%%%%%%%%%%%%%%%%%%%%%%%%%%%%%%%%%%%%%%%%%%%%%%
%%%%%%%%%%%%%%%%%%%%%%%%%%%%%%%%%%%%%%%%%%%%%%%%%%%%

%
\subsection{The order of phase transition}

In order to analyze the phase transition, 
it is useful to introduce the exponential 
integral function
\begin{eqnarray}
z^n_0\int^{\infty}_{z_0} dz
z^{-n-1} e^{-z}&=&
z^n_0\Gamma(-n,z_0).
\label{integral-fun}
\end{eqnarray}
This function diverges for $n\le 0$ as follows
\begin{eqnarray}
\lim_{z_0\rightarrow 0} 
z^n_0\int^{\infty}_{z_0} dz
z^{-n-1} e^{-z}
&=&\lim_{z_0\rightarrow 0} z^n_0\Gamma(-n,z_0)
\rightarrow ~~\mbox{diverge}.
\label{exp_integ}
\end{eqnarray}
In case $n=0$ it diverges logarithmically.
On the contrary, it converges to $1/n$ for $n>0$.
Hence, the analysis to find the order of phase transition
in the limit
$s\rightarrow s_0$ 
becomes a straightforward one. 
It is useful to adopt the following approximation,
\begin{eqnarray}
s&=&s_{\mbox{Hadrons}}+
\left.C^{\star}(T)\int^{\infty}_{v_0}
dv v^{-\alpha}e^{-v(s-s_0)}\right|_{s\rightarrow s_0},
\nonumber\\
s&\approx&\left.C^{\star}(T)\int^{\infty}_{v_0}
dv v^{-\alpha}e^{-v(s-s_0)}\right|_{s\rightarrow s_0}.
\label{analysis_trans}
\end{eqnarray}
Indeed, Eq.(\ref{analysis_trans}) leads 
to the following conclusions~\cite{Zakout2006},
\begin{eqnarray}
\mbox{for}~ \alpha>1:~ s &\rightarrow& ~\mbox{finite}~ 
(\mbox{a possible phase transition}),
\nonumber\\
\mbox{for}~\alpha\le 1:~ s &\rightarrow& ~\mbox{diverge}~ 
(\mbox{no phase transition}).
\end{eqnarray}
In the case $\alpha\le 1$, the fireball pressure diverges
in the limit $s=s_0$ and subsequently the phase transition
does not exist.
On the other hand, in the case $\alpha> 1$ 
the fireball isobaric pressure converges 
in the limit $s=s_0$, though it diverges as  
the bag's internal isobaric pressure exceeds
the hadronic external isobaric pressure 
$s_0>s$ and the phase transition to 
an explosive quark-gluon plasma 
takes place in the system. 

The order of the phase transition is determined 
by examining the continuation of the isobaric pressure 
outside and inside the Hagedorn's bag 
and its n-th derivative as well.
Its derivative with respect to temperature 
(or any thermodynamical ensemble)
reads 
\begin{eqnarray}
s'&=&
\left(C^{\star}(T)\right)'\int^{\infty}_{v_0} dv 
v^{-\alpha}e^{-v(s-s_0)}
-{C^{\star}}(T)(s'-s'_0)\int^{\infty}_{v_0} dv
v^{-\alpha+1}e^{-v(s-s_0)}.
\label{order-phase-trans}
\end{eqnarray}
The exponential integral in 
Eq.(\ref{order-phase-trans}) 
at the point of the phase transition leads to 
the following conditions
\begin{eqnarray}
(s'-s'_0)&\approx&\frac{\mbox{finite value}}{
\lim_{s\rightarrow s_0}
\int^{\infty}_{v_0}dv v^{-\alpha+1}e^{-v(s-s_0)}},
\nonumber\\
&=& 0, ~\mbox{for}~ 2\ge\alpha>1
~\mbox{(a higher order phase transition)},
\nonumber\\
&\neq& 0, ~\mbox{for}~  \alpha> 2
~\mbox{(a first order phase transition)}.
\end{eqnarray}
In the summary, first and second order phase transitions
take place under the following constraints
\begin{eqnarray}
\infty\ge\alpha>2, 
~~~(\mbox{first order phase transition}),
\end{eqnarray}
and
\begin{eqnarray}
2\ge\alpha>1+\frac{1}{2},
~~~(\mbox{second order phase transition}),
\end{eqnarray}
respectively.
The $n$-th order phase transition 
takes place 
whenever $\alpha$ takes the following value
\begin{eqnarray}
1+\frac{1}{n-1}\ge \alpha>1+\frac{1}{n},
~~~(\mbox{$n$-th order phase transition}).
\end{eqnarray}
Finally, there is no phase transition 
for the bag of an internal structure of
\begin{eqnarray}
\alpha\le 1, 
~~~(\mbox{no phase transition}).  
\end{eqnarray}
%%%%%%%%%%%%%%%%%%%%%%%%%%%%%%%%%%%%%%%%%%%%%%%%%%%%%%%%%%%
%%%%%%%%%%%%%%%%%%%%%%%%%%%%%%%%%%%%%%%%%%%%%%%%%%%%%%%%%%%
The Hagedorn bubbles considered in the present work are
color-singlet states with an internal structure 
of $\alpha=3$.
They may appear in RHIC and LHC 
as fireballs (i.e. the Hagedorn states).
This means that the gas of fireballs 
(i.e. the Hagedorn phase) undergoes 
a first order phase transition 
to an explosive quark-gluon plasma.
However, when the explosive quark-gluon plasma appears
it expands rapidly.
Nevertheless, in a context of an alternate scenario 
it is reasonable to assume that the colored bags
could appear in the system 
before the appearance of an explosive quark-gluon plasma.
The isobaric function can be written as follows
\begin{eqnarray}
f_{\mbox{Fireballs}}(T,s)/f_{\mbox{colored-bags}}(T,s)
\sim \int^{\infty}_{v_0} dv v^{-3}
e^{-v(s-s_0)}/
\int^{\infty}_{v_0} dv v^{-\frac{1}{2}}
e^{-v(s-s_0)}.
\end{eqnarray}
In this case the phase transition from the Hagedorn phase
to the gas of colored bags is a first order phase transition.
The surprise in this scenario is that the phase transition 
from the gas of colored bags to the explosive quark-gluon plasma 
is not possible, 
keeping in mind in this scenario 
we only consider a simple color group symmetry 
with no other associated symmetry.
This leads to the conclusion that the explosive
quark-gluon plasma must take place for the quark 
and gluon bags with some specific internal structure.

On the other hand, 
it is also possible to think about a gas 
of colored bags with a specific internal 
color-flavor correlation associated 
with additional configuration-space 
internal symmetry such as $O(N)$ or $SO(N)$ 
or even any other associated non-trivial effect.
Then the generalization of the isobaric function 
for a gas of colored bags becomes 
\begin{eqnarray}
f_{\mbox{exotic-bags}}(T,s)
\sim
\int^{\infty}_{v_0} dv v^{-\alpha_e}
e^{-v(s-s_0)},
\end{eqnarray}
with an internal structure 
$\frac{1}{2}<\alpha_e\le \frac{3}{2}$.
Then is this scenario, it is possible 
that the Hagedorn's gas 
$f(T,s)_{\mbox{Fireballs}}/f(T,s)_{\mbox{exotic-bags}}$
undergoes a higher order phase transition 
to a gas of non-color-singlet bags.
The resultant gas of non-color-singlet bags undergoes 
a higher order phase transition 
to an explosive quark-gluon plasma.

%%%%%%%%%%%%%%%%%%%%%%%%%%%%%%%%%%%%%%%%%%%%%%%%%%
%%%%%%%%%%%%%%%%%%%%%%%%%%%%%%%%%%%%%%%%%%%%%%%%%%

%%%%%%%%%%%%%%%%%%%%%%%%%%%%%%%%%%%%%%%%%%%%%%%%%%
%%%%%%%%%%%%%%%%%%%%%%%%%%%%%%%%%%%%%%%%%%%%%%%%%%
\section{The role of chiral phase transition
\label{chiral-restoration}}
In this section the role of the chiral phase transition 
is considered in detail. 
A comprehensive review of the chiral fields
can be found in Ref.~\cite{Ripka:1997a}.
In the context of the Gell-Mann L$\acute{e}$vy model, 
namely $\sigma$-model,
the quark is assumed coupled to the chiral field 
of linear $\sigma$-model. 
The model can also be studied in the context of 
the Nambu Jona-Lasinio model in a similar way.
The analysis of the chiral restoration in the context of
the Nambu Jona-Lasinio model and color deconfinement 
will be considered elsewhere~\cite{Zakout:in-progress2}.
We can extend the Lagrangian density of the 
$\sigma$-model 
of Refs.~\cite{Mcgovern:1990a,Mcgovern:1990b}
to include the flavor chemical potentials as follows   
%%%%%%%%
\begin{eqnarray}
{\cal L}&=&\overline{q}\left[i\gamma^{\mu}\partial_{\mu}
-m_{0}-g_{YM}\gamma^{\mu} A_{b\mu}\tau_b
-g_{\sigma}\left(\sigma_a \lambda_a+i\gamma_5\lambda_a\pi_a\right)
+i\frac{1}{\beta}
\left(\theta_{B}+\theta_S\cdots\right)\gamma^{0}\right]q
\nonumber\\
&+&
\frac{1}{2}\left[\partial_{\mu}\sigma_a\partial^{\mu}\sigma_a
+\partial_{\mu}\pi_a \partial^{\mu}\pi_a
\right]
-\frac{1}{4} F_{b\mu\nu} F_b^{\mu\nu}
-U(\sigma,\vec{\pi}),
\label{chiral_lagrangian1}
\end{eqnarray}
%%%%%%%
where the axial scalar and axial pseudo-scalar 
field potential is given by
%%%%%%%%
\begin{eqnarray}
U(\sigma,\vec{\pi})&=&\frac{\lambda^2_{\sigma}}{8}
\left[\sigma^2_a+\pi^2_a-f^2\right]^2
\nonumber\\
&+&\frac{\eta_{\sigma}}{12}\left[
\mbox{tr}_F
\left[(\sigma_a\lambda_a+i\pi_a\lambda_a)
(\sigma_a\lambda_a-i\pi_a\lambda_a)\right]^2
-\frac{1}{3}
\left[\mbox{tr}_F(\sigma_a\lambda_a+i\pi_a\lambda_a)
(\sigma_a\lambda_a-i\pi_a\lambda_a)\right]^2
\right]
\nonumber\\
&-&\frac{\kappa^2}{2}\left[\mbox{det}
\left(\sigma_a\lambda_a+i\pi_a\lambda_a\right)+
\mbox{det}\left(\sigma_a\lambda_a+i\pi_a\lambda_a\right)
\right].
\label{chiral_potential}
\end{eqnarray}
%%%%%%%%%%
The non-Abelian color-gluon field reads
\begin{eqnarray}
F_{b\mu\nu}&=&
\partial_{\mu} A_{b\nu}-\partial_{\nu} A_{b\mu}
-g_{YM}f_{bb'b''} A_{b'\mu} A_{b''\nu}.
\label{nonabelian_field}
\end{eqnarray}
%%%%%%%%%%%%%%%%%%%%%%%%%%%%%%%%%%%%%%%%%%%%%%%%%%%%%%%%%%%
The quark satisfies the color fundamental representation 
defined by the following group transformation
$U_{\mbox{fun}}=
\exp\left( \imath \theta_a \tau_a \right)$ where $\tau_a$
is a set of 
the fundamental symmetric group $SU(N_c)$ generators.
On the other hand, the gluon satisfies the
adjoint group representation defined by the transformation
$U_{\mbox{adj}}=\exp\left( \imath\phi_a T_a\right)$ 
where $T_a$ is the set of adjoint generators 
of the same group that generates the fundamental generators. 
Furthermore, the adjoint color parameters $\phi_a$ 
are related to the fundamental ones by the relation 
$\phi_a\propto (\theta_i-\theta_j)$ where the index $a$
runs over $1,\cdots,~(N^2_c-1)$ while $i,j$ indices 
run over $1,\cdots,~ N_c$.
The adjoint and fundamental indices are 
related by $a\equiv (ij)$.
The set of generators that commute with the Hamiltonian 
is retained with the correspondence conservative parameters 
set $\theta_i$. 
The details will be given 
elsewhere~\cite{Zakout:in-progress2}.
The last term in the Lagrangian 
%%%%%%%%
\begin{eqnarray}
V_{U1}=-\frac{\kappa^2}{2}\left[
\mbox{det}(\sigma_a\lambda_a+i\pi_a\lambda_a)
+
\mbox{det}(\sigma_a\lambda_a-i\pi_a\lambda_a)
\right],
\label{chiral_axial}
\end{eqnarray}
is $U_{\mbox{axial}}\left(1\right)$ symmetry breaking term. 
For two flavors $N_f=2$, we have $\eta_{\sigma}=0$.
The model can be simplified furthermore by retaining only one 
scalar $\sigma$ field and three pseudo-scalar $\pi_i$ fields 
and the constant $\kappa^2=0$ of 
the axial symmetry breaking term. 
The easiest way to analyze the chiral phase 
is to adopt the mean-field approximation by replacing the 
$\sigma$ and $\pi_i$ fields by the their expectation values or
condensations
$\sigma\approx\left<\sigma\right>$ 
and $\pi_i\approx\left<\pi_i\right>$, respectively.
The trivial example is the set 
for the symmetric nuclear matter $\left<\pi_i\right>=0$. 
Finally, we neglect the interaction between 
the gluons and quarks 
by setting the non-Abelian coupling constant to $g_{YM}=0$. 
The small coupling $g_{YM}$ 
can be treated perturbatively~\cite{Zakout:in-progress2}.
%%%%%%%%%%%%%%%%%%%%%%%%%%%%%%%%%%%%%%%%%%%%%%%%%%%%%%%%%%

The effective chiral Lagrangian reads,
\begin{eqnarray}
{\cal L}={\cal L}_{q\overline{q}}+{\cal L}_{g}
+{\cal L}_{\sigma}.
\label{total_chiral_lagrangian}
\end{eqnarray}
The quark's term becomes 
\begin{eqnarray}
{\cal L}_{q\overline{q}}&\approx&
\overline{q}\left[i\gamma^{\mu}\partial_{\mu}
-m^{*}_{q}(\sigma)
+i\frac{1}{\beta}\theta_{B}\gamma^{0}\right]q,
\nonumber\\
&\doteq&
\overline{q}\left[i\gamma^{\mu}\partial_{\mu}
-m^{*}_{q}(\left<\sigma\right>)
+i\frac{1}{\beta}\frac{\mu_B}{T}\gamma^{0}\right]q,
\label{chiral_quark_lagrangian}
\end{eqnarray}
where chiral constituent quark mass is given by
$m^{*}_{q}(\left<\sigma\right>)=
m_0+g_{\sigma}\left<\sigma\right>$, 
while the conservative charge parameter 
such as the baryonic charge is defined by
$\imath \theta_{B}=\frac{\mu_B}{T}$. 
Hereinafter, we shall neglect the chemical potential
$\mu_B$ as far we are interested in 
the phase transition along temperature axis for 
the diluted and hot nuclear matter. 
The hot hadronic matter at zero baryonic density 
simplifies the calculations drastically.
The gluon-Lagrangian part 
in the limit of zero coupling constant, 
$g_{YM}=0$, is reduced to 
\begin{eqnarray}
{\cal L}_{g}&\approx&
-\frac{1}{4} F^{a}_{\mu\nu} F^{a\mu\nu},
\label{gluon_lagrangian}
\end{eqnarray}
where the color-gluon field becomes
\begin{eqnarray}
F^a_{\mu\nu}&=&
\partial_{\mu} A^a_{\nu}-\partial_{\nu} A^a_{\mu}.
\end{eqnarray}
%%%%%%%%%%%%%%%%%%%%%%%%%%%%%%%%%%%%%%%%%%%%%%%%%%%%%%%%%%%%%
There is an additional term in the Lagrangian due 
the chiral interaction. 
This chiral-Lagrangian term is responsible to 
the chiral restoration phase transition at high temperature. 
The chiral-Lagrangian reads
\begin{eqnarray}
{\cal L}_{\sigma}&\approx&
\frac{1}{2}\partial_{\mu}
\sigma\partial^{\mu}\sigma-U(\sigma),
\nonumber\\
&\doteq&-U(\left<\sigma\right>).
\label{sigma_lagrangian_part}
\end{eqnarray}
Hereinafter, the mean field brackets are discarded 
and redefined as $\left<\sigma\right>=\sigma$.
The partition function for a gas of quarks and gluons 
and chiral source is given by a tensor product of 
the quark, the gluon and the chiral field
Fock spaces
\begin{eqnarray}
Z_{q\overline{q}g\sigma}(\beta,V;\sigma)&=&
Z_{q\overline{q}}(\beta,V;\sigma)
\cdot Z_{g}(\beta,V)
\cdot Z_{\sigma}(\beta,V;\sigma).
\label{total_chiral_partition}
\end{eqnarray}
However, the chiral potential 
is to be subtracted 
from the quark and gluon bag energy when 
the micro-canonical ensemble is computed. 
The whole idea is that the hadronic bags 
are embedded in the chiral field 
background and the constituent quarks 
are coupled to the external chiral field.
%
%%%%%%%%%%%%%%%%%%%%%%%%%%%%%%%%%%%%%%%%%%%%%%%%%%%
%%%%%%%%%%%%%%%%%%%%%%%%%%%%%%%%%%%%%%%%%%%%%%%%%%%
%%%%%%%%%%%%%%%%%%%%%%%%%%%%%%%%%%%%%%%%%%%%%%%%%%%

The color-singlet state is projected as follows
\begin{eqnarray}
Z_{\mbox{Singlet}\sigma}\left(\beta,V;\sigma\right)&=&
\int d\mu(g)
Z_{q\overline{q}g\sigma}(\beta,V;\sigma),
\nonumber\\
&=&
\int d\mu(g) 
Z_{q\overline{q}}(\beta,V;\sigma)
\cdot Z_{g}(\beta,V)
\cdot Z_{\sigma}(\beta,V;\sigma),
\nonumber\\
&=&
\left[\int d\mu(g) 
Z_{q\overline{q}}(\beta,V;\sigma)
\cdot Z_{g}(\beta,V)\right]
\cdot Z_{\sigma}(\beta,V;\sigma).
\label{singlet_total_chiral_partition}
\end{eqnarray}
The quark-antiquark partition function
with an internal color degree
of freedom is given by
\begin{eqnarray}
Z_{q\overline{q}}(\beta,V;\sigma)&=&
\exp\left[(2j+1)\frac{1}{N_c}
\mbox{tr}_c\int\frac{Vd^3\vec{p}}{(2\pi)^3}
\right.
\nonumber\\
&~&
\left.
\cdot\ln\left[1+
\left({\bf R}_{\mbox{fun}}(g)+{\bf R}^{*}_{\mbox{fun}}(g)\right)
e^{-\beta \sqrt{\vec{p}^2+m^{*}_q(\sigma)}}
+
e^{-2\beta \sqrt{\vec{p}^2+m^{*}_q(\sigma)}}\right]\right].
\label{chiral_qq_partition}
\end{eqnarray}
The gluon partition function is not modified 
by the chiral field and it reads
\begin{eqnarray}
Z_{g}(\beta,V)&=&\exp\left[
-\frac{(2j+1)}{N^2_c-1}\mbox{tr}_c
\int \frac{Vd^3\vec{p}}{(2\pi)^3}\ln\left[
1-{\bf R}_{\mbox{adj}}(g)e^{-\beta \sqrt{\vec{p}^2+m_g}}
\right]\right],
\label{chiral_gluon_partition}
\end{eqnarray}
where the gluon mass remains $m_g=0$ in the hadronic phase.
Finally, the $\sigma$-energy partition function 
is calculated from the $\sigma$-potential as follows 
\begin{eqnarray}
Z_{\sigma}(\beta,V;\sigma)&=&e^{-\beta V U(\sigma)}.
\label{sigma_partition}
\end{eqnarray} 
%
%%%%%%%%%%%%%%%%%%%%%%%%%%%%%%%%%%%%%%%%%%%%%%%%%%%%%%%%%%%%%
%%%%%%%%%%%%%%%%%%%%%%%%%%%%%%%%%%%%%%%%%%%%%%%%%%%%%%%%%%%%%
%
The micro-canonical ensemble is calculated by finding 
the inverse Laplace transform of the mixed-canonical ensemble
\begin{eqnarray}
Z(W,V;\sigma)=\frac{1}{2\pi i}
\int^{\beta_c+\infty}_{\beta_c-i\infty}
d\beta
e^{\beta W} Z_{\mbox{Singlet}\sigma}
\left(\beta,V;\sigma\right).
\label{micro_total_chiral_partition}
\end{eqnarray}
The $\sigma$-potential background 
is to be subtracted from the bag energy 
in order to scale the quark and gluon bag's 
energy correctly.
The $\sigma$-field is the scalar chiral field 
interaction among the bags and it is coupled 
to the constituent quarks. 
The quark and gluon bag's energy $\widetilde{W}$ 
is scaled and re-defined as follows
\begin{eqnarray}
\widetilde{W}=W-V U_{\sigma}(\sigma).
\label{scale_bag_chiral_energy}
\end{eqnarray}
Hence, the micro-canonical ensemble 
of the chiral quark and gluon bag is calculated as follows,
\begin{eqnarray}
Z(\widetilde{W},V;\sigma)&=&
\frac{1}{2\pi i}\int^{\beta_c+\infty}_{\beta_c-i\infty}
d\beta
e^{\beta \widetilde{W}+\beta V U_{\sigma(\sigma)}} 
Z_{\mbox{Singlet}\sigma}\left(\beta,V;\sigma\right),
\nonumber\\
&=&\frac{1}{2\pi i}
\int^{\beta_c+\infty}_{\beta_c-i\infty}
d\beta
e^{\beta \widetilde{W}} 
e^{\beta V U_{\sigma(\sigma)}}
Z_{\mbox{Singlet}\sigma}\left(\beta,V;\sigma\right).
\label{micro_net_chiral_partition}
\end{eqnarray}
The resultant micro-canonical ensemble given by 
Eq.(\ref{micro_net_chiral_partition})
is equivalent to start from the beginning
with a mixed-canonical ensemble as a tensor product
of the Fock spaces of chiral quark and gluon
projected in the color-singlet state as follows
\begin{eqnarray}
Z(\widetilde{W},V;\sigma)&=&\frac{1}{2\pi i}
\int^{\beta_c+\infty}_{\beta_c-i\infty}
d\beta 
e^{\beta \widetilde{W}} 
Z_{\mbox{Singlet}}\left(\beta,V;\sigma\right),
\label{micro_eq_chiral_partition}
\end{eqnarray}
where $\widetilde{W}$ is the quark and gluon bag energy 
while the chiral color-singlet 
mixed-canonical partition function 
is given by
\begin{eqnarray}
Z_{\mbox{Singlet}}\left(\beta,V;\sigma\right)
&=&
\int d\mu(g) 
Z_{q\overline{q}}(\beta,V;\sigma)
\cdot Z_{g}(\beta),
\nonumber\\
&=&
\int d\mu(g) 
Z_{q\overline{q}g}(\beta,V;\sigma).
\label{chiral_singlet_bag_partition}
\end{eqnarray}
%
%%%%%%%%%%%%%%%%%%%%%%%%%%%%%%%%%%%%%%%%%%%%%%%%%%%%%%%%%%%%%%
%%%%%%%%%%%%%%%%%%%%%%%%%%%%%%%%%%%%%%%%%%%%%%%%%%%%%%%%%%%%%%
%%%%%%%%%%%%%%%%%%%%%%%%%%%%%%%%%%%%%%%%%%%%%%%%%%%%%%%%%%%%%%
%
% 

It is demonstrated in sections~\ref{sect_fund_adj} 
and \ref{sect_micro_canonical}
that the non-chiral hadronic phase undergoes 
a third order phase transition from a hadronic gas
dominated by the low-lying mass spectrum to another 
hadronic gas dominated by 
the continuous Hagedorn mass spectrum. 
It is interpreted that low-lying mass 
spectrum corresponds 
to the discrete hadronic mass spectrum 
found experimentally 
and consisting meson, baryons 
and any exotic hadronic mass 
spectrum could be found 
in the data book\cite{databook2004}.
On the other hand, the Hagedorn states 
are the highly excited metastable hadronic 
states those are produced just above 
the discrete low-lying mass spectrum.
The density of states for the continuous 
Hagedorn mass spectrum is calculated 
from the micro-canonical ensemble. 
In the calculation of the color-singlet 
canonical ensemble, the multi-integrations 
over the colors are preformed. 
The standard procedure is to adopt  
the saddle points approximation. 
In the limit of chiral Hagedorn states,
the quark and gluon canonical ensemble
is approximated by expanding the exponential around the 
color saddle points up to the quadratic term similar to 
the same procedure preformed to the non-chiral Hagedorn 
states.
%
%%%%%%%%%%%%%%%%%%%%%%%%%%%%%%%%%%%%%%%%%%%%%%%%%%%%%%%%%%%%%%
%%%%%%%%%%%%%%%%%%%%%%%%%%%%%%%%%%%%%%%%%%%%%%%%%%%%%%%%%%%%%%
The quadratic expansion 
of the chiral quark ensemble given 
by Eq.(\ref{chiral_qq_partition})
with respect to the color saddle points reads,
\begin{eqnarray}
\ln Z_{q\overline{q}}(\beta,V,\sigma)&=&
\Phi^{(0)}_{q\overline{q}\sigma}-\frac{1}{2}
\Phi^{(2)}_{q\overline{q}\sigma}
\sum^{N_c}_{i=1}\theta^2_i.
\label{chiral_qq_zeroth_quad}
\end{eqnarray}
The coefficients of the zeroth 
and the quadratic terms read, 
respectively,
\begin{eqnarray}
\Phi^{(0)}_{q\overline{q}\sigma}&=&
2 N_c\int \frac{V d^{3}\vec{p}}{(2\pi)^3} 
2\ln\left[1+e^{-\beta 
\sqrt{\vec{p}^2+{m^{*}_{q}}^2(\sigma)}}\right],
\nonumber\\
&=&
\frac{V}{\beta^3}\frac{2N_c}{\pi^2}
\int^{\infty}_0 dx x^2 \ln\left(
1+e^{ -\sqrt{x^2+\beta^2 {m^*_{q}}^2(\sigma)} }
\right),
\nonumber\\
&=&\frac{V}{\beta^3} 
\widetilde{\Phi}^{(0)}_{q\overline{q}\sigma}%
\left(\beta m^{*}_{q}(\sigma)\right),
\nonumber\\
&=&\frac{V}{\beta^3} 
\widetilde{\Phi}^{(0)}_{q\overline{q}\sigma}%
\left(\overline{\beta} m^{*}_{q}(\sigma)\right),
\label{chiral_qq_zeroth}
\end{eqnarray}
and 
\begin{eqnarray}
\Phi^{(2)}_{q\overline{q}\sigma}
&=&2 N_c \frac{1}{N_c}
\int \frac{V d^{3}\vec{p}}{(2\pi)^3} 
\frac{2 e^{-\beta\sqrt{\vec{p}^2+{m^{*}_{q}}^2(\sigma)}
}}
{\left(1+e^{-\beta \sqrt{\vec{p}^2+{m^{*}_{q}}^2(\sigma)}}
\right)^2},
\nonumber\\
&=&
\frac{V}{\beta^3}\frac{2}{\pi^2}
\int^{\infty}_0 dx x^2 
\frac{ 
e^{ -\sqrt{x^2+\beta^2 
{m^*_{\sigma}}^2(\sigma)}} 
}
{\left(1+e^{ -\sqrt{x^2+\beta^2 
{m^*_{\sigma}}^2(\sigma)} }\right)^2},
\nonumber\\
&=&\frac{V}{\beta^3} 
\widetilde{\Phi}^{(2)}_{q\overline{q}\sigma}
\left(\beta m^{*}_{q}(\sigma)\right),
\nonumber\\
&=&\frac{V}{\beta^3} 
\widetilde{\Phi}^{(2)}_{q\overline{q}\sigma}
\left(\overline{\beta} m^{*}_{q}(\sigma)\right),
\label{chiral_qq_quad}
\end{eqnarray}
where  
$m_{\sigma}^{*}(\sigma)=\left(m_0+g_{\sigma}\sigma\right)$.
The first order term is not needed in the expansion 
due to the saddle point constraint of extremization.

%%%%%%%%%%%%%%%%%%%%
%%%%%%%%%%%%%%%%%%%%
The same thing can be performed for the gluon-part 
given by Eq.(\ref{chiral_gluon_partition}).
The only difference is that, 
in the gluon ensemble, 
the expansion is carried over the adjoint color 
variables rather the fundamental ones
in the chiral quark-antiquark ensemble.
The gluonic quadratic color expansion 
is identical to that found in the non-chiral Hagedorn 
as follows
\begin{eqnarray}
\ln Z_{g}(\beta)=
\Phi_g^{(0)}-\frac{1}{2}\Phi_g^{(2)}
\sum^{N_c}_{n=1}\sum^{N_c}_{m=1}
(\theta_n-\theta_m)^2,
\label{gluon_zeroth_quad}
\end{eqnarray}
where
\begin{eqnarray}
\Phi^{(0)}_{g}&=&\frac{V}{\beta^3} \widetilde{\Phi}^{(0)}_{g},
~~~\mbox{where}~~
\widetilde{\Phi}^{(0)}_{g}=2(N^2_c-1)\frac{\pi^2}{90},
\nonumber\\
\Phi^{(2)}_{g}&=&\frac{V}{\beta^3}\widetilde{\Phi}^{(2)}_{g},
~~~\mbox{where}~~
\widetilde{\Phi}^{(2)}_{g}=\frac{1}{3}.
\end{eqnarray}

%%%%%%%%%%%%%%%%%%%%%%%%%%%%%%%%%%%%%%%%%%%%%%%%%%%%%%%%%%%%%
%%%%%%%%%%%%%%%%%%%%%%%%%%%%%%%%%%%%%%%%%%%%%%%%%%%%%%%%%%%%%
%%%%%%%%%%%%%%%%%%%%%%%%%%%%%%%%%%%%%%%%%%%%%%%%%%%%%%%%%%%%%
%%%%%%%%%%%%%%%%%%%%%%%%%%%%%%%%%%%%%%%%%%%%%%%%%%%%%%%%%%%%%
The resultant mixed-canonical ensemble that generates 
the density of states for the color-singlet 
chiral Hagedorn Eq.(\ref{chiral_singlet_bag_partition})
is approximated to
\begin{eqnarray}
Z_{\mbox{Singlet}}(\beta,V;\sigma)
&\approx&\int d\mu(g)_{\mbox{saddle}}
Z_{q\overline{q}}(\beta,V;\sigma) Z_{g}(\beta,V),
\label{singlet_chiral_bag_saddle}
\end{eqnarray}
where $\int d\mu(g)_{\mbox{saddle}}$ is the invariance
measure given by the Gaussian-like saddle point method
and Eq.(\ref{invariance_measure_saddle}).
After evaluating the multi-integrations over 
the color variables using the saddle points 
approximation, the chiral color-singlet 
mixed-canonical ensemble is reduced to
\begin{eqnarray}
Z_{(II)q\overline{q} g}(\beta,V;\sigma)&=&
{\cal N}_{q\overline{q} g}
\left(
\frac{\beta^3/V}{ 2 N_c\tilde{\Phi}^{(2)}_g+
\widetilde{\Phi}^{(2)}_{q\overline{q}\sigma}
(\overline{\beta} m^{*}_{\sigma})
}
\right)^{\frac{N^2_c-1}{2}}
e^{\frac{V}{\beta^3}\left[
\widetilde{\Phi}^{(0)}_{q\overline{q}\sigma}
(\overline{\beta} m^{*}_{\sigma})
+\widetilde{\Phi}^{(0)}_{g}\right]}.
\label{canonical_chiral_bag_ensemble}
\end{eqnarray} 
where the pre-factor normalization coefficient
${\cal N}_{q\overline{q} g}$
is identical to the non-chiral 
mixed-canonical ensemble given 
by Eq.(\ref{norm_qqg}).
%%%%%%%%%%%%%%%%%%%%%%%%%%%%%%%%%%%
%%%%%%%%%%%%%%%%%%%%%%%%%%%%%%%%%%%%%%%%%%%%%%%%%%%%%%

The chiral color-singlet micro-canonical ensemble 
is found by calculating the inverse Laplace transform 
of the mixed-canonical ensemble as follows,
\begin{eqnarray}
Z_{(II)q\overline{q} g}(W,V;\sigma)
=
\frac{{\cal N}_{q\overline{q} g}}{2\pi\imath}
\int^{\beta_c+\imath\infty}_{\beta_c-\imath\infty} 
&d\beta& e^{\beta W}
\beta^{\frac{3}{2}(N^2_c-1)}
\left(
\frac{1/V}{ 2 N_c\widetilde{\Phi}^{(2)}_g
+\widetilde{\Phi}^{(2)}_{q\overline{q}\sigma}
(\overline{\beta} m^{*}_{\sigma})}
\right)^{\frac{N^2_c-1}{2}}
\nonumber\\
&\times&
\exp\left[\frac{V}{\beta^3}
\left(\widetilde{\Phi}^{(0)}_{q\overline{q}\sigma}
(\overline{\beta} m^{*}_{\sigma})
+
\widetilde{\Phi}^{(0)}_{g}\right)\right].
\label{microcanonical_chiral_bag_Laplace}
\end{eqnarray}
The inverse Laplace transform is evaluated using 
the steepest descent method.
The Laplace stationary-saddle point is determined 
by extremizing the exponential under 
the integral in 
Eq.(\ref{microcanonical_chiral_bag_Laplace}) 
with respect to the Laplace transform variable $\beta$.
Unfortunately, the solution of the Laplace-saddle point
is found a transcendental one 
and can not be written in a closed form. 
However, we are concerned by the solution near 
the chiral restoration phase transition 
where the constituent quark masses 
approach their current ones.
In the limit of small constituent quark masses such 
as light flavors, 
the Laplace saddle point is found by iteration.
In this kind of transcendental problem, 
the solution converges rapidly.
The saddle point solution is found as follows,   
\begin{eqnarray}
\beta^{(1)}_0&=&\left[\frac{3V}{W}
\left(\widetilde{\Phi}^{(0)}_{q\overline{q}\sigma}
({\overline{\beta} m^{*}_{\sigma}}\approx 0)
+
\tilde{\Phi}^{(0)}_{g}\right)\right]^{\frac{1}{4}},
\nonumber\\
\beta^{(2)}_0&=&\left[\frac{3V}{W}
\left(\widetilde{\Phi}^{(0)}_{q\overline{q}\sigma}
(\beta^{(1)}_0 m^{*}_{\sigma})
+
\tilde{\Phi}^{(0)}_{g}\right)\right]^{\frac{1}{4}},
\nonumber\\
&\vdots&
\nonumber\\
\beta_0&\approx& \beta^{\left(I_n\right)}_{0},
~~~\mbox{and}~~
\overline{\beta}\approx\beta^{\left(I_n-1\right)},
\end{eqnarray}
where $I_n$ is the number of iteration.
In the present kind of transcendental equation, 
we assume that the value 
$\beta_0=\beta^{(2)}_0$ 
(i.e. $I_n=2$ and 
$\overline{\beta}=\beta_0^{(1)}$) 
is a sufficient approximation in the limit below 
but close to the chiral restoration phase transition 
with small constituent quark masses 
like up and down flavors.
This approximation is also justified 
for the strangeness.
%%%%%%%%%%%%%%%%%%%%%%%%%%%%%%%%%%%%%%%%%%%%%%%%%%%%%%%%%%
The micro-canonical ensemble takes the following form,
\begin{eqnarray}
Z_{(II)q\overline{q}g}(W,V;\sigma)&=&
\frac{1}{2\sqrt{2\pi}}
{\cal N}_{q\overline{q} g}
N_{\widetilde{\Phi}}
V^{-\frac{(N^2_c-2)}{8}} W^{-\frac{(3N^2_c+2)}{8}}
\exp\left[
\frac{4}{3}
\left(\tilde{\Phi}^{(0)}_{q\overline{q} g\sigma}
(\overline{\beta} m^{*}_{\sigma})
\right)^{\frac{1}{4}}
W^{3/4} V^{1/4}
\right],
\nonumber\\
\label{microcanonical_chiral_bag_ensemble}
\end{eqnarray}
for the chiral color-singlet bag of quarks and gluons.
The pre-exponential coefficient 
$N_{\tilde{\Phi}}$ is dimensionless and reads
\begin{eqnarray}
N_{\widetilde{\Phi}}&=&
\frac{
\left(
\widetilde{\Phi}^{(0)}_{q\overline{q} g\sigma}
(\overline{\beta} m^{*}_{\sigma})
\right)^{\frac{3N^2_c-2}{8}}
}{
\left(
2N_c\widetilde{\Phi}^{(2)}_g
+\widetilde{\Phi}^{(2)}_{q\overline{q}\sigma}
(\overline{\beta}m^{*}_{\sigma})
\right)^{\frac{N^2_c-1}{2}}
}.
\label{micro_chiral_norm}
\end{eqnarray}
The neutralness of the pre-factor coefficient 
$N_{\tilde{\Phi}}$  
does not generate a power 
function with respect 
to the bag's energy or volume. In this sense,
its variation with respect to the chiral constituent 
quark mass does not play a significant role 
in the chiral restoration phase transition 
and subsequently this variation is neglected.
Therefore, it is reasonable to ignore 
the chiral effect in $N_{\tilde{\Phi}}$ 
and to approximate it to the following:
\begin{eqnarray}
N_{\widetilde{\Phi}}&\approx&
\frac{
\left(
\widetilde{\Phi}^{(0)}_{q\overline{q} g\sigma}
(0)
\right)^{\frac{3N^2_c-2}{8}}
}{
\left(
2N_c\widetilde{\Phi}^{(2)}_g
+\widetilde{\Phi}^{(2)}_{q\overline{q}\sigma}
(0)
\right)^{\frac{N^2_c-1}{2}}
},
\nonumber\\
&\approx&
\frac{
\left(
2\cdot 2 N_c \cdot 7\pi^2/720
+
2 \cdot (N^2_c-1)\pi^2/90
\right)^{(3N^2_c-2)/8}
}{
\left(
2N_c/6+1/3
\right)^{(N^2_c-1)/2}
}.
\label{micro_chiral_0_norm}
\end{eqnarray}
In the contrary, the exponential plays a significant 
role in the chiral phase transition. 
It is decisive to determine 
the scalar chiral $\sigma$-mean 
field and the order of the chiral phase transition.
It must be written explicitly as a function 
of the chiral constituent quark mass as follows
\begin{eqnarray}
\widetilde{\Phi}^{(0)}_{q\overline{q} g\sigma}
(\overline{\beta}m^{*}_{\sigma})&=&
\widetilde{\Phi}^{(0)}_{q\overline{q}\sigma}
(\overline{\beta}m^{*}_{\sigma})
+\widetilde{\Phi}^{(0)}_{g}.
\label{chiral_qqgs_exp1}
\end{eqnarray}
The explicit expression with respect 
to the chiral constituent quark mass reads
\begin{eqnarray}
\widetilde{\Phi}^{(0)}_{q\overline{q}\sigma}
(\overline{\beta}m^{*}_{\sigma})&=&
\frac{2N_c}{\pi^2}\int^{\infty}_{0} dx x^2 
\ln\left[
1+e^{-\sqrt{x^2+(\overline{\beta}{m^{*}_{\sigma}})^2}}
\right],\nonumber\\
&=&
\frac{2N_c}{3\pi^2}\int^{\infty}_{0} dx
\frac{x^4}{\sqrt{x^2+(\overline{\beta}{m^{*}_{\sigma}})^2}} 
\frac{1}{\left[
1+e^{-\sqrt{x^2+(\overline{\beta} {m^{*}_{\sigma}})^2}}
\right]}.
\label{chiral_qqs_exp1}
\end{eqnarray}
%%%%%%%%%%%%%%%%%%%%%%%%%%%%%%%%%%%%%%%%%%%%%%%%%%%%%%%%%%%%%%%%%%
%
%%%%%%%%%%%%%%%%%%%%%%%%%%%%%%%%%%%%%%%%%%%%%%%%%%%%%%%%%%%%%%%%%%
The variation of the exponential with respect 
to the scalar chiral $\sigma$-mean field is essential 
for calculating the scalar chiral $\sigma$-field. 
This can be done by extremizing the isobaric pressure.
The explicit variation of 
$\widetilde{\Phi}^{(0)}_{q\overline{q}\sigma}
(\overline{\beta}m^{*}_{\sigma})$ 
with respect to the scalar chiral $\sigma$-field 
takes the following form,
\begin{eqnarray}
\frac{\delta}{\delta \sigma} 
\left(\widetilde{\Phi}^{(0)}_{q\overline{q}\sigma}
(\overline{\beta}m^{*}_{\sigma})\right)&=&
-\left(\frac{2N_c}{\pi^2}\int^{\infty}_{0} dx x^2 
\frac{1}{\left[
e^{\sqrt{x^2+(\overline{\beta}{m^{*}_{\sigma}})^2}}+1
\right]}
\frac{\overline{\beta}^2 m^{*}_{\sigma}}
{\sqrt{x^2+(\overline{\beta}m^{*}_{\sigma})^2}}\right)
\frac{\delta m^{*}_{\sigma}}{\delta\sigma}.
\label{chiral_variation}
\end{eqnarray}
%%%%%%%%%%%%%%%%%%%%%%%%%%%%%%%%%%%%%%%%%%%%%%%%%%%%%%%%
%%%%%%%%%%%%%%%%%%%%%%%%%%%%%%%%%%%%%%%%%%%%%%%%%%%%%%%%
%
%
%
%
The variable transformations $x=W/V$ and $v=V$ transform 
the independent variable set as follows 
$\{W,V\}\rightarrow \{x,v\}$.
However, this transformation set simplifies 
the analysis of phase transition in terms of 
the bag's energy density and volume 
rather than the bag's energy and volume. 
Nonetheless, according to the standard MIT bag model, 
the bag's mass is given by subtracting 
the bag's energy constant as follows $W=m-Bv$,
where $BV$ is interpreted as the bag's vacuum energy.
The micro-canonical ensemble as a function 
of two independent variables 
$\{x,v\}$ is displayed as follows
\begin{eqnarray}
Z_{(II)q\overline{q}g}(x,v;\sigma)
&\equiv&
\frac{1}{2\sqrt{2\pi}}
{\cal N}_{q\overline{q} g}
N_{\widetilde{\Phi}}
v^{-\frac{N^2_c}{2}} x^{-\frac{3N^2_c+2}{8}}
\exp\left[
\frac{4}{3}
\left(
\widetilde{\Phi}^{(0)}_{q\overline{q} g\sigma}
(\overline{\beta}m^{*}_{\sigma})
\right)^{\frac{1}{4}}
x^{3/4} v
\right].
\label{chiral_x_v_trans}
\end{eqnarray}
The Hagedorn's density of states in the context 
of MIT bag model with a sharp surface boundary
becomes,
\begin{eqnarray}
\rho_{\mbox{Bags}}&=&\delta\left(m-4Bv\right) 
Z_{(II)q\overline{q}g}(x,v;\sigma),
\nonumber\\
&=&\frac{1}{v}\delta\left(x-3B\right) 
Z_{(II)q\overline{q}g}(x,v;\sigma).
\label{bag_density}
\end{eqnarray}
The vantage of point adopting the standard bag model 
with a sharp surface boundary condition 
is simplifying the model drastically.
However, in the realistic physical situations, 
it is expected that the bag surface boundary has 
an extended surface in the extreme hot bath rather 
than a sharp one.
Furthermore, it would be reasonable 
to think that the bag's 
surface boundary is distorted and becomes a fuzzy one.
It is argued that bag's surface boundary is important 
in the determination of the order of phase transition
and the existence of the tri-critical point in
the phase transition diagram~\cite{Bugaev:2007a,Zakout2006}.  
Nonetheless, the surface distortion may cause a mechanical 
instability. 
Such a mechanical instability is ignored here. 
%%%%%%%%%%%%%%%%%%%%%%%%%%%%%%%%%%%%%%%%%%%%%%%%%%%%%%%%%%
%%%%%%%%%%%%%%%%%%%%%%%%%%%%%%%%%%%%%%%%%%%%%%%%%%%%%%%%%%%%%%
%
The Hagedorn's isobaric pressure is calculated 
by integrating the degenerate single-particle ensemble function 
over the mass and volume distribution function. 
It is reasonable to assume that
the gas of Hagedorn states is satisfying 
the Maxwell-Boltzmann statistics in order
to simplify the model analysis.
The Hagedorn isobaric function is integrated 
over the continuous mass and volume variables
as follows 
\begin{eqnarray}
f_{\mbox{Fireballs}}(T,s)&=& 
\int^{\infty}_{v_0}\int^{\infty}_{m_0} d v d m e^{-vs}
\rho_{\mbox{Bags}}(m,v)\varphi_{\mbox{MB}}(T;m).
\label{fireball_chiral_isobaric1}
\end{eqnarray}
As far the initial Hagedorn mass is relatively heavy,
it is adequate to adopt the following approximation,
\begin{eqnarray}
f_{\mbox{Fireballs}}(T,s)
&=& 
\int^{\infty}_{v_0}\int^{\infty}_{m_0} d v d m e^{-vs}
\rho_{\mbox{Bags}}(m,v) 
\left(\frac{m T}{2\pi}\right)^{3/2} e^{-m/T},
\nonumber\\
&=& 
\int^{\infty}_{v_0}\int^{\infty}_{x_0} d v d x e^{-vs}
\delta(x-3 B) Z_{(II)q\overline{q}g}(x,v;\sigma) 
\left(\frac{\left(x+B\right) v T}{2\pi}\right)^{3/2} 
e^{-\left(x+B\right) v/T}.
\nonumber\\
\label{fireball_chiral_isobaric2}
\end{eqnarray}
The integration over the $\delta$-function 
eliminates the energy density integration 
and reduces the isobaric function to 
an integration only over the 
Hagedorn's excluded volume as follows
\begin{eqnarray}
f_{\mbox{Fireballs}}(T,s)
&=&
\int^{\infty}_{v_0} d v e^{-vs}
\left.Z_{(II)q\overline{q}g}(x,v;\sigma)\right|_{x=3B} 
\left(\frac{4B T}{2\pi}\right)^{3/2} v^{3/2} 
e^{-\left(4 B/T\right) v}.
\label{fireball_chiral_vol}
\end{eqnarray}
In order to simplify the analysis, 
the chiral isobaric function 
for the Hagedorn gas is written 
in short as follows
\begin{eqnarray}
f_{\mbox{Fireballs}}(T,s)&=& 
C'(T)\int^{\infty}_{v_0} d v 
v^{-\frac{(N^2_c-3)}{2}}
e^{-v\left(s-s_0\right)},
\label{fireball_chiral_asym}
\end{eqnarray}
where the integral's pre-factor function reads
\begin{eqnarray}
C'(T)&=& 
\frac{1}{2\sqrt{2\pi}}
{\cal N}_{q\overline{q} g}
N_{\widetilde{\Phi}}
\left(3B\right)^{-\frac{3N^2_c+2}{8}}
\left(\frac{4B T}{2\pi}\right)^{3/2}.
\label{fireball_chiral_pre_fact}
\end{eqnarray}
It is worth to note that 
Eq.(\ref{fireball_chiral_asym})
differs from Eq.(\ref{fireball-isobaric-fun}) 
where the former equation includes 
the scalar chiral $\sigma$-field
unlike the later one.
Furthermore, 
the Hagedorn bag's internal isobaric
pressure is given by
\begin{eqnarray}
p_{\mbox{Internal}}/T&=&s_{0},
\nonumber\\
&=&
\frac{4}{3}
\left(
\widetilde{\Phi}^{(0)}_{q\overline{q} g\sigma}
(\overline{\beta}m^{*}_{\sigma})
\right)^{\frac{1}{4}}
\left(3B\right)^{3/4}-4B/T.
\label{fireball_chiral_intern_pr}
\end{eqnarray}
%%%%%%%%%%%%%%%%%%%%
%%%%%%%%%%
%%%%%%%%%%%%%%%%%%%%%%%%%%%%%%%%%%%%%%%%%%%%
%%%%%%%%%%%%%%%%%%%%%%%%%%%%%%%%%%%%%%%%%%%%

On the other hand, 
the chiral low-laying mass spectrum is realized as 
the discrete hadronic mass spectrum found experimentally
and displayed in the data book~\cite{databook2004}. 
The scalar chiral $\sigma$-mean field potential 
interacts and couples to the hadrons and generates 
the hadron masses. 
At highly extreme conditions, the chiral symmetry 
will be restored and the hadrons with light flavors
will dissolve to massless states
while the scalar chiral $\sigma$-potential vanishes. 
Under the above assumption the density of states for 
the low-lying mass spectrum is generalized to take 
into consideration the chiral discrete mass spectrum
as follows,  
\begin{eqnarray}
\rho_{\mbox{Low-Lying}}(m,v,T,\Lambda)&=&
\sum^{\mbox{Baryons}}_{\mbox{B}}  
\left(2J_{\mbox{B}}+1\right)
D_{\mbox{FD}}(\Lambda_{\mbox{B}}) 
\delta\left(m-m^{*}_{\mbox{B}}(\sigma)\right)
\delta\left(v-v_H\right)
\nonumber\\
&+&
\sum^{\mbox{Mesons}}_{\mbox{M}} 
\left(2J_{\mbox{M}}+1\right)
D_{\mbox{BE}}(\Lambda_{\mbox{M}})
\delta\left(m-m^{*}_{\mbox{M}}(\sigma)\right)
\delta\left(v-v_H\right).\nonumber\\
\label{chiral_low_lying_1}
\end{eqnarray}
The sums run over the baryon and the meson mass spectra
those are satisfying 
Fermi-Dirac and Bose-Einstein statistics, respectively.
The coupling of the scalar chiral $\sigma$-field
to the constituent quark generates the hadron's mass 
as follows 
\begin{eqnarray}
M^{*}_{\mbox{Hadron}}&=& g_{H\sigma}\sigma+n_q m_{q0},
\nonumber\\
&\approx&n_{q}\cdot g^{\mbox{input}}_{q\sigma}\cdot \sigma,
\label{chiral_low_lying_mass}
\end{eqnarray}
where $n_q=2,3$ for meson and baryon, respectively, 
and $m_{q0}$ is the quark current mass
and it is usually massless 
for light flavors $m_{q0}\approx 0$.
In general, the coupling constants $g_{H\sigma}$ 
or $g^{\mbox{input}}_{q\sigma}$ 
are determined by the phenomenology in order 
to fit the experimental hadronic mass spectrum. 
However, the above procedure is boring and produces 
a tremendous number of fitting parameters.
In order to reduce the number of phenomenological
parameters $g_{H\sigma}$ and 
to include the effect of the discrete hadronic mass spectrum, 
we assume the effective discrete hadronic mass spectra 
for baryons and mesons are, respectively, generated by
\begin{eqnarray}
m^{*}_{\mbox{B}}(\sigma)&=&
\left(M_{\mbox{B}}-M_{\mbox{Nucleon}}\right)
+3g_{q\sigma} \sigma,
\nonumber\\
m^{*}_{\mbox{M}}(\sigma)&=&
\left(M_{\mbox{M}}-\frac{2}{3}M_{\mbox{Nucleon}}\right)
+2g_{q\sigma} \sigma.
\label{chiral_low_lying_hadrons}
\end{eqnarray}
In the computational calculations, a special consideration 
is given for light mesons such as pions as far they 
are replaced by the scalar $\sigma$-mean field. 
Furthermore, the $\omega$ meson is replaced 
by the vector mean field while 
the other non-strange hadrons
are left as hadron particles.
The generalization of the isobaric function 
of the discrete low-lying hadronic mass spectrum,
namely the chiral low-lying isobaric function, 
is given by 
\begin{eqnarray}
f_{\mbox{Low-Lying}}(T,s)
&=&
\sum_{\mbox{B}}^{\mbox{Baryons}}
\left[
\varphi_{{\mbox{FD}}}
\left(T,s;m^{*}_{\mbox{B}}(\sigma),\Lambda^{\star}_{\mbox{B}}
\right)
+
\varphi_{{\mbox{FD}}}
\left(
T,s;m^{*}_{\mbox{B}}(\sigma),{\overline{\Lambda}}^{\star}_{\mbox{B}}
\right)\right]
\nonumber\\
&+&
\sum_{\mbox{M}}^{\mbox{Mesons}}
\left[\varphi_{{\mbox{BE}}}
\left(
T,s;m^{*}_{\mbox{M}}(\sigma),\Lambda^{\star}_{\mbox{M}}
\right)
+
\varphi_{{\mbox{BE}}}
\left(
T,s;m^{*}_{\mbox{M}}(\sigma),{\overline{\Lambda}}^{\star}_{\mbox{M}}
\right)\right].
\label{chiral_low_lying_fun}
\nonumber\\
\end{eqnarray}
The ensemble functions for single-particle species obeying the
Fermi-Dirac
and Bose-Einstein statistics read, respectively, 
\begin{eqnarray}
\varphi_{\mbox{FD}}
\left(T,s;m^{*}_{\mbox{B}}(\sigma),\Lambda^{\star}_{\mbox{B}}
\right)=
\left(2J_{\mbox{B}}+1\right)\int\frac{d^{3}k}{(2\pi)^{3}}
\ln\left[
1+\Lambda_{\mbox{B}} e^{-vs}
e^{-\frac{1}{T}\sqrt{k^{2}+{m^{*2}_{\mbox{B}}}(\sigma)}}
\right],
\label{quant_stats_fd_sigma}
\end{eqnarray}
and
\begin{eqnarray}
\varphi_{\mbox{BE}}\left(
T,s;m^{*}_{\mbox{M}}(\sigma),
\Lambda^{\star}_{\mbox{M}})\right)= 
-\left(2J_{\mbox{M}}+1\right)\int\frac{d^{3}k}{(2\pi)^{3}}
\ln\left[1-\Lambda_{\mbox{M}} e^{-vs}
e^{-\frac{1}{T}\sqrt{k^{2}+m_{\mbox{M}}^{*2}(\sigma)}}\right].
\label{quant_stats_be_sigma}
\end{eqnarray}
%%%%%%%%%%%%%%%%%%%%%%%%%%%%%%%%%%%%%%%%%%%%%%%%%%%
%
%
%
%
The pressure $p=T s^{*}$ is determined by finding 
the isobaric extreme right-hand singularity as follows,
\begin{eqnarray}
s^{*}&=&f_{\mbox{Fireballs}}(T,s^{*}) 
+f_{\mbox{Low-Lying}}(T,s^{*})
+f_{\sigma}(T,s^{*}),
\nonumber\\
&=& 
 f_{\mbox{Fireballs}}(T,s^{*}) 
+f_{\mbox{Low-Lying}}(T,s^{*})-U(\sigma),
\label{chiral_low_lying_singularity}
\end{eqnarray}
where the isobaric term of the scalar chiral $\sigma$-field 
is given by $f_{\sigma}(T,s^{*})=-U(\sigma)$.

The value of the scalar chiral $\sigma$-mean field 
is determined by extremizing the chiral isobaric 
pressure as follows 
\begin{eqnarray}
\frac{\partial}{\partial \sigma} s^{*}=0.
\label{chiral_pressure_extremum}
\end{eqnarray}
The chiral variation of the Hagedorn's isobaric function reads
\begin{eqnarray}
\frac{\partial}{\partial \sigma}
f_{\mbox{Fireballs}}(T,s^{*}) &=&
C'(T)\int^{\infty}_{v_0} dv v^{-\alpha+1} 
e^{-v\left(s-s_0\right)}
\frac{\partial s_0}{\partial \sigma} 
+\left(\cdots\right) \frac{\partial s^{*}}{\partial \sigma}.
\label{chiral_fireball_variation}
\end{eqnarray}
The chiral variation of the bag's internal pressure reads,
\begin{eqnarray} 
\frac{\partial p_{\mbox{Internal}}}{\partial\sigma}=
T \frac{\partial s_{0}}{\partial\sigma},
\label{chiral_internal_variation}
\end{eqnarray}
where 
\begin{eqnarray}
\frac{\partial s_0}{\partial \sigma}&=&
\frac{1}{3}
\left(3B/\widetilde{\Phi}^{(0)}_{q\overline{q} g\sigma}
(\overline{\beta}m^{*}_{\sigma})
\right)^{3/4} 
\times
\frac{\delta \widetilde{\Phi}^{(0)}_{q\overline{q} g\sigma}
(\overline{\beta}m^{*}_{\sigma})}{\delta \sigma}.
\end{eqnarray}
The asymptotic approximation of the chiral variation
of the Hagedorn's isobaric function reads
\begin{eqnarray}
\frac{\partial}{\partial \sigma}
f_{\mbox{Fireballs}}(T,s^{*}) &\propto&
-m^{*}_{q}\int^{\infty}_{v_0} 
dv v^{-\alpha+1} e^{-v\left(s-s_0\right)}
+\left(\cdots\right) \frac{\partial s^{*}}{\partial \sigma}.
\label{chiral_fireball_vary_asym}
\end{eqnarray}
%%%%%%%%%%%%%%%%%%%%%
%%%%%%%%%%%%%%%%%%%%%
The variation of the chiral isobaric 
function of the discrete low-lying mass 
spectrum with respect to the
scalar chiral $\sigma$-field reads
\begin{eqnarray}
\frac{\partial}{\partial \sigma}f_{\mbox{Low-Lying}}(T,s^{*})
&=&
\sum_{\mbox{B}}^{\mbox{Baryons}}
\left[
\varphi^{(\sigma)}_{{\mbox{FD}}}
\left(T,s^{*};m^{*}_{\mbox{B}}(\sigma),\Lambda^{\star}_{\mbox{B}}
\right)
+
\varphi^{(\sigma)}_{{\mbox{FD}}}
\left(
T,s^{*};m^{*}_{\mbox{B}}(\sigma),
{\overline{\Lambda}}^{\star}_{\mbox{B}}
\right)\right]
\nonumber\\
&+&
\sum_{\mbox{M}}^{\mbox{Mesons}}
\left[\varphi^{(\sigma)}_{{\mbox{BE}}}
\left(
T,s^{*};m^{*}_{\mbox{M}}(\sigma),\Lambda^{\star}_{\mbox{M}}
\right)
+
\varphi^{(\sigma)}_{{\mbox{BE}}}
\left(
T,s;m^{*}_{\mbox{M}}(\sigma),
{\overline{\Lambda}}^{\star}_{\mbox{M}}
\right)\right]
\nonumber\\
&+& 
\left(\cdots\right)\frac{\partial s^{*}}{\partial\sigma}.
\label{chiral_low_lying_vary}
\end{eqnarray}
The explicit expressions used in the above 
equations are displayed as follows
\begin{eqnarray}
\frac{\partial}{\partial\sigma}
\varphi_{\mbox{FD}} 
\left(T,s^*;m^{*}_{\mbox{B}}(\sigma),\Lambda^{\star}_{\mbox{B}}
\right)
&=&
\varphi^{(\sigma)}_{\mbox{FD}} 
\left(T,s^*;m^{*}_{\mbox{B}}(\sigma),\Lambda^{\star}_{\mbox{B}}
\right)
+
\left(\cdots\right) \frac{\partial s^{*}}{\partial\sigma},
\nonumber\\
\varphi^{(\sigma)}_{\mbox{FD}}
\left(T,s^*;m^{*}_{\mbox{B}}(\sigma),\Lambda^{\star}_{\mbox{B}}
\right)
&=&
-\left(2 J_{\mbox{B}}+1\right) g_{q\sigma}
\int\frac{dk k^2}{2\pi^2}
\frac{\frac{m^{*}_q(\sigma)}{T}/
\sqrt{k^{2}+{m^{*2}_{\mbox{B}}}(\sigma)}}
{\left[\Lambda^{-1}_{\mbox{B}} e^{vs^*}
e^{\frac{1}{T}\sqrt{k^{2}+{m^{*2}_{\mbox{B}}}(\sigma)}}+1\right]
},
\end{eqnarray}
and
\begin{eqnarray}
\frac{\partial}{\partial\sigma}
\varphi_{\mbox{BE}} 
\left(T,s^*;m^{*}_{\mbox{M}}(\sigma),\Lambda^{\star}_{\mbox{M}}
\right)
&=&
\varphi^{(\sigma)}_{\mbox{BE}} 
\left(T,s^*;m^{*}_{\mbox{M}}(\sigma),\Lambda^{\star}_{\mbox{M}}
\right)
+
\left(\cdots\right) \frac{\partial s^{*}}{\partial\sigma},
\nonumber\\
\varphi^{(\sigma)}_{\mbox{BE}}
\left(T,s^*;m^{*}_{\mbox{M}}(\sigma),\Lambda^{\star}_{\mbox{M}}
\right)
&=&
-\left(2 J_{\mbox{M}}+1\right) g_{q\sigma}
\int\frac{dk k^2}{2\pi^2}
\frac{\frac{m^{*}_q(\sigma)}{T}/
\sqrt{k^{2}+{m^{*2}_{\mbox{M}}}(\sigma)}}
{\left[\Lambda^{-1}_{\mbox{M}} e^{vs^*}
e^{\frac{1}{T}\sqrt{k^{2}+{m^{*2}_{\mbox{M}}}(\sigma)}}-1\right]
},
\end{eqnarray}
where the result
$\partial m^*_{q}(\sigma)/\partial\sigma=g_{q\sigma}$ 
is adopted. 
%%%%%%%%%%%%%%%%%%%%%%%%%%%%%%%%%%%%%%%%%%%%%%%%%%%%%%%%%%%%%%%
Finally, the variation of 
the scalar chiral $\sigma$-potential 
is  written for the simplicity as follows
\begin{eqnarray}
-\frac{\partial U(\sigma)}{\partial\sigma}
&\equiv&
M^{2}_{\sigma}(\sigma)\sigma+(\cdots),
\label{chiral_sigma_vary}
\end{eqnarray}
where $M_{\sigma}(\sigma)$ is interpreted 
as the effective mass of 
the scalar chiral $\sigma$-field particle. 
The dots term is the remaining terms 
those do not affect 
our final interpretation and conclusion in
the chiral restoration phase transition.

The scalar chiral $\sigma$-mean field condensate 
is determined by extremizing the total isobaric pressure 
(i.e. grand potential)
in the following way:
\begin{eqnarray}
T\frac{\partial s^{*}}{\partial \sigma}&=&0,
\nonumber\\
-\frac{\partial U(\sigma)}{\partial\sigma}
&=&-\frac{\partial}{\partial \sigma}\left[
f_{\mbox{Fireballs}}(T,s^{*}) 
+ 
f_{\mbox{Low-Lying}}(T,s^{*})
\right].
\label{chiral_discrete_hagedorn_extremum}
\end{eqnarray}
The Eq.(\ref{chiral_discrete_hagedorn_extremum}) 
is approximated and simplified to 
\begin{eqnarray}
M^{2}_{\sigma}(\sigma)\sigma
&=&-\frac{\partial}{\partial \sigma}\left[
f_{\mbox{Fireballs}}(T,s^{*}) 
+ 
f_{\mbox{Low-Lying}}(T,s^{*})
\right].
\label{chiral_discrete_hagedorn_extremum2}
\end{eqnarray}
It has been pointed out that the hadronic phase exhibits 
the Gross-Witten point in a way that the gas of 
the discrete low-lying mass spectrum undergoes
a third order phase transition from 
the discrete low-lying hadronic mass spectrum phase 
to a hadronic phase dominated 
by the continuous Hagedorn mass spectrum. 
This leaves three possibilities in order to study 
the chiral phase transition. 
In the case (I), we have a mixed low-lying and Hagedorn 
phases near the Gross-Witten point. 
In this case, the Gross-Witten
and chiral restoration phase transitions 
overlap around the Gross-Witten point, 
although the chiral restoration phase transition 
is a cross-over one unlike the third order 
Gross-Witten phase transition.
Moreover, the both transitions 
take place below and far away from
the deconfinement phase transition 
to an explosive quark-gluon plasma.
In the second case (II), the chiral restoration 
phase transition takes place in the hadronic phase 
dominated by 
the discrete low-lying hadronic mass spectrum particles.
In this scenario (i.e. case (II))
the chiral restoration phase transition is 
a smooth cross-over and is located below 
the Gross-Witten point and  
far away from the phase transition 
to an explosive quark-gluon plasma.
In this case (II),
the scalar chiral $\sigma$-mean field is determined 
by solving the following equation of extremization, 
\begin{eqnarray}
M^{2}_{\sigma}(\sigma)\sigma&\approx&
-\frac{\partial U(\sigma)}{\partial \sigma},
\nonumber\\
&=&-\frac{\partial}{\partial \sigma}\left[
f_{\mbox{Low-Lying}}(T,s^{*}) 
\right].
\label{chiral_discrete_extremum1}
\end{eqnarray}
This equation produces a smooth cross-over 
chiral restoration phase transition where 
the scalar chiral $\sigma$-mean field 
decreases smoothly.
This class of equation has been studied 
extensively in context of the Walecka model 
where the Van der Waals effect is neglected.  
When the bag's excluded volume is ignored, 
we get the standard $\sigma$-model as follows 
\begin{eqnarray}
M^{2}_{\sigma}(\sigma)\sigma
&\approx&-\frac{\partial}{\partial \sigma}\left[
f_{\mbox{Low-Lying}}(T,s^{*}) 
\right],\nonumber\\
&=&
2\left(J_{\mbox{B}}+1\right) g_{q\sigma}
\sum_{\mbox{B}}
\int\frac{dk k^2}{2\pi^2}
\frac{\frac{m^{*}_q(\sigma)}{T}/
\sqrt{k^{2}+{m^{*2}_{\mbox{B}}}(\sigma)}}
{\left[
e^{\frac{1}{T}\sqrt{k^{2}+{m^{*2}_{\mbox{B}}}(\sigma)}}+1\right]
}
\nonumber\\
&+&
2\left(J_{\mbox{M}}+1\right) g_{q\sigma} 
\sum_{\mbox{M}}\int\frac{dk k^2}{2\pi^2}
\frac{\frac{m^{*}_q(\sigma)}{T}/
\sqrt{k^{2}+{m^{*2}_{\mbox{M}}}(\sigma)}}
{\left[
e^{\frac{1}{T}\sqrt{k^{2}+{m^{*2}_{\mbox{M}}}(\sigma)}}-1\right]
},
\label{chiral_discrete_extremum2}
\end{eqnarray}
where the fugacity are set to 
$\Lambda_{\mbox{B}}=1$, $\Lambda_{\mbox{M}}=1$ 
as far the present analysis focuses 
in the region near zero baryonic 
chemical potential along the temperature axis.
%%%%%%%%%%%%%%%%%%%%%%%%%%%%%%%%%%%%%%%%%%%%%%%
In this case, the chiral restoration phase transition
takes place before the Gross-Witten point 
in contrary to the case (I) where in the case (I) 
the chiral restoration phase transition
overlaps the the Gross-Witten point and
is located far away from the deconfinement phase 
transition to an explosive quark-gluon plasma.
Hence, the case (II) sounds to fail to produce 
the continuous Hagedron threshold 
since it generates massless hadron masses before 
the continuous Hagedorn mass spectrum is reached.
Since in this scenario, 
the Hagedorn mass threshold is much higher 
than the maximum discrete mass spectrum, 
the continuous Hagedorn states production 
is denied.
The self-consistent phase transition 
from the hadronic phase 
to an quark-gluon plasma is not possible 
in the context of case (II)
because the gas of bags will never 
be produced unlike the cases (I) and (III).
Hence, the Gibbs construction becomes essential
in case (II) if we want to study 
the phase transition to quark-gluon plasma. 
It is hard to image an explosive quark-gluon 
plasma phase transition in this case. 
However, the case (I) sounds to be more physical 
than the case (II) since in the former case
the chiral restoration phase transition overlaps 
the third order Gross-Witten point. 
This overlapping mechanism virtually make 
the Hagedorn phase difficult to be detected 
directly experimentally~\cite{Bugaev:2008a}.
Furthermore, in this case the chiral restoration 
phase transition is supposed to take place 
prior to the color-deconfinement phase transition 
to an explosive quark-gluon plasma. 
Technically speaking, 
it will be hard to distinguish the color-deconfinement 
from the chiral restoration phase transition.
Nonetheless, the indirect detection and analysis 
can distinguish the explosive quark-gluon plasma 
from the static hadronic 
phase~\cite{Stoecker:2007a,Betz:2007b}.
Finally, the third case (III)
is the most important case. 
Indeed, in this case the chiral phase transition 
takes place in a hadronic phase dominated by 
the Hagedorn's gas. 
The chiral restoration phase transition 
takes place above the Gross-Witten point 
and utmost simultaneous the point 
of the deconfinement phase transition to 
an explosive quark-gluon plasma. 
The scalar chiral $\sigma$-mean field 
is determined by extremizing 
the isobaric pressure as follows
\begin{eqnarray}
-\frac{\partial U(\sigma)}{\sigma}&\approx&
M^{2}_{\sigma}(\sigma)\sigma, \nonumber\\
&\doteq&-\frac{\partial}{\partial \sigma}\left[
f_{\mbox{Fireballs}}(T,s^{*}) 
\right].
\label{chiral_fireball_extremum}
\end{eqnarray}
Hence, the hadronic matter dominated 
by the fireballs 
(i.e. the Hagedorn states) 
leads to the following equation
\begin{eqnarray}
M^{2}_{\sigma}(\sigma) \sigma&\doteq&
\left(\cdots\right)
m^{*}_{q}\cdot
\int^{\infty}_{v_0} dv v^{-\alpha+1} e^{-v\left(s-s_0\right)}.
\label{Hagedorn-chiral-asym1}
\end{eqnarray}
%%%%%%%%%%%%%%%%%%%%%%%%%%%%%%
Eq.(\ref{Hagedorn-chiral-asym1}) has very important 
characteristics which has been discussed in details 
in section~ \ref{sect_VdW} and Ref.~\cite{Zakout2006}.
At the point just below the phase transition 
from the Hagedorn phase to 
an explosive quark-gluon plasma,
it is shown that 
Eq.(\ref{Hagedorn-chiral-asym1}) 
has the following characteristic properties,
\begin{eqnarray}
M^{2}_{\sigma}(\sigma) \sigma
&\propto&
m^{*}_{q}\cdot
\lim_{s\rightarrow s_0}\left(s-s_0\right)^{\alpha-2}
\Gamma\left(-\alpha+2,v_0(s-s_0)\right),
\nonumber\\
&\sim&\mbox{finite} 
~~~\mbox{for}~~ \alpha>2,
\nonumber\\
&\sim& \infty ~~~\mbox{for}~~ 
\alpha\le 2 \rightarrow \sigma=0,
\label{Hagedorn-chiral-solution}
\end{eqnarray}
where $\Gamma(-n,x)$ is the exponential integral function 
of order $n$ where $n>0$. 
The integral on the right hand side 
of Eq.(\ref{Hagedorn-chiral-asym1})
diverges whenever the bag's internal isobaric pressure 
$s_0$ exceeds the external one $s$.
The phase transition from the hadronic 
phase to an explosive quark-gluon plasma 
takes place in the limit 
$(s-s_0)\rightarrow 0$  
when the bag's internal isobaric pressure $s_0$ reaches
the external isobaric hadronic pressure $s$ from below.
The solution $\sigma=0$ is the only non-trivial 
solution for the scalar chiral $\sigma$-mean field 
condensate when the explosive quark-gluon plasma
takes place subsequent the point $(s-s_0)=0^{-}$. 
The abrupt vanishing of 
the scalar chiral $\sigma$-mean field 
indicates evidently that the chiral symmetry 
restoration phase transition takes place 
below or utmost at the same point of the phase 
transition to an explosive quark-gluon plasma.
Actually, the point $s=s_0$ decides 
the order of chiral phase transition 
where the chiral phase transition takes 
place simultaneously with the deconfinement 
phase transition to an explosive quark-gluon plasma. 
However, if the cross-over chiral restoration phase 
transition takes place prior to the deconfinement 
phase transition then it takes place  
far away from the explosive 
quark-gluon plasma phase transition. 
%%%%%%%%%%%%%%%%%%%%%%%%%%%%%%%%%%%%%%%%%%%%%%%%%%%%%%%

The same analysis done in section~\ref{sect_VdW}
to determining the order of phase transition
can be carried out for the chiral symmetry 
restoration phase transition.
As the isobaric ensemble reaches 
the limit $(s-s_0)=0^{+}$ 
just below the point of the phase transition,
the scalar chiral $\sigma$-mean field solution  
is finite for the gas of bags with an internal 
structure of $\alpha>2$ 
while it must vanish for $\alpha\le 2$ 
in order to get the physical solution. 
This means that for the gas of bags 
with internal structures of $\alpha\le 2$
the chiral restoration phase transition
takes place below the point of the deconfinement 
phase transition to an explosive quark-gluon plasma. 
In this case, the order of the phase transition 
will be a rapid cross-over transition.
Furthermore, the chiral restoration phase 
transition must be completed far away 
the point of the phase transition
to an explosive quark-gluon plasma is reached.
If the smooth cross-over phase transition 
is not completed prior the deconfinement point,
then the chiral restoration is prohibited 
for $\alpha\le 2$.
On the other hand, the case is completely 
different for the gas of bags with 
the internal structure of $\alpha>2$.
In this case, the gas of bags undergoes 
the chiral restoration phase transition 
simultaneously at the same point of the 
explosive quark-gluon plasma phase transition.
The order of the chiral restoration 
phase transition is found 
of a second order for $\alpha=3$ 
while of a higher order for $3\ge\alpha>2$.
Finally, the chiral phase transition is of the 
first order for the gas of bags with
an internal structure of $\alpha>3$.
%%%%%%%%%%%%%%%%%%%%%%%%%%%%%%%%%%%%%%%%%%%%%%%

In the context of a model consisting hadronic 
matter dominated by the fireballs with an internal 
structure of $\alpha=3$ (i.e. the color-singlet bag),  
the scenario of the phase transition 
is summarized as follows:
The gas of the discrete low-lying hadronic mass 
spectrum particles undergoes 
a third order phase transition
to the Hagedorn phase at the Gross-Witten point.
The chiral restoration phase transition likely 
takes place in the hadronic matter that 
is dominated by the continuous Hagedorn states. 
The chiral restoration phase transition 
is located above the critical 
Gross-Witten point and it takes place 
simultaneously with the point 
of the phase transition to an explosive 
quark-gluon plasma.
The chiral restoration phase transition 
is of the second order and it takes place 
just below or utmost simultaneously the point 
of the phase transition 
to an explosive quark-gluon plasma.
Finally, the Hagedorn phase undergoes 
a first order phase transition to 
an explosive quark-gluon plasma. 
Nonetheless, this scenario does not 
exclude the possibility for 
the sharp cross-over chiral restoration 
phase transition to take place 
subsequent Gross-Witten point 
but far away from the point 
of the phase transition 
to an explosive quark-gluon plasma.
However, in the case that 
the smooth cross-over chiral restoration 
phase transition is not completed yet 
in the Hagedorn phase, 
then the second order chiral restoration 
phase transition must take place simultaneously 
with the point of phase transition 
to an explosive quark-gluon plasma. 
Indeed, in this scenario it seems likely 
that the chiral restoration phase transition 
takes place in the continuous Hagedorn 
spectrum phase rather than in the 
discrete low-lying mass spectrum phase.
The continuity of the Hagedorn mass spectrum makes 
the Hagedorn phase domain narrow in the hot bath. 
The narrowness of the Hagedorn phase range besides
being the restoration phase transition 
lain above Gross-Witten point 
and taken place utmost simultaneously 
with the deconfinement phase transition 
makes the continuous Hagedorn states 
difficult to be distinguished 
from the quark-gluon plasma.
It makes also more hard to distinguish 
the chiral restoration phase transition 
from the deconfinement phase transition. 
For example it has been argued that the heavy 
Hagedorn states have large widths \cite{Bugaev:2008a}.
Nevertheless, this scenario, however, 
is inconclusive although it explains 
the existence of the Gross-Witten point, 
the chiral restoration,
and deconfinement and even the existence 
of the tri-critical point 
for some class of bag's internal structures. 
The existence of 
the tri-critical point is essential 
in the multi-processes scenario.
%
%%%%%%%%%%%%%%%%%%%%%%%%%%%%%%%%%%%%%%%%%%
%%%%%%%%%%%%%%%%%%%%%%%%%%%%%%%%%%%%%%%%%%

The other alternative scenarios for 
a hadronic gas dominated 
by the fireballs can also be imagined. 
The fireball's internal 
structure $\alpha$ varies 
under extreme conditions due 
to the color-flavor correlations and 
the modification of the bag's volume fluctuations.
In the case that the bag's internal structure 
is of order $\alpha>3$, 
then both the chiral restoration 
and the color deconfinement  
are of first order phase transitions 
if they are taken place simultaneously. 
In contrary, the chiral restoration phase transition 
is found of a higher order for the gas of bags 
with the internal structure of $3\ge\alpha>2$.
The phase transition to an explosive quark-gluon plasma
is of the first order for $\alpha>2$ while 
of a higher order for $2\ge\alpha>1$.

In the class of scenarios 
where the chiral restoration and deconfinement 
phase transitions overlap, 
the $n$-th order chiral restoration
phase transition takes place simultaneously 
with the first order phase transition 
to an explosive quark-gluon plasma. 
Nonetheless, the Gross-Witten point remains 
a third order phase transition point.
However, the simultaneous phase transition 
to the chiral restoration and to 
an explosive quark-gluon plasma is not conclusive.
If the fireball has 
the internal structure of $2\ge\alpha>1$, 
then the deconfinement phase transition 
is of $n$-th order 
while the chiral phase transition
must take place far away from 
the point of phase transition 
to an explosive quark-gluon plasma. 
Moreover, the gas of hadronic bags 
with the internal structure of $\alpha\le 1$ 
never undergo the phase 
transition to an explosive quark-gluon plasma. 
It is interesting to note that 
much of the above scenarios 
are precursors that the smooth cross-over
chiral restoration phase transition is taken place below
and far away from the point of the phase transition 
to an explosive quark-gluon plasma 
and moreover as the chiral restoration phase transition 
approaches 
the point of the phase transition to quark-gluon plasma 
the cross-over restoration phase transition 
becomes more sharp and rapid.
%
%
%
%%%%%%%%%%%%%%%%%%%%%%%%%%%%%%%%%%
%%%%%%%%%%%%%%%%%%%%%%%%%%%%%%%%%%
%%%%%%%%%%%%%%%%%%%%%%%%%%%%%%%%%%

%%%%%%%%%%%%%%%%%%%%%%%%%%%%%%%%%
%%%%%%%%%%%%%%%%%%%%%%%%%%%%%%%%%
%%%%%%%%%%%%%%%%%%%%%%%%%%%%%%%%%
\section{Conclusions\label{sect_conclusion}}

In order to study 
the chiral restoration and deconfinement 
phase transitions in QCD,
the hadronic density of states must be known for 
the entire energy domain below the point of the 
phase transition to the true deconfined quark-gluon plasma.
The theoretical procedure to find the hadronic density 
of states is carried out by computing 
the micro-canonical ensemble.
The canonical ensemble for the color-singlet bag 
of constituent particles with the underlying symmetric 
group $SU(N_c)$ (and $U(N_c)$)
has been considered in detail.
The color structure has been attracted much attention
in order to understand the confinement/deconfinement in 
QCD~\cite{Elze:1983a,Elze84a,Elze985a,
Gorenstein:1983a,Skagerstam:1984a,Dumitru2004a,
Dumitru:2004b,Dumitru:2005b,Zakout2006}
or even the extended gauge field theories such 
as AdS/CFT~\cite{Sundborg:2000a,Aharony2003a,Schnitzer2004a}.
However, sometime ago, Gross and Witten have argued 
the existence of the tri-critical Gross-Witten point.
This point is thought to play a significant
role in the phase transition mechanism. 
It has been derived using
the spectral density of color eigenvalues method originally
introduced by Brezin {\em et. al.}~\cite{Brezin1978a}.
Recently, the Gross-Witten critical point
and the deconfinement mechanism has been reviewed
in the context of the Brezin {\em et. al.}~\cite{Brezin1978a}
method to study the quark-gluon plasma 
in QCD~\cite{ Lang:1981a,Skagerstam:1984a,Azakov1987a,
Dumitru2004a,Dumitru:2004b,Dumitru:2005b}
and the black hole formation 
in AdS/CFT~\cite{Sundborg:2000a,Aharony2003a,Aharony2005a,
Aharony:2005b,Alvarez-Gaume2005}.

In the present work, we have introduced an alternate 
and simpler method to derive the color-singlet 
canonical ensemble for the asymptotic large thermal 
running parameter $\lambda/N^2_c$.
Furthermore, we have demonstrated in details how to locate
the Gross-Witten critical point for the phase transition.
This novel method suits the realistic and 
complicate physical situations.
For the small thermal running parameter 
$\lambda/N^2_c$, the saddle points
are distributed uniformly over the entire color circle range 
$|\theta_i|\le\pi$.
Since the saddle points are distributed uniformly 
over the entire range, 
the Vandermonde determinant contributes to the action 
as an additional effective potential term. 
The integral of the resultant ensemble 
has been evaluated trivially.
However, this procedure fails when the saddle points 
congregate around the origin rather 
than distribute uniformly 
over the entire color circle range $|\theta_i|\le\pi$
and the Vandermonde effective potential 
develops a virtual singularity. 
In this case, a further consideration 
must be taken into account in order to regulate 
the action involving the Vandermonde effective potential  
and finally to evaluate the canonical ensemble correctly.
Therefore, this behavior indicates that the solution changes 
its analytical function characteristic 
and a subsequent phase transition takes place.
The solution of the asymptotic large $\lambda/N^2_c$ 
is evaluated using the Gaussian-like saddle points method.
The action is expanded around the stationary
Fourier color variables 
where only the quadratic terms are retained
while the Vandermonde determinant is regulated 
in a nontrivial way.
Fortunately, in this procedure the stationary 
Fourier color points are found dominant around the origin 
and this simplifies the problem drastically.
In spite of the action complexity due to the realistic
physical situation involved, it will be always 
an easy way to find the quadratic expansion around 
the saddle color points and 
the resultant integration over the color variables 
is evaluated using the standard Gaussian quadrature.
%
%%%%%%%%%%%%%%%%%%%%%%%%%%%%%%

The critical point for the Hagedorn phase transition
is determined in the midway of the interpolation
between the two different analytical solutions 
for the small- and the large-$\lambda/N^2_c$, respectively.
The Gaussian-like saddle points method
is compared with the spectral density of color eigenvalues
method (i.e. spectral density method, see for example
Refs.~\cite{Brezin1978a,GrossWitten1980a,Aharony2003a})
and the exact numerical solution as well. 
It is found that the exact numerical solution
fits precisely the results 
of the Gaussian-like saddle points method
for the large thermal running parameter
$\lambda/N^2_c>\lambda_0/N^2_c$ 
and furthermore it is found in a good agreement 
with the spectral density of color eigenvalues method.
When the large-$\lambda/N^2_c$ is extrapolated 
to the small-$\lambda/N^2_c<\lambda_0/N^2_c$, 
the solution becomes slippery 
and when the $\lambda/N^2_c$ reaches 
some critical value, 
it is deflected to increase. 
This solution is concave up and has a minimum
at the critical point $\lambda_{(II)\mbox{min}}$. 
This point is the threshold point 
for the acceptable physical solution (II).
%%%%%%%%%%%%%%%%%%%%%%%%%%%%%%%%%%%%%%%%%%%%%%%%%%%%%%%%%%%%%%%%%%%%%
%%%%%%%%%%%%%%%%%%%%%%%%%%%%%%%%%%%%%%%%%%%%%%%%%%%%%%%%%%%%%%%%%%%%%
The minimum point for the extrapolation 
of the asymptotic large-$\lambda/N^2_c$ solution 
is actually the minimal threshold point.
The extreme left hand side interval for 
the asymptotic large-$\lambda/N^2_c$ solution 
is presumed to start from the threshold 
point $\lambda_{(II)\mbox{min}}$.
The extrapolation down below this threshold point 
is unphysical and is declined as a solution.
This means that beyond the threshold point, 
the asymptotic large-$\lambda/N^2_c$ 
solution (II) is deflected and changes 
its analyticity in order to match and satisfy 
the asymptotic small-$\lambda/N^2_c$ solution (I).
The solution (I) is found in agreement with 
the exact numerical results for the small $\lambda/N^2_c$.
The asymptotic small-$\lambda/N^2_c$ solution is found
be adopting standard approximations.
The action is expanded to $\lambda$-power expansion 
and then is evaluated using the group orthogonality 
over the full color range. 
The ultimate limit of the approximation
validity is given by the point 
$\lambda_{(I)\mbox{max}}$
and beyond this point the solution will be broken.
The action structure is simplified drastically 
in the range $\lambda\le\lambda_{(I)\mbox{max}}$.
For example, the quantum statistics can be approximated always
to Maxwell-Boltzmann statistics and so on.
However, the asymptotic small- and large-$\lambda/N^2_c$ 
solutions (i.e. solutions (I) and (II), respectively) 
may split by a small additional constant. 
This constant is simply the approximation 
redundant in particular when both solutions 
are extrapolated far away from their asymptotic limits.
The critical point $\lambda_0/N^2_c$
for the phase transition from the asymptotic
small-$\lambda/N^2_c$ to large-$\lambda/N^2_c$ solutions 
is located near or above 
$\lambda\ge\lambda_{(II)\mbox{min}}$ 
but below 
$\lambda\le\lambda_{(I)\mbox{max}}$.
It is roughly the midway interpolation between
the small and large solutions.
The advantage of the Gaussian-like saddle points 
method is the simplicity 
in the sense it can deal with a complicated physical
problem such as the deconfinement 
phase transition in QCD.

%%%%%%%%%%%%%%%%%%%%%%%%%%%%%%%%%%%%%%%%%
%%%%%%%%%%%%%%%%%%%%%%%%%%%%%%%%%%%%%%%%%
%%%  Fig.
%%%  Figure
%%%  Hagedorn_illus.eps
%%%  Hagedorn_illus_higher.eps
%%%
%%%  fig_hagedorn_illus
%%%  fig_hagedorn_illus_higher
%%%
%%%%%%%%%%%%%%%%%%%%%%%%%%%%%%%%%%%%%%%%%
%%%%%%%%%%%%%%%%%%%%%%%%%%%%%%%%%%%%%%%%%
The QCD-phase transition is studied in the context 
of the color-singlet state of quark and gluon bags.
It is found that the density of states 
for low-lying masses is a discrete spectrum. 
The low-lying density of states is determined 
by the known hadronic mass spectrum particles that 
are found experimentally and are available 
in the data book~\cite{databook2004}.
The Hagedorn states appear just above 
the low-lying known mass spectrum. 
The mass spectrum for the Hagedorn states is continuous. 
The gas dominated by the low-lying mass spectrum particles
undergoes a third order phase transition to a hadronic 
phase dominated by the Hagedorn states.
When the system is thermally excited beyond 
the Hagedorn phase, the hadronic phase 
undergoes another phase transition 
to an explosive quark-gluon plasma.

\begin{figure}
\includegraphics{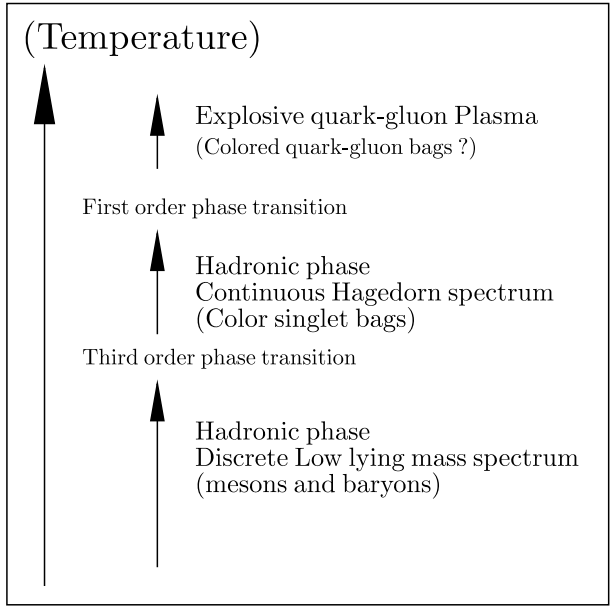}%
\caption{\label{fig_hagedorn_illus}
The phase transition scenario in the context of the present model. 
At low temperature,
the diluted nuclear matter is dominated
by the low-lying hadronic mass spectrum such pions, nucleons 
etc $\cdots$.
At the Hagedorn critical temperature, the system undergoes 
a third order phase transition
to system dominated by the highly excited hadronic states
known as Hagedorn states or fireballs.
When the system is thermally excited to higher temperature, 
the hadronic phase undergoes a first order phase transition 
to an explosive quark-gluon plasma.}
\end{figure}

\begin{figure}
\includegraphics{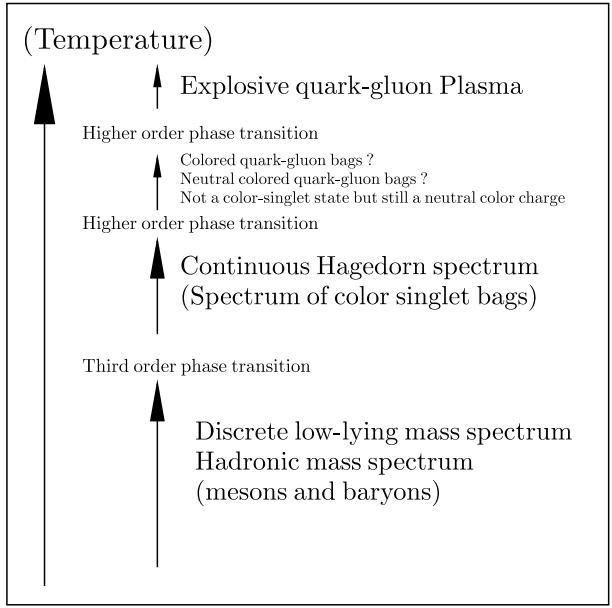}%
\caption{\label{fig_hagedorn_illus_higher}
Same as Fig.\ref{fig_hagedorn_illus} but with an alternate scenario
for the phase transition.
The hadronic phase dominated by Hagedorn states undergoes
a higher order phase transition to another phase dominated
by highly excited neutral color bound quark-gluon bags 
(i.e. bound state but with a color non-singlet state). 
At higher temperature, the gas of colored quark and gluon bags 
undergoes 
a higher order phase transition
to an explosive real deconfined quark-gluon plasma. 
In this scenario, the phase transition 
to an explosive quark-gluon plasma takes place
only through more complicated multi-processes.}
\end{figure}

The scenario for the phase transition is depicted
in Fig.(\ref{fig_hagedorn_illus}).
In the diluted nuclear matter when the temperature increases 
and reaches the critical one, 
the gas of the discrete low-lying hadronic 
mass spectrum particles undergoes 
a third order phase transition to a gas of continuous 
Hagedorn states
(i.e. the high-lying hadronic states).
Furthermore, when the system is thermally excited above this temperature, 
the gas of Hagedorn states undergoes another first order or higher order
phase transition to an explosive quark-gluon plasma.
Moreover, the scenario for the phase transition 
can be extended to consider
into account the multiple intermediate processes.
An alternate scenario is depicted in
Fig.(\ref{fig_hagedorn_illus_higher}). 
The gas of Hagedorn states undergoes 
a higher order phase transition to a gas of neutral colored bags 
or even non-singlet bags.
These bags are not color-singlet states but carry neutral 
color charge and this charge is fixed 
by the effective color chemical potentials. 
Subsequently, the gas of neutral colored bags undergoes 
a higher order phase transition to a gas of colored bags. 
The non-singlet bags are not colored bags
and not true deconfined colors 
but rather bags with specific internal color structures. 
In the colored bags, the quarks and gluons are still bounded in
finite size blobs and  they carry total color charge.
The phase transition to the colored bags 
with conserved color charges 
might be associated with breaking 
the group symmetry $SU(N_c)$ to $U(1)^{N_c-1}$.   
These colored bags become unstable and when the system 
is perturbed slightly and thermally excited, the system eventually 
undergoes another higher order phase transition to a real deconfined
quark-gluon plasma.

\begin{figure}
\includegraphics{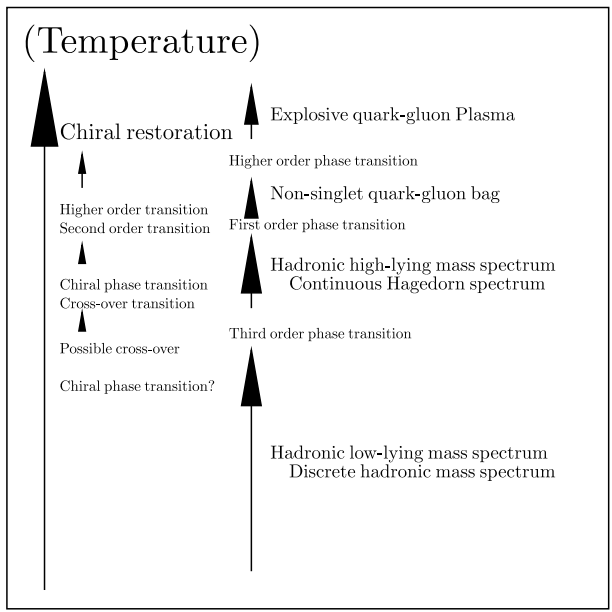}%
\caption{\label{fig_hagedorn_chiral}
The consistent chiral and deconfinement phase transition scenario.}
\end{figure}

The possible consistent color-deconfinement 
and chiral restoration 
phase transition scenarios are depicted 
in Fig.(\ref{fig_hagedorn_chiral}).
It is illustrated that the discrete low-lying 
hadronic mass spectrum undergoes
a third order phase transition 
to the Hagedorn phase.
The gas of continuous Hagedorn states 
with internal structures $\alpha>2$ undergoes 
a first order phase transition 
to an explosive quark-gluon plasma.
However, it is possible in some alternate scenarios 
that the Hagedorn phase undergoes a higher order phase 
transition to the gas of non-singlet bags 
(i.e. with $\alpha\le 2$)
and the subsequent gas 
of exotic states undergoes 
a higher order phase transition 
to an explosive quark-gluon plasma. 
The chiral phase transition 
for the gas of Hagedorn states 
with the internal structure $\alpha=3$
is found of the second order and 
it takes place utmost simultaneously 
the deconfinement first order phase transition 
to an explosive quark-gluon plasma.
However, in the case that the chiral phase 
transition persists to take place 
in the gas of discrete low-lying hadronic
mass spectrum states prior to the Hagedorn phase 
then the chiral phase transition 
will be a smooth cross-over one. 
In this case, it will be hard to reach 
the Hagedorn threshold production 
and subsequently the Hagedorn states 
will not be produced and no explosive quark-gluon 
plasma will be generated. The quark-gluon plasma 
in this scenario can be found by the Gibbs construction.
It is possible in other scenarios 
that the the chiral phase transition  
coincides the Gross-Witten point 
and it will be  a cross-over transition.
It seems that the chiral restoration 
phase transition likely takes
place utmost in the Hagedorn phase 
just below the deconfinement phase transition 
or at least coincides the Gross-Witten point.
In some scenarios, it is likely that 
the chiral phase transition coincides the Gross-Witten point
while in other scenarios the chiral restoration phase transition
is likely takes place simultaneously with the deconfinement
phase transition to an explosive quark-gluon plasma.
Indeed, the Hagedorn bag's internal structure $\alpha$
is found essential in the phase transition diagram.
Furthermore, it is also possible 
the Hagedorn phase undergoes 
a higher order phase transition 
through multi-processes internal-structure 
phase transition while 
the chiral restoration phase transition 
is a smooth cross-over transition 
and takes place in the continuous Hagedorn phase.
The order of the chiral phase transition becomes
sharper and of a lower order as its point approaches
the deconfinement point to an explosive quark-gluon plasma. 
The QCD phase transition diagram 
is proved very rich and non-trivial.

%%%%%%%%%%%%%%%%%%%%%%%%%
%%%%%%%%%%%%%%%%%%%%%%%%%%%%%%%%%%%%%%%%%%%%%%%%%%%
%%%%%%%%%%%%%%%%%%%%%%%%%%%%%%%%%%%%%%%%%%%%%%%%%%%
%%%%%%%%%%%%%%%%%%%%%%%%%%%%%%%%%%%%%%%%%%%%%%%%%%%
%%%%%%%%%%%%%
%%%%%%%%%%%%%
%%%%%%%%%%%%%
%%%%%%%%%%%%%
%%%%%%%%%%%%%
%%%%%%%%%%%%%%%%%%%%%%%%%%%%%%%%%%%%%%%%%%%%%%%%%%%%%%%%%%%%%
%%%%%%%%%%%%%%%%%%%%%%%%%%%%%%%%%%%%%%%%%%%%%%%%%%%%%%%%%%%%%
%%%%%%%%%%%%%%%%%%%%%%%%%%%%%%%%%%%%%%%%%%%%%%%%%%%%%%%%%%%%%
%%%%%%%%%%%%%%%%%%%%%%%%%%%%%
%%%%%%%%%%%%%%%%%%%%%%%%%%%%%%%%%%%%%%%%%%%%%%%%%%%%%%%%%%%%%
%%%%%%%%%%%%%%%%%%%%%%%%%%%%%%%%%%%%%%%%%%%%%%%%%%%%%%%%%%%%%
%%%%%%%%%%%%%%%%%%%%%%%%%%%%%%%%%%%%%%%%%%%%%%%%%%%%%%%%%%%%%
%%%%%%%%%%%%%%%%%%%%%%%%%%%%%%%%%%%%%%%%%%%%%%%%%%%%%%%%
%%%%%%%%%%%%%%%%%%%%%%%%%%%%%%%%%%%%%%%%%%%%%%%%%%%%%%%%
%%%%%%%%%%%%%%%%%%%%%%%%%%%%%%%%%%%%%%%%%%%%%%%%%%%%%%%%
%%%%%%%%%%%%%%%%%%%%%%%%%%%%%%%%%%%%%%%%%%%%%%%%%%%%%%%%
%%%%%%%%%%%%%%%%%%%%%%%%%%%%%%%%%%%%%%%%%%%%%%%%%%%%%%%%%%
%%%%%%%%%%%%%%%%%%%%%%%%%%%%%%%%%%%%%%%%%%%%%%%%%%%%%%%%%%
%%%%%%%%%%%%%%%%%%%%%%%%%%%%%%%%%%%%%%%%%%%%%%%%%%%%%%%%%%

\begin{acknowledgments}
I. Z. gratefully acknowledges support from Frankfurt institute
for advanced studies.
He is indebted to Walter Greiner and Horst St\"ocker for
their encouragements and discussions and Melissa Franklin
and John Huth for their support during
his visit to Harvard University and
George Brandenburg for the hospitality in Laboratory
for Particle Physics and Cosmology at Harvard University.
The authors thank J. Schaffner-Bielich for the collaboration
in the earlier stages of the present work 
and H. T. Elze.
The authors also thank the Frankfurt Center for Scientific Computing.
This work was supported by Alexander von Humboldt foundation
and  Harvard University.
\end{acknowledgments}

%%%%%%%%%%%%%%%%%%%%%%%%%%%
\bibliography{gg_ext_x1}

\begin{thebibliography}{53}
\expandafter\ifx\csname natexlab\endcsname\relax\def\natexlab#1{#1}\fi
\expandafter\ifx\csname bibnamefont\endcsname\relax
  \def\bibnamefont#1{#1}\fi
\expandafter\ifx\csname bibfnamefont\endcsname\relax
  \def\bibfnamefont#1{#1}\fi
\expandafter\ifx\csname citenamefont\endcsname\relax
  \def\citenamefont#1{#1}\fi
\expandafter\ifx\csname url\endcsname\relax
  \def\url#1{\texttt{#1}}\fi
\expandafter\ifx\csname urlprefix\endcsname\relax\def\urlprefix{URL }\fi
\providecommand{\bibinfo}[2]{#2}
\providecommand{\eprint}[2][]{\url{#2}}

\bibitem[{\citenamefont{Hagedorn}(1965)}]{Hagedorn:1965a}
\bibinfo{author}{\bibfnamefont{R.}~\bibnamefont{Hagedorn}},
  \bibinfo{journal}{Nuovo Cim. Suppl.} \textbf{\bibinfo{volume}{3}},
  \bibinfo{pages}{147} (\bibinfo{year}{1965}).

\bibitem[{\citenamefont{Frautschi}(1971)}]{Frautschi:1971a}
\bibinfo{author}{\bibfnamefont{S.~C.} \bibnamefont{Frautschi}},
  \bibinfo{journal}{Phys. Rev.} \textbf{\bibinfo{volume}{D3}},
  \bibinfo{pages}{2821} (\bibinfo{year}{1971}).

\bibitem[{\citenamefont{Kapusta}(1981)}]{Kapusta1981a}
\bibinfo{author}{\bibfnamefont{J.~I.} \bibnamefont{Kapusta}},
  \bibinfo{journal}{Phys. Rev.} \textbf{\bibinfo{volume}{D23}},
  \bibinfo{pages}{2444} (\bibinfo{year}{1981}).

\bibitem[{\citenamefont{Kapusta}(1982)}]{Kapusta:1982a}
\bibinfo{author}{\bibfnamefont{J.~I.} \bibnamefont{Kapusta}},
  \bibinfo{journal}{Nucl. Phys.} \textbf{\bibinfo{volume}{B196}},
  \bibinfo{pages}{1} (\bibinfo{year}{1982}).

\bibitem[{\citenamefont{Redlich and Turko}(1980)}]{Redlich80a}
\bibinfo{author}{\bibfnamefont{K.}~\bibnamefont{Redlich}} \bibnamefont{and}
  \bibinfo{author}{\bibfnamefont{L.}~\bibnamefont{Turko}}, \bibinfo{journal}{Z.
  Phys.} \textbf{\bibinfo{volume}{C5}}, \bibinfo{pages}{201}
  (\bibinfo{year}{1980}).

\bibitem[{\citenamefont{Turko}(1981)}]{Turko:1981a}
\bibinfo{author}{\bibfnamefont{L.}~\bibnamefont{Turko}},
  \bibinfo{journal}{Phys. Lett.} \textbf{\bibinfo{volume}{B104}},
  \bibinfo{pages}{153} (\bibinfo{year}{1981}).

\bibitem[{\citenamefont{Gross and Witten}(1980)}]{GrossWitten1980a}
\bibinfo{author}{\bibfnamefont{D.~J.} \bibnamefont{Gross}} \bibnamefont{and}
  \bibinfo{author}{\bibfnamefont{E.}~\bibnamefont{Witten}},
  \bibinfo{journal}{Phys. Rev.} \textbf{\bibinfo{volume}{D21}},
  \bibinfo{pages}{446} (\bibinfo{year}{1980}).

\bibitem[{\citenamefont{Brezin et~al.}(1978)\citenamefont{Brezin, Itzykson,
  Parisi, and Zuber}}]{Brezin1978a}
\bibinfo{author}{\bibfnamefont{E.}~\bibnamefont{Brezin}},
  \bibinfo{author}{\bibfnamefont{C.}~\bibnamefont{Itzykson}},
  \bibinfo{author}{\bibfnamefont{G.}~\bibnamefont{Parisi}}, \bibnamefont{and}
  \bibinfo{author}{\bibfnamefont{J.~B.} \bibnamefont{Zuber}},
  \bibinfo{journal}{Commun. Math. Phys.} \textbf{\bibinfo{volume}{59}},
  \bibinfo{pages}{35} (\bibinfo{year}{1978}).

\bibitem[{\citenamefont{Lang et~al.}(1981)\citenamefont{Lang, Salomonson, and
  Skagerstam}}]{Lang:1981a}
\bibinfo{author}{\bibfnamefont{C.~B.} \bibnamefont{Lang}},
  \bibinfo{author}{\bibfnamefont{P.}~\bibnamefont{Salomonson}},
  \bibnamefont{and} \bibinfo{author}{\bibfnamefont{B.~S.}
  \bibnamefont{Skagerstam}}, \bibinfo{journal}{Nucl. Phys.}
  \textbf{\bibinfo{volume}{B190}}, \bibinfo{pages}{337} (\bibinfo{year}{1981}).

\bibitem[{\citenamefont{Skagerstam}(1984)}]{Skagerstam:1984a}
\bibinfo{author}{\bibfnamefont{B.~S.} \bibnamefont{Skagerstam}},
  \bibinfo{journal}{Z. Phys.} \textbf{\bibinfo{volume}{C24}},
  \bibinfo{pages}{97} (\bibinfo{year}{1984}).

\bibitem[{\citenamefont{Azakov et~al.}(1987)\citenamefont{Azakov, Salomonson,
  and Skagerstam}}]{Azakov1987a}
\bibinfo{author}{\bibfnamefont{S.~I.} \bibnamefont{Azakov}},
  \bibinfo{author}{\bibfnamefont{P.}~\bibnamefont{Salomonson}},
  \bibnamefont{and} \bibinfo{author}{\bibfnamefont{B.~S.}
  \bibnamefont{Skagerstam}}, \bibinfo{journal}{Phys. Rev.}
  \textbf{\bibinfo{volume}{D36}}, \bibinfo{pages}{2137} (\bibinfo{year}{1987}).

\bibitem[{\citenamefont{Hallin and Persson}(1998)}]{Hallin1998a}
\bibinfo{author}{\bibfnamefont{J.}~\bibnamefont{Hallin}} \bibnamefont{and}
  \bibinfo{author}{\bibfnamefont{D.}~\bibnamefont{Persson}},
  \bibinfo{journal}{Phys. Lett.} \textbf{\bibinfo{volume}{B429}},
  \bibinfo{pages}{232} (\bibinfo{year}{1998}), \eprint{arXiv:hep-ph/9803234}.

\bibitem[{\citenamefont{Dumitru et~al.}(2004)\citenamefont{Dumitru, Hatta,
  Lenaghan, Orginos, and Pisarski}}]{Dumitru2004a}
\bibinfo{author}{\bibfnamefont{A.}~\bibnamefont{Dumitru}},
  \bibinfo{author}{\bibfnamefont{Y.}~\bibnamefont{Hatta}},
  \bibinfo{author}{\bibfnamefont{J.}~\bibnamefont{Lenaghan}},
  \bibinfo{author}{\bibfnamefont{K.}~\bibnamefont{Orginos}}, \bibnamefont{and}
  \bibinfo{author}{\bibfnamefont{R.~D.} \bibnamefont{Pisarski}},
  \bibinfo{journal}{Phys. Rev.} \textbf{\bibinfo{volume}{D70}},
  \bibinfo{pages}{034511} (\bibinfo{year}{2004}),
  \eprint{arXiv:hep-th/0311223}.

\bibitem[{\citenamefont{Dumitru
  et~al.}(2005{\natexlab{a}})\citenamefont{Dumitru, Lenaghan, and
  Pisarski}}]{Dumitru:2004b}
\bibinfo{author}{\bibfnamefont{A.}~\bibnamefont{Dumitru}},
  \bibinfo{author}{\bibfnamefont{J.}~\bibnamefont{Lenaghan}}, \bibnamefont{and}
  \bibinfo{author}{\bibfnamefont{R.~D.} \bibnamefont{Pisarski}},
  \bibinfo{journal}{Phys. Rev.} \textbf{\bibinfo{volume}{D71}},
  \bibinfo{pages}{074004} (\bibinfo{year}{2005}{\natexlab{a}}),
  \eprint{arXiv:hep-ph/0410294}.

\bibitem[{\citenamefont{Dumitru
  et~al.}(2005{\natexlab{b}})\citenamefont{Dumitru, Pisarski, and
  Zschiesche}}]{Dumitru:2005b}
\bibinfo{author}{\bibfnamefont{A.}~\bibnamefont{Dumitru}},
  \bibinfo{author}{\bibfnamefont{R.~D.} \bibnamefont{Pisarski}},
  \bibnamefont{and}
  \bibinfo{author}{\bibfnamefont{D.}~\bibnamefont{Zschiesche}},
  \bibinfo{journal}{Phys. Rev.} \textbf{\bibinfo{volume}{D72}},
  \bibinfo{pages}{065008} (\bibinfo{year}{2005}{\natexlab{b}}),
  \eprint{arXiv:hep-ph/0505256}.

\bibitem[{\citenamefont{Nambu et~al.}(1982)\citenamefont{Nambu, Bambah, and
  Gross}}]{Nambu:1982a}
\bibinfo{author}{\bibfnamefont{Y.}~\bibnamefont{Nambu}},
  \bibinfo{author}{\bibfnamefont{B.}~\bibnamefont{Bambah}}, \bibnamefont{and}
  \bibinfo{author}{\bibfnamefont{M.}~\bibnamefont{Gross}},
  \bibinfo{journal}{Phys. Rev.} \textbf{\bibinfo{volume}{D26}},
  \bibinfo{pages}{2875} (\bibinfo{year}{1982}).

\bibitem[{\citenamefont{Bambah}(1984)}]{Bambah:1983a}
\bibinfo{author}{\bibfnamefont{B.}~\bibnamefont{Bambah}},
  \bibinfo{journal}{Phys. Rev.} \textbf{\bibinfo{volume}{D29}},
  \bibinfo{pages}{1323} (\bibinfo{year}{1984}).

\bibitem[{\citenamefont{Gross}(1983)}]{GrossM:1983a}
\bibinfo{author}{\bibfnamefont{M.}~\bibnamefont{Gross}},
  \bibinfo{journal}{Phys. Rev.} \textbf{\bibinfo{volume}{D27}},
  \bibinfo{pages}{432} (\bibinfo{year}{1983}).

\bibitem[{\citenamefont{Jaimungal and Paniak}(1998)}]{Jaimungal:1997a}
\bibinfo{author}{\bibfnamefont{S.}~\bibnamefont{Jaimungal}} \bibnamefont{and}
  \bibinfo{author}{\bibfnamefont{L.~D.} \bibnamefont{Paniak}},
  \bibinfo{journal}{Nucl. Phys.} \textbf{\bibinfo{volume}{B517}},
  \bibinfo{pages}{622} (\bibinfo{year}{1998}), \eprint{arXiv:hep-th/9710044}.

\bibitem[{\citenamefont{Gattringer et~al.}(1997)\citenamefont{Gattringer,
  Paniak, and Semenoff}}]{Gattringer:1996a}
\bibinfo{author}{\bibfnamefont{C.~R.} \bibnamefont{Gattringer}},
  \bibinfo{author}{\bibfnamefont{L.~D.} \bibnamefont{Paniak}},
  \bibnamefont{and} \bibinfo{author}{\bibfnamefont{G.~W.}
  \bibnamefont{Semenoff}}, \bibinfo{journal}{Annals Phys.}
  \textbf{\bibinfo{volume}{256}}, \bibinfo{pages}{74} (\bibinfo{year}{1997}),
  \eprint{arXiv:hep-th/9612030}.

\bibitem[{\citenamefont{Schnitzer}(2004)}]{Schnitzer2004a}
\bibinfo{author}{\bibfnamefont{H.~J.} \bibnamefont{Schnitzer}},
  \bibinfo{journal}{Nucl. Phys.} \textbf{\bibinfo{volume}{B695}},
  \bibinfo{pages}{267} (\bibinfo{year}{2004}), \eprint{arXiv:hep-th/0402219}.

\bibitem[{\citenamefont{Sundborg}(2000)}]{Sundborg:2000a}
\bibinfo{author}{\bibfnamefont{B.}~\bibnamefont{Sundborg}},
  \bibinfo{journal}{Nucl. Phys.} \textbf{\bibinfo{volume}{B573}},
  \bibinfo{pages}{349} (\bibinfo{year}{2000}), \eprint{arXiv:hep-th/9908001}.

\bibitem[{\citenamefont{Aharony et~al.}(2004)\citenamefont{Aharony, Marsano,
  Minwalla, Papadodimas, and Van~Raamsdonk}}]{Aharony2003a}
\bibinfo{author}{\bibfnamefont{O.}~\bibnamefont{Aharony}},
  \bibinfo{author}{\bibfnamefont{J.}~\bibnamefont{Marsano}},
  \bibinfo{author}{\bibfnamefont{S.}~\bibnamefont{Minwalla}},
  \bibinfo{author}{\bibfnamefont{K.}~\bibnamefont{Papadodimas}},
  \bibnamefont{and}
  \bibinfo{author}{\bibfnamefont{M.}~\bibnamefont{Van~Raamsdonk}},
  \bibinfo{journal}{Adv. Theor. Math. Phys.} \textbf{\bibinfo{volume}{8}},
  \bibinfo{pages}{603} (\bibinfo{year}{2004}), \eprint{arXiv:hep-th/0310285}.

\bibitem[{\citenamefont{Aharony et~al.}(2005)\citenamefont{Aharony, Marsano,
  Minwalla, Papadodimas, and Van~Raamsdonk}}]{Aharony2005a}
\bibinfo{author}{\bibfnamefont{O.}~\bibnamefont{Aharony}},
  \bibinfo{author}{\bibfnamefont{J.}~\bibnamefont{Marsano}},
  \bibinfo{author}{\bibfnamefont{S.}~\bibnamefont{Minwalla}},
  \bibinfo{author}{\bibfnamefont{K.}~\bibnamefont{Papadodimas}},
  \bibnamefont{and}
  \bibinfo{author}{\bibfnamefont{M.}~\bibnamefont{Van~Raamsdonk}},
  \bibinfo{journal}{Phys. Rev.} \textbf{\bibinfo{volume}{D71}},
  \bibinfo{pages}{125018} (\bibinfo{year}{2005}),
  \eprint{arXiv:hep-th/0502149}.

\bibitem[{\citenamefont{Aharony et~al.}(2006)\citenamefont{Aharony, Minwalla,
  and Wiseman}}]{Aharony:2005b}
\bibinfo{author}{\bibfnamefont{O.}~\bibnamefont{Aharony}},
  \bibinfo{author}{\bibfnamefont{S.}~\bibnamefont{Minwalla}}, \bibnamefont{and}
  \bibinfo{author}{\bibfnamefont{T.}~\bibnamefont{Wiseman}},
  \bibinfo{journal}{Class. Quant. Grav.} \textbf{\bibinfo{volume}{23}},
  \bibinfo{pages}{2171} (\bibinfo{year}{2006}), \eprint{arXiv:hep-th/0507219}.

\bibitem[{\citenamefont{Alvarez-Gaume et~al.}(2005)\citenamefont{Alvarez-Gaume,
  Gomez, Liu, and Wadia}}]{Alvarez-Gaume2005}
\bibinfo{author}{\bibfnamefont{L.}~\bibnamefont{Alvarez-Gaume}},
  \bibinfo{author}{\bibfnamefont{C.}~\bibnamefont{Gomez}},
  \bibinfo{author}{\bibfnamefont{H.}~\bibnamefont{Liu}}, \bibnamefont{and}
  \bibinfo{author}{\bibfnamefont{S.~R.} \bibnamefont{Wadia}},
  \bibinfo{journal}{Phys. Rev.} \textbf{\bibinfo{volume}{D71}},
  \bibinfo{pages}{124023} (\bibinfo{year}{2005}),
  \eprint{arXiv:hep-th/0502227}.

\bibitem[{\citenamefont{Muller and Rafelski}(1982)}]{Muller:1982b}
\bibinfo{author}{\bibfnamefont{B.}~\bibnamefont{Muller}} \bibnamefont{and}
  \bibinfo{author}{\bibfnamefont{J.}~\bibnamefont{Rafelski}},
  \bibinfo{journal}{Phys. Lett.} \textbf{\bibinfo{volume}{B116}},
  \bibinfo{pages}{274} (\bibinfo{year}{1982}).

\bibitem[{\citenamefont{Zakout et~al.}(2007)\citenamefont{Zakout, Greiner, and
  Schaffner-Bielich}}]{Zakout2006}
\bibinfo{author}{\bibfnamefont{I.}~\bibnamefont{Zakout}},
  \bibinfo{author}{\bibfnamefont{C.}~\bibnamefont{Greiner}}, \bibnamefont{and}
  \bibinfo{author}{\bibfnamefont{J.}~\bibnamefont{Schaffner-Bielich}},
  \bibinfo{journal}{Nucl. Phys.} \textbf{\bibinfo{volume}{A781}},
  \bibinfo{pages}{150} (\bibinfo{year}{2007}).

\bibitem[{\citenamefont{Elze et~al.}(1983)\citenamefont{Elze, Greiner, and
  Rafelski}}]{Elze:1983a}
\bibinfo{author}{\bibfnamefont{H.~T.} \bibnamefont{Elze}},
  \bibinfo{author}{\bibfnamefont{W.}~\bibnamefont{Greiner}}, \bibnamefont{and}
  \bibinfo{author}{\bibfnamefont{J.}~\bibnamefont{Rafelski}},
  \bibinfo{journal}{Phys. Lett.} \textbf{\bibinfo{volume}{B124}},
  \bibinfo{pages}{515} (\bibinfo{year}{1983}).

\bibitem[{\citenamefont{Elze et~al.}(1984)\citenamefont{Elze, Greiner, and
  Rafelski}}]{Elze84a}
\bibinfo{author}{\bibfnamefont{H.~T.} \bibnamefont{Elze}},
  \bibinfo{author}{\bibfnamefont{W.}~\bibnamefont{Greiner}}, \bibnamefont{and}
  \bibinfo{author}{\bibfnamefont{J.}~\bibnamefont{Rafelski}},
  \bibinfo{journal}{Z. Phys.} \textbf{\bibinfo{volume}{C24}},
  \bibinfo{pages}{361} (\bibinfo{year}{1984}).

\bibitem[{\citenamefont{Elze and Greiner}(1986{\natexlab{a}})}]{Elze985a}
\bibinfo{author}{\bibfnamefont{H.~T.} \bibnamefont{Elze}} \bibnamefont{and}
  \bibinfo{author}{\bibfnamefont{W.}~\bibnamefont{Greiner}},
  \bibinfo{journal}{Phys. Rev.} \textbf{\bibinfo{volume}{A33}},
  \bibinfo{pages}{1879} (\bibinfo{year}{1986}{\natexlab{a}}).

\bibitem[{\citenamefont{Elze and Greiner}(1986{\natexlab{b}})}]{Elze86a}
\bibinfo{author}{\bibfnamefont{H.~T.} \bibnamefont{Elze}} \bibnamefont{and}
  \bibinfo{author}{\bibfnamefont{W.}~\bibnamefont{Greiner}},
  \bibinfo{journal}{Phys. Lett.} \textbf{\bibinfo{volume}{B179}},
  \bibinfo{pages}{385} (\bibinfo{year}{1986}{\natexlab{b}}).

\bibitem[{\citenamefont{Elze et~al.}(1987)\citenamefont{Elze, Miller, and
  Redlich}}]{Elze86b}
\bibinfo{author}{\bibfnamefont{H.~T.} \bibnamefont{Elze}},
  \bibinfo{author}{\bibfnamefont{D.~E.} \bibnamefont{Miller}},
  \bibnamefont{and} \bibinfo{author}{\bibfnamefont{K.}~\bibnamefont{Redlich}},
  \bibinfo{journal}{Phys. Rev.} \textbf{\bibinfo{volume}{D35}},
  \bibinfo{pages}{748} (\bibinfo{year}{1987}).

\bibitem[{\citenamefont{Gorenstein et~al.}(1983)\citenamefont{Gorenstein,
  Lipskikh, Petrov, and Zinovev}}]{Gorenstein:1983a}
\bibinfo{author}{\bibfnamefont{M.~I.} \bibnamefont{Gorenstein}},
  \bibinfo{author}{\bibfnamefont{S.~I.} \bibnamefont{Lipskikh}},
  \bibinfo{author}{\bibfnamefont{V.~K.} \bibnamefont{Petrov}},
  \bibnamefont{and} \bibinfo{author}{\bibfnamefont{G.~M.}
  \bibnamefont{Zinovev}}, \bibinfo{journal}{Phys. Lett.}
  \textbf{\bibinfo{volume}{B123}}, \bibinfo{pages}{437} (\bibinfo{year}{1983}).

\bibitem[{\citenamefont{Gorenstein et~al.}(1981)\citenamefont{Gorenstein,
  Petrov, and Zinovev}}]{Gorenstein:1981a}
\bibinfo{author}{\bibfnamefont{M.~I.} \bibnamefont{Gorenstein}},
  \bibinfo{author}{\bibfnamefont{V.~K.} \bibnamefont{Petrov}},
  \bibnamefont{and} \bibinfo{author}{\bibfnamefont{G.~M.}
  \bibnamefont{Zinovev}}, \bibinfo{journal}{Phys. Lett.}
  \textbf{\bibinfo{volume}{B106}}, \bibinfo{pages}{327} (\bibinfo{year}{1981}).

\bibitem[{\citenamefont{Gorenstein et~al.}(1982)\citenamefont{Gorenstein,
  Petrov, Shelest, and Zinovev}}]{Gorenstein:1982a}
\bibinfo{author}{\bibfnamefont{M.~I.} \bibnamefont{Gorenstein}},
  \bibinfo{author}{\bibfnamefont{V.~K.} \bibnamefont{Petrov}},
  \bibinfo{author}{\bibfnamefont{V.~P.} \bibnamefont{Shelest}},
  \bibnamefont{and} \bibinfo{author}{\bibfnamefont{G.~M.}
  \bibnamefont{Zinovev}}, \bibinfo{journal}{Theor. Math. Phys.}
  \textbf{\bibinfo{volume}{52}}, \bibinfo{pages}{843} (\bibinfo{year}{1982}).

\bibitem[{\citenamefont{Gorenstein et~al.}(1984)\citenamefont{Gorenstein,
  Lipskikh, and Zinovev}}]{Gorenstein:1984a}
\bibinfo{author}{\bibfnamefont{M.~I.} \bibnamefont{Gorenstein}},
  \bibinfo{author}{\bibfnamefont{S.~I.} \bibnamefont{Lipskikh}},
  \bibnamefont{and} \bibinfo{author}{\bibfnamefont{G.~M.}
  \bibnamefont{Zinovev}}, \bibinfo{journal}{Z. Phys.}
  \textbf{\bibinfo{volume}{C22}}, \bibinfo{pages}{189} (\bibinfo{year}{1984}).

\bibitem[{\citenamefont{Gorenstein et~al.}(1998)\citenamefont{Gorenstein,
  Greiner, and Yang}}]{Gorenstein:1998a}
\bibinfo{author}{\bibfnamefont{M.~I.} \bibnamefont{Gorenstein}},
  \bibinfo{author}{\bibfnamefont{W.}~\bibnamefont{Greiner}}, \bibnamefont{and}
  \bibinfo{author}{\bibfnamefont{S.~N.} \bibnamefont{Yang}},
  \bibinfo{journal}{J. Phys.} \textbf{\bibinfo{volume}{G24}},
  \bibinfo{pages}{725} (\bibinfo{year}{1998}).

\bibitem[{\citenamefont{Auberson
  et~al.}(1986{\natexlab{a}})\citenamefont{Auberson, Epele, and
  Mahoux}}]{Auberson:1986a}
\bibinfo{author}{\bibfnamefont{G.}~\bibnamefont{Auberson}},
  \bibinfo{author}{\bibfnamefont{L.}~\bibnamefont{Epele}}, \bibnamefont{and}
  \bibinfo{author}{\bibfnamefont{G.}~\bibnamefont{Mahoux}},
  \bibinfo{journal}{Nuovo Cim.} \textbf{\bibinfo{volume}{A94}},
  \bibinfo{pages}{339} (\bibinfo{year}{1986}{\natexlab{a}}).

\bibitem[{\citenamefont{Gorenstein et~al.}(2005)\citenamefont{Gorenstein,
  Gazdzicki, and Greiner}}]{Gorenstein:2005a}
\bibinfo{author}{\bibfnamefont{M.~I.} \bibnamefont{Gorenstein}},
  \bibinfo{author}{\bibfnamefont{M.}~\bibnamefont{Gazdzicki}},
  \bibnamefont{and} \bibinfo{author}{\bibfnamefont{W.}~\bibnamefont{Greiner}},
  \bibinfo{journal}{Phys. Rev.} \textbf{\bibinfo{volume}{C72}},
  \bibinfo{pages}{024909} (\bibinfo{year}{2005}).

\bibitem[{\citenamefont{Auberson
  et~al.}(1986{\natexlab{b}})\citenamefont{Auberson, Epele, Mahoux, and
  Simao}}]{Auberson:1986b}
\bibinfo{author}{\bibfnamefont{G.}~\bibnamefont{Auberson}},
  \bibinfo{author}{\bibfnamefont{L.}~\bibnamefont{Epele}},
  \bibinfo{author}{\bibfnamefont{G.}~\bibnamefont{Mahoux}}, \bibnamefont{and}
  \bibinfo{author}{\bibfnamefont{F.~R.~A.} \bibnamefont{Simao}},
  \bibinfo{journal}{J. Math. Phys.} \textbf{\bibinfo{volume}{27}},
  \bibinfo{pages}{1658} (\bibinfo{year}{1986}{\natexlab{b}}).

\bibitem[{\citenamefont{Balian and Bloch}(1970)}]{Balian:1970a}
\bibinfo{author}{\bibfnamefont{R.}~\bibnamefont{Balian}} \bibnamefont{and}
  \bibinfo{author}{\bibfnamefont{C.}~\bibnamefont{Bloch}},
  \bibinfo{journal}{Ann. Phys.} \textbf{\bibinfo{volume}{60}},
  \bibinfo{pages}{401} (\bibinfo{year}{1970}).

\bibitem[{\citenamefont{Balian and Bloch}(1974)}]{Balian70b}
\bibinfo{author}{\bibfnamefont{R.}~\bibnamefont{Balian}} \bibnamefont{and}
  \bibinfo{author}{\bibfnamefont{C.}~\bibnamefont{Bloch}},
  \bibinfo{journal}{Ann. Phys.} \textbf{\bibinfo{volume}{E84}},
  \bibinfo{pages}{559} (\bibinfo{year}{1974}).

\bibitem[{\citenamefont{Balian and Bloch}(1971)}]{Balian71a}
\bibinfo{author}{\bibfnamefont{R.}~\bibnamefont{Balian}} \bibnamefont{and}
  \bibinfo{author}{\bibfnamefont{C.}~\bibnamefont{Bloch}},
  \bibinfo{journal}{Ann. Phys.} \textbf{\bibinfo{volume}{64}},
  \bibinfo{pages}{271} (\bibinfo{year}{1971}).

\bibitem[{\citenamefont{Eidelman et~al.}(2004)}]{databook2004}
\bibinfo{author}{\bibfnamefont{S.}~\bibnamefont{Eidelman}} \bibnamefont{et~al.}
  (\bibinfo{collaboration}{Particle Data Group}), \bibinfo{journal}{Phys.
  Lett.} \textbf{\bibinfo{volume}{B592}}, \bibinfo{pages}{1}
  (\bibinfo{year}{2004}).

\bibitem[{\citenamefont{Ripka}(1997)}]{Ripka:1997a}
\bibinfo{author}{\bibfnamefont{G.}~\bibnamefont{Ripka}},
  \emph{\bibinfo{title}{Quarks bound by chiral fields: The quark-structure of
  the vacuum and of light mesons and baryons}} (\bibinfo{publisher}{Clarendon
  Press, 205 p}, \bibinfo{address}{Oxford, UK}, \bibinfo{year}{1997}).

\bibitem[{\citenamefont{Zakout}()}]{Zakout:in-progress2}
\bibinfo{author}{\bibfnamefont{I.}~\bibnamefont{Zakout}}, \bibinfo{note}{work
  in progress}.

\bibitem[{\citenamefont{Mcgovern and
  Birse}(1990{\natexlab{a}})}]{Mcgovern:1990a}
\bibinfo{author}{\bibfnamefont{J.~A.} \bibnamefont{Mcgovern}} \bibnamefont{and}
  \bibinfo{author}{\bibfnamefont{M.~C.} \bibnamefont{Birse}},
  \bibinfo{journal}{Nucl. Phys.} \textbf{\bibinfo{volume}{A506}},
  \bibinfo{pages}{367} (\bibinfo{year}{1990}{\natexlab{a}}).

\bibitem[{\citenamefont{Mcgovern and
  Birse}(1990{\natexlab{b}})}]{Mcgovern:1990b}
\bibinfo{author}{\bibfnamefont{J.~A.} \bibnamefont{Mcgovern}} \bibnamefont{and}
  \bibinfo{author}{\bibfnamefont{M.~C.} \bibnamefont{Birse}},
  \bibinfo{journal}{Nucl. Phys.} \textbf{\bibinfo{volume}{A506}},
  \bibinfo{pages}{392} (\bibinfo{year}{1990}{\natexlab{b}}).

\bibitem[{\citenamefont{Bugaev}(2007)}]{Bugaev:2007a}
\bibinfo{author}{\bibfnamefont{K.~A.} \bibnamefont{Bugaev}},
  \bibinfo{journal}{Phys. Rev.} \textbf{\bibinfo{volume}{C76}},
  \bibinfo{pages}{014903} (\bibinfo{year}{2007}),
  \eprint{arXiv:hep-ph/0703222}.

\bibitem[{\citenamefont{Bugaev et~al.}(2008)\citenamefont{Bugaev, Petrov, and
  Zinovjev}}]{Bugaev:2008a}
\bibinfo{author}{\bibfnamefont{K.~A.} \bibnamefont{Bugaev}},
  \bibinfo{author}{\bibfnamefont{V.~K.} \bibnamefont{Petrov}},
  \bibnamefont{and} \bibinfo{author}{\bibfnamefont{G.~M.}
  \bibnamefont{Zinovjev}} (\bibinfo{year}{2008}), \eprint{arXiv:0801.4869
  [hep-ph]}.

\bibitem[{\citenamefont{Stoecker et~al.}(2006)\citenamefont{Stoecker, Betz, and
  Rau}}]{Stoecker:2007a}
\bibinfo{author}{\bibfnamefont{H.}~\bibnamefont{Stoecker}},
  \bibinfo{author}{\bibfnamefont{B.}~\bibnamefont{Betz}}, \bibnamefont{and}
  \bibinfo{author}{\bibfnamefont{P.}~\bibnamefont{Rau}}, \bibinfo{journal}{PoS}
  \textbf{\bibinfo{volume}{CPOD2006}}, \bibinfo{pages}{029}
  (\bibinfo{year}{2006}), \eprint{arXiv:nucl-th/0703054}.

\bibitem[{\citenamefont{Betz et~al.}(2008)\citenamefont{Betz, Rau, and
  Stocker}}]{Betz:2007b}
\bibinfo{author}{\bibfnamefont{B.}~\bibnamefont{Betz}},
  \bibinfo{author}{\bibfnamefont{P.}~\bibnamefont{Rau}}, \bibnamefont{and}
  \bibinfo{author}{\bibfnamefont{H.}~\bibnamefont{Stocker}},
  \bibinfo{journal}{Int. J. Mod. Phys.} \textbf{\bibinfo{volume}{E16}},
  \bibinfo{pages}{3082} (\bibinfo{year}{2008}), \eprint{arXiv:0707.3942
  [hep-th]}.

\end{thebibliography}
%%%%%%%%%%%%%%%%%%%%%%%%%%%

%%%%%%%%%%%%%%%%%%%%%%%%%%%%%%%%%%%
%% Fig1
%% Figure : stationary solution
%%          and the ultimate limit
%% Fig.: westen_station_a.eps fig_w_station_a
%%
%%%%%%%%%%%%%%%%%%%%%%%%%%%%%%%%%%%

%%%%%%%%%%%%%%%%%%%%%%%%%%%%%%%%%%%
\newpage
%%%%%%%%%%%%%%%%%%%%%%%%%%%%

%%%%%%%%%%%%%%%%%%%%%%%%%%%%%%%%%%%%%%%%%%
%%%%%%%%%%%%%%%%%%%%%%%%%%%%%%%%%%%%%%%%%%
%% Figures                             %%%
%% Figure: westen_fun.eps              %%%
%%  fig_w_fun                          %%%
%% fig_w_fun_smallb                    %%%
%% fig_w_fun_match                     %%%
%% Figure: westen_fun_smallb.eps       %%%
%% Figure: westen_fun_smallc.eps       %%%
%%%%%%%%%%%%%%%%%%%%%%%%%%%%%%%%%%%%%%%%%%
%%%%%%%%%%%%%%%%%%%%%%%%%%%%%%%%%%%%%%%%%%
\newpage 
%%%%%%%%%%%%%

%%%%%%%%%%%%%
\newpage

%%%%%%%%%%%%%
\newpage
%%%%%%%%%%%%%%%%%%%%%%%%%%%%%%%%%

%%%%%%%%%%%%%%%%%%%%%%%%%%%%%%%%%%%%%%%%%%%%%%%%%%%%%%%%%%%%%%
%%%%%%%%%%%%%%%%%%%%%%%%%%%%%%%%%%%%%%%%%%%%%%%%%%%%%%%%%%%%%%
%%  Fig.
%%  Figure
%%  westen_adj_a.eps
%%  westen_adj_smallb.eps
%%  westen_adj_smallc.eps
%%  fig_w_adj_a
%%  fig_w_adj_smallb (0-0.5)
%%  fig_w_adj_smallc (0-1)
%%
%%%%%%%%%%%%%%%%%%%%%%%%%%%%%%%%%%%%%%%%%%%%%%%%%%%%%%%%%%%%%%
%%%%%%%%%%%%%%%%%%%%%%%%%%%%%%%%%%%%%%%%%%%%%%%%%%%%%%%%%%%%%%
%%%%%%%%%%%%%

\newpage
%%%%%%%%%%%%%

%%%%%%%%%%%%%%%%%%%%%%%%%%%%%
\newpage
%%%%%%%%%%%%%%%%%%%%%%%%%
%%%%%%%%%%%%%%%%%%%%%%%%%%%%%

\newpage
%%%%%%%%%%%%%%%%%%%%%%%%%%%%%%%%%%%%%%%%%%%%%%%%%%%%%%%%%%%%

%%%%%%%%%%%%%%%%%%%%%%%%%%%%%%%%%%%%%%%%%%%%
%%%%%%%%%%%%%%%%%%%%%%%%%%%%%%%%%%%%%%%%%%%%
%%%
%%%  Figure
%%%  Fig
%%%
%%%  westen_fun_adj.eps
%%%  westen_fun_adj_match.eps
%%%  westen_fun_adj_small.eps
%%%  fig_w_fun_adj
%%%  fig_w_fun_adj_small
%%%  fig_w_fun_match
%%%
%%%
%%%%%%%%%%%%%%%%%%%%%%%%%%%%%%%%%%%%%%%%%%%%%
%%%%%%%%%%%%%%%%%%%%%%%%%%%%%%%%%%%%%%%%%%%%%
\newpage
%%%%%%%%%%%%%%%%%%%%%%%%%%%%%%%%%%%%%%%%%%%%%%%%%%%%%%%%%%%%%%%%%%%%%%%%%
%%%%%%%%%%%%%%%%%%%%%%%%%%%%%%%%%%%%%%%%%%%%%%%%%%%%%%%%%%%%%%%%%%%%%%%%%

%%%%%%%%%%%%%%%%%%%%%%%%%%%%%
\newpage
%%%%%%%%%%%%%%%%%%%%%%%%%

%%%%%%%%%%%%%%%%%%%%%%%%%%%%%
\newpage
%%%%%%%%%%%%%%%%%%%%%%%%%%%%%
%
%%%%%%%%%%%%%%%%%%%%%%%%%%%%%

\newpage
%%%%%%%%%%%%%%%%%%%%%%%%%%%%%

%%%%%%%%%%%%%%%%%%%%%%%%%%%%%
%%%%%%%%%%%%%%%%%%%%%%%%%%%%%
\newpage
%%%%%%%%%%%%%%%%%%%%%%%%%

%%%%%%%%%%%%%%%%%%%%%%%%%%%%%
%%%%%%%%%%%%%%%%%%%%%%%%%%%%%

\newpage
%%%%%%%%%%%%%%%%%%%%%%%%%%%%%

\end{document}